\documentclass[11pt, onehalf, phd]{osudiss-2} 

\usepackage{graphicx} 
\usepackage{lipsum}
\usepackage{amsmath,amsfonts,amsthm}
\usepackage{commath}
\usepackage{amssymb}
\usepackage{verbatimbox}
\usepackage[utf8]{inputenc}
\usepackage{bm} 
\usepackage{booktabs} 
\usepackage[titletoc]{appendix}

\usepackage{bookmark} 
\hypersetup{colorlinks=true,urlcolor=blue,linkcolor=blue, citecolor=blue, breaklinks} 
\usepackage[all]{hypcap}

\usepackage[numbers,square,sort&compress]{natbib}

\usepackage{subcaption}
\usepackage[export]{adjustbox}
\PassOptionsToPackage{obeyspaces}{url}

\usepackage{thesis-commands}
\graphicspath{{fig/}}

\author{Lipei Du}
\title{Hydrodynamic Description of the Baryon-charged Quark-Gluon Plasma}
\advisorname{Ulrich Heinz}
\degree{Doctor of Philosophy}
\member{Richard Furnstahl}
\member{Michael Lisa}
\member{Jay Gupta} 
\authordegrees{M.S.}
\graduationyear{2021}
\unit{Graduate Program in Physics}

\begin{document}

\frontmatter


\begin{abstract}
One of the primary goals of nuclear physics is studying the phase diagram of Quantum Chromodynamics, where a hypothetical critical point serves as a landmark. A systematic model-data comparison of heavy-ion collisions at center-of-mass energies between 1 and 100 GeV per nucleon is essential for locating the critical point and the phase boundary between the deconfined quark-gluon plasma and the confined hadron resonance gas. At these energies the net baryon density of the system can be high and critical fluctuations can become essential in the presence of the critical point. Simulating their dynamical evolution thus becomes an indispensable part of theoretical modeling. 

In this thesis we first present the (3+1)-dimensional relativistic hydrodynamic code {\sc BEShydro}, which solves the equations of motion of second-order Denicol-Niemi-Molnar-Rischke theory, including bulk and shear viscous components as well as baryon diffusion current. We then study the effects caused by the baryon diffusion on the longitudinal dynamics and on the phase diagram trajectories of fluid cells at different space-time rapidities of the system, and how they are affected by critical dynamics near the critical point. We finally explore the evolution of non-hydrodynamic slow processes describing long wavelength critical fluctuations near the critical point, by extending the conventional hydrodynamic description by coupling it to additional explicitly evolving slow modes, and their back-reaction to the bulk matter properties. 
\end{abstract}
\clearpage 
\tableofcontents 


\mainmatter
\chapter{Introduction}
\label{ch:intro}

\section{The early universe and the Big Bang}

\begin{figure}[!htb]
\centering
\includegraphics[width=\linewidth]{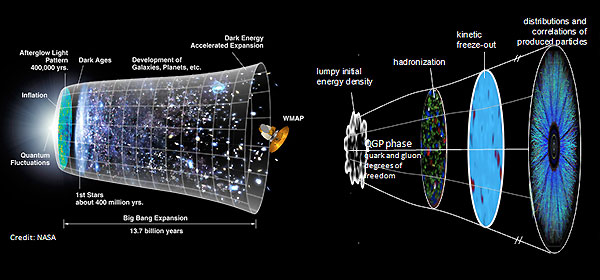}\par 
\caption{{\sl Left:} An illustration of the expansion of the universe from the Big Bang to the present day with a timeline of different stages. The structure of the universe we see today is a reflection of structures created much earlier. {\sl Right:} A graphical representation of the expanding system created in heavy-ion collisions from the Little Bang to the final detectable free-streaming particles. As an analogy to studying the early universe, nuclear physicists seek for fingerprints of the early structure of the created system from the particle tracks observed in detectors.  Figure taken from Ref.~\cite{universegrows} where the credit of the left figure is given as NASA.}
\label{fig:earlyuniverse}
\end{figure}

According to the prevailing cosmological model -- the Big Bang theory -- our universe started from a ``Big Bang singularity'' about 13.7 billion years ago (see the left panel of Fig.~\ref{fig:earlyuniverse}) \cite{Agashe:2014kda,Heinz:2004qz}. The singularity had almost infinite energy density and temperature, and during the first few microseconds of the early universe, quarks, anti-quarks and gluons which are the fundamental building blocks of today's observable universe were in a deconfined phase, called quark-gluon plasma (QGP). Our universe was permeated by this phase of matter until its temperature decreased below the pseudo-critical temperature $T_{pc}\approx155\,$MeV, at which the color-charged quarks, anti-quarks and gluons got confined into colorless hadrons (``hadronization'').

%
%

Hadronization of the QGP was followed shortly after by primordial nucleon synthesis  which happened about 3 minutes into life of our universe. When the temperature decreased to about 100\,KeV, small atomic nuclei started to form (``Big Bang nucleosynthesis''). During the stage of nucleosynthesis, all unstable hadrons decayed, anti-particles got annihilated, and only a tiny fraction of excess protons, neutrons and electrons were left, with the surviving neutrons being bound inside small atomic nuclei. After that the chemical composition ceased to change and thus the ``chemical freeze-out'' of the early universe was reached. 

Following the Big Bang nucleosynthesis was the ``photon epoch'' which lasted for about 370~000 years. During this epoch the temperature was still so high that the typical photon energy exceeded the binding of electrons to nuclei, and thus the universe remained ionized and was filled with a plasma of nuclei, electrons and photons. Because of the ionized environment, photons interacted frequently with nuclei and electrons and thus could not travel freely. Consequently the universe was opaque to electromagnetic radiation.

Finally, at the end of the photon epoch, when the temperature dropped to about 3000 K, hydrogen and helium nuclei were able to capture electrons and form stable, electrically neutral atoms (``recombination''), and the universe started to become transparent to photons. Photons were no longer able to stay in thermal equilibrium with the matter, and thus the ``thermal freeze-out'' was reached. The photons of the cosmic microwave background (CMB) radiation started to decouple and stream freely ever since. Today's astronomical observations can only trace back to the moment when the electromagnetic radiation decoupled but not earlier because of the opacity of the early universe. Further extrapolation to times before 370~000 years after the Big Bang requires strong help from cosmological theory. The QGP that filled the very early universe cannot be accessed by observations. Fortunately, a QGP with almost identical thermodynamic properties can be recreated by high energy nuclear collisions in the laboratory -- the Little Bang -- the theoretical description of whose dynamical evolution is the main topic of this thesis (see the right panel of Fig.~\ref{fig:earlyuniverse}).

\section{Heavy-ion collisions and the Little Bang}

To recreate the QGP matter in the laboratory, physicists need the help of particle colliders which can provide unprecedented energy per constituent to highly compressed beams of fully ionized atomic nuclei, with mass numbers ranging from $A=1$ (protons) to $A=238$ (Uranium)~\cite{Shiltsev:2019rfl}. There are two hadron colliders in operation today, the Large Hadron Collider (LHC) at CERN and the Relativistic Heavy-Ion Collider (RHIC) at Brookhaven National Laboratory (BNL), both of which collide either protons or heavy ions, or protons with ions \cite{hiccern,hicrhic,doerhic}. RHIC is the only collider dedicated for heavy ion research and can accelerate almost any ion species whereas the LHC runs with ions only about one month per year \cite{doerhic}.

%
\begin{figure}[!t]
\centering
\includegraphics[width=\linewidth]{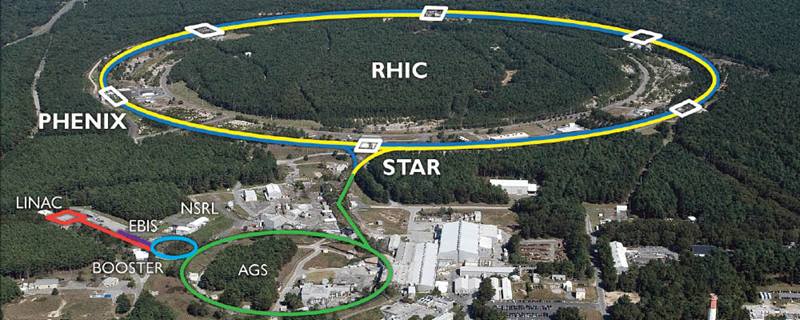}\par 
\caption{Aerial picture of the RHIC facility at BNL, and the layout of the collider, injector complex and the principal running and completed experiments. Figure taken from Ref.~\cite{doerhic}.}
\label{fig:rhicillus}
\end{figure}%
%

Fig.~\ref{fig:rhicillus} shows an aerial picture of the RHIC facility at BNL and its injector complex. RHIC has two storage rings that collide heavy ions or protons circulating in opposite directions, which get accelerated through several stages of boosters before reaching the two rings (yellow and blue circles in the figure). These two rings cross at six interaction points, and the two principal experiments in operation, STAR (for Solenoidal Tracker at RHIC) and PHENIX (for Pioneering High Energy Nuclear Interaction eXperiment), are located at two of these points, as indicated in the figure.  Near the interaction points, particles in one high energy beam can collide with the ones from the other beam, with each nucleus-nucleus collision creating hundreds to thousands of charged particles. These particles reach and leave tracks in the detectors surrounding the interaction regions.

%
\begin{figure}[!htbp]
\centering
\includegraphics[width=\linewidth]{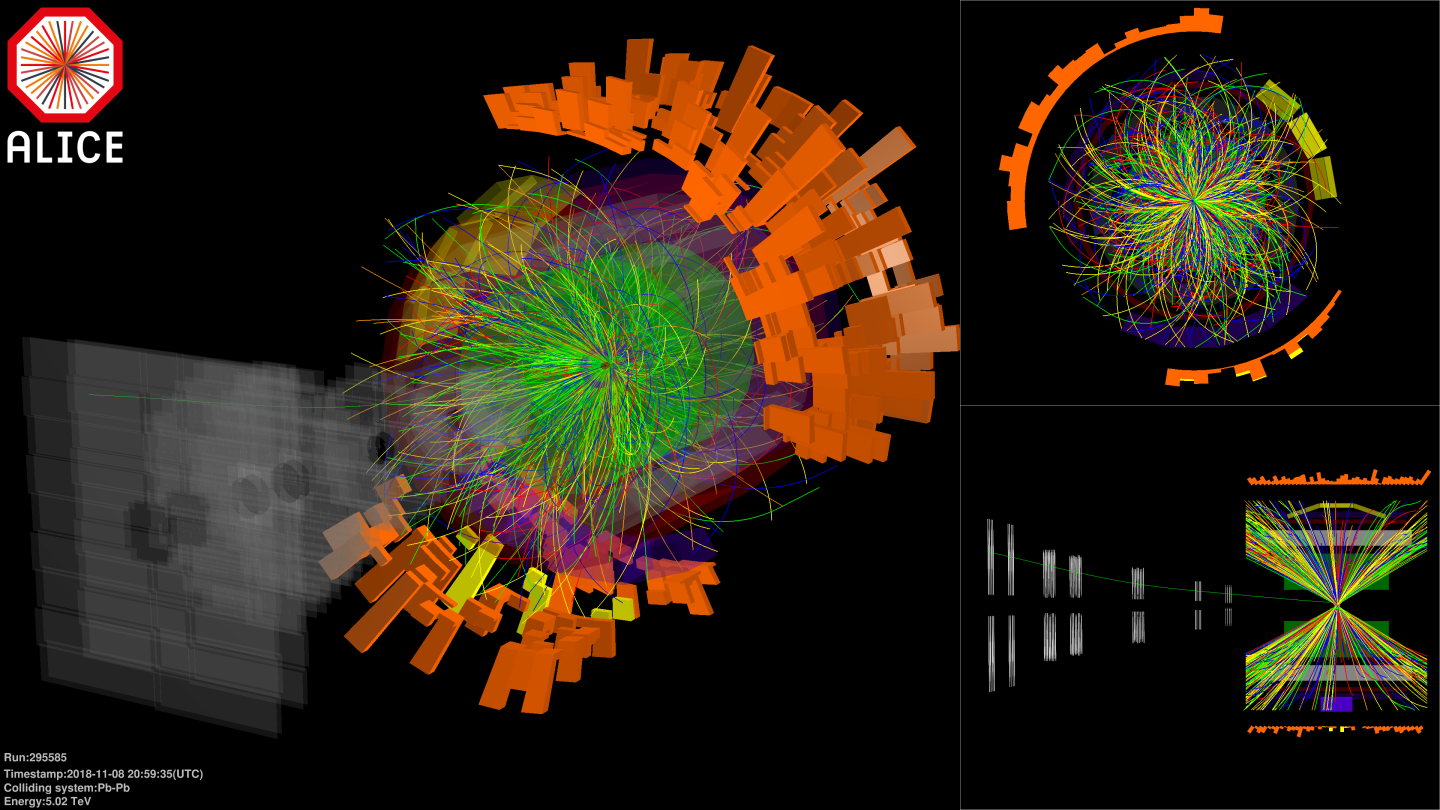}\par 
\caption{Event display of the first Pb-Pb collisions at the center-of-mass energy of 5.02~TeV per nucleon pair in 2018, where the curved lines represent the trajectories of showers of particles through the ALICE detector. Figure taken from Ref.~\cite{hiccernalice}.}
\label{fig:detector}
\end{figure}%
%

Fig.~\ref{fig:detector} shows the particle showers produced by the first Pb-Pb collisions at top LHC energy as reconstructed by the ALICE (A Large Ion Collider Experiment) Collaboration.  ALICE, one of eight detector experiments at the LHC, is optimized for studying heavy-ion collisions \cite{hiccernalice}. 
With the trajectories left by the charged particles in the detectors and the help of various additional particle identification (PID) methods, the momenta and species of the final particles can be determined, and thus observables such as particle spectra can be obtained. Through these observables of the final particles, the nuclear physicists can study the properties of the created systems and even trace back to the very early stage of their evolution. This is an analogy to how the cosmologists study the very early universe through observations of today's universe. We shall discuss the phenomenology of heavy-ion collisions and a few observables of interest to this thesis in greater detail in Ch.~\ref{ch:pheno}.

\section{QCD phase diagram and the RHIC Beam Energy Scan}\label{sec:phase_diagram}

\begin{figure}[!tbp]
\centering
\includegraphics[width=0.65\linewidth]{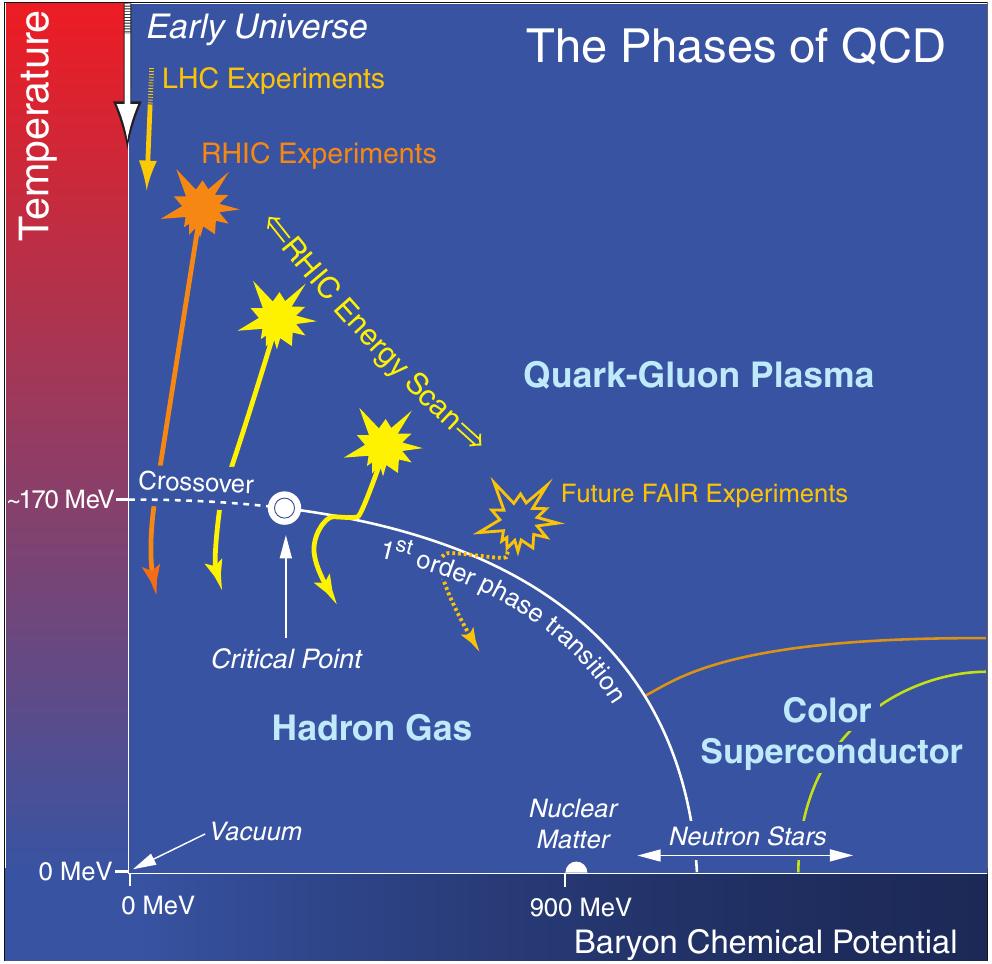}\par 
\caption{The QCD phase diagram with trajectories indicating nuclear collisions at various beam energies carried out at different colliders. Figure taken from Ref.~\cite{osti_1296778}.}
\label{fig:phase_diagram}
\end{figure}

The strong interaction, as one of the four known fundamental interactions, is described by the theory of Quantum Chromodynamics (QCD) in the Standard Model of particle physics \cite{Agashe:2014kda}. Matter consisting of constituents whose interaction is governed by the strong force manifests itself in different phases in the QCD phase diagram (see, e.g., the reviews \cite{Stephanov:2004wx, Fukushima:2010bq, BraunMunzinger:2009zz}). The aforementioned QGP, which existed during the early universe and is reproduced in heavy-ion collisions, is one of those phases of QCD matter. Our modern understanding of the QCD phase diagram, illustrated in Fig.~\ref{fig:phase_diagram}, is based on perturbative QCD calculations at asymptotically large temperature and densities, first-principles Lattice QCD calculations, model calculations and empirical nuclear physics \cite{Stephanov:2004wx, kogut_stephanov_2003}. The QCD phase diagram is expressed in the space of thermodynamic parameters (usually on a plane with temperature $T$ and baryon chemical potential $\mu$ as the axes), where different states have well defined thermodynamic properties. In fact, any point in the phase diagram represents a stable thermodynamic state, characterized by various thermodynamic properties, including entropy density, pressure, baryon density, etc.

%
%

Studying the QCD phase diagram is one of the primary goals of nuclear physics \cite{osti_1296778}.  First principles calculations from Lattice QCD show that, at zero $\mu$, the phase transition from a deconfined QGP phase to a confined hadron resonance gas (HRG) phase \cite{Huovinen:2009yb,Andronic:2017pug,Vovchenko:2020lju} from high to low temperature is a dramatic but smooth crossover \cite{Aoki:2006we, Bazavov:2009zn, Borsanyi:2010cj, Bazavov:2011nk}. At large $\mu$, calculations of phase transition using Lattice QCD are not yet available, since the techniques suffer from the ``sign problem" \cite{PHILIPSEN201355,Ratti:2018ksb}, while a lot of efforts are spent on extending the physics at zero $\mu$ to regions of non-zero $\mu$. Nevertheless, theoretical models \cite{Ejiri:2008xt,Asakawa:1989bq,Barducci:1989wi,Barducci:1989eu} indicate that at large chemical potential the phase transition is first order \cite{Stephanov:2004wx, Stephanov:2007fk, Fukushima:2010bq}, and this implies that a critical point  (i.e., a second-order phase transition) exists at non-zero chemical potential \cite{Stephanov:1998dy,Stephanov:1999zu}, at the end of the first order phase transition line. Confirming the existence and finding the location of the hypothetical QCD critical point has attracted tremendous amount of attention over the last two decades \cite{Rajagopal:1999cp,Busza:2018rrf,Bzdak:2019pkr}. There are other phases at even higher chemical potentials, such as color superconductivity \cite{Alford:2007xm} and neutron star matter \cite{Lattimer:2000nx,Lattimer:2012nd,Baiotti:2016qnr}. In this thesis we are interested in the QGP phase and its transition to HRG, which can be studied experimentally through heavy-ion collisions \cite{Bzdak:2019pkr,Luo:2020pef,Luo:2017faz}.

It is helpful to first summarize some general features of the first- and second-order phase transitions \cite{Stephanov:2004wx, Stephanov:2007fk,Bzdak:2019pkr}. Along the first-order phase transition line, given particular $(T, \mu)$, the system can coexist in two distinct phases at the same time (in our case the QGP and HRG phases) whereas the densities of conserved quantities, such as energy and baryon densities, are discontinuous. By moving along the phase transition line, changing $(T, \mu)$ towards their critical values $(T_c, \mu_c)$, the two distinct phases become more and more similar, and finally turn into one indistinguishable phase, at the critical point. At this special point, the discontinuities and thermodynamic barrier between the two phases vanish, which results in large fluctuations, and the system is characterized by critical phenomena, such as singularities in thermodynamic susceptibilities. These singularities are a result of long-range correlations among the thermal fluctuations in the system. Remarkably, near critical points, materials of completely different microscopic properties can have universal critical behavior, and these materials are categorized into a common universality class.

Universality classes can be further classified by associated static and dynamical critical phenomena. In terms of static phenomena, based on dimensionality and symmetry of the order parameter, the QCD critical point belongs to the static universality class of the 3-dimensional Ising model \cite{Guida:1996ep, zinn2002quantum, Berges:1998rc, Halasz:1998qr}, just like the critical point associated with the liquid-gas transition of a normal fluid. The concept of universality class was extended to dynamical critical phenomena by Hohenberg and Halperin in Ref.~\cite{RevModPhys.49.435}, and according to this Hohenberg-Halperin classification, QCD belongs to the dynamical model H \cite{Son:2004iv}. Within model H, the dynamic critical exponent is predicted to be $z\approx3$, which controls the phenomenon of critical slowing down \cite{Berdnikov:1999ph}, where the equilibration time grows with the critical correlation length, as $\tau_\mathrm{eq}\sim\xi^z$. This is a very important phenomenon near the critical point, which controls the dynamical evolution of the matter created in heavy-ion collision as it evolves through in the critical region and thus is phenomenologically relevant when searching for criticality of QCD. Dynamical behavior includes relaxation times, responses to perturbations and transport coefficients, etc. We shall discuss this in more detail in Ch.~\ref{ch:diffcp} and Ch.~\ref{ch.fluctuations} in this thesis. The order parameter of QCD matter, conventionally noted as the $\sigma$ field, is not easy to determine, as it controlled by a combination of fluctuations of the chiral condensate, baryon density, and energy-momentum densities \cite{Son:2004iv}. 

The main method for mapping the QCD phase diagram is to carry out heavy-ion collisions at various beam energies, among which the collisions at lower energies are expected to scan higher chemical potential regions of the diagram \cite{Luo:2017faz, Busza:2018rrf, Bzdak:2019pkr}. Such collisions have been carried out at different experimental facilities, such as the Large Hadron Collider (LHC) at CERN and the Relativistic Heavy-Ion Collider (RHIC) at Brookhaven National Laboratory, and large sets of data have been accumulated. One of the most promising signatures of the QCD critical point is a non-monotonic beam energy dependence of higher-order cumulants of the fluctuations in the net proton production yields \cite{Stephanov:1998dy, Stephanov:1999zu, Hatta:2003wn, Stephanov:2008qz, Stephanov:2011pb, Luo:2017faz}. This is based on the idea that these observables are more sensitive to the correlation length of fluctuations of the order parameter which, in the thermodynamic limit, diverges at the critical point \cite{RevModPhys.49.435}. Fireballs created in heavy-ion collisions at different beam energies should freeze out with correlation lengths that depend non-monotonically on the collision energy \cite{Stephanov:1998dy, Stephanov:1999zu}, and this should be reflected in the net baryon cumulants (see Sec.~\ref{sec:cumucrit}). Strongly motivated by this, a Beam Energy Scan (BES) program has been carried out at RHIC during the last decade \cite{Mohanty:2011nm,Schmah:2013vea,McDonald:2015tza,Odyniec:2013aaa,Tlusty:2018rif}. During a first campaign that ended in 2011 (BES-I), Au-Au collisions were studied at collision energies $\sqrt{s_\mathrm{NN}}$ from 200 GeV down to 7.7 GeV (BES-I) \cite{Adamczyk:2017iwn, Bzdak:2019pkr, Luo:2020pef}. A second campaign, BES-II, with significantly increased beam luminosity has been completed this year, after having explored collision energies down to $\sqrt{s_\mathrm{NN}}=3.0$\,GeV in fixed-target mode. 

Besides RHIC and LHC, there are various other experimental programs for heavy-ion collisions, including the earlier NA49 \cite{Afanasev:1999iu,Gazdzicki:2004ef} and ongoing NA61/SHINE \cite{Abgrall:2014xwa,Gazdzicki:2008kk} experiments at the Super Proton Synchrotron (SPS) \cite{Satz:2004zd}, the High Acceptance DiElecton Spectrometer (HADES) which is installed at SIS18 at the GSI Helmholtzzentrum f\"ur Schwerionenforschung (Germany) \cite{Agakishiev:2009am}. There are other planned experiments, such as the Compressed Matter Experiment (CBM) at the Facility for Antiproton and Ion Research (FAIR) \cite{Ablyazimov:2017guv,Spiller:2006gj} and the Multi Purpose Detector (MPD) at NICA (Nuclotron-based Ion Collider fAcility) at the Joint Institute for Nuclear Research (Dubna, Russia) \cite{Sissakian:2009zza} which are both in advanced stages of construction, as well as the planned CEE (CSR External Target) at High Intensity Heavy-Ion Accelerator Facility (HIAF) in China \cite{Yang:2013yeb}, and a possible future heavy-ion program at J-PARC (Japan Proton Accelerator Research Complex) which presently is a high intensity proton accelerator facility \cite{Sakaguchi:2019xjv}.

\section{Theoretical status and challenges}\label{sec:theochall}

Identifying possible signals of the hypothetical QCD critical point experimentally in heavy-ion collisions is highly non-trivial, because of the dynamical nature of the fireballs created in the collisions. Within their short lifetimes of several dozen yoctoseconds ($10^{-24}$ s) the fireballs' energy density decreases rapidly by collective expansion, from initially hundreds of GeV/fm$^3$ to well below 1~GeV/fm$^3$ at final freeze-out (see e.g. Refs.~\cite{Heinz:2013th,Busza:2018rrf}). The rapid dynamical evolution of the thermodynamic environment keeps the system permanently out of thermal equilibrium and thus critical fluctuations never reach their thermodynamic equilibrium distributions. In addition, in those parts of the fireball which pass through the quark-hadron phase transition close to the QCD critical point, the dynamics of critical fluctuations is affected by ``critical slowing-down'' \cite{Berdnikov:1999ph}. This is both a curse and a blessing: If critical fluctuations would relax quickly to thermal equilibrium, all memory of critical dynamics might have been erased from the hadronic freeze-out distributions by the time the hadron yields and momenta decouple. If, on the other hand, the dynamical evolution of fluctuations is slowed in the vicinity of the critical point, some signals of critical dynamics may survive until freeze-out but they will then most definitely not feature their equilibrium characteristics near the critical point \cite{Berdnikov:1999ph}.

%
\begin{figure}[!t]
\centering
\includegraphics[width= \linewidth]{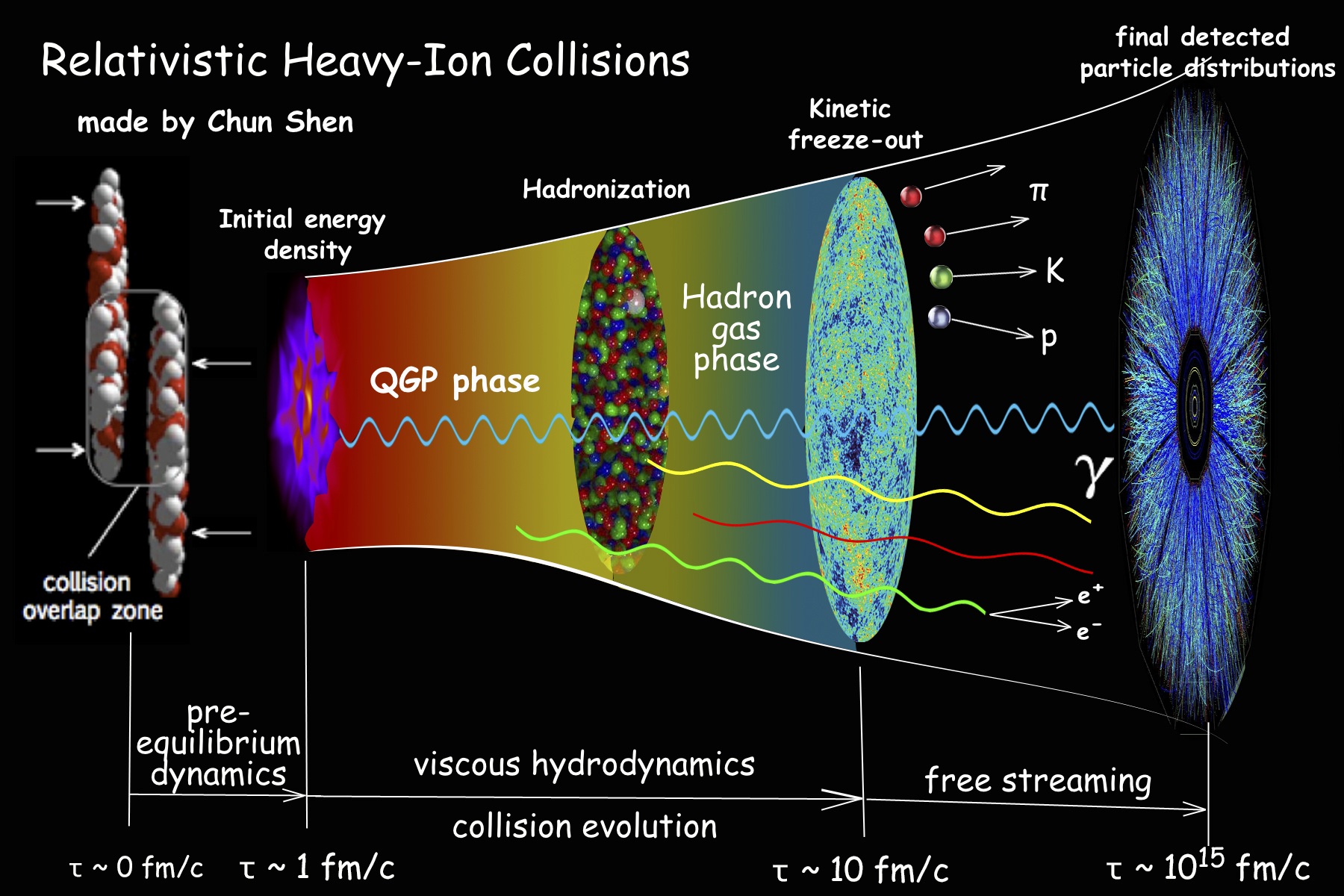}\par 
\caption{Various evolution stages of relativistic heavy-ion collisions from left to right, with corresponding physics descriptions and associated time scales at the bottom. Figure taken from Ref.~\cite{Shen:2014vra}.}
\label{fig:littbang}
\end{figure}%
%

Thus, to confirm or exclude the critical point via systematic model-data comparison, reliable dynamical simulations of off-equilibrium critical fluctuations and the associated final particle cumulants, on top of a well-constrained comprehensive dynamical description of the bulk medium at various beam energies, are indispensable \cite{Nahrgang:2011mv, Nahrgang:2011mg, Mukherjee:2015swa, Akamatsu:2016llw, Murase:2016rhl, Herold:2016uvv, Stephanov:2017ghc, Hirano:2018diu, Singh:2018dpk, Herold:2018ptm, Nahrgang:2018afz, Yin:2018ejt}. Recently, the Hydro+/++ framework \cite{Stephanov:2017ghc,An:2019csj} was developed for incorporating off-equilibrium fluctuations and critical slowing-down into hydrodynamic simulations, and some practical progress within simplified settings has since been made using this framework \cite{Rajagopal:2019xwg, Du:2020bxp}. On the other hand, while a fully developed and calibrated (2+1)-dimensional multistage description of heavy-ion collisions shown in Fig.~\ref{fig:littbang} (including initial conditions + pre-hydrodynamic dynamics + viscous hydrodynamics + hadronic afterburner) exists (see, e.g., the most recent versions described in \cite{Nijs:2020ors, Nijs:2020roc, Everett:2020xug, Everett:2020yty}) and has met great phenomenological success at top RHIC and LHC energies \cite{Heinz:2013th, Busza:2018rrf, Bzdak:2019pkr}, such a comprehensive and fully validated framework is still missing for collisions at the lower BES energies. 

Compared to high-energy collisions at LHC and top RHIC energies, collisions at BES energies introduce a number of additional complications \cite{Shen:2020gef, Shen:2020mgh}. These include (i) a much more complex, intrinsically (3+1)-dimensional and temporally extended nuclear interpenetration stage and its associated dynamical  deposition of energy and baryon number \cite{Shen:2017bsr, Du:2018mpf, Martinez:2019jbu, Martinez:2019rlp, Shen:2017ruz}, (ii) the need to account for and properly propagate conserved charge currents for baryon number and strangeness \cite{Shen:2017ruz, Denicol:2018wdp, Monnai:2019hkn, Du:2019obx, Greif:2017byw, Fotakis:2019nbq, Shen:2020jwv}, (iii) a consistent treatment of singularities in the thermodynamic properties associated with the critical point \cite{RevModPhys.49.435, RevModPhys.50.85}, and (iv) the aforementioned off-equilibrium nature of critical dynamics \cite{Stephanov:2017ghc, Nahrgang:2018afz, An:2019csj, Rajagopal:2019xwg, Du:2020bxp, Nahrgang:2020yxm}. The situation is made even more complicated by the back-reaction of the non-equilibrium critical fluctuation dynamics on the bulk evolution of the medium. This back-reaction causes a potential dilemma: On the one hand, locating the critical point requires reliable calculations at various beam energies of critical fluctuations on top of a well-constrained bulk evolution of the fireball medium; on the other hand, the back-reaction of the off-equilibrium critical fluctuations from a critical point whose location is yet to be determined onto the medium evolution might interfere with the calibration of the latter and turn it into an impossibly complex iterative procedure whose convergence cannot be guaranteed. 

Guidance on how (and perhaps even whether) to incorporate critical effects when constraining the bulk dynamics is direly needed, not least since dynamical simulations for low beam energies are computationally very expensive. Some effects on the bulk medium evolution arising from singularities in the thermodynamic properties of the QCD matter \cite{RevModPhys.49.435, RevModPhys.50.85}, by adding a critical point to the QCD Equation of State (EoS) \cite{Nonaka:2004pg, Parotto:2018pwx, Monnai:2019hkn, Monnai:2021kgu, Stafford:2021wik}  and/or explicitly including critical scaling of its transport coefficients, have been explored, with special attention to the bulk viscous pressure since critical fluctuations of the chiral order parameter, which couples to the baryon mass, can be directly related to a peak of the bulk viscosity near the critical point \cite{Karsch:2007jc, Moore:2008ws, NoronhaHostler:2008ju, Denicol:2009am}. The authors of \cite{Monnai:2016kud} showed that critical effects on the bulk viscous pressure have non-negligible phenomenological consequences for the rapidity distributions of hadronic particle yields, implying that critical effects might indeed play an important role in the calibration of the bulk medium. More investigations on how significant the critical effects can have on the bulk dynamics are certainly needed, which also motivates our exploration on baryon transport near the critical point in Ch.~\ref{ch:diffcp}.

\section{Overview of the thesis}

In this thesis, we attempt to tackle a few theoretical challenges discussed in the previous section. Regarding the bulk dynamics of fireballs created at BES, we study the dynamical initialization and hydrodynamic evolution at non-zero baryon density. Especially, we explore the effects from baryon diffusion current on the longitudinal evolution, both away from and near the critical point. We also study the off-equilibrium dynamics of critical fluctuations in an expanding fireball. This thesis is organized as follows. In Ch.~\ref{ch:pheno}, we briefly discuss a few observables of great interest to the thesis, which have constraining power on bulk evolution, especially for the longitudinal dynamics, and which are sensitive to criticality in heavy-ion collisions. In Ch.~\ref{ch:multistage}, we focus on different physics stages of a multistage framework for heavy-ion collisions at low beam energies. Ch.~\ref{ch:numerics} is devoted to the numerical simulation and validations of a (3+1)-dimensional hydrodynamic code at non-zero net baryon density, \bes, which can be used to simulate a fireball evolution at BES. Then evolution of a baryon-charged medium undergoing Gubser flow is demonstrated in Ch.~\ref{ch.gubser}. We then focus on effects from the baryon diffusion current on the baryon evolution in the longitudinal direction, and see how the evolution can be affected by the presence of a critical point in Ch.~\ref{ch:diffcp}. After these discussions on the bulk evolution of a fireball, we then explore the off-equilibrium dynamics of critical fluctuations and its back-reaction on the bulk evolution in Ch.~\ref{ch.fluctuations}. We then conclude and provide some outlook for studies of heavy-ion collisions in regions of high charge density in Ch.~\ref{ch:concl}.

In this thesis we use natural units, $\hbar\eq{}c\eq{}k_B\eq{}1$, which are the reduced Planck constant, the speed of light and the Boltzmann constant, respectively. We also use the Milne coordinates, $x^\mu\eq{}(\tau,x,y,\eta_s)$, where $\tau$ and $\eta_s$ are the (longitudinal) proper time and space-time rapidity, respectively, related to the Cartesian coordinates via $t\eq{}\tau\cosh\eta_s,\, z\eq{}\tau\sinh\eta_s$. We employ the mostly-minus convention of the metric tensor $g^{\mu\nu}\eq{}\text{diag}(+1,-1,-1,-1/\tau^2)$. In relativistic heavy-ion collisions, we also often use $1\,\mathrm{GeV}\eq{}10^3\,\mathrm{MeV}\eq{}10^9\,\mathrm{eV}$ as our natural energy scale, $1\,\mathrm{fm}\eq{}10^{-15}\,\mathrm{m}$ as the natural length scale, and $1\,\mathrm{fm}/c\approx0.33\times10^{-23}\,\mathrm{s}$ as a typical time scale. It is also convenient to use $\hbar{}c\eq{}0.197\,\mathrm{GeV}\cdot\mathrm{fm}\,\simeq200\,\mathrm{MeV}\cdot\mathrm{fm}\eq{}1$ to convert the dimension of energy to that of inverse length, e.g., $0.197\,\mathrm{GeV}\eq{}1\,\mathrm{fm}^{-1}$. Using the Boltzmann constant $k_B\eq{}8.6173\times10^{-14}\,\mathrm{GeV}\cdot\mathrm{K}^{-1}\eq{}1$, one can get $1\,\mathrm{GeV}\eq{}1.1605\times10^{13}\,\mathrm{K}$; thus a typical temperature in QGP, $0.3\,\mathrm{GeV}\simeq3.5\times10^{12}\,\mathrm{K}$, which is about $10^5$ times hotter than the center of the Sun.

\chapter{Phenomenology of heavy-ion collisions}
\label{ch:pheno}

\section{Kinematics of high energy nuclear collisions}\label{sec:kinematics}

In high energy nuclear collisions experiments \cite{hiccern,hiccernalice,hicrhic,doerhic}, bunches of ions first get accelerated to very high speed (close to the speed of light) along the beam direction in an accelerator, and then they have a chance to collide either with a fixed target or with ions from another beam flying in the opposite direction. Depending on the overlap of the colliding nuclei in the plane transverse to the beam direction and their collision energies, hundreds to thousands of charged particles are produced, and can be recorded with the detectors surrounding the interaction regions where the collisions occur. The experimentalists can change nuclear isotopes and their collision energies. The detectors measure the momenta of produced charged particles (neutral particles are more difficult to measure), and some use particle identification (PID) methods to also measure their energies and masses.

As mentioned in Ch.~\ref{ch:intro}, the systems produced in high energy nuclear collisions have extremely small size with very short lifetime, and one can only extract their properties from the information of measurable final particles. In other words, quantitative extraction of the systems properties needs a systematic model-data comparison. For this purpose, we first need to specify the coordinate system and kinematic quantities, etc., with which we can quantify the colliding system, throughout various evolution stages, for both experimental measurements and theoretical descriptions -- the topic we shall focus on in this section. 

%
\begin{figure}[!htb]
\centering
\includegraphics[width=0.5\linewidth]{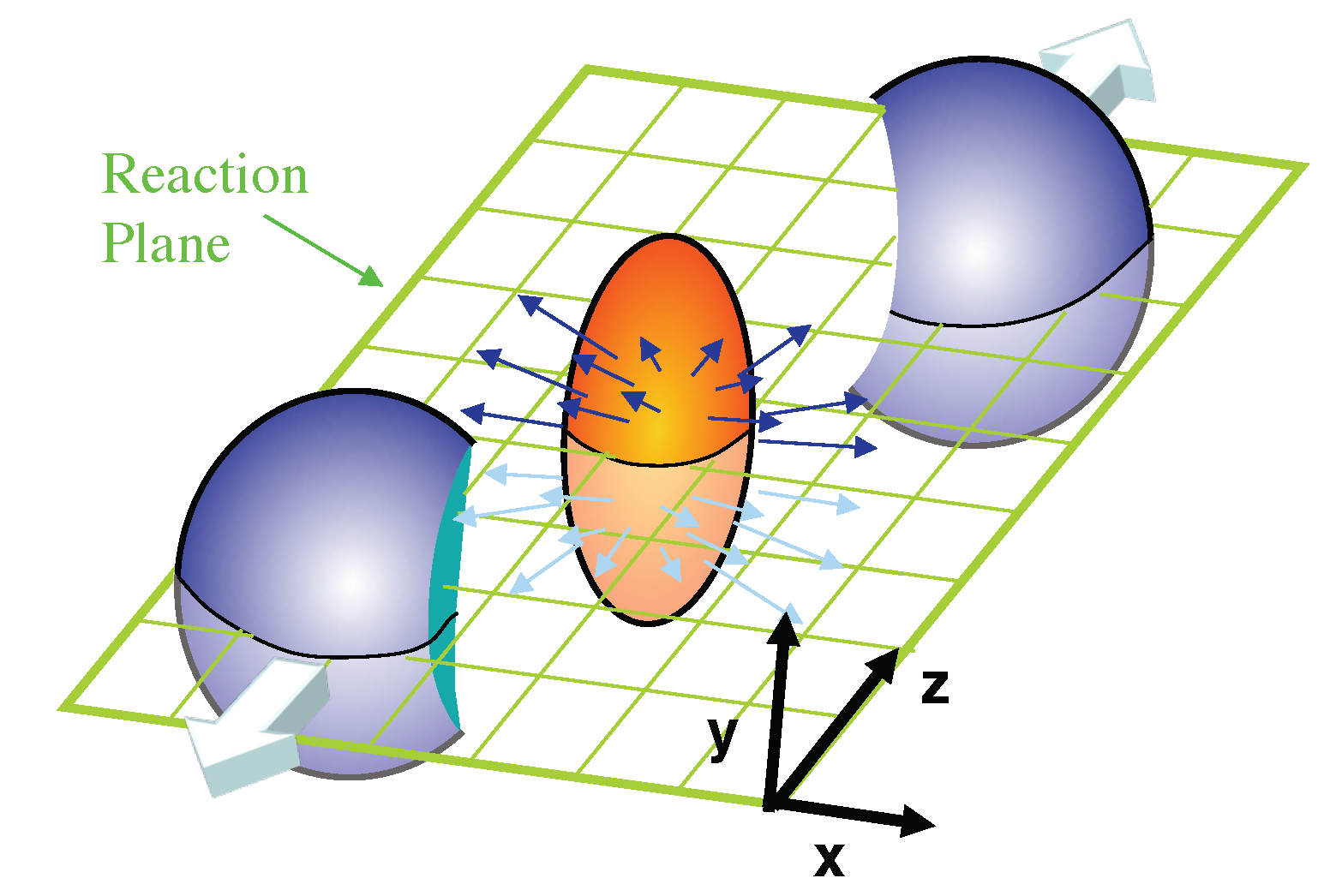}\includegraphics[width=0.5\linewidth]{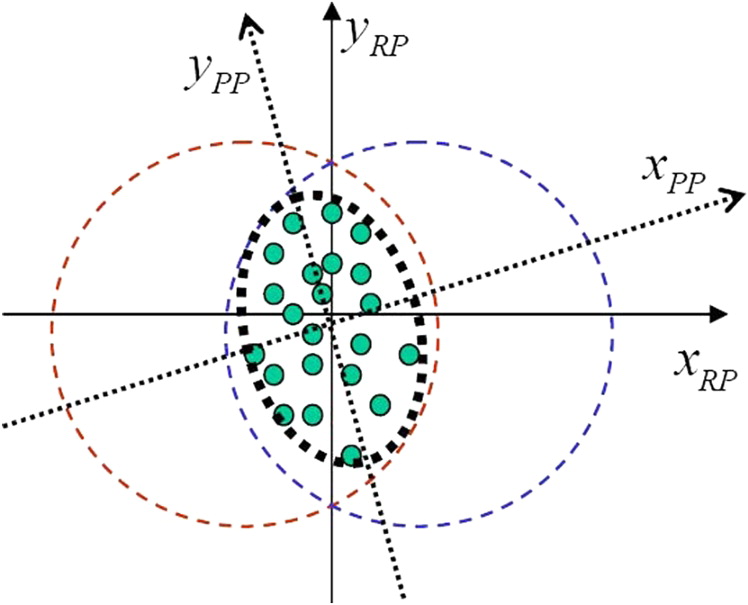}\par 
\caption{{\sl Left:} Schematic plot of a non-central nuclear collision. The orange object indicates the expanding fireball, produced in the geometrical overlap region with an almond shape in the transverse plane, whose short axis lying in the reaction plane. Figure taken from the internet with unknown credit. {\sl Right:}  Illustration of the overlap region in the transverse plane, where nucleons in the colliding nuclei have fluctuating positions. The short axis of the colliding region defining the participant plane, denoted as $x_\mathrm{PP}$, may differ from that of the reaction plane, denoted as $x_\mathrm{RP}$. Figure taken from Ref.~\cite{Voloshin:2007pc}.}
\label{fig:reaction_plane}
\end{figure}%
%

Conventionally, the $z$-axis of the coordinate system is chosen to be along the beam direction (longitudinal direction); the nucleus moving in the positive $z$-direction is called the projectile, the other one the target nucleus. The direction pointing from the center of the projectile  to that of the target defines the $+x$-direction, and then the $y$-direction is fixed following the convention for a right-handed coordinate system (see the left panel of Fig.~\ref{fig:reaction_plane}).\footnote{
	Note that in the left panel of Fig.~\ref{fig:reaction_plane} a left-handed coordinate system is plotted, and thus the $x$-axis should be reversed to make it right-handed.
}
The plane spanned by the $x$- and $z$-axes is called the reaction plane (RP), while the $x$-$y$ plane is usually referred to as the transverse plane; thus the axis labels are often denoted with a subscript ``RP'' (see the right panel of Fig.~\ref{fig:reaction_plane}). However, because of the intrinsic fluctuations of the positions of the nucleons inside the colliding nuclei at the time of impact, the overlap region (between the two dashed circles) and the colliding zone among the nucleons (illustrated by the green points) may not match perfectly, as illustrated in the right panel of Fig.~\ref{fig:reaction_plane}; thus the major axis of the created system defines different coordinates in the transverse plane, the so-called participant plane (PP) coordinates, and their axes are labeled with the subscript ``PP'' in Fig.~\ref{fig:reaction_plane}.

To characterize the collision system further through the produced particles, at relativistic energies it is convenient to introduce coordinates which transform simply between different longitudinal reference frames. For a particle with four-momentum $p^\mu=(E_p, \bp)=(E_p, p_x, p_y, p_z)$, where $E_p$ denotes the energy and $\bp$ the momentum of the particle, one can define its rapidity as:
\begin{equation}
y=\frac{1}{2}\ln\left(\frac{E_p+p_z}{E_p-p_z}\right)=\frac{1}{2}\ln\left(\frac{1+v_z}{1-v_z}\right)\,,
\end{equation}
where $v_z=p_z/E_p$ is the $z$-component of its velocity, and $E_p=\sqrt{m^2+\bp^2}$ with $m$ being its mass. 
For on-shell particles, given their energy and rapidity, one has
\begin{equation}
E_p=m_T\cosh{}y\,,\quad p_z=m_T\sinh{}y\,,
\end{equation}
where $m_T$ is the transverse mass, defined as
\begin{equation}
m_T^2=m^2+p_T^2\,.
\end{equation}
Here $p_T=\sqrt{p_x^2+p_y^2}$ is the transverse momentum. Rapidity is convenient because it transforms in a simple way under Lorentz transformation along the beam direction: For a particle with rapidity $y_A$ in a reference frame $A$, its rapidity reads $y_B=y_A+\delta y$ in a different reference frame $B$ that moves in the $-z$ direction with rapidity $\delta y$ with respect to $A$.

If only the direction of $\bp$ for a particle is measured, with no information on $p=|\bp|$ or its mass (which are needed for PID), then pseudo-rapidity is convenient, defined by
\begin{equation}
\eta=\frac{1}{2}\ln\left(\frac{p+p_z}{p-p_z}\right)=\ln\left(\cot\frac{\theta}{2}\right)\,,
\end{equation}
where $\theta$ is the polar angle between $\bp$ and the beam direction $z$, $p_z=p\cos\theta$.
When particles have  momenta that are large compared to their mass, $E_p\approx{}|\bp|$ and pseudo-rapidity approaches rapidity. Note that
\begin{equation}
|\bp|=p_T\cosh\eta\,,\quad p_z=p_T\sinh\eta\,.
\end{equation}
Pseudo-rapidity is very useful because it is easy to measure, while a measurement of rapidity requires PID. For example, the charged particle multiplicity is usually reported as a function of pseudo-rapidity, i.e., $dN_\mathrm{ch}/d\eta$, which requires no PID, i.e.~no reconstruction of particle masses.

%
\begin{figure}[!htb]
\centering
\includegraphics[width=\linewidth]{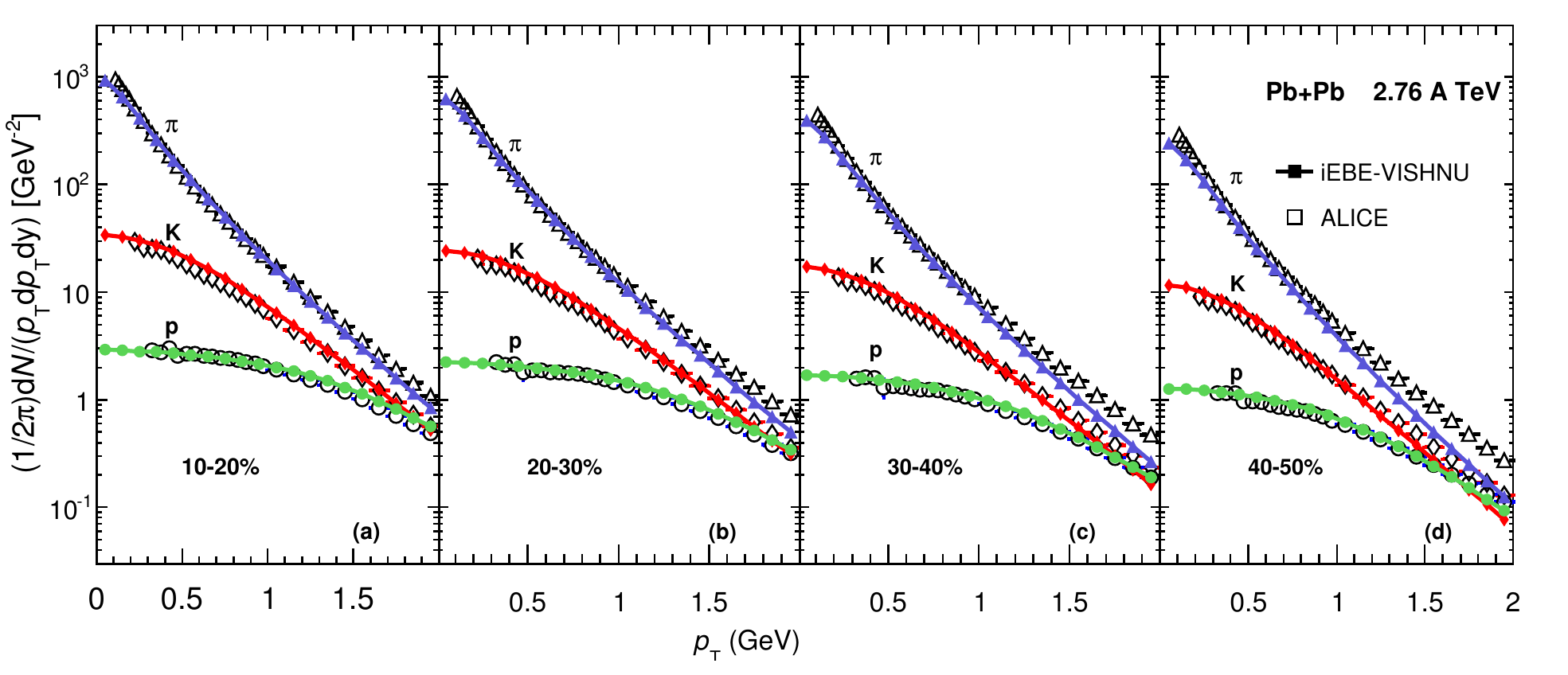}\par 
\caption{Comparison for $p_T$ spectra of  identified hadrons between simulations and experimental measurements by ALICE, for Pb+Pb collisions at $\snn=2.76\,$TeV, for four different centrality classes. Figure taken from Ref.~\cite{Xu:2016hmp}.}
\label{fig:identified_spectra}
\end{figure}%
%

The system created in heavy-ion collisions then can be characterized by the multiplicities and momentum spectra of the final (identified or charged) hadrons.  Using the kinematic quantities and coordinates introduced above, one can construct the following Lorentz invariant spectrum for detected particles:
\begin{equation}\label{eq:inv_spect}
E_p\frac{dN}{d^3\bp}=\frac{dN}{d^2 \bm p_Tdy}=\frac{dN}{p_Tdp_Td\phi_pdy}=\frac{dN}{m_Tdm_Td\phi_pdy}\,,
\end{equation}
where $\phi_p$ is the azimuthal angle of an emitted particle with respect to $x$-axis in transverse polar coordinates with $\bm p_T=(p_T\cos\phi_p, p_T\sin\phi_p)$. Eq.~\eqref{eq:inv_spect} defines the fundamental single-particle observable which model simulations must reproduce. An example is shown in Fig.~\ref{fig:identified_spectra}, where the azimuthal dependence in $\phi_p$ has been integrated out. The invariant spectra of identified particles are important for describing relativistic collisions, also because they are invariant under Lorentz boosts, i.e.~they are identical in the projectile, target or center-of-mass rest frames. The laboratory frame (lab frame for short) is the frame where the detector is at rest, while the center-of-mass frame is where the total momentum of the colliding system is zero, i.e., their center-of-mass is at rest. These two frames do not generally coincide, except in a collider facility colliding equal-mass nuclei with equal energies of both beams.

%
\begin{figure}[!htb]
\centering
\includegraphics[width=0.67\linewidth]{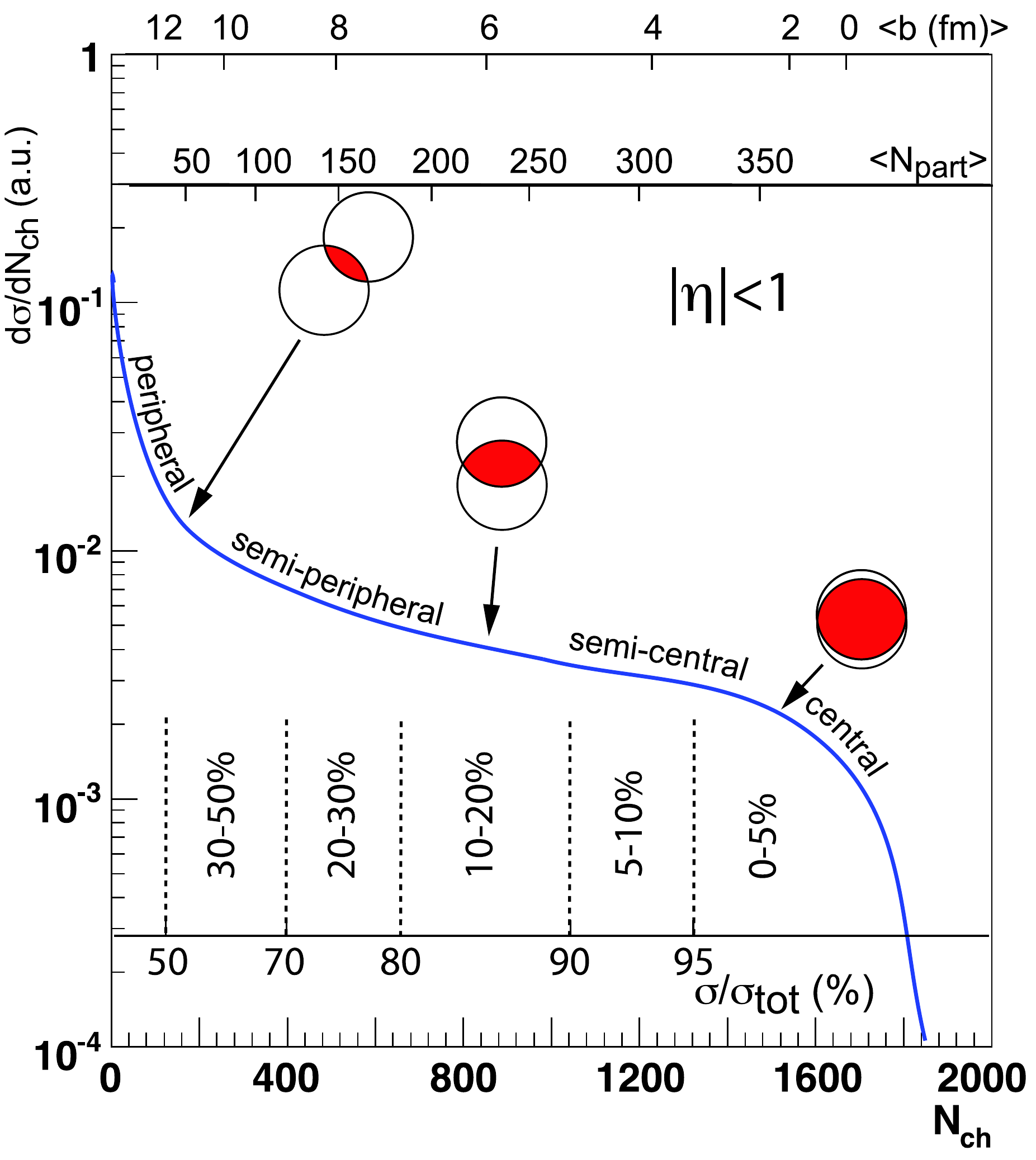}\par 
\caption{An illustration of Glauber modeling in high energy nuclear collisions. In the figure, the events are binned into various centrality classes, and for each of them, the final particle multiplicity at mid-rapidity ($|\eta|<1$) can be related to an average impact parameter $\langle b\rangle$ and an average number of participants $\langle N_\mathrm{part}\rangle$. Figure taken from Ref.~\cite{Miller:2007ri}.}
\label{fig:centrality_glauber}
\end{figure}%
%

Nuclear collisions can be further categorized into different collision centralities, e.g., central and peripheral collisions which, in collisions of large nuclei, are directly related to the impact parameter of the collisions \cite{Miller:2007ri}. The impact parameter is the distance between the nuclear centers of two colliding nuclei, and thus determines the overlap of them. The most central collisions correspond to maximum overlap between the two colliding nuclei and thus have the smallest impact parameters and the largest number of nucleons participating in the collision (also called participants). Unfortunately, the impact parameter is not experimentally measurable, but one can categorize collisions using other experimental observables, such as charged particle multiplicity $N_\mathrm{ch}$ and transverse energy $E_T$ of the final particles. These are strongly correlated with impact parameter, although due to event-by-event fluctuations of the nucleon positions in the colliding nuclei, the relation is not exact and only holds on average. In minimum bias collisions, where all possible collisions are allowed, the  events can be grouped into different centrality classes, by binning the charged particle multiplicities: Starting from the events with maximum multiplicities, the top 10\%{} events are categorized as 0-10\%{} centrality class, and so on. Within the Glauber model \cite{Miller:2007ri}, the centrality classes can be directly connected to the averaged impact parameter and participants, etc. (see to horizontal axes in Fig.~\ref{fig:centrality_glauber}).

\section{Bulk properties and collectivity}

To map the QCD phase diagram by carrying out heavy-ion collisions at various beam energies and using the measurements to extract the transport properties of the produced QCD matter, a calibrated multistage theoretical framework is needed (see, e.g., Refs.~\cite{Hirano:2012kj,Heinz:2013wva}) which can accurately describe the bulk evolution of the system. Due to the smaller Lorentz contraction and the violation of boost-invariance at low beam energies, the space-time evolution in the longitudinal direction becomes rather complicated. Constraining the longitudinal dynamics and disentangling features arising from the matter deposition at the initial stage and from transport during the hydrodynamic stage is essential, for which the (pseudo-)rapidity distributions are the most important experimental observables to use. In this section, we briefly discuss a few examples of such observables, which have constraining power on bulk dynamics at low collision energies.

\subsection{Longitudinal distributions and baryon stopping}\label{sec:barystop}

At low beam energies, after the two colliding nuclei have passed through each other, some of the incoming valence quarks have been decelerated, doping the matter produced near mid-rapidity with net baryon charge. Longitudinal boost-invariance is no longer a good approximation and thus the longitudinal dynamics deviates from simple Bjorken flow \cite{Shen:2020gef, Shen:2020mgh}. This is strikingly different from the systems produced at top RHIC and LHC energies, which are usually well described as baryon-neutral fluids undergoing Bjorken expansion. As a consequence of this complication, it becomes essential to constrain the dynamics of net baryon density, especially in the longitudinal direction. Exactly for this reason we shall discuss the rapidity distributions of protons, anti-protons and net protons, which reflect directly (although not trivially) the longitudinal evolution of the non-zero net baryon density.

\begin{figure}[!htbp]
\centering
\includegraphics[width=0.53\linewidth]{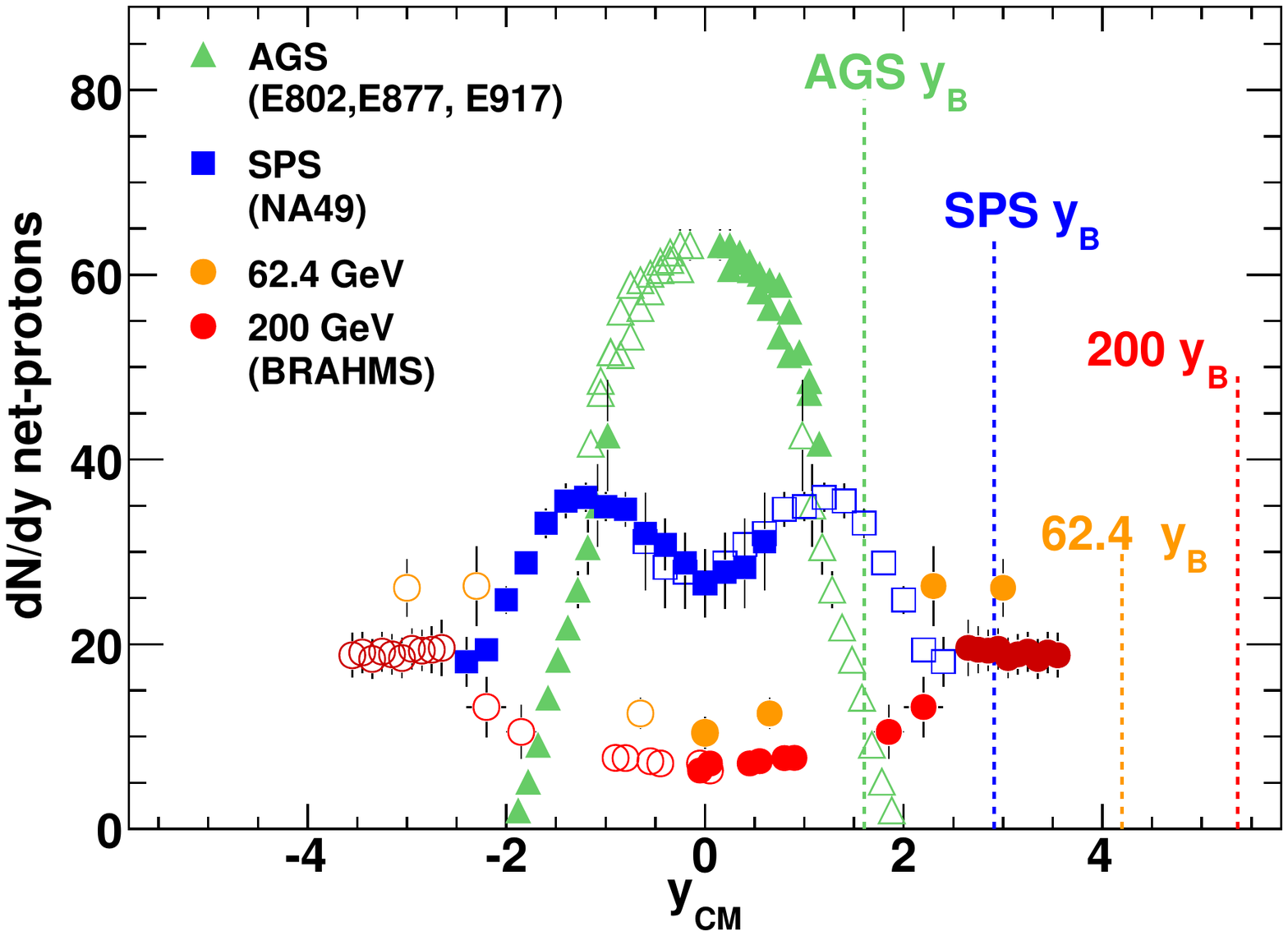}\hspace{0.15cm}\includegraphics[width=0.45\linewidth]{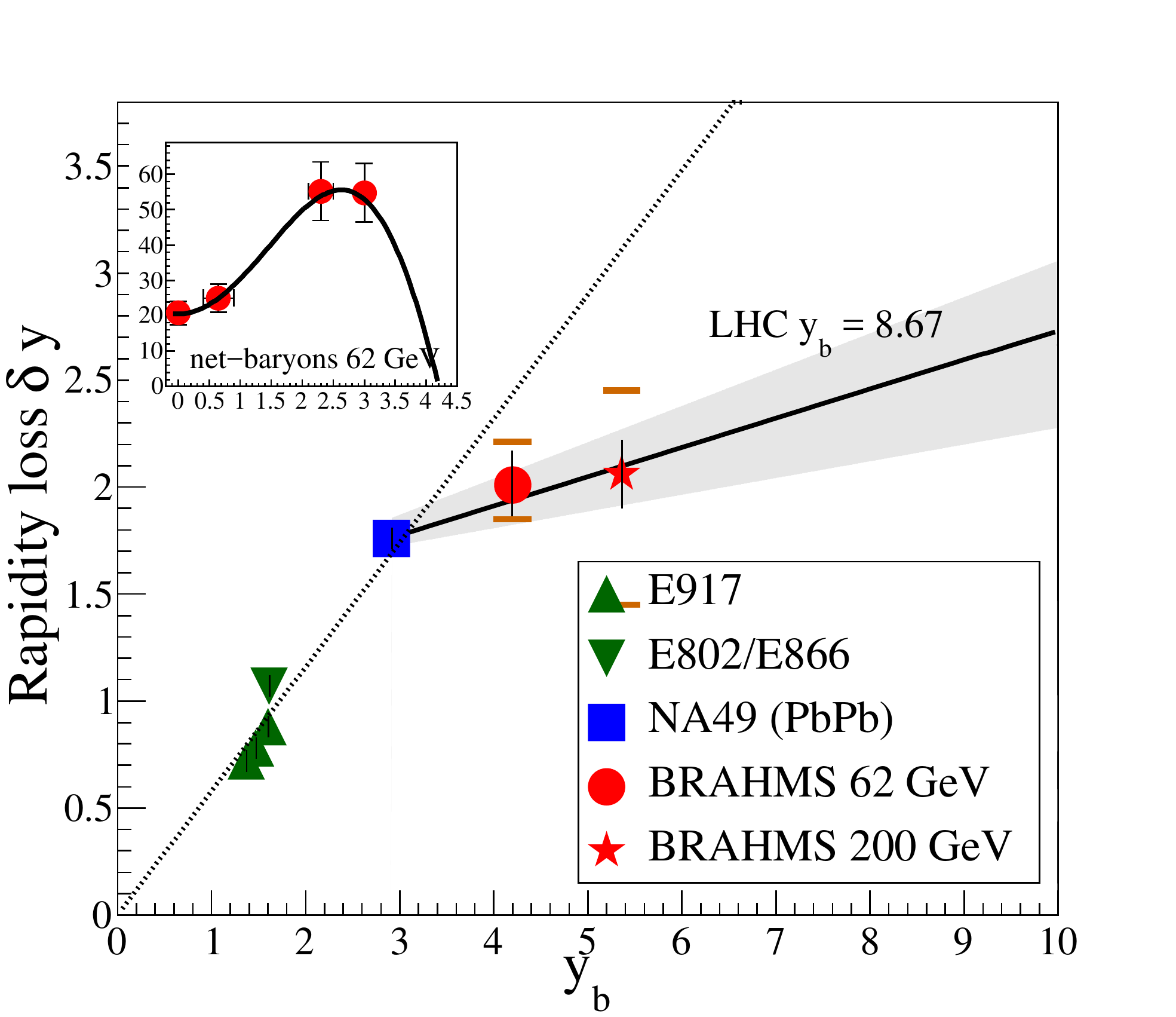}\par 
\caption{{\sl Left:} Rapidity-distributions of net protons, from AGS (Au+Au at $\snn\eq5$  GeV), SPS (Pb+Pb at $\snn\eq17$  GeV), and BRAHMS (Au+Au at $\snn\eq62.4$  and 200 GeV). Figure taken from Ref.~\cite{Videbaek:2009zy}. {\sl Right:}  Mean rapidity loss for net baryons at AGS, SPS, and RHIC as a function of beam rapidity. Figure taken from Ref.~\cite{Arsene:2009aa}.}
\label{fig:baryon_dist}
\end{figure}

Recent measurements of rapidity densities of identified particles, including protons and anti-protons, at RHIC BES energies have been reported for the mid-rapidity region ($|y|<0.1$) by the STAR collaboration at RHIC \cite{Adamczyk:2017iwn}; they have been used to test systematic model calculations (see, e.g., Ref.~\cite{Shen:2020jwv}). However, data at only mid-rapidity cannot provide enough constraining power for modeling the (3+1)-D evolution. Fortunately, some earlier experiments by the BRAHMS and PHOBOS collaborations provide particle multiplicity densities covering a large ranged rapidity, which can be used to calibrate longitudinal baryon evolution at low collision energies \cite{Shen:2017ruz, Shen:2020jwv}. Some rapidity-distributions of net protons at various beam energies are shown in the left panel of Fig.~\ref{fig:baryon_dist}, where data are taken from AGS (Au+Au at $\snn\eq5$  GeV, $y_b\eq1.7$), SPS (Pb+Pb at $\snn\eq17$  GeV, $y_b\eq2.9$), and BRAHMS at RHIC (Au+Au at $\snn\eq62.4$  and 200 GeV, $y_b\eq4.2$ and $5.4$, respectively), with $y_b\approx\ln(\snn/m_p)$ being the beam rapidity given $m_p$ the nucleon mass.

%
%

The left panel of Fig.~\ref{fig:baryon_dist} indicates that indeed, because the QCD matter is baryon-charged, the yield of protons is larger than that of antiproton, and that the rapidity distribution of net protons has a non-trivial shape. Generally speaking,  for the very low beam energy provided by the AGS, the rapidity density of net protons has a single peak at mid-rapidity. However, as the collision energy increases, two peaks separated by a dip at $y=0$ emerge, and the locations of the maximum rapidity densities become more and more separated, while the net baryon number around mid-rapidity decreases. This is generally described by the phenomenon called lack of baryon stopping and onset of nuclear transparency, i.e., the inability of colliding nuclei to completely stop each other as the collision energy increases \cite{Busza:1983rj}. From the perspective of multistage hybrid models, we note that the final net proton rapidity density distribution combines the effects from  initial baryon stopping and subsequent hydrodynamic baryon transport \cite{Shen:2017ruz, Denicol:2018wdp, Du:2018mpf, Monnai:2019hkn, Du:2019obx}. 

To characterize baryon stopping, the following mean rapidity loss  \cite{Videbaek:1995mf} was proposed:
\begin{equation}
\delta y=y_\mathrm{b}-\frac{2}{N_\mathrm{part}}\int_0^{y_\mathrm{b}} y\frac{dN^{B-\bar B}}{dy}dy\,.
\end{equation}
Here $N_\mathrm{part}$ is the number of participants, $dN^{B-\bar B}/dy$ the net baryon rapidity density, and thus the second term on the right hand represents the average rapidity per baryon for the forward-moving final particles. We note that in the left panel of Fig.~\ref{fig:baryon_dist}, only the net proton number is measured while the neutrons lack of electric charge and thus are not measurable and some other hadrons carrying baryon number may not be measured, either. Thus a conversion from the measured net proton distribution to the inferred net baryon distribution is needed, which brings in systematic uncertainties. Additionally, in some of the experiments the net proton number is measured at only a few rapidities, and thus some interpolation and extrapolation is required to obtain a continuous $dN^{p-\bar p}/dy$  distribution, that covers the entire region from mid-rapidity to beam rapidity. The resulting mean rapidity loss of net baryons is shown in the right panel of Fig.~\ref{fig:baryon_dist}, where the inset shows a fit to experimental data at $\snn\eq62.4$ GeV using a 3rd order polynomial in $y^2$. From the figure, one can see that the rapidity loss increases linearly with the beam rapidity at AGS and SPS energies, whereas the linearity appears to break down at higher energies \cite{Arsene:2009aa}.

\subsection{Collective flows}

To constrain the structure in the transverse plane, it is useful to look at the harmonic flows, which describe the anisotropy in the azimuthal distribution of the final particles (see reviews \cite{Heinz:2013th,Jeon:2015dfa}). The transverse flow is driven by the pressure gradient; flow anisotropies are resulted from geometric anisotropies and fluctuations in the initial energy and pressure distribution in the transverse plane. Differences in such flows between protons and anti-protons should be helpful to constrain the initial baryon distribution and baryon transport coefficients (see, e.g., Refs.~\cite{Denicol:2018wdp,Shen:2020jwv}).

As mentioned in Sec.~\ref{sec:kinematics}, experimentalists typically measure the spectra of hadrons (within different centrality classes, at various beam energies, and for certain collision systems, etc.). From these spectra one can obtain other observables which quantify the system, including yields (Sec.~\ref{sec:barystop}) and anisotropic flows, etc., differentially in or integrated over transverse momentum $p_T$. Here we focus on the anisotropic flows, which are defined in terms of the azimuthal Fourier components of the invariant momentum spectrum, for the particle species $i$, as follows:
\begin{equation}
\frac{dN_i}{p_Tdp_Td\phi_pdy}=\frac{1}{2\pi}\frac{dN_i}{p_Tdp_Tdy}\left(1+2\sum_{n\geq1}v^{(i)}_n(y,p_T)\cos[n(\phi_p-\Psi_n^{(i)}(y,p_T))]\right)\,.
\end{equation}
Here $v^{(i)}_n(y,p_T)\geq0$ is called the $n$th-order differential anisotropic flow coefficient, $\Psi_n^{(i)}(y,p_T)$ the $n$th-order differential flow plane, both of which depend on the rapidity $y$ and the transverse momentum $p_T$. Formally, the anisotropic flow coefficients can be obtained as
\begin{equation}
v^{(i)}_n(y,p_T)=\av{\cos[n(\phi_p-\Psi_n^{(i)}(y,p_T))]}\,,
\end{equation}
where the average $\av{\cdots}$ is taken over the invariant momentum spectrum. Sometimes the $p_T$-integrated anisotropic flows are also reported, for particles with all transverse momenta, defined as
\begin{equation}
\frac{dN_i}{d\phi_pdy}=\frac{1}{2\pi}\frac{dN_i}{dy}\left(1+2\sum_{n\geq1}v^{(i)}_n(y)\cos[n(\phi_p-\Psi_n^{(i)}(y))]\right)\,,
\end{equation}
where the $p_T$-integrated spectrum is given by
\begin{equation}
\frac{dN_i}{d\phi_pdy}=\int_0^\infty p_Tdp_T \frac{dN_i}{p_Tdp_Td\phi_pdy}\,.
\end{equation}
Usually these coefficients are measured in the mid-rapidity region, and thus the rapidity-dependence is dropped from the expression, but we note that the anisotropic flows at different rapidity can be essential for constraining the longitudinal structure of the fireball at low beam energies.

%
\begin{figure}[!htb]
\centering
\includegraphics[width=1.07\linewidth]{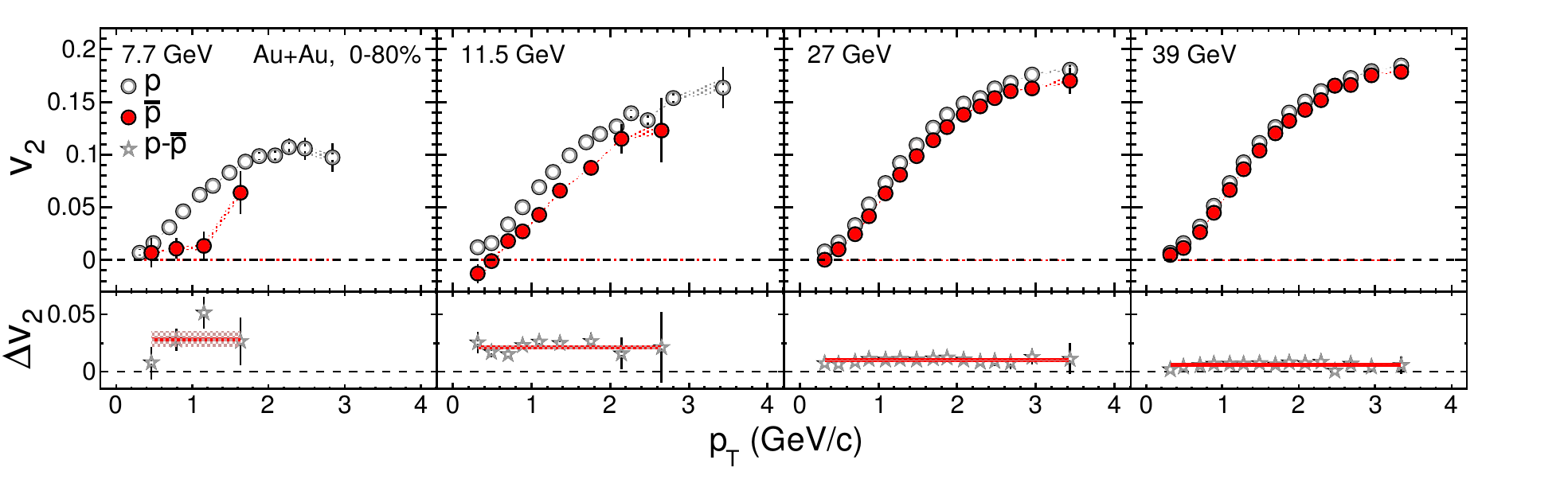}\par 
\caption{The $p_T$-differential elliptic flow of protons and anti-protons produced in Au+Au collisions within $0\%$-$80\%$. The lower panels show the difference of $v_2$ between protons and anti-protons, while the red bands are obtained from a constant fit to that ratio. Figure taken from Ref.~\cite{Adamczyk:2013gv}.}
\label{fig:baryon_v2}
\end{figure}%
%

The second order anisotropic flow coefficient $v_2$ is called elliptic flow, which is directly driven by the gradient of pressure corresponding to the initial geometric ellipticity, i.e., the almond-shaped overlap region shown in Fig.~\ref{fig:reaction_plane}. Similarly, there is a mapping between higher-order flow coefficients and the corresponding initial eccentricities at the same order, with a mixture of contributions from other orders as well. At BES energies, the fireball is not neutral, i.e., additional constraints are required to understand the evolution of its charge distributions. In this thesis, we are interested in the distribution of baryon density, and thus we expect the anisotropic flows of protons and anti-protons, together with their difference, to be able to provide information on the distribution of baryon number in the transverse plane. 

The difference of $v_2$ between particles and their corresponding anti-particles produced at different beam energies and within various centralities have been measured systematically by the STAR collaboration \cite{Adamczyk:2013gv, Adamczyk:2013gw, Adamczyk:2015fum}. The $p_T$-differential $v_2$ of protons and anti-protons at four beam energies are shown in Fig.~\ref{fig:baryon_v2}. The figure indicates that at BES energies protons have a larger elliptic flow than anti-protons and that the difference decreases with increasing beam energy, but does not have a strong $p_T$ dependence. Such a difference may be due to the distribution of baryon density or arising from other sources, such as interactions in the hadronic afterburner which are more important at lower beam energies \cite{Adamczyk:2013gv}. Theoretical modeling shows that larger baryon diffusion can reduce the elliptic flow of protons and enhance that of anti-protons by changing the transverse baryon distribution through dissipative corrections at particlization \cite{Denicol:2018wdp}.

%
\begin{figure}[!htb]
\centering
\hspace{-0.12cm}\includegraphics[width=0.455\linewidth]{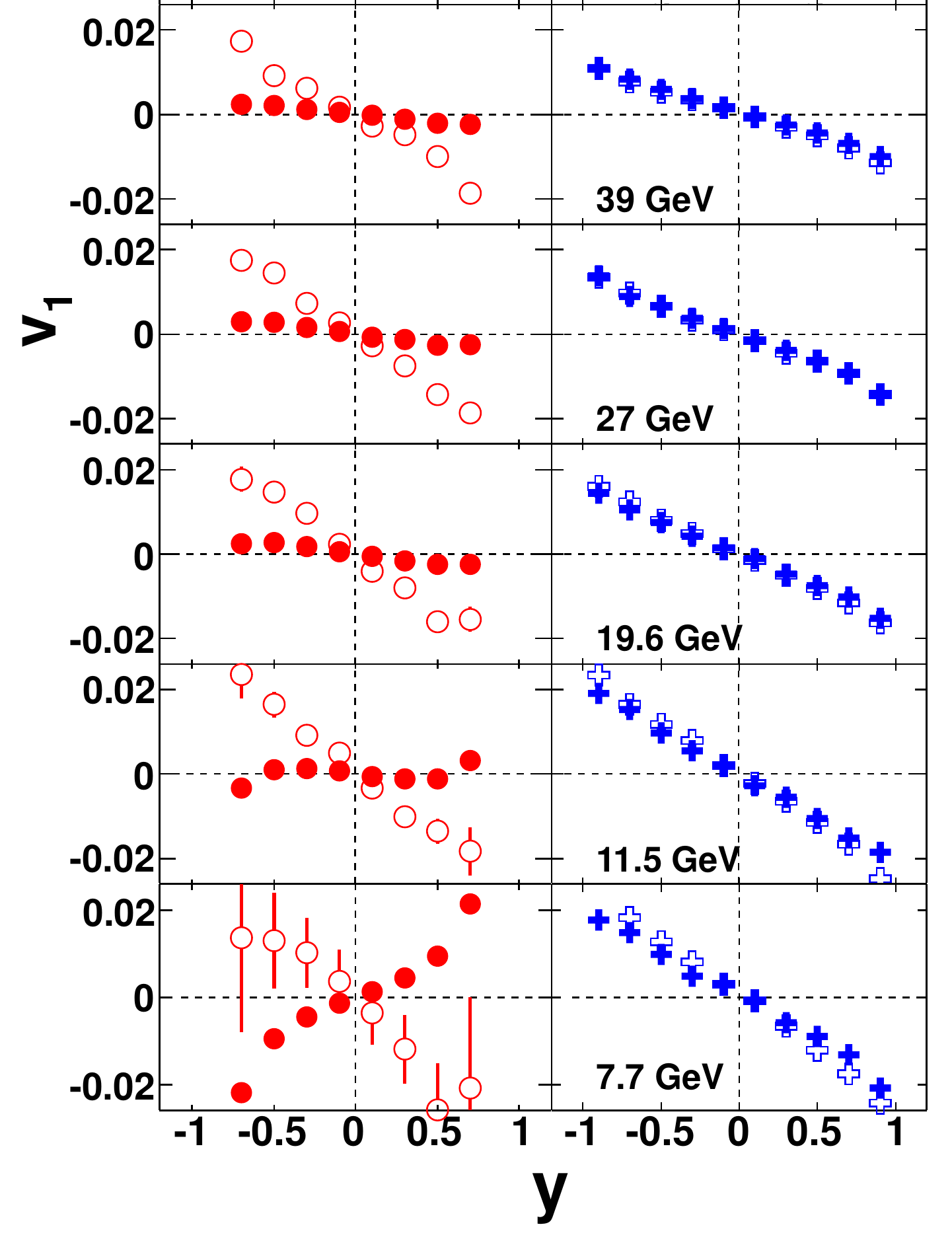}\hspace{0.25cm}\includegraphics[width=0.5\linewidth]{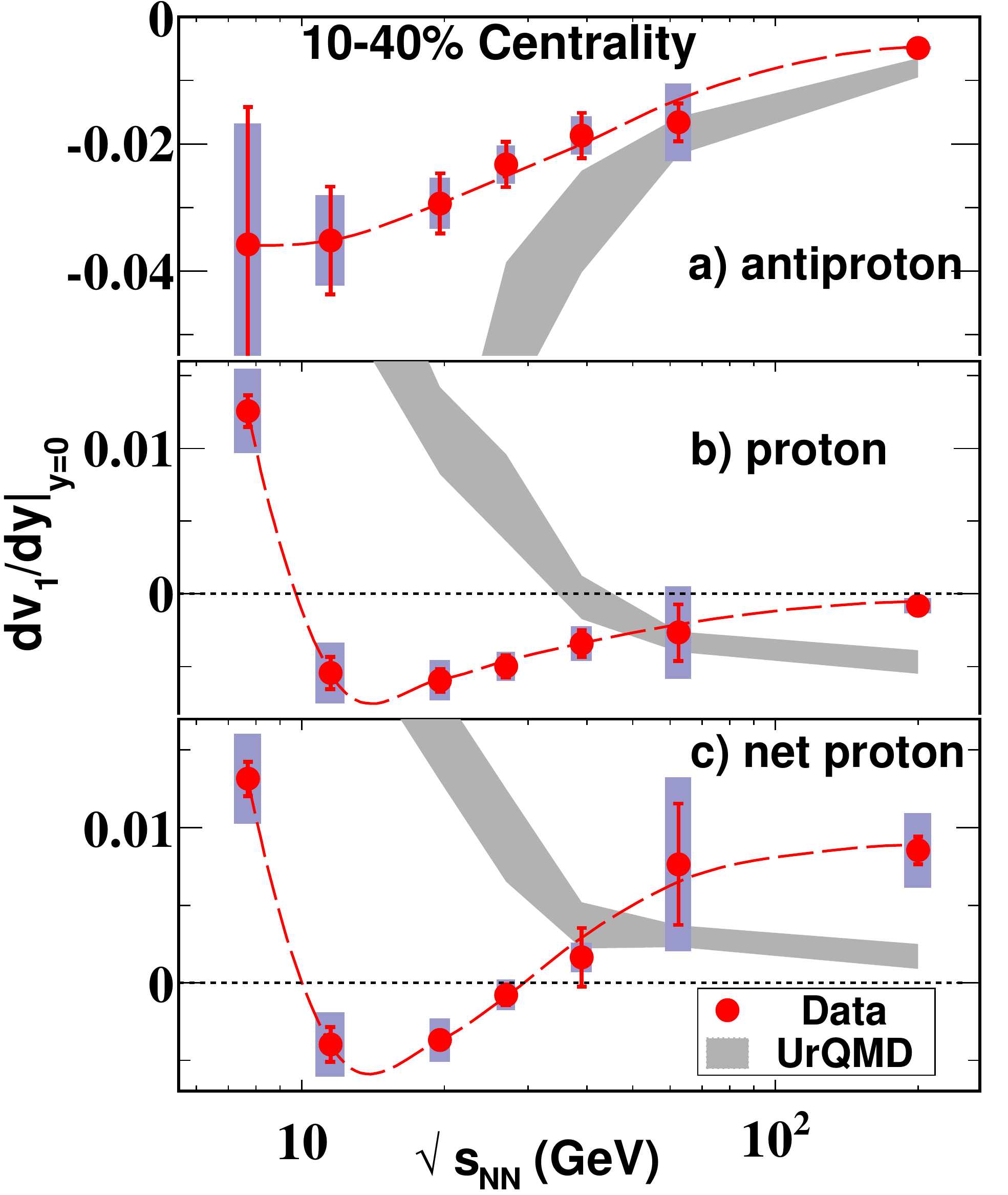}\par 
\caption{{\sl Left:} Directed flow $v_1$ as a function of rapidity $y$ for protons (red dot), anti-protons (red circle), $\pi^+$ (blue open cross), and $\pi^-$ (blue solid cross) from Au+Au collisions at intermediate centrality $10\%$-$40\%$. {\sl Right:}  The slope at mid-rapidity of directed flow at different beam energies for anti-protons (a), protons (b) and net protons (c). The left figure adapted and the right taken from Ref.~\cite{Adamczyk:2014ipa}.}
\label{fig:v1_dist_BES}
\end{figure}%
%

The first order anisotropic flow coefficient $v_1$, also called directed flow, describes collective sidewards deflection of the final particles (see a recent review \cite{Singha:2016mna}). Measurements of directed flow for identified hadrons at $|y|<1$ in Au+Au collisions at BES have been reported by the STAR collaboration \cite{Adamczyk:2014ipa}. As shown in the left panel of Fig.~\ref{fig:v1_dist_BES}, at intermediate centrality of $10\%$-$40\%$, the slopes of $v_1(y)$ for $\pi^\pm$, protons and anti-protons are negative for all energies, except for protons at 7.7 GeV \cite{Adamczyk:2014ipa}. Besides, the slopes for $\pi^+$ and $\pi^-$ are quite close with slight difference at low beam energies, whereas a large difference in percentage between protons and anti-protons can be seen. Indeed, given the statistics it is not obvious whether the sign of the $v_1(y)$ slope for protons is negative at collision energies above 7.7 GeV. With a cubic fit $v_1(y)=F_1y+F_3y^3$, Ref.~\cite{Adamczyk:2014ipa} plotted the slope of $v_1(y)$ at mid-rapidity, $dv_1(y)/dy|_{y=0}=F_1$, in the right panel of Fig.~\ref{fig:v1_dist_BES}. From that plot, we do see that the slope for protons is small but indeed negative, except at 7.7 GeV.

As noted in Ref.~\cite{Adamczyk:2014ipa}, different models with various mechanisms have been used to explain the slope of $v_1(y)$, among which we are interested in the hydrodynamic description. Within such a framework, $v_1(y)$ around mid-rapidity can provide some constraints on the early stage of fireballs expansion. Ref.~\cite{Bozek:2010bi} proposed a tilted structure for the initial distribution of the fireball in the reaction plane, whose corresponding pressure gradient can drive a directed flow with a negative slope for charged particles over a broad pseudo-rapidity range, consistent with the experimental data (for 200 GeV Au+Au and Cu+Cu collisions \cite{Abelev:2008jga}).  However, a recent calculation of $v_1(y)$ with such a tilted structure in the initial distribution of energy density gave positive slopes for $v_1(y)$ of protons at various beam energies, which are opposite to the experimental data. It was also proposed that the minimum in the slope of the directed flow may be a signature of a first-order phase transition \cite{Stoecker:2004qu}. Simultaneously fitting the directed flow of identified particles at BES, along with the elliptic flow and longitudinal distributions of net protons, would provide strong constraints on the initial structure of the fireball.

\section{Cumulants and criticality}\label{sec:cumucrit}

In this section, we focus on cumulants of the baryon number, which are believed to be sensitive to the QCD critical point and thus could provide signatures of it  in the near future when the data of the BES-II campaign are analyzed \cite{Bzdak:2019pkr,Nahrgang:2016ayr,Luo:2017faz}. As a measurable and directly related observable, the net proton number cumulants can be obtained from protons and anti-protons multiplicities on an event-by-event basis.

\subsection{Baryon number cumulants}\label{sec:bcumul}

The baryon susceptibilities, $\chi_j$, calculated from the equation of state can be related to the cumulants, $\kappa_j$, of the fluctuations of the net baryon number $N$. Noting the relationship between the partition function and pressure for a grand canonical system
\begin{equation}
Z=\frac{\exp(Vp(T,\mu))}{T}\,,
\end{equation}
and that between the baryon density and pressure
\begin{equation}
n=\frac{\langle N\rangle}{V}=\left(\frac{\partial p}{\partial \mu}\right)_T\,,
\end{equation}
one can connect the net baryon number cumulants and the susceptibilities \cite{Gupta:2011wh,Bzdak:2019pkr,Luo:2017faz}
\begin{equation}\label{eq:chicum}
\chi_j\equiv\left(\frac{\partial^j (p/T^4)}{\partial(\mu/T)^j}\right)_T=\frac{1}{VT^3}\kappa_j\equiv\frac{1}{VT^3}\left(\frac{\partial^j \ln Z}{\partial(\mu/T)^j}\right)_T\,,
\end{equation}
where $V$ is the volume of the thermal system. More specifically, the cumulants at the first few orders are given as
\begin{equation}
\kappa_1 = \langle N\rangle\,,\quad \kappa_2 = \langle(\delta N)^2\rangle\,,\quad \kappa_3 = \langle(\delta N)^3\rangle\,,\quad \kappa_4 = \langle(\delta N)^4\rangle-3\langle(\delta N)^2\rangle^2\,,
\end{equation}
where $\delta N=N-\langle N\rangle$ denotes the fluctuation of the net baryon number,  $N$ being the event-by-event net baryon number and $\langle N\rangle$ the event-averaged number. Eq.~\eqref{eq:chicum} indicates that if the system is in equilibrium, then the cumulants of the observed event-by-event distribution of net-baryon number in heavy-ion collisions should be directly comparable to the baryon susceptibilities calculated from Lattice QCD -- a possible test of the QCD in the non-perturbative domain (see, e.g., Ref.~\cite{Gupta:2011wh}). The problem with this is, however, that the dynamically evolving heavy-ion collision fireballs are never really in thermal equilibrium.

By taking ratios between the susceptibilities or the cumulants, one can cancel out $T$ and $V$ (also volume fluctuations to some extent) and obtain \cite{Bzdak:2019pkr,Nahrgang:2016ayr,Luo:2017faz}
\begin{equation}\label{eq:centrlm}
\frac{\sigma^2}{M}=\frac{\chi_2}{\chi_1}=\frac{\kappa_2}{\kappa_1}\,,\quad S\sigma=\frac{\chi_3}{\chi_2}=\frac{\kappa_3}{\kappa_2}\,,\quad \kappa\sigma^2=\frac{\chi_4}{\chi_2}=\frac{\kappa_4}{\kappa_2}\,,
\end{equation}
where $M, \sigma^2, S, \kappa$ are central moments of net baryon number, including mean, variance, skewness and  kurtosis, respectively. Explicitly, they are defined in terms of the baryon number, as follows:
\begin{equation}
\sigma^2=\langle(\delta N)^2\rangle\,,\quad S=\frac{\langle(\delta N)^3\rangle}{\langle(\delta N)^2\rangle^{3/2}}\,,\quad \kappa=\frac{\langle(\delta N)^4\rangle}{\langle(\delta N)^2\rangle^{2}}-3\,,
\end{equation}
and  the skewness and kurtosis describe how the shape of a probability distribution deviates from the Gaussian distribution, where the former gives the asymmetry and the later the ``tailedness'' of the distribution. For a distribution, which has larger tail to the left (right), the skewness is negative (positive). A distribution, which is more (less) concentrated around its mean, is called a leptokurtic (platykurtic) distribution and its kurtosis is positive (negative). Sometimes, $S$, $\kappa$ and higher order cumulants are called non-Gaussian fluctuations.

\subsection{Equilibrium critical fluctuations}\label{sec:critifluc}

The cumulants of fluctuations are especially important in the context of critical point searching, as the most characteristic feature of criticality is the enhancement and divergence of fluctuations \cite{Stephanov:2008qz,Stephanov:2011pb,Bzdak:2019pkr}. With varying order parameter $\sigma(\bm x)$, and its fluctuation $\delta\sigma(\bm x)=\sigma(\bm x)-\av{\sigma(\bm x)}$, one can introduce the volume integrated fluctuation $\sigma_V\equiv\int{}d^3\bm x\,\delta\sigma(\bm x)$. The second cumulant of this extensive quantity is 
\begin{equation}\label{eq:kapp2}
\kappa_2[\sigma_V]=\av{\sigma_V^2}=V\int{}d^3\bm x\av{\delta\sigma(\bm x)\delta\sigma(0)}\approx VT\xi^2\,,
\end{equation}
where $\xi$ is the correlation length and, in the last step, the thermodynamic limit $V\gg\xi^3$ has been assumed, together with the following Gaussian approximation of the two point correlation:
\begin{equation}\label{eq:twoptcorre}
\av{\delta\sigma(\bm x)\delta\sigma(\bm y)}=\frac{T}{4\pi|\bm x-\bm y|}\exp\left(-\frac{|\bm x-\bm y|}{\xi}\right)\,.
\end{equation}
Eq.~\eqref{eq:kapp2} indicates that the second-order cumulant diverges with the correlation length as $\kappa_2\sim\xi^2$, which can be viewed as the divergence of the integral of $1/|\bm x-\bm y|$ being cut off at $|\bm x-\bm y|\sim\xi$ (see Eq.~\eqref{eq:twoptcorre}). Here, it is also important to note that the divergence of fluctuations at the critical point is a collective and macroscopic phenomenon, which involves the correlation among fluctuations of many d.o.f. in a correlation volume $V_\xi\sim\xi^3$. In other words, the divergence of fluctuations is not referring to that of the {\it magnitude} of the fluctuations of local d.o.f. but to the divergence of the {\it range of the correlations} for the fluctuations.

Similarly, the higher order non-Gaussian fluctuations are found to be \cite{Stephanov:2008qz,Stephanov:2011pb}
\begin{eqnarray}
\kappa_3[\sigma_V]&=&\av{\sigma_V^3}=2\tilde\lambda_3VT^{3/2}\xi^{9/2}\,,\label{eq:higherc1}\\
\kappa_4[\sigma_V]&=&\av{\sigma_V^4}_\mathrm{c}\equiv\av{\sigma_V^4}-3\av{\sigma_V^2}^2=6(2\tilde\lambda_3^2-\tilde\lambda_4)VT^2\xi^7\,,\label{eq:higherc2}
\end{eqnarray}
where the subscript ``c'' in the last line means  ``connected'', and $\tilde\lambda_3, \tilde\lambda_4$ are some coupling constants in the effective action functional of $\sigma$ field \cite{Stephanov:2008qz,Stephanov:2011pb}. Eqs.~(\ref{eq:higherc1}, \ref{eq:higherc2}) indicate that higher-order cumulants, which describe the shape of the probability distribution of $\sigma$, are more sensitive to the correlation length than the second-order one which describes the width of the distribution \cite{Stephanov:2008qz,Stephanov:2011pb,Bzdak:2019pkr}. As noted in Sec.~\ref{sec:phase_diagram}, the order parameter $\sigma$ field has a mixture of contents, which, at large chemical potential   where the hypothetical QCD critical point is expected, is dominated by the net baryon density. Thus the fluctuation of $\sigma$ near the critical point can affect  other measurable quantities, for example, the fluctuations of charged particles \cite{Stephanov:2008qz}. In fact, the coupling to the critical mode  is isospin blind, and the cumulants of the fluctuations of protons, anti-protons and net protons (as well as for neutral particles) have similar patterns near the critical point \cite{Bzdak:2019pkr}.

Stephanov showed that the cumulants of multiplicity fluctuations of pions and protons are expected to diverge as \cite{Stephanov:2008qz}
\begin{equation}
\omega_3(N)_\sigma\sim\xi^{9/2}\,,\quad\omega_4(N)_\sigma\sim\xi^7\,,
\end{equation}
where the subscript $\sigma$ indicates that only critical mode contribution is considered, $N$ is the multiplicity of pions or protons, and $\omega_k\equiv\av{(\delta N)^k}_\mathrm{c}/\langle N\rangle$ are the normalized cumulants of multiplicity, with $\langle N\rangle$ being the mean total multiplicity. Introducing the normalized cumulants of multiplicity can cancel out the dependence on volume $V$ of the numerator and denominator. Note that since protons have much larger mass than pions, the critical effects on their multiplicity fluctuations are expected to be much stronger than for pions \cite{Stephanov:2008qz}, and thus people are more interested in proton multiplicity fluctuations when searching for the critical point experimentally (see, e.g., Refs.~\cite{Bzdak:2019pkr,Adam:2020unf,Abdallah:2021fzj}). We also note that the discussions in this subsection are for an infinite equilibrium system; some complications from off-equilibrium and other sources of fluctuations shall be briefly mentioned in the next subsection. 

%
\begin{figure}[!tb]
\centering
\hspace{-0.12cm}\includegraphics[width=0.35\linewidth]{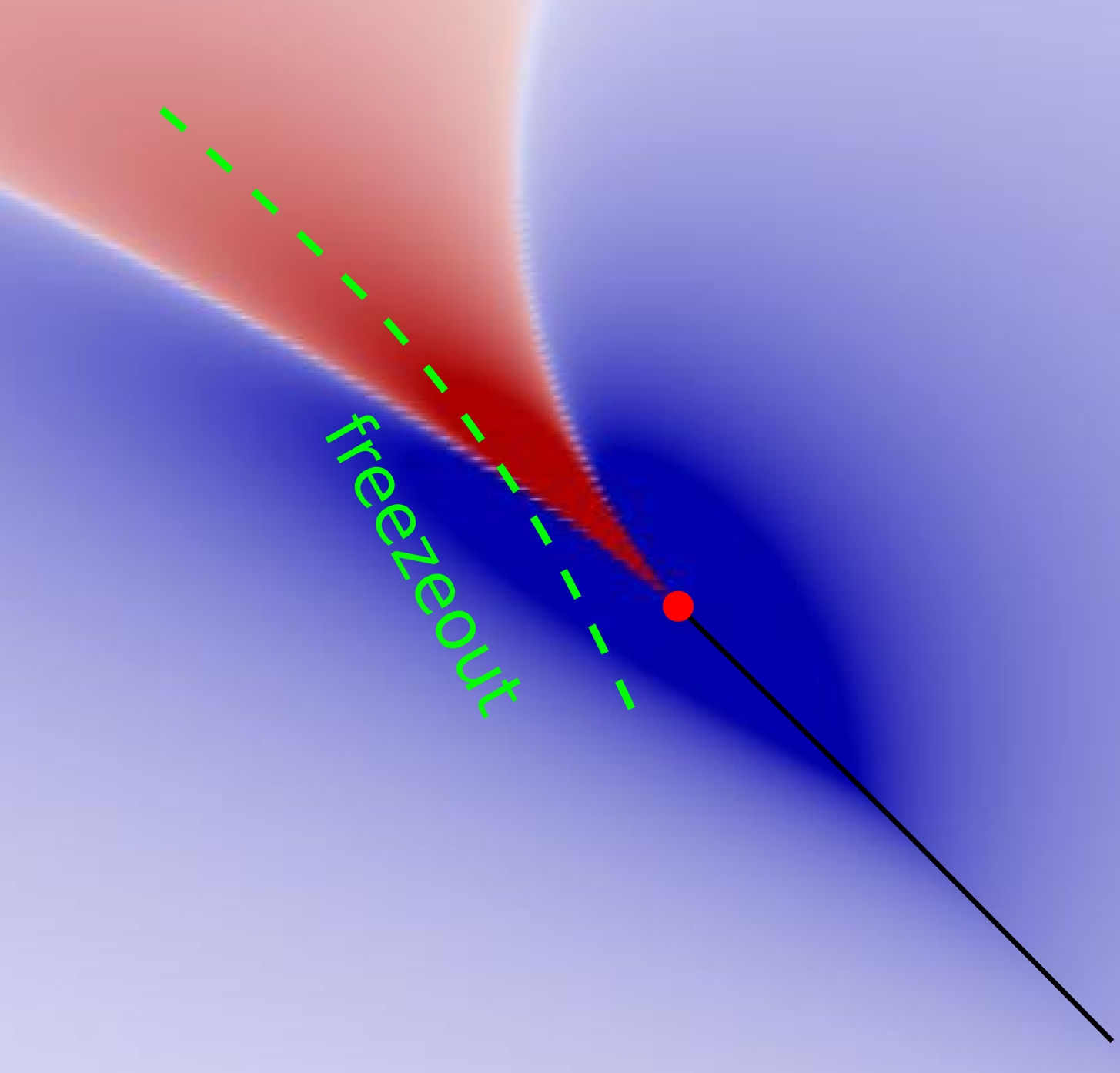}\hspace{0.25cm}\includegraphics[width=0.56\linewidth]{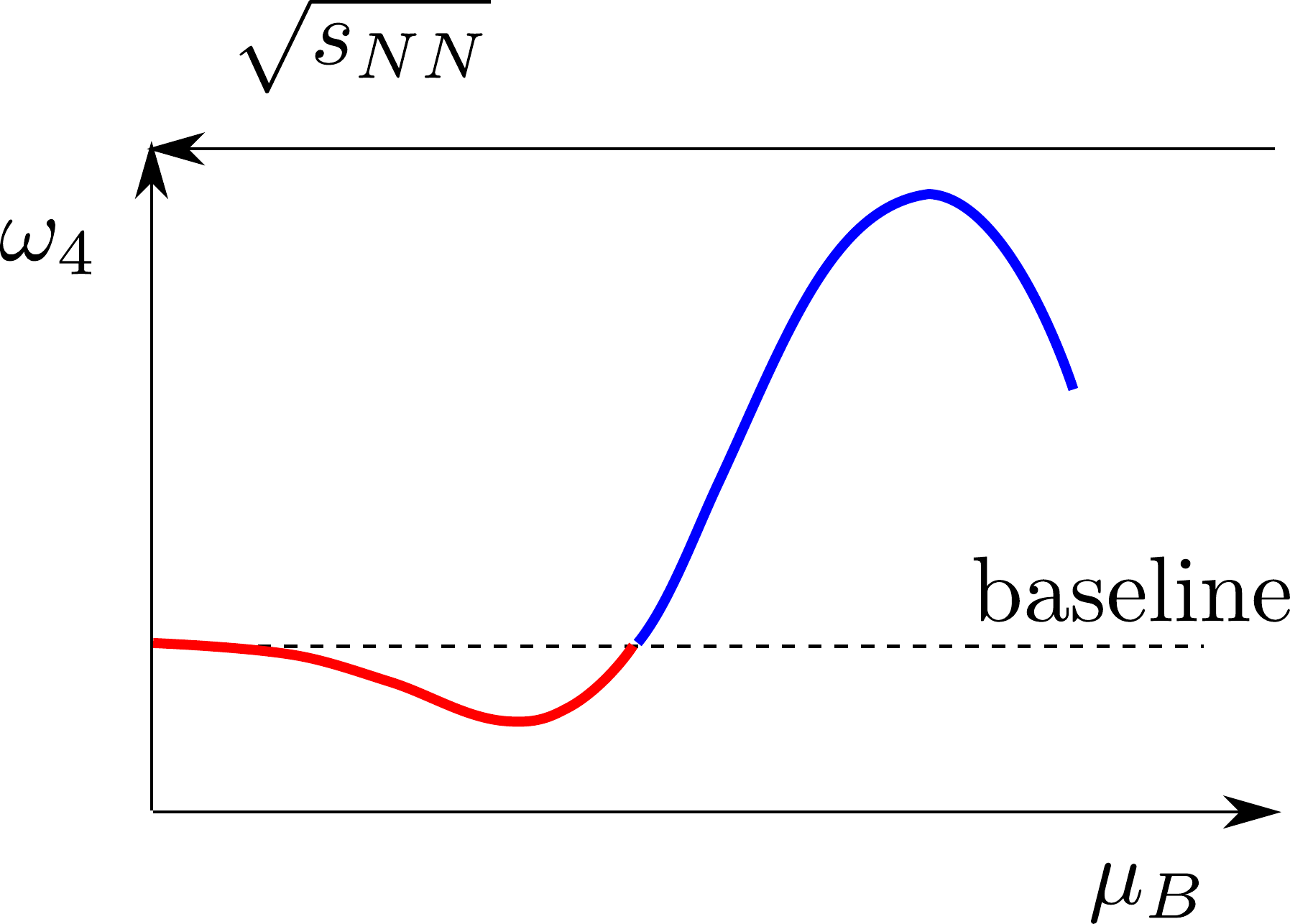}\par 
\caption{{\sl Left:} Density plot of the fourth order cumulant of the order parameter near the critical point, by mapping the EoS of the 3D Ising model to that of QCD. The green dashed line indicates the freeze-out line, from whose left end to the right, chemical freeze-out is expected to happen, in heavy-ion collisions from high to low beam energies. {\sl Right:}  Normalized fourth order cumulant of the proton multiplicity as a function of chemical potential along the freeze-out line in the left panel. Experimentally, increasing the chemical potential can be achieved by lowering the collision energies, as indicated by the upper arrow of energy pointing to the left.  Figures taken from Ref.~\cite{Bzdak:2019pkr}.}
\label{fig:fz_out_omega}
\end{figure}%
%

\subsection{Fluctuation cumulants in heavy-ion collisions}

As discussed in the previous subsection, the higher order cumulants of the multiplicity fluctuations of net protons (as well as protons and anti-protons) are more sensitive to the enhancement in the correlation length. Along the freeze-out curve in the phase diagram, as shown in the left panel of Fig.~\ref{fig:fz_out_omega}, from the left to the right end, the system first approaches and then moves away from the critical point, and thus the correlation length first increases and then decreases. With this picture in mind, one can expect the higher order cumulants to increase and then decrease in systems which freeze out with that increase from low to high chemical potentials. In heavy-ion collisions, tuning the chemical potential can be achieved by changing the collision energy: At lower energies, more baryon number can be stopped into the produced QCD matter, and thus its chemical potential increases. Based on these arguments, when decreasing or increasing the beam energy in the RHIC BES campaign, the higher order cumulants of net protons should show a non-monotonic behavior (see sketch in the right panel of Fig.~\ref{fig:fz_out_omega}) \cite{Stephanov:1998dy,Stephanov:1999zu,Stephanov:2008qz,Stephanov:2011pb,Bzdak:2019pkr}. This is one of the most important motivations for the BES studies at RHIC and else. Again, the non-monotonic behavior of these cumulants as a function of beam energy $\snn$ as an experimental signature of a critical point is based on the assumption that the fluctuations are at equilibrium at freeze-out \cite{Stephanov:2008qz}, which is likely not the case in reality. 

%
\begin{figure}[!t]
\centering
\includegraphics[width=0.9\linewidth]{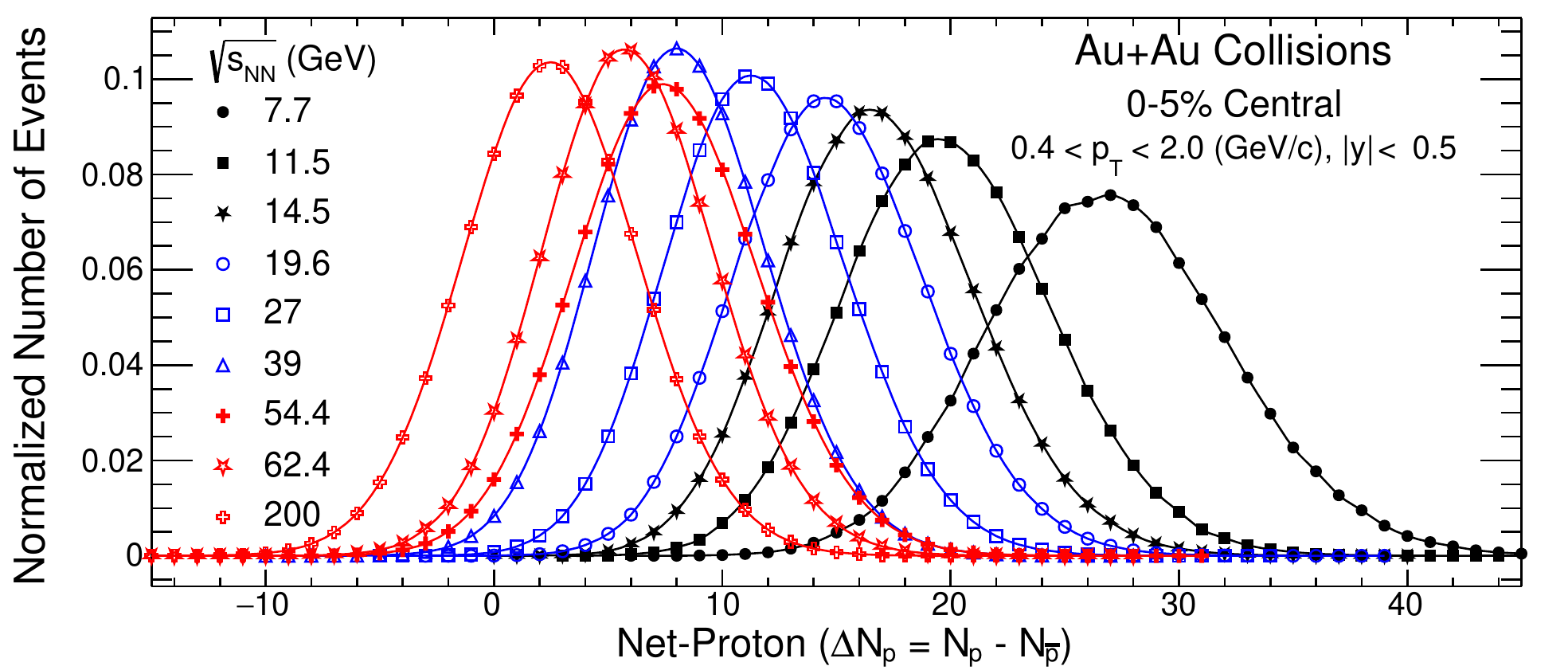}\par 
\caption{Event-by-event distributions of net proton number for central ($0\%$-$5\%$) Au+Au collisions at BES energies, normalized by the total event numbers at each energy \cite{Adam:2020unf,Abdallah:2021fzj}. Figure taken from Ref.~\cite{Adam:2020unf}}
\label{fig:ebe_multiplicity}
\end{figure}%
%

Recently, the STAR collaboration reported the measurements of the cumulant ratios (central moments) listed in Eq.~\eqref{eq:centrlm} of the net proton number at BES energies (from 7.7 GeV to 200 GeV)  \cite{Adam:2020unf,Abdallah:2021fzj}, obtained from data taken from 2010 to 2017, as part of the BES-I program. The event-by-event distribution of net proton multiplicity (mid-rapidity $|y|<0.5$) at BES energies is shown in Fig.~\ref{fig:ebe_multiplicity}, from which the cumulants can be obtained. As noted, near the critical point, the high-order cumulants are expected to be enhanced, and thus the distributions of net proton cumulants in the figure would deviate from a Gaussian distribution (see Sec.~\ref{sec:bcumul}), if at some collision energies the system freezes out close to the critical point.

\begin{figure}[!t]
\centering
\includegraphics[width=0.9\linewidth]{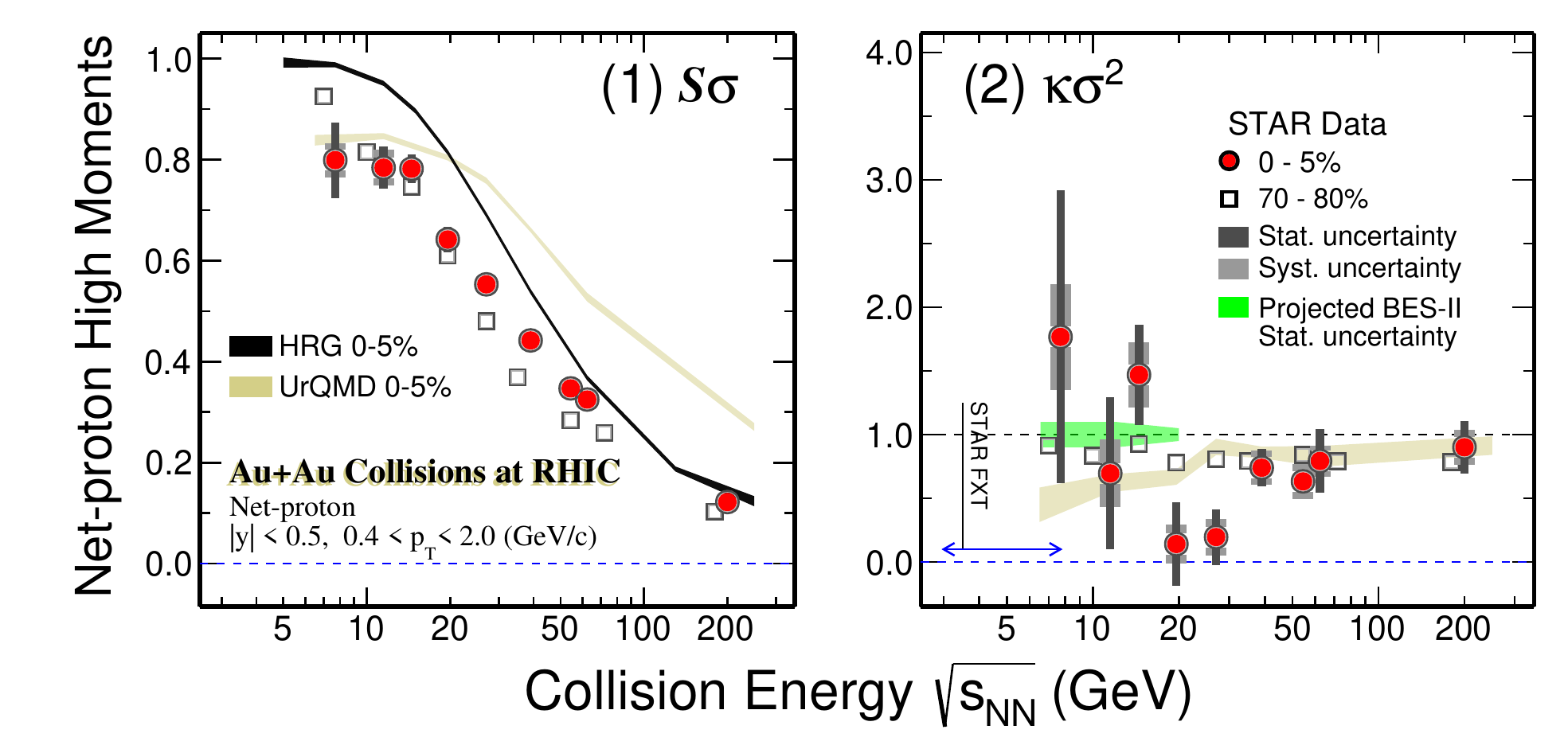}\par 
\caption{$S\sigma$ and $\kappa\sigma^2$ of the net proton distributions measured in Au+Au collisions at different collision energies, at two centralities, including central ($0\%$-$5\%$, filled circles) and peripheral ($70\%$-$80\%$, open squares) collisions. Some baseline calculations without any critical dynamics are also presented, including {\sc UrQMD} and HRG calculations \cite{Adam:2020unf,Abdallah:2021fzj}. Figure taken from Ref.~\cite{Adam:2020unf}.}
\label{fig:bes_cumulants}
\end{figure}

In Refs.~\cite{Adam:2020unf,Abdallah:2021fzj}, the STAR collaboration presented the 3rd and 4th order moments, $S\sigma$ and $\kappa\sigma^2$, which characterize the shape of the net proton multiplicity distribution functions, at various collision energies, as shown in Fig.~\ref{fig:bes_cumulants}. There is a hint from the figure that $\kappa\sigma^2$ may vary non-monotonically as a function of the collision energy, whereas $S\sigma$ does not favor such a trend.  The experimental measurements are also compared to the HRG model which assumes an ideal gas of hadron resonances in thermal equilibrium, and the {\sc UrQMD} model, which is  a hadronic transport model; neither of these two models includes critical dynamics and neither of them can explain the energy dependence of the cumulant $\kappa\sigma^2$. Note that there are still large uncertainties in the data points at low beam energies, which will be reduced with data taken during the second phase of BES \cite{Luo:2017faz,Bzdak:2019pkr}.

Finally, we note that to confirm and eventually locate (or exclude) the hypothetical QCD critical point, a systematic model-data comparison at various beam energies is required, as a number of real-life complexities may affect the thermal equilibrium discussions in this section. For example, the trend and sign change of $\omega_4(N)$ as a function of chemical potential shown in Fig.~\ref{fig:fz_out_omega} are based on a non-universal mapping between the EoS of a 3D Ising model and that of QCD matter \cite{Stephanov:2008qz,Stephanov:2011pb}, and the mapping itself has large uncertainties (see, e.g., Refs.~\cite{Parotto:2018pwx, Pradeep:2019ccv}). Furthermore, the discussion in Sec.~\ref{sec:critifluc} is based on critical fluctuations in systems at complete thermal equilibrium, while in realistic systems produced in heavy-ion collisions, the critical fluctuations do not have time to relax to their equilibrium values because of the rapid expansion of the system and critical slowing down (see, e.g., Refs.~\cite{Rajagopal:2019xwg,Stephanov:2017ghc}). This is also the motivation for our study in Ch.~\ref{ch.fluctuations}. Besides, the experimental measurements show the cumulants as a function of beam energies, a mapping of which to chemical potentials and temperatures in the phase diagram where the critical point sits, is also highly non-trivial. Last but not least, we note that there are other dynamical effects in heavy-ion collisions that can contribute to the cumulants measured in experiments, including volume fluctuations, finite-size effects, hadronic rescatterings, initial stage fluctuations, etc. (see, e.g., the review \cite{Jeon:2003gk} for more discussion).

%
%

\section{Other observables}

In the previous sections of this chapter, we have briefly discussed several observables, including longitudinal distributions of particle multiplicities and collective flows which are sensitive to the system's bulk dynamics, and fluctuation cumulants of net protons which are expected to be sensitive to criticality. Because of the complicated dynamical effects involved in searching for the critical point, a systematic comparison between experimental measurements and theoretical dynamical calculations for particle cumulants at various beam energies are crucial. For this purpose, the off-equilibrium dynamics of critical fluctuations needs to be simulated on top of a well-calibrated bulk dynamic evolution. The calibration requires observables with adequate constraining power. The aforementioned ones are some examples, but many more measurements and observables not mentioned above exist, at RHIC and also other facilities \cite{Luo:2017faz,Bzdak:2019pkr,Aggarwal:2010cw}. Because of the much more complicated dynamics at low beam energies, and the associated more demanding computational cost, when compared with collisions at the LHC and at top RHIC energies, the additional constraining power of these additional data should be explored. Most of this work must be left to the future. Fortunately, additional facilities and efforts are under way to explore the high baryon density region of the QCD phase diagram. For discussion of some of those outside the scope of this thesis, we recommand Refs.~\cite{Mohanty:2011nm,Schmah:2013vea,McDonald:2015tza,Odyniec:2013aaa,Tlusty:2018rif,Afanasev:1999iu,Gazdzicki:2004ef,Abgrall:2014xwa,Gazdzicki:2008kk,Satz:2004zd,Agakishiev:2009am}.

\chapter{Multistage description of heavy-ion collisions at BES energies}
\label{ch:multistage}

In Sec.~\ref{sec:theochall} we briefly mentioned the ingredients of a multistage framework for heavy-ion collisions at low beam energies, which, compared to high-energy collisions at LHC and top RHIC energies, introduces a number of additional complications. In this chapter, we shall describe the physics of the different stages of the framework in greater detail. In Sec.~\ref{sec:dyna_init}, we discuss the (3+1)-dimensional and temporally extended nuclear interpenetration stage and the construction of source terms describing the dynamical  deposition of energy and baryon number  for the hydrodynamic description. In Sec.~\ref{sec:relat_hydro}, we shall demonstrate the hydrodynamic equations which properly propagate conserved charge currents for baryon number, electric charge and strangeness. We then focus on intrinsic fluctuations at equilibrium in the hydrodynamic regime and near the critical point in Sec.~\ref{sec:fluct}. Then Sec.~\ref{transcoeff} and Sec.~\ref{subsec-eos} are devoted to the transport coefficients and Equation of State at non-zero chemical potentials, which describe microscopic properties of the bulk medium, and their singularities near the critical point. In Sec.~\ref{sec:part_hadr}, we discuss the particlization which converts the fluid into particles and the kinetic description for the produced particles. 

This chapter also includes the new ingredients of the multistage framework that we developed to adapt it for BES energies. It is largely based on material previously published in Refs.~\cite{Du:2018mpf, Du:2019obx, Du:2020bxp, du2021baryon}.

\section{Dynamical initialization}\label{sec:dyna_init}

\subsection{Pre-equilibrium dynamics}
\label{sec1}

At RHIC BES energies, the dynamics of the pre-equilibrium stage and the effects resulting from a nonzero net baryon current become critical components of the dynamical evolution of the collision fireball (see recent reviews \cite{Shen:2020gef,Shen:2020mgh}). Hybrid models of heavy-ion collisions, consisting of multiple stages describing different physics, have recently received intensive attention. In many approaches (see Table I in \cite{Oliinychenko:2015lva}), a hydrodynamic stage describing the evolution of quark-gluon plasma is initialized with output from some pre-equilibrium evolution model on a surface of constant (proper) time. Recently, dynamical initialization models in which the pre-equilibrium matter is converted to fluid gradually while the colliding nuclei are passing through each other were proposed in, e.g., Refs.~\cite{Shen:2017ruz, Okai:2017ofp, Shen:2017bsr, Akamatsu:2018olk,Du:2018mpf}.\footnote{%
	While hybrid modeling of heavy-ion collisions at RHIC BES has led to a recent surge of activity in (3+1)D dynamical initialization, similar ideas were already developed more than two decades ago (see e.g., Refs.~\cite{KAJANTIE1982203,KAJANTIE1983261,KAJANTIE1983152}). Indeed, heavy-ion collisions were carried out at low beam energies before the operation of RHIC, and a lot of work had been done to model such systems, similar to what we are trying to accomplish for BES energies nowadays. This reminds us that a lot can be learnt by studying the historic works of the pioneers of this field.
}
An incomplete list of recent works which attempted to include dynamical initialization in hybrid models can be found in Table \ref{tab:dynamical}.

In this section, we study specifically the dynamical initialization of hydrodynamics from {\sc UrQMD} \cite{Bass:1998ca,Bleicher:1999xi}. Our approach has many similarities with \cite{Akamatsu:2018olk} (where JAM was used instead of {\sc UrQMD}) but, different from \cite{Akamatsu:2018olk} and similar to \cite{Denicol:2018wdp}, it uses dissipative hydrodynamics, including evolution of the baryon diffusion current. We will here focus on differences between the initial conditions obtained from our approach and that of \cite{Shen:2017bsr,Denicol:2018wdp}, and on the dynamical effects of baryon number diffusion which were studied in \cite{Denicol:2018wdp} but not in \cite{Akamatsu:2018olk}.

\begin{table*}[!tb]
\centering
\caption{Some dynamical initialization models for hydrodynamics at low energies, where different smearing kernels are employed.}
\label{tab:dynamical}
\begin{tabular}{lll} 
 \hline\hline
 References & Dynamics & Smearing kernel \\ 
 \hline
 Kajantie {\it et al.}, 1982 \cite{KAJANTIE1982203,KAJANTIE1983261,KAJANTIE1983152} & Collision geometry & $\delta$-function\\
 Naboka {\it et al.}, 2015 \cite{Naboka:2014eha} & Relaxation model of $T^{\mu\nu}$ & Gaussian with $R_x$ and $R_y$ \\
 Shen {\it et al.}, 2018 \cite{Shen:2017bsr, Shen:2017ruz} & MC-Glauber + string & Gaussian with $\sigma_\perp$ and $\sigma_\eta$ \\
 Du {\it et al.}, 2018 \cite{Du:2018mpf} & Modified {\sc UrQMD} & Lorentz invariant kernel \\
 Akamatsu {\it et al.}, 2018 \cite{Akamatsu:2018olk} & JAM & Lorentz invariant kernel \\
 Kanakubo {\it et al.}, 2018 \cite{Kanakubo:2018vkl,Kanakubo:2019ogh} & PYTHIA & Gaussian with $\sigma_\perp$ and $\sigma_\eta$ \\
 \hline\hline
\end{tabular}
\end{table*}

\subsection{Dynamical sources}
\label{sec2}

During the interpenetration stage of the two nuclei, we describe the medium created in the collision as a superposition of freshly produced, still un-thermalized particles and an approximately thermalized dissipative fluid. In the conservation laws for energy, momentum and net baryon number, hydrodynamic source currents describe the conversion of particles into fluid via thermalization:
\begin{equation}
\label{eq1}
   d_\mu T^{\mu\nu}_{\mathrm{fluid}} 
   = J_{\mathrm{source}}^\nu(x) \equiv -d_\mu T^{\mu\nu}_{\mathrm{particle}}(x) \;,\quad 
   d_\mu N^\mu_{\mathrm{fluid}} = \rho_{\mathrm{B,}{\textrm{source}}}(x)  \equiv -d_\mu N^\mu_{\mathrm{particle}}(x)\;,
\end{equation}
where $d_\mu$ stands for the covariant derivative (see Eqs.~(\ref{hydro_eqs_T},\ref{hydro_eqs_n}) below). The particle contributions on the right hand side are obtained from {\sc UrQMD} \cite{Bass:1998ca,Bleicher:1999xi}, a kinetic model based on hadronic degrees of freedom that describes the initial collision stage in terms of the decay of strings and resonances created in the primary collisions between nucleons as the colliding nuclei interpenetrate each other. With the exception of leading baryons carrying at least one of the incoming valence quarks, particles produced in these decays are not allowed to rescatter but assumed to become part of the fluid after free-streaming for a formation time $\tau_{\mathrm{form}}$ which encapsulates in a single, species-independent number both their formation and thermalization, in their own rest frame. Leading baryons are allowed to scatter multiple times if the secondary collision occurs within their formation time, until the nuclei have completely passed through each other; then they, too, become part of the fluid.

The energy-momentum tensor and net baryon current of the particles thus produced are given by \cite{Oliinychenko:2015lva}
\begin{equation}
\label{eq:cartesians}
   T^{\mu\nu}_{\mathrm{particle}}(t, \br) 
   = \sum_i \frac{p_i^\mu p_i^\nu}{p_i^0} K(\br{-}\br_i(t), \bp_i)\,\Theta_i\;, \quad
   N^\mu_{\textrm{particle}}(t, \br) 
   = \sum_i b_i \frac{p_i^\mu}{p_i^0} K(\br{-}\br_i(t), \bp_i)\,\Theta_i\;,
\end{equation}
where $p_i^0{\,=}\sqrt{m_i^2{\,+\,}\bp_i^2}$, $\br_i(t)=\br_{i0}+\bigl(\bp_i/p_i^0\bigr)(t{\,-\,}t_{i0})$ is the free-streaming trajectory of a particle produced at space-time point $x_{i0}^\mu=(t_{i0},\br_{i0})$. Here $K(\br{-}\br_i(t),\bp_i)$ and $\Theta_i{\,\equiv\,}\Theta(t_{\mathrm{form},i}{\,-\,}t_{\mathrm{rf}})$ are a spatial smearing kernel, assumed to be Gaussian, and a step-like temporal switching function in the rest frame (rf) of the particle, respectively:
\begin{equation}\label{eq:cartesiank}
   K(\br,\bp_i) = \frac{\gamma_i}{\left(2\pi\sigma^2\right)^{3/2}}
   \exp\left(-\frac{\br^2 + (\br\cdot\bu_i)^2}{2\sigma^2}\right),\qquad	
   \Theta(t_{\textrm{form}}{\,-\,}t)
   =\frac{1}{2}\left[\tanh\left(\frac{t_{\textrm{form}}-t}{\Delta \tau_{\textrm{th}}}\right)+1\right].
\end{equation}
In the first expression $\gamma_i=1/\sqrt{1{-}\bp_i^2/(p_i^0)^2}$ and $\bu_i = \bp_i/m_i$ are the Lorentz-contraction factor and spatial part of the four-velocity of a particle with mass $m_i$ and momentum $\bp_i$ in the lab frame. The temporal switching function describes the disappearance and absorption by the fluid of particle $i$ around time $t_\mathrm{rf}=t_{\mathrm{form},i}=t_{i0,\mathrm{rf}}+\tau_\mathrm{form}$ in the particle rest frame; it approaches a step function $\theta(t_{\textrm{form}}{-}t)$ when $\Delta\tau_{\mathrm{th}}\to 0$. We here use a non-zero $\Delta\tau_{\mathrm{th}}{\,=\,}0.5$\,fm/$c$ to avoid large dissipative effects arising from large temporal gradients of the hydrodynamic source terms in Eq.~(\ref{eq1}).

%
\begin{figure}[!t]
\centering
\includegraphics[width=0.45\linewidth]{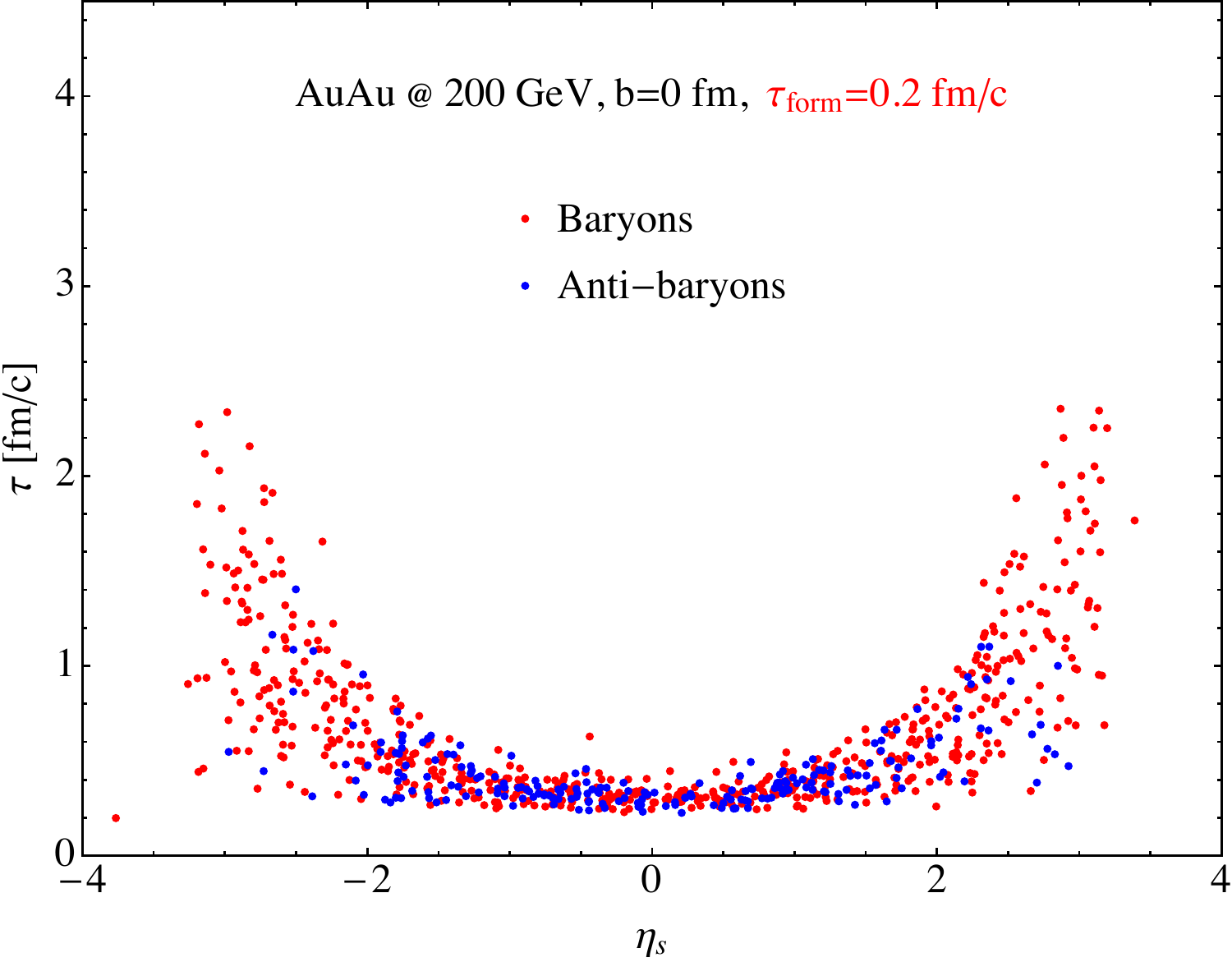}\quad 
\includegraphics[width=0.46\linewidth]{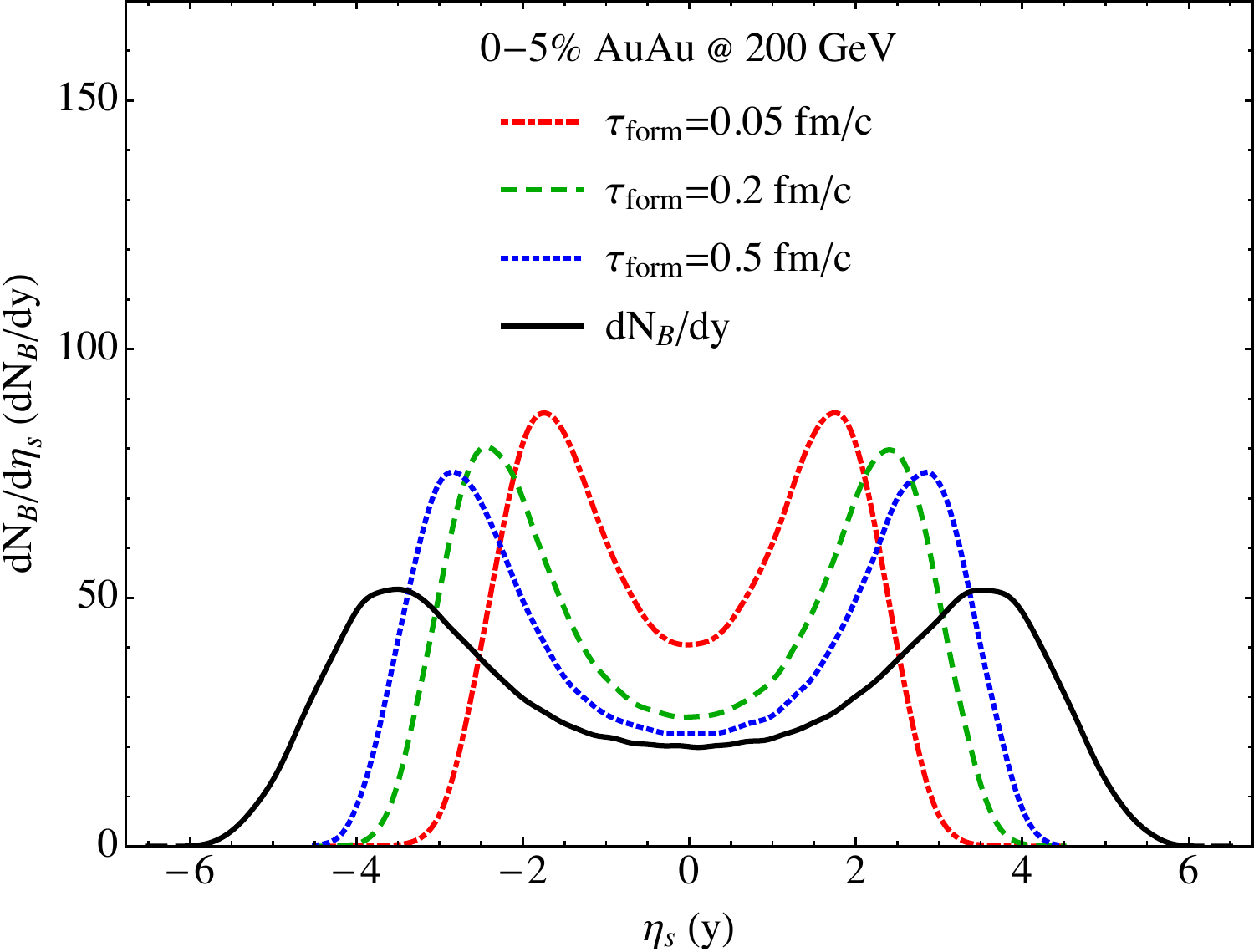}\par 
\caption{{\sl Left:} $\tau{-}\eta_s$ distribution of the produced particles at formation time (mesons not 
	     shown). 
	     {\sl Right:} Space-time rapidity distribution (dashed lines) and rapidity distribution (solid line) of 
	     net baryon number from the pre-equilibrium stage, for different formation times.}
\label{fig:partscatter}
\end{figure}%
%

The left panel of Fig.~\ref{fig:partscatter} shows a scatter plot in the $(\tau,\eta_s)$ plane of the baryons and antibaryons created by {\sc UrQMD} at their thermalization time $t_{\mathrm{form},i}$, using $\tau_\mathrm{form}=0.2$\,fm/$c$. It shares qualitative features with Fig.~4b in Ref.~\cite{Shen:2017bsr}. In the right panel, the green dashed line converts this information into an initial space-time-rapidity ($\eta_s$) distribution for the hydrodynamic evolution. One observes a large difference between the initial rapidity ($y$, black line) and space-time-rapidity ($\eta_s$) distributions (colored lines), and this difference depends sensitively on the formation time $\tau_\mathrm{form}$. In the dynamical string-fragmentation model of Ref.~\cite{Shen:2017bsr} the initial $y$ and $\eta_s$ distributions for net baryons are much closer to each other, both at $\snn=200$ and 19.6\,$A$\,GeV (see Fig.~7 in \cite{Shen:2017bsr}). In our model, at 19.6\,$A$\, GeV the two peaks near projectile and target rapidities merge into a single peak around midrapidity, for both the $y$ and $\eta_s$ distributions, with the width of the $\eta_s$ distribution depending strongly on the choice of $\tau_\mathrm{form}$ but being generically much smaller than that of the rapidity distribution. -- Different initial net-baryon $\eta_s$ distributions correspond to different initial space-time distributions of the baryon chemical potential $\mu/T$ whose gradients drive the baryon diffusion current. How the final baryon momentum distributions are affected by the ensuing differences in hydrodynamic evolution is an interesting question.

%
\begin{figure}[!htbp]
\centering
\includegraphics[width= 0.8\linewidth]{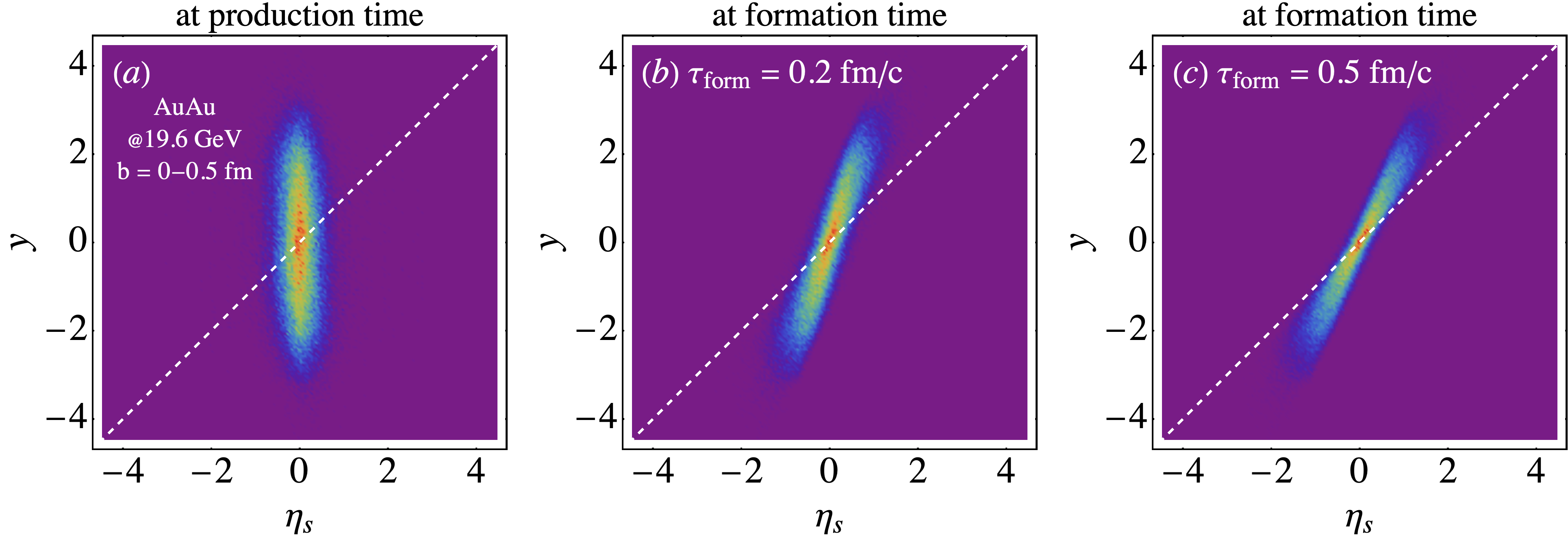}\\
\includegraphics[width= 0.8\linewidth]{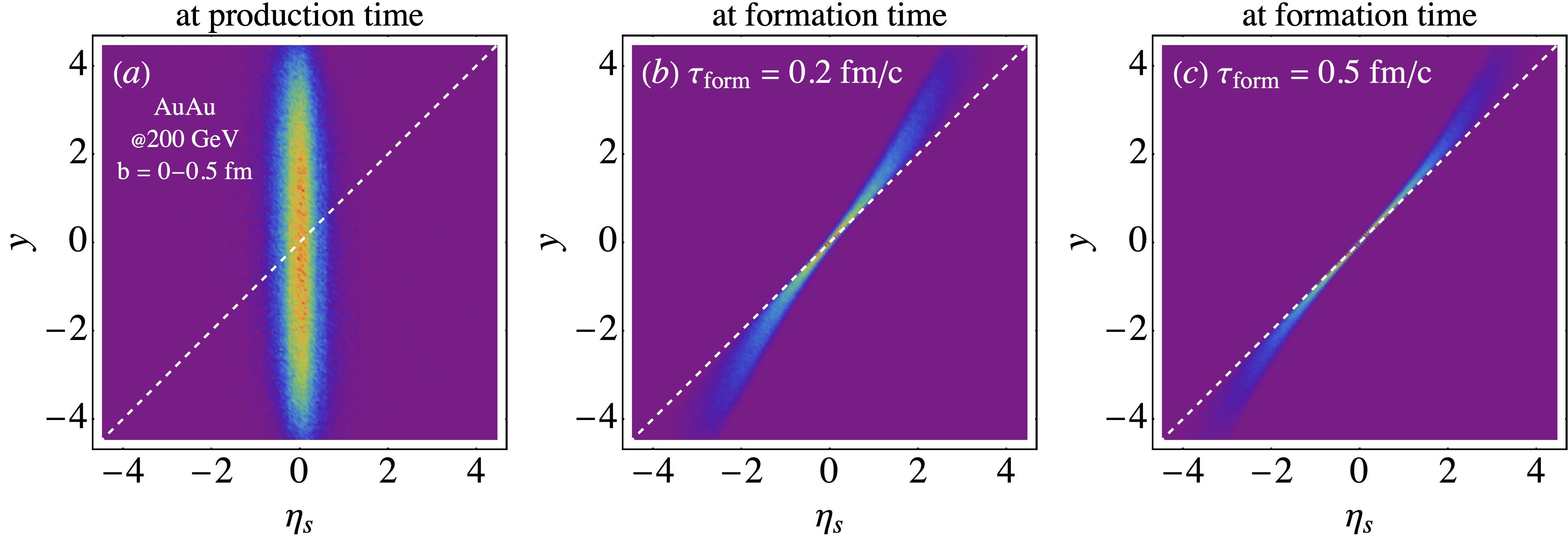}\par 
\caption{Density distributions in $\eta_s$-$y$ plane for particles produced in central Au-Au collisions at $\snn=19.6\,$ GeV (upper panels) and 200\,GeV (lower panels) from the Modified-{\sc UrQMD} model. The left most column shows the distributions of particles at production time, while the right two columns are for particles at two different formation times, with $\tau_\mathrm{form}=0.2\,$fm$/c$ (middle) and 0.5\,fm$/c$ (right), respectively.
}
\label{fig:yeta_dist_all}
\end{figure}%
%

%
\begin{figure}[!htbp]
\centering
\includegraphics[width= 0.8\linewidth]{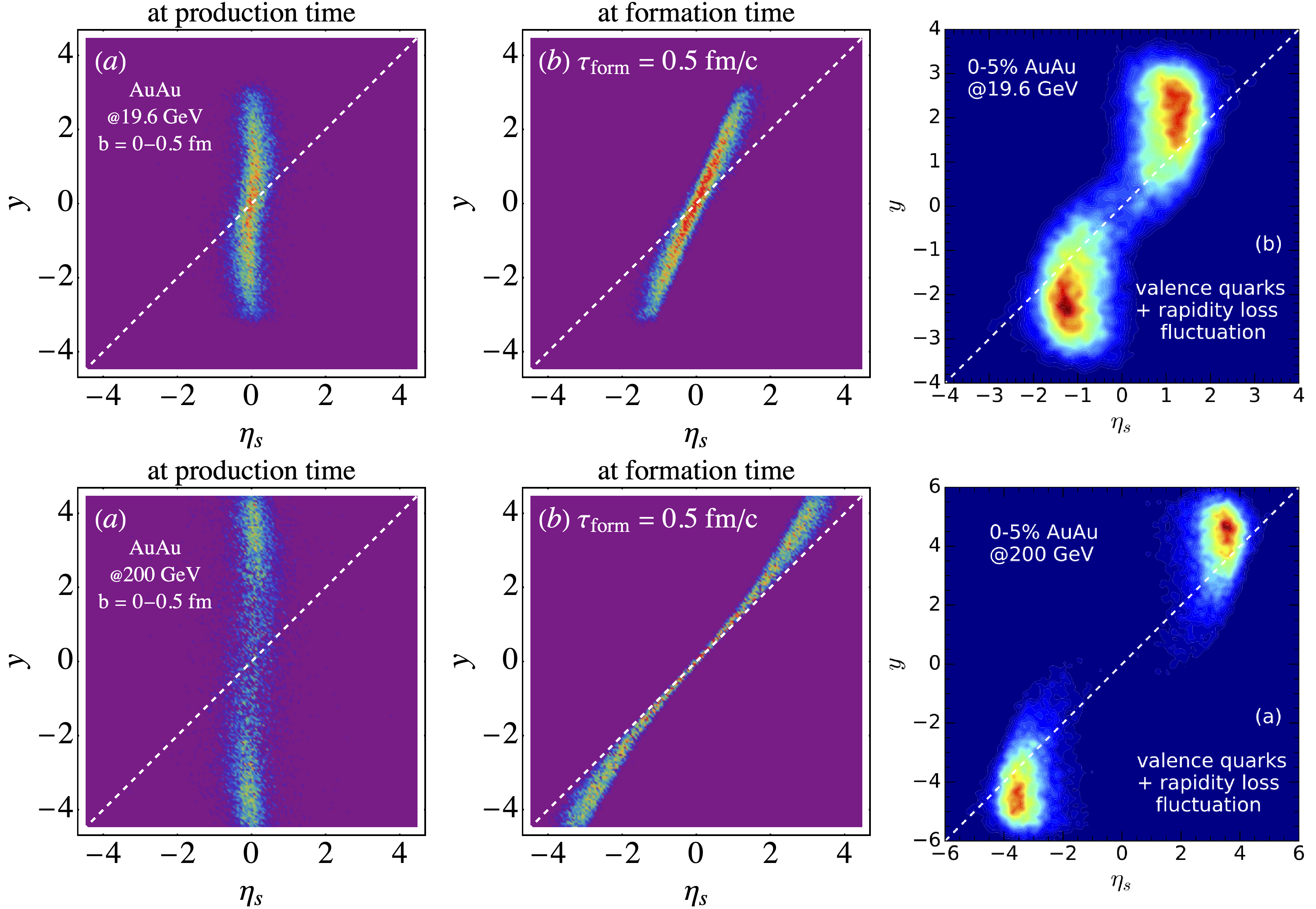}\par 
\caption{Similar to Fig.~\ref{fig:yeta_dist_all}, but for baryon number distributions in $\eta_s$-$y$ plane at the two collision energies. The left two columns are again from the Modified-{\sc UrQMD} model, where the formation time is chosen as $\tau_\mathrm{form}=0.5\,$fm$/c$ (middle column). The right most column is from the dynamical string-fragmentation model \cite{Shen:2017bsr} for comparison.
}
\label{fig:yeta_baryon}
\end{figure}%
%

We further illustrate this point by plotting in Fig.~\ref{fig:yeta_dist_all} the particle density distributions in $\eta_s$-$y$ plane for central Au-Au collisions at $\snn=19.6\,$ GeV (upper panels) and 200\,GeV (lower panels) from the Modified-{\sc UrQMD} model described above. These particle distributions should be a reflection of the corresponding distribution of the initial energy deposition. From the left column which shows particle distributions at production time, we can see that the initial energy deposition strongly breaks the boost-invariance at both top and low  RHIC energies. From the right two columns, where the particle distributions are shown at their formation time during which the particles stream  freely, we see that the initial distribution evolves towards boost-invariance by longitudinal free-streaming. Consequently, when the formation time increases the initial distribution approaches boost-invariance, and this evolution is faster at higher collision energies, as one sees by comparing the upper panels and the lower ones. Finally, we also show the $\eta_s$-$y$ distributions for baryon number in Fig.~\ref{fig:yeta_baryon}, which provides a different perspective for Fig.~\ref{fig:partscatter}. Baryon distributions share the features observed in Fig.~\ref{fig:yeta_dist_all}, while the two peaks do not emerge at low beam energy after formation time, which is different from the dynamical string-fragmentation model (i.e., the MC-Glauber\,+\,String model \cite{Shen:2017bsr, Shen:2017ruz} listed in Table~\ref{tab:dynamical}). From the discussion, we can see that there is large uncertainty in hydrodynamic initial conditions related to the formation time.

\begin{figure}[!tb]
\centering
\includegraphics[width=0.35\linewidth]{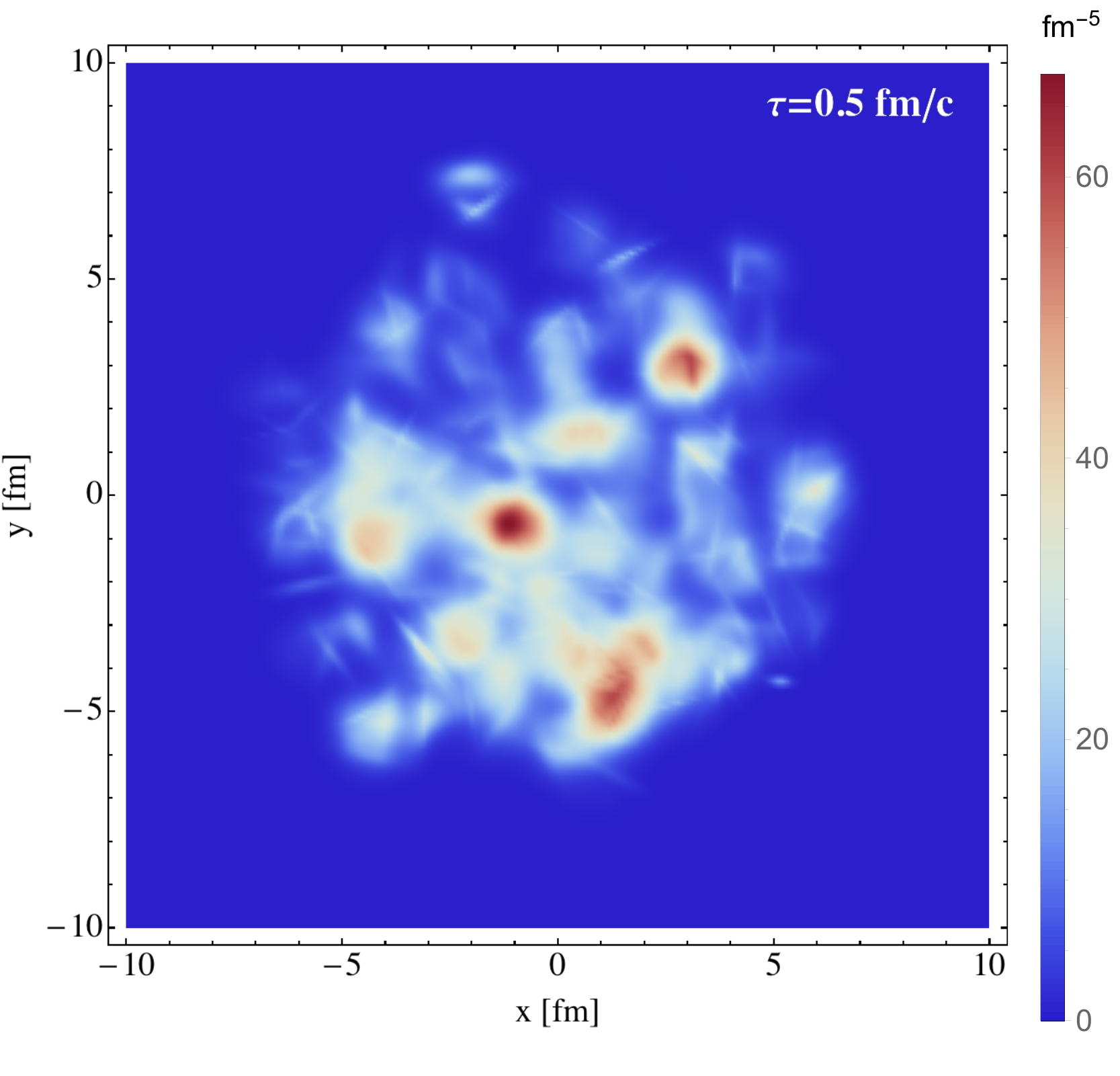}\quad \quad
\includegraphics[width=0.35\linewidth]{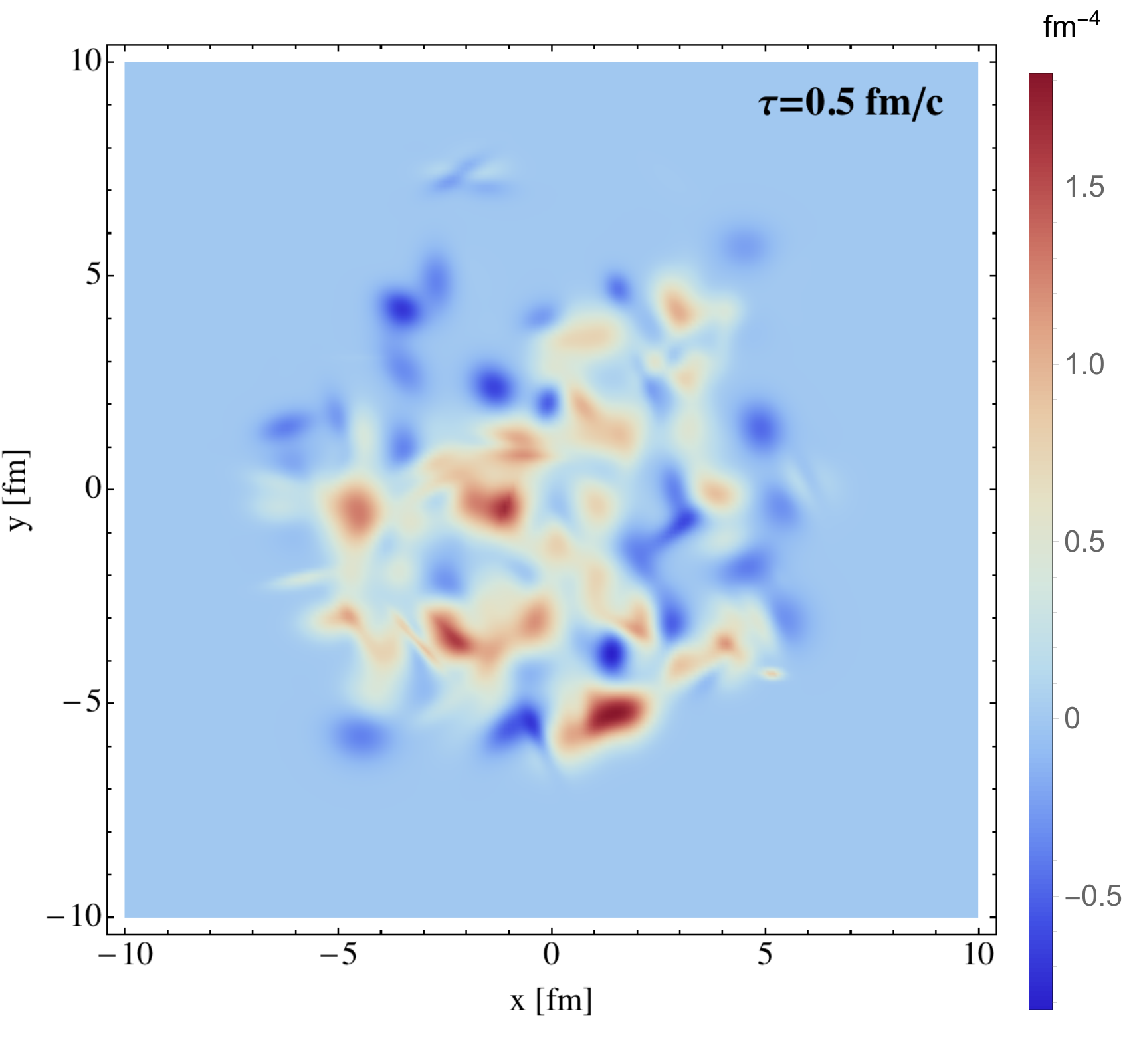}\par 
\vspace{0.2cm}
\hspace{-0.4cm}\includegraphics[width=0.35\linewidth]{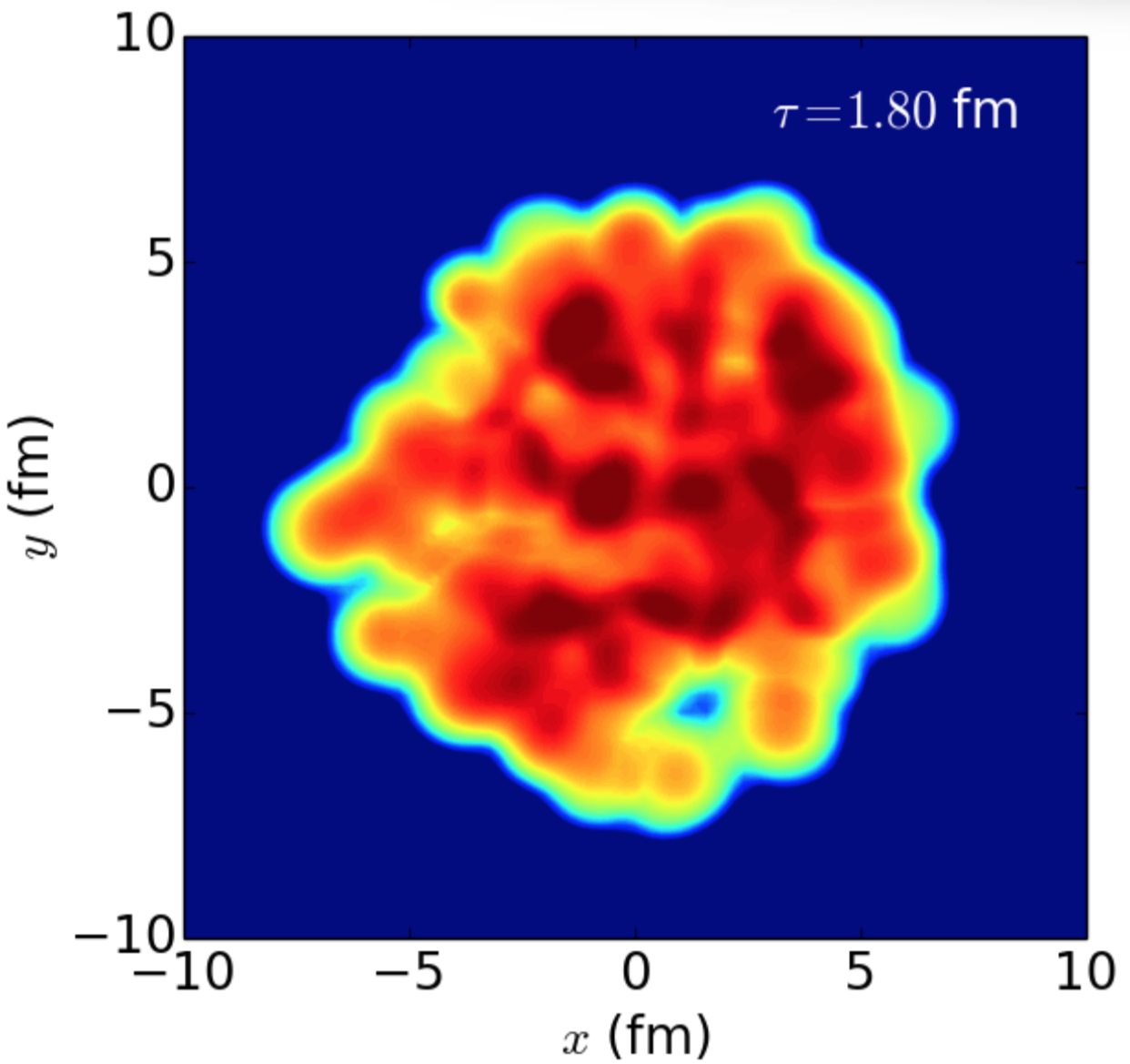}\quad \quad 
\includegraphics[width=0.35\linewidth]{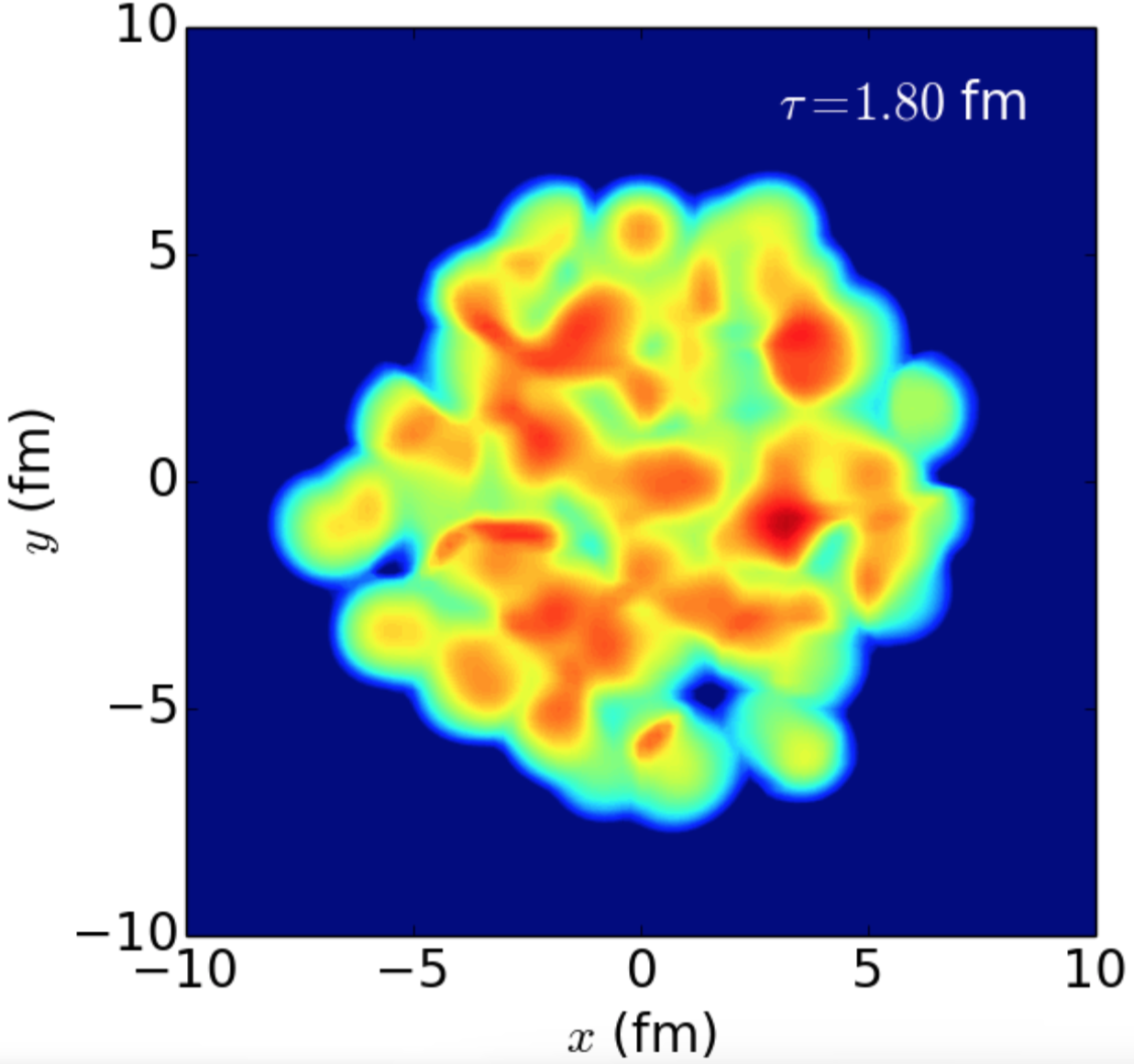}\par 
\caption{{\sl Upper panels:} Transverse density distributions of (left) $J^\tau_{\mathrm{source}}$ and (right) $\rho_{\mathrm{B,}{\textrm{source}}}$ defined in Eq.~\eqref{eq1} calculated using Eqs.~\eqref{eq:cartesians} and \eqref{eq:cartesiank}.
	     {\sl Lower panels:} Corresponding distributions of energy and baryon densities from Ref.~\cite{Shen:2017bsr}.}
\label{fig:trans_compare}
\end{figure}

We also note that, different from Ref.~\cite{Shen:2017bsr} where the fluctuations in the transverse plane of net baryon and energy densities are correlated with each other and across rapidities by their common string breaking origin, such correlations are not visible in our model, due to the effects of individual transverse and longitudinal motion of the produced particles before becoming part of the fluid (see Fig.~\ref{fig:trans_compare}). 

%
%

As illustrated above, because of the violation of boost-invariance, at low beam energies the space-time distribution of the initial condition is decoupled from the energy-momentum distribution. Another complication comes from the smearing kernel used to construct the dynamical sources from partons or strings. Above we use a smearing kernel in Cartesian coordinates which has Lorentz contraction, and then convert the obtained source terms into Milne coordinates for hydrodynamic evolution \cite{du2020ds}. However, since most of the hydrodynamic codes are written in Milne coordinates, other authors often employ smearing kernels that are Gaussian distributions (without Lorentz contraction) in this coordinate system. The choice of smearing kernel was not an issue for constructing initial conditions at ultra-relativistic collisions, where boost-invariant Bjorken expansion is generally assumed intially. However, in the case of constructing time-dependent initial conditions which are non-trivial in longitudinal direction as well, the choice of smearing kernel can be of phenomenological relevance. 

For example, a 3D initial condition at a fixed proper time $\tau_0$ is constructed from the AMPT model as follows \cite{Pang:2012he}:
\begin{equation}
  T^{\mu\nu} (\tau_{0},x,y,\eta_{s}) = K\sum_{i}
  \frac{p^{\mu}_{i}p^{\nu}_{i}}{p^{\tau}_{i}}\frac{1}{\tau_{0}\sqrt{2\pi\sigma_{\eta_{s}}^{2}}}\frac{1}{2\pi\sigma_{r}^{2}}  \exp\left[-\frac{(x-x_{i})^{2}+(y-y_{i})^{2}}{2\sigma_{r}^{2}} - \frac{(\eta_{s}-\eta_{i s})^{2}}{2\sigma_{\eta_{s}}^{2}}\right],
  \label{eq-tmunu-pang}
\end{equation}
where $K$ is a normalization factor, $\sigma_{r}$ and $\sigma_{\eta_{s}}$ the smearing widths in the transverse plane and longitudinal direction, respectively. Here $p^\mu_i=(p^{\tau}_{i},\,p^x,\,p^y,\,p^\eta)$ is the four-momentum of a particle $i$ produced from AMPT, and $(x_i,\,y_i,\,\eta_i)$ its spatial coordinates. Note that here $\mu=(\tau, x, y, \eta_s)$ are the four components of the Milne coordinates. Similar smearing kernels are used for most of such studies, when initial conditions are constructed, where Gaussian distributions are assumed for both transverse and longitudinal directions, without Lorentz contraction. We find this less motivated than our ansatz (\ref{eq:cartesians},\ref{eq:cartesiank}) since the processes that lead to the initially produced particles becoming part of the fluid happen in the particle rest frame, not in the global frame.

\begin{figure}[!tb]
        \centering
        \includegraphics[width= 0.47\textwidth]{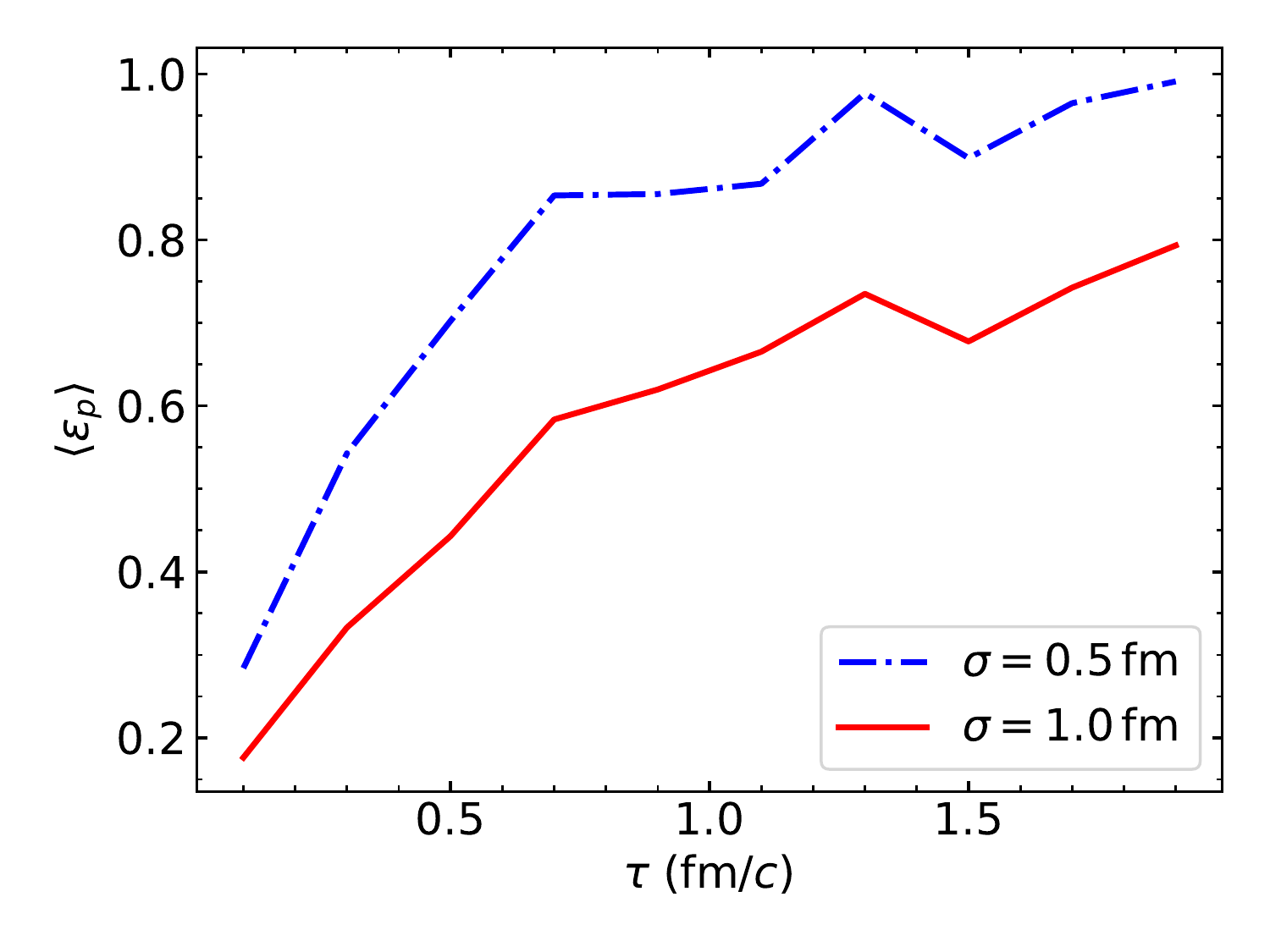}
        \includegraphics[width= 0.47\textwidth]{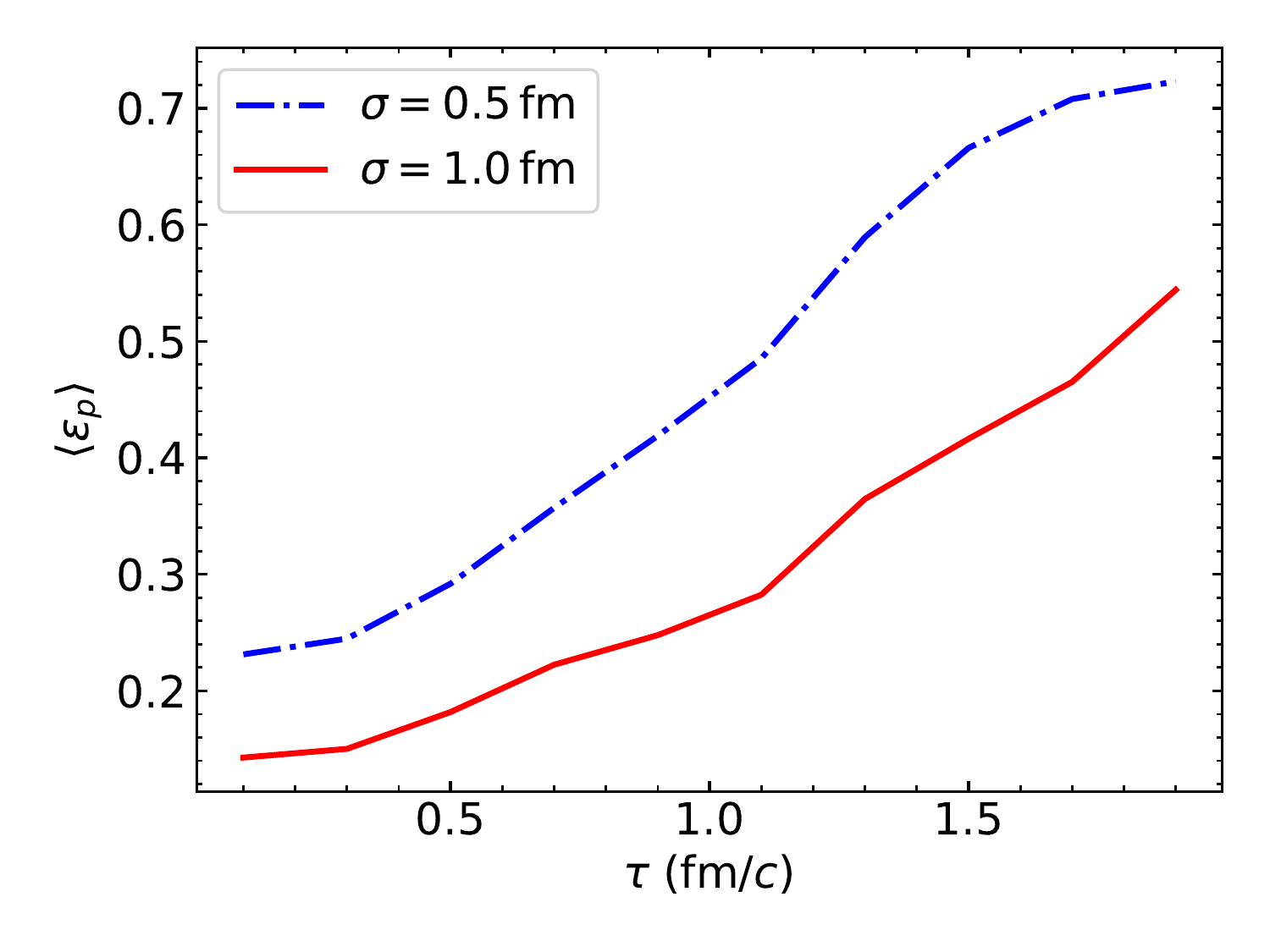}
        \caption{Comparison of transverse momentum anisotropy at $\eta_s=1.5$ with a smearing kernel in Cartesian (left) and Milne (right) coordinates, where all smearing widths are set to be the same, i.e., $\sigma$ in the plot.}\label{fig:anisop_comp}
\end{figure}

To compare the kernel in Eq.~\eqref{eq-tmunu-pang}  and the one in Eqs.~\eqref{eq:cartesians} and \eqref{eq:cartesiank}, we calculate $T^{\mu\nu} (\tau,x,y,\eta_{s})$ with the same particle list, and then compute the corresponding transverse momentum anisotropy, defined as
\begin{equation}
	\langle\epsilon_p \rangle=\sqrt{\frac{\langle T^{xx}-T^{yy} \rangle^2+\langle 2T^{xy} \rangle^2}{\langle T^{xx}+T^{yy}  \rangle^2}}\,,
\end{equation}
where the averaged is taken over the entire transverse plane ($K=1$ in Eq.~\eqref{eq-tmunu-pang} for this comparison). The evolution of  $\langle\epsilon_p \rangle$ at  $\eta_s=1.5$ for these two different smearing kernels are shown in Fig.~\ref{fig:anisop_comp}, where both quantitative and qualitative differences can be observed between these two cases.\footnote{%
    The code we developed to compute the dynamical sources can be downloaded from \url{https://github.com/LipeiDu/part2s}, which is called {\sc part2s}, whose name is from its purpose to convert PARTticles TO Source terms.}
Apparently, when the smearing kernel with Lorentz contraction is employed, the momentum anisotropy generated at early times is larger, and more fluctuations can be seen. These differences can result in different phenomenological results, such as anisotropic flows, and thus can affect the values of transport properties (e.g., shear viscosity) that one extracts from model-data comparisons.

\section{Relativistic hydrodynamics}\label{sec:relat_hydro}
%
%

Over the last decade, second-order dissipative relativistic fluid dynamics (RFD) \cite{ISRAEL1976310, rspa.1977.0155, Israel:1979wp} has developed into a powerful and phenomenologically very successful tool for the description of the dynamical evolution of the hot and dense matter created in relativistic heavy-ion collisions \cite{Heinz:2005bw, Song:2008si, Song:2007fn, Romatschke:2007mq, Schenke:2010nt, Gale:2012rq, Martinez:2010sc, Florkowski:2010cf, Martinez:2012tu, Florkowski:2014bba, Karpenko:2013wva, Molnar:2009tx, Denicol:2012cn}. The initial development of dissipative RFD ignored the evolution of conserved currents such as net baryon number and strangeness because the community's attention was focused on experiments performed at the highest available collision energies at RHIC at Brookhaven National Laboratory and the LHC at CERN, at which the colliding atomic nuclei are largely transparent to each other, creating a system of approximately zero net baryon number, strangeness, and isospin charge near midrapidity in the center of mass frame. Only with the BES program at RHIC \cite{Aggarwal:2010cw}, in which heavy-ion collisions were studied at lower collision energies where some of the incoming baryon charge gets stopped near midrapidity, became the need urgent for including the dynamics of the baryon number and other conserved charge currents in the hydrodynamic description. These developments are also relevant for the theoretical description of future experiments at NICA \cite{Sissakian:2009zza} and FAIR \cite{SPILLER2006305,CHATTOPADHYAY2014267} and some other facilities mentioned in Sec.~\ref{sec:phase_diagram}.

In this section, we review the hydrodynamic evolution of systems with non-zero conserved charges (such as baryon number). We develop  \code{} \cite{Du:2019obx} to solve these equations and shall provide a detailed description of its structure and performance in Ch.~\ref{ch:numerics}.

\subsection{Conservation laws}

Hydrodynamics is a macroscopic theory describing the space-time evolution of the 14 components of the energy-momentum tensor $T^{\mu\nu}(x)$ and the net (baryon) charge current $N^\mu(x)$.\footnote{%
     For $N_C>1$ conserved charges, the number of evolved quantities increases to $10{\,+\,}4N_C$. We note that recently some efforts have been spent on the hydrodynamic description of multiple charges, including baryon charge, electric charge and strangeness (see Refs.~\cite{Greif:2017byw, Fotakis:2019nbq, Rose:2020sjv}). See more discussion in Sec.~\ref{sec:coef_multi} on the transport coefficients and Sec.~\ref{sec:eos_multi} on Equation of State involving multiple conserved charges.}
Five evolution equations arise from the conservation laws for energy, momentum, and the baryon charge \cite{Molnar:2009tx}:
\begin{eqnarray}
d_{\mu}T^{\mu\nu} &\equiv&\frac{1}{\sqrt{g}}\partial_{\mu}(\sqrt{g}T^{\mu\nu})+\Gamma^{\nu}_{\mu\lambda}T^{\mu\lambda}=0\;,
\label{hydro_eqs_T}\\
d_{\mu}N^{\mu} &\equiv&\frac{1}{\sqrt{g}}\partial_{\mu}(\sqrt{g}N^{\mu})=0\;.
\label{hydro_eqs_n}
\end{eqnarray}
Here $d_\mu$ ($\mu=0,1,2,3$) stands for the covariant derivative in a general system of space-time coordinates, with metric tensor $g^{\mu\nu}$ defined with negative signature (``mostly minus'' convention $(+,-,-,-)$), $g\equiv-\det{(g_{\mu\nu})}$, and the Christoffel symbols (see, e.g., \cite{carroll2004spacetime})
\begin{equation}
    \Gamma^{\mu}_{\alpha\beta}\equiv \frac{1}{2} g^{\mu\nu} \bigl(\partial_{\beta}g_{\alpha\nu}
         +\partial_{\alpha}g_{\nu\beta}
         -\partial_{\nu}g_{\alpha\beta}\bigr)
    =\Gamma^{\mu}_{\beta\alpha}\;.
\end{equation}

The 14 independent components of $T^{\mu\nu}$ and $N^\mu$ are more physically defined in terms of the hydrodynamic decomposition of these tensors \cite{landau2013fluid},
\begin{eqnarray}
    T^{\mu\nu} &=& \ed u^{\mu}u^{\nu}-(\peq+\Pi)\Delta^{\mu\nu}+\pi^{\mu\nu}\;,\label{eq-dec-T}\\
    N^{\mu} &=& \n u^{\mu}+\V^{\mu}\;. \label{eq-dec-N}
\end{eqnarray}
Here the flow 4-velocity $u^\mu(x)$, with $u^\mu u_\mu=1$, is defined as the time-like eigenvector of the energy-momentum tensor,
\begin{equation}
    T^{\mu\nu}u_\nu=\ed u^\mu\;,
\end{equation}
and specifies the local rest frame (LRF) of the fluid at point $x$ (the so-called ``Landau frame''). The tensors $u^{\mu}u^{\nu}$ and $\Delta^{\mu\nu} \equiv g^{\mu\nu} - u^{\mu}u^{\nu}$ then are projectors on the temporal and spatial directions in the LRF. $\ed$ and $\n$ are the energy and net baryon density in the LRF which can be obtained as the following projections of $T^{\mu\nu}$ and $N^\mu$:
\begin{equation}
    \ed = u_\mu T^{\mu\nu} u_\nu\;,\qquad \n = u_\mu N^{\mu}\;.\label{eq-landau}
\end{equation}
From these, the local equilibrium pressure $\peq$ is obtained through the EoS $\peq=\peq(\ed,\n)$. The shear stress $\pi^{\mu\nu}$, the bulk viscous pressure $\Pi$, and the baryon diffusion current $n^\mu$ are dissipative flows describing deviations from local equilibrium.

Using the decomposition (\ref{eq-dec-T},\ref{eq-dec-N}) the conservation laws (\ref{hydro_eqs_T},\ref{hydro_eqs_n}) can be brought into the physically intuitive form \cite{Jeon:2015dfa} 
\begin{eqnarray}
 D\n &=& -\n\theta -\nabla_\mu n^\mu\;,
\label{eq:vhydro-N}
\\
 D\ed &=& -(\ed{+}\peq{+}\Pi)\theta + \pi_{\mu\nu}\sigma^{\mu\nu}\;,
\label{eq:vhydro-E}
\\
  (\ed{+}\peq{+}\Pi)\, Du^\mu &=& \nabla^\mu(\peq{+}\Pi)
    - \Delta^{\mu\nu} \nabla^\sigma\pi_{\nu\sigma} + \pi^{\mu\nu} Du_\nu\;.
\label{eq:vhydro-u}
\end{eqnarray}
Here $D=u_\mu d^\mu$ denotes the time derivative in the LRF, $\theta=d_\mu u^\mu$ is the scalar expansion rate, $\nabla^\mu=\partial^{\langle\mu\rangle}$ (where generally $A^{\langle\mu\rangle} \equiv \Delta^{\mu\nu}A_\nu$) denotes the spatial gradient in the LRF, and $\sigma^{\mu\nu} = \nabla^{\langle\mu}u^{\nu\rangle}$ (where generally $B^{\langle\mu\nu\rangle} \equiv \Delta^{\mu\nu}_{\alpha\beta} B^{\alpha\beta}$, with the traceless spatial projector $\Delta^{\mu\nu}_{\alpha\beta} \equiv \frac{1}{2} (\Delta^\mu_\alpha \Delta^\nu_\beta + \Delta^\nu_\alpha \Delta^\mu_\beta) - \frac{1}{3} \Delta^{\mu\nu} \Delta_{\alpha\beta}$) is the shear flow tensor. While these equations clearly exhibit the physics in the LRF (which varies from point to point), \code\ solves the conservation laws (\ref{hydro_eqs_T},\ref{hydro_eqs_n}) in a fixed global computational frame. Their explicit form in the global frame is discussed in Sec.~\ref{sec2.1.3}.

The 5 conservation laws (\ref{eq:vhydro-N}-\ref{eq:vhydro-u}) are sufficient to determine the energy and baryon density, $\ed$ and $\n$, together with the 3 independent components of the flow velocity $u^\mu$, as long as the shear stress $\pi^{\mu\nu}$, the bulk viscous pressure $\Pi$, and the baryon diffusion current $n^\mu$ vanish.\footnote{\label{fn2}%
     Note that the shear stress is traceless, $\pi^\mu_\mu=0$, and both $\pi^{\mu\nu}$ and $n^\mu$ have only spatial components in the LRF, $u_\mu \pi^{\mu\nu}=\pi^{\mu\nu} u_\nu = u_\mu n^\mu =0$. $\pi^{\mu\nu}$, $\Pi$, and $n^\mu$ thus describe 5+1+3=9 dissipative degrees of freedom.}
Their evolution is not directly constrained by conservation laws but controlled by the competition between microscopic scattering processes (which drive the system towards local equilibrium and the dissipative flows to zero) and the macroscopic expansion (which drives the system away from equilibrium and the dissipative flows away from zero). Their evolution is thus controlled by both micro- and macroscopic physics. One way to obtain their evolution equations is DNMR theory \cite{Denicol:2012cn, Denicol:2010xn, Molnar:2009tx} which uses the method of moments of the Boltzmann equation and which we employ here. 

Expressed through its natural variables, i.e.\ the temperature $T$ and baryon chemical potential $\mu$, the equilibrium pressure $\peq(\ed,\n)=\peq(T,\mu)$ is recognized as the grand-canonical thermodynamic potential for a system with temperature $T(\ed,\n)$ and chemical potential $\mu(\ed,\n)$. In principle, $T$ and $\mu$ are not needed for the hydrodynamic evolution, but they may be required to compute certain signatures of the evolving fluid (such as the spectrum of electromagnetic radiation emitted during its evolution or the spectrum of hadrons into which it decays at the end of the life of the quark-gluon plasma phase), and in \code\ the driving force for net baryon number diffusion is formulated in terms of the gradient of $\mu/T$ rather than that of the net baryon density. Also, the transport coefficients controlling the evolution of the dissipative flows are most naturally expressed as functions of $T$ and $\mu$ since they are defined as response functions of the thermal equilibrium system described by the potential $\peq(T,\mu)$. Different versions of the EoS $\peq(T,\mu)$ or $\peq(\ed,\n)$ used in \code\ will be described in Sec.~\ref{subsec-eos}.

\subsection{Evolution equations for the dissipative flows}
\label{sec2.1.2}

In \code\ the dissipative flows are evolved with DNMR theory \cite{Molnar:2009tx, Denicol:2010xn, Denicol:2012cn}. While the equations of motion in this theory are derived from the Boltzmann equation which is applicable only to weakly coupled systems \cite{Arnold:2002zm}, the hydrodynamic description is an effective theory which is generic and applicable also in the strong coupling regime where the Boltzmann equation is not valid \cite{Baier:2007ix}. When applying the DNMR equations to the fluid produced in nuclear collisions, which appears to be strongly coupled, one must replace its material properties, i.e. the EoS and transport coefficients, by those for real QCD matter.

\begin{figure}[!htbp]
\begin{center}
\includegraphics[width= \textwidth]{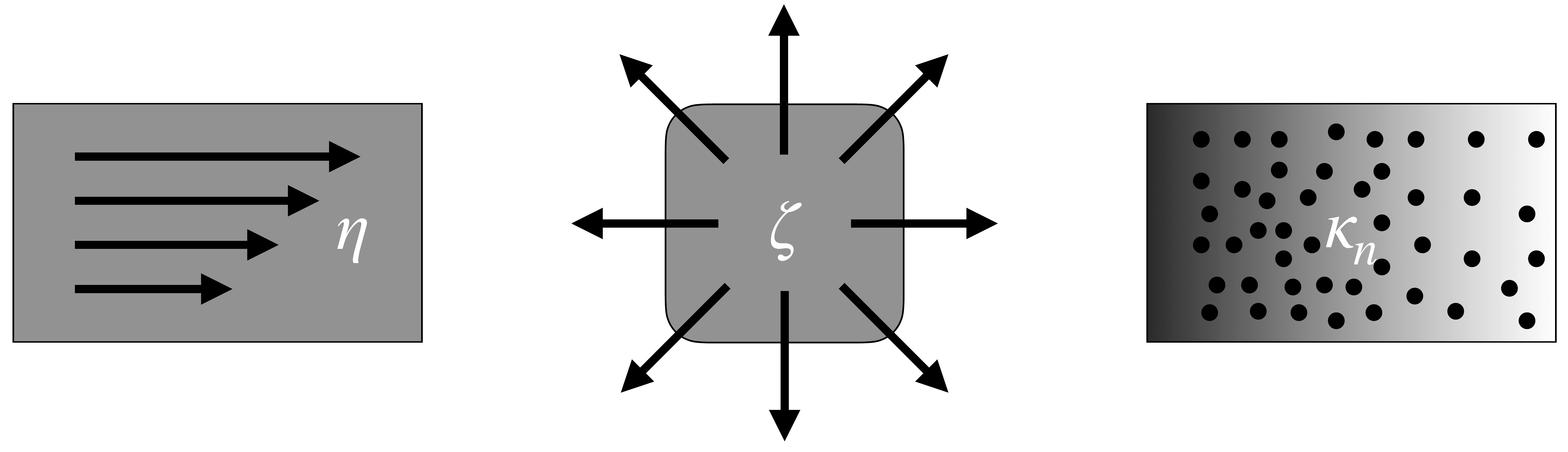}
\caption{Illustration of transport coefficients associated with three viscous effects. {\sl Left:} Shear viscosity $\eta$ measures the resistance to flow gradients and reduces the buildup of flow anisotropy. {\sl Middle:} Bulk viscosity $\zeta$ measures the resistance to isotropic compression or expansion and reduces the buildup of radial flow. {\sl Right:} Baryon diffusion coefficient $\kappa_n$ measures the strength of the response of baryon diffusion current to the gradient of the ratio between chemical potential and temperature.}
\label{fig:visco_illu}
\end{center}
\end{figure}

In the framework of DNMR theory, the dissipative transport equations are given by the following relaxation equations:
\begin{eqnarray}
    \tau_{\Pi }D{\Pi}+\Pi &=& \Pi_{\mathrm{NS}} +\mathcal{J}+\mathcal{K}+\mathcal{R}\;,\label{eq-relax-Pi}\\
    \tau_{n}(Dn)^{\left\langle \mu \right\rangle}+n^{\mu} &=& n^{\mu}_{\mathrm{NS}}+\mathcal{J}^{\mu }+\mathcal{K}^{\mu }+\mathcal{R}^{\mu }\;,\label{eq-relax-n}\\
    \tau _{\pi }(D\pi)^{\left\langle \mu \nu \right\rangle }+\pi ^{\mu \nu }
    &=& \pi^{\mu \nu }_{\mathrm{NS}}+\mathcal{J}^{\mu \nu }+\mathcal{K}^{\mu \nu }+\mathcal{R}^{\mu \nu }\;.\label{eq-relax-pi}
\end{eqnarray}
Here $(Dn)^{\langle\mu\rangle}\equiv \Delta^{\mu\nu} Dn_\nu$ and $(D\pi)^{\langle\mu\nu\rangle} \equiv \Delta^{\mu\nu}_{\alpha\beta} D\pi^{\alpha\beta}$, ensuring that all terms are purely spatial in the LRF and, where applicable, traceless. $\tau _{\Pi }$, $\tau_{n}$, and $\tau _{\pi}$ are the relaxation times for $\Pi$, $\V^\mu$, and $\pi^{\mu\nu}$, respectively. They control how fast the dissipative flows relax to their Navier-Stokes limits \cite{Jeon:2015dfa}:
\begin{eqnarray}
    \Pi_{\mathrm{NS}} &=& -\zeta \theta\;,\label{eq-ns-Pi}\\
    n^{\mu}_{\mathrm{NS}} &=& \kappa_n \nabla^{\mu}\left(\frac{\mu}{T}\right)\;,\label{eq-ns-n}\\
    \pi ^{\mu \nu }_{\mathrm{NS}} &=& 2\eta \sigma ^{\mu \nu }\;,\label{eq-ns-pi}
\end{eqnarray}
where $\zeta$, $\kappa_n$, and $\eta$ are the bulk viscosity, baryon diffusion coefficient, and shear viscosity, respectively, describing the first-order response of the dissipative flows to their driving forces, the (negative of the) scalar expansion rate $\theta$, the spatial gradient of $\mu/T$ in the LRF, $\nabla^{\mu}(\mu/T)$, and the shear flow tensor  $\sigma ^{\mu\nu}$, respectively, which drive the system away from local equilibrium. An illustration of these dissipative effects can be found in Fig.~\ref{fig:visco_illu}.

The scalar, vector and tensor source terms on the r.h.s.~of Eqs.~(\ref{eq-relax-Pi}-\ref{eq-relax-pi}), $\mathcal{J},\; \mathcal{K},\; \mathcal{R},\; \mathcal{J}^{\mu},\; \mathcal{K}^{\mu},\; \mathcal{R}^{\mu},\; \mathcal{J}^{\mu \nu },\; \mathcal{K}^{\mu \nu }$ and $\mathcal{R}^{\mu \nu }$, contain terms of second order in the small parameters Knudsen number (ratio between a characteristic microscopic and macroscopic time or length scale of the fluid) and inverse Reynolds number (ratio between dissipative quantities and local equilibrium values). According to the notation established in \cite{Denicol:2012cn}, the Navier-Stokes and $\mathcal{K}$ terms on the r.h.s.~of Eqs.~(\ref{eq-relax-Pi}-\ref{eq-relax-pi}) are of first and second order in the Knudsen number(s), respectively, the  $\mathcal{J}$ terms are of order Knudsen number times inverse Reynolds number, and the $\mathcal{R}$ terms are of second order in the inverse Reynolds number(s). Their explicit expressions can be found in Ref.~\cite{Denicol:2012cn}. Following the arguments in Ref.~\cite{Bazow:2016yra}, we here include only a subset of the $\mathcal{J}$ terms. As \code\ is ultimately designed for precision studies of relativistic heavy-ion collisions, one should perhaps not put too much blind trust into these arguments and rather check their validity; on the other hand, adding the missing second-order source terms to the code at a later time should be straightforward (even if additional code stability tests may be needed). Future code updates will include additional terms as required by specific applications.

As implemented in the code, the thus simplified relaxation equations for the dissipative flows read
\begin{eqnarray}
    \tau_{\Pi}D\Pi+\Pi &=&
        -\zeta\theta -\delta_{\Pi\Pi}\Pi\theta
        +\lambda_{\Pi\pi}\pi^{\mu\nu}\sigma_{\mu\nu}\;,
\label{eq-Pi-simple}
\\
    \tau _{n} D n^\mu+n^{\mu } &=&\text{ } \kappa_n \nabla^{\mu }\left(\frac{\mu}{T}\right) -\tau_n n_{\nu }\omega ^{\nu \mu }-\delta_{nn}n^{\mu }\theta \nonumber\\
      &-&\lambda _{nn}n_{\nu }\sigma ^{\mu \nu } - \tau_n n^\nu u^\mu D u_\nu\;,
\label{eq-n-simple}
\\
    \tau _{\pi }D\pi^{\mu \nu}+\pi ^{\mu \nu }&
        =&2\eta \sigma ^{\mu \nu }
        +2\tau_{\pi}\pi^{\langle\mu}_{\lambda}\omega^{\nu\rangle\lambda}
        -\delta_{\pi\pi}\pi^{\mu\nu}\theta\nonumber\\
        &-&\tau_{\pi\pi}\pi^{\lambda\langle\mu}\sigma^{\nu\rangle}_{\lambda}
        +\lambda_{\pi\Pi}\Pi\sigma^{\mu\nu}
        \nonumber \\
        &-&\tau_{\pi}( \pi ^{\lambda \mu }u^{\nu }+\pi ^{\lambda \nu}u^{\mu }) Du_{\lambda }\;; 
\label{eq-pi-simple}
\end{eqnarray}
here $\omega^{\mu\nu}=\frac{1}{2}(\nabla^\mu u^\nu-\nabla^\nu u^\mu)$ is the vorticity tensor. The additional transport coefficients $\delta_{\Pi\Pi}$, $\lambda_{\Pi\pi}$, $\tau_n$ etc. will be discussed in Sec.~\ref{transcoeff}. In Eqs.~(\ref{eq-n-simple}) and (\ref{eq-pi-simple}) we removed the transversality constraints on the l.h.s. by using footnote \ref{fn2} and
\begin{equation}
(D{n})^{\left\langle \mu \right\rangle} = \Delta^\mu_\nu D n^\nu = D n^\mu + u^\mu n^\nu D u_\nu\;,
\end{equation}
as well as its analog for $\pi^{\mu\nu}$, moving the extra terms as additional source terms to the r.h.s.

It is worth pointing out that in Eqs.~(\ref{eq-Pi-simple})-(\ref{eq-pi-simple}) we have followed Ref.~\cite{Denicol:2018wdp} in ignoring terms describing the direct influence of baryon diffusion, $n^\mu$, on the evolution of the shear and bulk viscous stresses, $\pi^{\mu\nu}$ and $\Pi$. Baryon evolution still affects the evolution of the system indirectly through the EoS. In this approach it has been shown \cite{Denicol:2018wdp, Du:2018mpf} that, while dissipative baryon diffusion effects directly influence the net-proton distributions, its indirect effects on the distributions of mesons and charged hadrons are negligible. It might be interesting to study to which extent second-order couplings between baryon diffusion and viscous stresses can modify this conclusion.

\subsection{Evolution equations in Milne coordinates}
\label{sec2.1.3}

Up to this point the formalism is completely general. For application to ultra-relativistic heavy-ion collisions we need the specific form of the evolution equations in Milne coordinates $x^{\mu}=(\tau,x,y,\eta_s)$ which are best adapted to the relativistic collision kinematics and subsequent almost boost-invariant longitudinal flow pattern \cite{PhysRevD.27.140}. In terms of Cartesian coordinates $(t,x,y,z)$ the longitudinal proper time $\tau$ and space-time rapidity $\eta_s$ are defined as
\begin{equation}
    \tau=\sqrt{t^2-z^2}\;,\qquad \eta_{s}=\frac{1}{2}\ln{\left(\frac{t+z}{t-z}\right)}\;.
\end{equation}
The mid-rapidity point $z=\eta_s=0$ at $\tau=0$ defines the collision point in the global (computational) frame. In Milne coordinates the metric tensor is
\begin{equation}
    g^{\mu\nu} =\mathrm{diag}\bigl(1,-1,-1,-1/\tau^2\bigr)\;,
\end{equation}
the fluid four-velocity is $u^\mu =(u^\tau,u^x,u^y,u^\eta)$, and the four-derivative is $\partial_{\mu}=(\partial_\tau,\partial_x,\partial_y,\partial_\eta)$.\footnote{%
        In all sub- and superscripts $\eta$ is short for $\eta_s$.}
The metric has the following non-vanishing Christoffel symbols:
\begin{equation}
    \Gamma^{\eta}_{\tau\eta}=\Gamma^{\eta}_{\eta\tau}=\frac{1}{\tau}\;,\qquad \Gamma^{\tau}_{\eta\eta}=\tau\;.\label{eq-chiris}
\end{equation}
Plugging them into Eqs.~(\ref{hydro_eqs_T},\ref{hydro_eqs_n}) we obtain the conservation laws in Milne coordinates:
\begin{eqnarray}
{\partial }_{\mu }T^{\mu\tau}& =&-\frac{1}{\tau}(T^{\tau\tau}+\tau^{2}T^{\eta\eta})\, , \label{relEqs_Tmut}\\
{\partial }_{\mu }T^{\mu x}& =&-\frac{1}{\tau}T^{\tau x}\, , 
\\ 
{\partial }_{\mu }T^{\mu y}&=&-\frac{1}{\tau}T^{\tau y}\, , 
\\ 
{\partial }_{\mu }T^{\mu \eta }&=&-\frac{3}{\tau}T^{\tau \eta}\ ,
\label{dT0i1}
\\
\partial_\mu N^{\mu}&=&-\frac{1}{\tau} N^\tau\;.
\label{relEqs_n}
\end{eqnarray}

Introducing the convective time derivative $d \equiv u^\mu \partial_\mu$, the relaxation equations can be written as
\begin{eqnarray}
    d\Pi&=&-\frac{\zeta}{\tau_\Pi}\theta-\frac{\Pi}{\tau_\Pi}-I_\Pi \label{relEqs_Pi}
\;,\\
    d \V^\mu &=& \frac{\kappa_n}{\tau_\V} \nabla^{\mu} \left(\frac{\mu}{T}\right) - \frac{\V^{\mu}}{\tau_{\V}} - I_\V^\mu 
    - G_\V^\mu\label{relEqs_nmu}\,,
    \\
    d\pi^{\mu\nu}&=&\frac{2\eta}{\tau_{\pi}}\sigma^{\mu\nu}-\frac{\pi^{\mu\nu}}{\tau_{\pi}}-I^{\mu\nu}_{\pi}-G^{\mu\nu}_{\pi}\,,
    \label{relEqs_pi}
\end{eqnarray}
with the shorthand notations $G^{\mu\nu}_{\pi}\equiv u^{\alpha}\Gamma^{\mu}_{\alpha\beta} \pi^{\beta\nu} + u^{\alpha}\Gamma^{\nu}_{\alpha\beta}\pi^{\beta\mu}$ and $G^\mu_n = u^\alpha \Gamma^\mu_{\alpha\beta} n^\beta$ for the geometrical source terms obtained when splitting the covariant LRF time derivative $D$ in Eqs.~(\ref{eq-Pi-simple})-(\ref{eq-pi-simple}) into the convective time derivatives $d$ and a remainder (for example, $D n^\mu \equiv u^\alpha d_\alpha n^\mu = u^\alpha (\partial_\alpha n^\mu + \Gamma^\mu_{\alpha\beta} n^\beta)$). The $I$-terms are explicitly
\begin{eqnarray}
I_{\Pi}&\equiv& \frac{\delta_{\Pi\Pi}}{\tau_\Pi}\Pi\theta-\frac{\lambda_{\Pi\pi}}{\tau_\Pi}\pi^{\mu\nu}\sigma_{\mu\nu}\;,\label{IPiterms}
\\
I^\mu_n &\equiv& I_1^\mu+ \frac{\delta _{nn}}{\tau _{n}} I_2^\mu+ I_3^\mu+ \frac{\lambda _{nn}}{\tau _{n}} I_4^\mu\;, \label{Interms}\\
I^{\mu\nu}_{\pi}&\equiv& I^{\mu\nu}_{1}
+\frac{\delta_{\pi\pi}}{\tau_\pi}I^{\mu\nu}_{2}
-I^{\mu\nu}_{3}
+\frac{\tau_{\pi\pi}}{\tau_\pi}I^{\mu\nu}_{4}
-\frac{\lambda_{\pi\Pi}}{\tau_\pi}\Pi\sigma^{\mu\nu}\;,\label{Ipiterms}
\end{eqnarray}
with 
\begin{eqnarray}
\label{I1-I4-mu}
  &&\!\!\!\!\!\!
    I_1^\mu = u^\mu n^\nu Du_\nu,\quad 
    I_2^\mu = n^{\mu}\theta,\quad
    I_3^\mu = n_{\nu}\omega^{\nu\mu},\quad
    I_4^\mu = n_{\nu}\sigma^{\nu\mu};
\\
\label{I1-I3-munu}
  &&\!\!\!\!\!\!
    I_{1}^{\mu\nu} = \left(u^\mu\pi^{\nu\lambda}+u^\nu\pi^{\mu\lambda}  
                     \right) Du_{\lambda},\quad 
    I_{2}^{\mu\nu} = \theta \pi^{\mu\nu},\quad 
    I_{3}^{\mu\nu} = \omega^\mu_{\ \lambda}\pi^{\lambda\nu}
                    +\omega^\nu_{\ \lambda}\pi^{\lambda\mu},\qquad
\\
\label{I4-munu}
  &&\!\!\!\!\!\!
    I_{4}^{\mu\nu} = \frac{1}{2}
    \left(\pi^{\mu\lambda}\sigma^{\ \nu}_\lambda
         +\pi^{\nu\lambda}\sigma^{\ \mu}_\lambda\right) 
   -\frac{1}{3}\Delta^{\mu\nu}\pi^{\alpha\beta}\sigma_{\beta\alpha}.
\end{eqnarray}

The conservation laws (\ref{relEqs_Tmut}-\ref{relEqs_n}) together with the dissipative transport equations (\ref{relEqs_Pi}-\ref{relEqs_pi}) constitute the equations of motion of the relativistic hydrodynamic system encoded in \code.

%
\section{Hydrodynamic fluctuations}
\label{sec:fluct}
%

The hydrodynamic equations discussed in the previous section are deterministic, in the sense that once specific initial conditions are given, the time evolution of hydrodynamic quantities are determined with no randomness. However, the thermodynamic quantities in the equations are  defined on hydrodynamic cells, by coarse-graining the d.o.f. on smaller scales. In other words, the hydrodynamic quantities are intrinsically stochastic and the corresponding microscopic system fluctuates among the members of the statistical ensemble; the deterministic quantities described in the previous section are the ensemble-averaged values for those stochastic ones \cite{landau2013statistical,Kapusta:2011gt}. Fluctuations in hydrodynamics are usually negligible when the d.o.f. is large, but for heavy-ion collisions with particle d.o.f. at $O(10^2-10^4)$, such fluctuations are essential and can be measurable, especially when the systems pass through the critical region. For this reason, description of hydrodynamic fluctuations becomes a key ingredient of theoretical modeling of heavy-ion collisions at BES energies, where one searches for the critical point.

To take into account the thermal fluctuations of hydrodynamic variables, one can add stochastic noise terms to the energy-momentum tensor and net baryon current, which drive the thermal fluctuations \cite{landau2013statistical,Kapusta:2011gt}:
\begin{equation}
	\tilde T^{\mu\nu} \equiv T^{\mu\nu}+\tilde S^{\mu\nu} \,,\quad \tilde N^{\mu} \equiv N^{\mu}+\tilde I^{\mu} \,,
\end{equation}
where $\tilde S^{\mu\nu}$ and $\tilde I^{\mu}$ are the noise terms, and $\tilde{A}$ denotes stochastic quantities. The stochastic hydrodynamic equations then become
\begin{equation}\label{eq:stoconseq}
	d_\mu\tilde T^{\mu\nu} =0 \,,\quad d_\mu\tilde N^{\mu} =0\,.
\end{equation}
The amplitude of the noise terms $(\tilde S^{\mu\nu}, \tilde I^{\mu})$ is proportional to the dissipative transport coefficients according to the fluctuation-dissipation theorem \cite{landau2013statistical}:
\begin{equation}\label{eq:fdt}
\begin{gathered}
\av{\tilde S^{\mu\nu}(x)}=\av{\tilde I^\lambda(x)}=0, \quad \av{\tilde S^{\mu\nu}(x)\tilde I^\lambda(x^\prime)}=0, \quad \av{\tilde I^\mu(x)\tilde I^\nu(x^\prime)}= 2\kappa_n\Delta^{\mu\nu}\delta^{(4)}(x-x^\prime)\,,\\
\av{\tilde S^{\mu\nu}(x)\tilde S^{\lambda\kappa}(x^\prime)}= 2T\left[ \eta\, (\Delta^{\mu\kappa}\Delta^{\nu\lambda}+\Delta^{\mu\lambda}\Delta^{\nu\kappa})+\left(\zeta-\frac{2}{3} \eta\right) \Delta^{\mu\nu}\Delta^{\lambda\kappa} \right] \delta^{(4)}(x-x^\prime)\,,
\end{gathered}
\end{equation}
where $\eta$, $\zeta$ and $\kappa_n$ are shear viscosity, bulk viscosity and baryon diffusion coefficient, respectively, which are assumed to be functions of ensemble-averaged hydrodynamic quantities.

Two approaches exist to solve these stochastic hydrodynamic equations \eqref{eq:stoconseq} and \eqref{eq:fdt}. The first approach directly implements those stochastic equations, by sampling these noise terms according to \eqref{eq:fdt} and then solving the equations~\eqref{eq:stoconseq} (see, e.g., Refs.~\cite{Singh:2018dpk,Sakai:2018sxp,Murase:2016rhl,Sakai:2020pjw}). The other one is called hydro-kinetic approach, which derives a set of deterministic evolution equations for the correlation functions of the fluctuations first, and then solves these evolution equations deterministically together with the hydrodynamic equations (see Refs.~\cite{Stephanov:2017ghc, Akamatsu:2018vjr, An:2019osr,An:2019csj,Martinez:2018wia,Rajagopal:2019xwg,Du:2020bxp}). In this thesis, we shall employ the second approach which is deterministic and is able to propagate off-equilibrium dynamics of these correlation function, and then embed these hydro-kinetic equations in the \bes{} code to obtain the \bes+ framework \cite{Du:2020bxp} (described in Ch.~\ref{ch.fluctuations}).

In this section, after briefly discussing hydrodynamic fluctuations and their correlations in Sec.~\ref{sec:hydrofluct}, we focus in Sec.~\ref{sec:equifluct} on the equilibrium value of the slowest mode near the QCD critical point. Out-of-equilibrium fluctuations and their equations of motion are studied in Ch.~\ref{ch.fluctuations} where we also compare the different models used in this work with each other and with those introduced in Refs.~\cite{Stephanov:2017ghc, Akamatsu:2018vjr, Rajagopal:2019xwg}.

\subsection{Stochastic noise}\label{sec:hydrofluct}
%

We begin by introducing one- and two-point functions of hydrodynamic quantities, following the treatment in Refs.~\cite{Stephanov:2017ghc,An:2019osr,An:2019csj}. One-point functions of hydrodynamic quantities $\Psi_m(t, \bm x)$ are denoted as $\langle\Psi_m(t, \bm x)\rangle$ where $m$ labels energy density, net baryon density, components of the flow velocity, etc., and $\langle\dots\rangle$ indicates the ensemble average over the classically statistically fluctuating thermal ensemble describing the fluid cell at point $(t,\bm x)$. The ensemble-averaged quantities $\langle\Psi_m(t, \bm x)\rangle$ vary only slowly  with $\bm x$ on a scale of inhomogeneity $\ell$ and evolve deterministically according to hydrodynamic evolution equations \cite{Stephanov:2017ghc, An:2019osr,An:2019csj}. 

Next we define equal-time two-point functions (also referred to as (equal-time) correlation functions or correlators) 
in the local fluid rest frame (LRF),\footnote{%
    \label{fn1}  
    In expanding fluids the LRF is a function of space and time, and formulating the equal-time condition in the LRF covariantly requires some care \cite{An:2019osr, An:2019csj}. We will write the equal-time (in the LRF) correlators non-covariantly, using LRF coordinates. For a covariant treatment see
    \cite{An:2019osr, An:2019csj}.
    }
\begin{eqnarray}
\label{eq:correlator}
    &&G_{mn}(t,\bm x_1, \bm x_2)
    = \langle\delta\Psi_m(t, \bm x_1)\delta\Psi_n(t, \bm x_2)\rangle
\\\nonumber
    &&\quad\equiv \langle\Psi_m(t, \bm x_1) \Psi_n(t, \bm x_2)\rangle - \langle\Psi_m(t, \bm x_1)\rangle\langle\Psi_n(t, \bm x_2)\rangle\,,
\end{eqnarray}
where
\begin{equation}
    \delta\Psi_m(t, \bm x) \equiv \Psi_m(t, \bm x)-\langle\Psi_m(t, \bm x)\rangle\,,
\label{eq:fluctuation}
\end{equation}
are the fluctuations of $\Psi_m(t, \bm x)$ \cite{Stephanov:2017ghc,An:2019osr,An:2019csj}. In this thesis our interest will focus on the ``slowest critical mode'' \cite{Stephanov:2017ghc}, i.e. the slowest-evolving correlator associated with critical fluctuations near the QCD critical point, and we therefore mostly suppress the matrix subscripts $m,n$ from here onward. 

Expressing the correlator (\ref{eq:correlator}) as a function of the mid-point $\bm x \equiv (\bm x_1{+}\bm x_2)/2$ and separation $\Delta\bm x \equiv (\bm x_1{-}\bm x_2)$ we can write 
\begin{equation}
     G(x, \Delta\bm x)=\Bigl\langle
     \delta\Psi\bigl(t, {\bm x}{+}{\textstyle\frac{1}{2}}\Delta\bm x \bigr)
     \delta\Psi\bigl(t, {\bm x}{-}{\textstyle\frac{1}{2}}
     \Delta\bm x\bigr)\Bigr\rangle
\end{equation}
where $x \equiv (t, \bm x)$ \cite{Stephanov:2017ghc, An:2019osr, An:2019csj}. (Remember that $\Delta\bm{x}$ is the two-point separation in the LRF.)   

The following discussion relies on a separation of scales between the (microscopic) correlation length $\xi$ and the (macroscopic) hydrodynamic (in-)homogeneity length $\ell$: Assuming that approximate thermal equilibrium is reached over regions of size $\ell$ (i.e. a length scale over which the macroscopic hydrodynamic quantities can be considered approximately constant), this length $\ell$ will also define the range of ${\bm x}$ over which $G(x, \Delta\bm x)$, considered as a function of the mid-point coordinate, will vary appreciably. As a function of the separation $\Delta{\bm x}$, on the other hand, the correlator $G(x, \Delta\bm x)$ typically falls off exponentially, with the decay length given by the correlation length $\xi$. We will follow \cite{Stephanov:2017ghc, An:2019osr, An:2019csj} and assume $\xi\ll\ell$, i.e. $G(x, \Delta\bm x)$ varies much more rapidly with $\Delta{\bm x}$ than with ${\bm x}$. In this so-called {\it thermodynamic limit} $V \sim \ell^3 \gg \xi^3$ the fluctuations are small, and their probability distribution is approximately Gaussian, a fact which will be used for calculating the equilibrium value of the correlators \cite{Stephanov:2017ghc,landau2013statistical} ({\it cf.} Eq.~(\ref{eq:phiQeq}) below).

Given the scale separation $\xi\ll\ell$ it is convenient to introduce the mixed Fourier transform (i.e. Wigner transform) with respect to the separation vector $\Delta\bm x$ for the correlator \cite{Stephanov:2017ghc,An:2019osr,An:2019csj}:
\begin{eqnarray}
  W_{\bm Q}(x) =  \int d^3(\Delta \bm x)\, G(x, \Delta\bm x)\, e^{i\bm Q \cdot \Delta\bm x}  = \int_{\Delta \bm x}
  \Bigl\langle
     \delta\Psi\bigl(t, {\bm x}{+}{\textstyle\frac{1}{2}}\Delta\bm x \bigr)
     \delta\Psi\bigl(t, {\bm x}{-}{\textstyle\frac{1}{2}}
     \Delta\bm x\bigr)\Bigr\rangle
    e^{i\bm Q\cdot \Delta\bm x}\,. 
\label{eq:wigner}
\end{eqnarray}
$W_{\bm Q}(x)$ can be thought of as a mode distribution function, similar to the particle phase-space distribution function $f(x,\bm Q)$ in kinetic theory, where the mode index $\bm Q$ specifies the wave vector of the mode in the LRF \cite{Stephanov:2017ghc}.\footnote{%
    As shown in \cite{Stephanov:2017ghc, An:2019osr}, a specific linear combination of the matrix elements $W^{mn}_{\bm Q}(x)$ can be identified with the phase-space distribution function of phonons with momentum $\bm Q$ at point $x$ and shown to satisfy a Boltzmann equation.}
The scale separation implies an infrared momentum cutoff for $\bm Q$: $\ell^{-1} \ll Q \sim \xi^{-1}$, with $Q$ being the magnitude of $\bm Q$ (see Sec.~\ref{sec:nonequi}).

%
\subsection{Equilibrium critical fluctuations}
\label{sec:equifluct}

In the {\sc hydro+} framework \cite{Stephanov:2017ghc} attention is focused on the most slowly evolving correlator associated with critical fluctuations. Near the QCD critical point, Ref.~\cite{Stephanov:2017ghc} identifies the slowest mode as the diffusion of fluctuations in the entropy per baryon at fixed pressure, $\delta(s/n)$.\footnote{%
    As shown in \cite{Stephanov:2017ghc, Akamatsu:2018vjr} the fluctuation of $s/n$ is a diffusive eigenmode of linearized hydrodynamics whose evolution decouples from that of other hydrodynamic fluctuations.}
In this work we consider a special partial-equilibrium state where, except for the slowest mode, all other fluctuations have achieved a sufficient degree of local equilibrium on length scales $\ell$ \cite{Stephanov:2017ghc} that they can be described by dissipative fluid dynamics. Since the slowest mode needs more time to thermalize it will be described dynamically. This special case captures the situation near a critical point where the correlation length $\xi$ becomes large and the relaxation rates for different fluctuation modes, scaling with different critical exponents, are suppressed by different powers of $\xi$. 

The relaxation rate for $\delta(s/n)$ satisfies $\Gamma_Q \propto (\lambda_T/\cp)\,Q^2$, with $Q$ being the wave number \cite{Stephanov:2017ghc,Akamatsu:2018vjr}. $\delta(s/n)$ is the slowest critical mode near the QCD critical point since the heat capacity $\cp$ is the most rapidly divergent equilibrium susceptibility, diverging  quadratically with the correlation length, $\cp \propto \xi^2$, while the heat conductivity $\lambda_T$ in the numerator diverges only linearly, $\lambda_T\propto\xi$ (see, e.g., Refs. \cite{Stephanov:2017ghc,An:2019csj}). Thus the relaxation rate for $Q\sim \xi^{-1}$ is $\Gamma_\xi\propto \xi^{-3}$, which is the most strongly suppressed relaxation rate of all critical modes. Consequently, the specific entropy fluctuations $\delta(s/n)$ are the first to fall out of equilibrium when the system passes through the critical region \cite{Akamatsu:2018vjr}.

$\delta(s/n)$ can be written in terms of the fluctuations of the hydrodynamic variables $\delta\Psi_m(x)$, and its correlator
\begin{equation}
    \phi_{\bm Q}(x){\,\sim\!}\int_{\Delta \bm x} \!\!
    \Bigl\langle
     \delta\frac{s}{n}\bigl(t, {\bm x}{+}{\textstyle\frac{1}{2}}\Delta\bm x \bigr)
     \delta\frac{s}{n}\bigl(t, {\bm x}{-}{\textstyle\frac{1}{2}}
     \Delta\bm x\bigr)\Bigr\rangle\,
     e^{i\bm Q \cdot\Delta\bm x}\
\label{eq:phiQ}
\end{equation}
is therefore some combination of the matrix elements $W^{mn}_{\bm Q}(x)$. The normalization of $\phi_{\bm Q}$ is irrelevant (see discussions after Eq.~(\ref{eq:deltas})) and therefore left arbitrary. $\phi_{\bm Q}$ are the non-hydrodynamic slow degrees of freedom (d.o.f.) near the QCD critical point, added in {\sc hydro+} as additional dynamical variables \cite{Stephanov:2017ghc}.\footnote{%
    Even though they represent the fluctuations of the single thermodynamic field $s/n$, we refer to them in the plural as {\it critical slow modes} since they are labeled by a continuous spectral index $\bm Q$ representing their wave number in the LRF.}
In Sec.~\ref{sec:quasieos}, we will see how these additional d.o.f. affect the bulk properties of the system, e.g. its entropy and pressure. 

Due to the separation of scales $\ell^{-1} \ll Q \sim \xi^{-1}$, the values of the hydrodynamic fields $e$, $n$, $u^\mu$ are approximately constant over the range $|\Delta \bm x|\lesssim 1/|\bm Q|$ over which the integral (\ref{eq:phiQ}) receives non-vanishing contributions, and the spatial correlator in the integrand can be taken as approximately translation invariant. Assuming also spatial isotropy in the local rest frame this implies \cite{Stephanov:2017ghc} that the dependence of the equilibrium correlator $\bar\phi_{\bm Q}$ on $\bm Q$ and $\xi$ involves only the magnitude $Q=|\bm Q|$ and must enter in the critical regime near the critical point through a universal scaling function $f_2$,
\begin{eqnarray}
\label{phi_eq0}
    \bar\phi_{\bm Q} &=& \int_{\Delta \bm x} \left\langle\delta \frac sn\left(\Delta\bm x\right)\delta \frac sn\left(\bm 0\right)\right\rangle e^{i\bm Q \cdot\Delta\bm x}
\\
    &=& \bar\phi_0 f_2(Q\xi,\Theta) \label{phi_eq}
\end{eqnarray}
where $\Theta=\Theta(e,n)$ is a scaling variable \cite{Stephanov:2017ghc} and both sides of the equation (although not explicitly indicated) depend on the position $x=(t,\bm x)$ through the local values of $e$ and $n$. The equilibrium value $\bar\phi_0$ of the static mode is obtained as the $Q{\,\to\,}0$ limit (the hydrodynamic limit) of the first of these equations (Eq. (\ref{phi_eq0})),
\begin{equation}
    \bar\phi_0(x) = V\left\langle\left(\delta \frac sn(x)\right)^2\right\rangle = \frac{\cp}{n^2}(x)\,, 
\label{eq:phiQeq}
\end{equation}
where the homogeneity volume around the point $x$, $V{\,\sim\,}\ell^3$, arises from integrating out $\Delta \bm x$ in Eq.~(\ref{phi_eq0}) and then using the Gaussian probability distribution of fluctuations in the thermodynamic limit \cite{landau2013statistical, Stephanov:2017ghc}.  
Here the heat capacity at constant pressure $c_p$ is given by \cite{Akamatsu:2018vjr}
\begin{equation}
    \cp = nT\left(\frac{\partial (s/n)}{\partial T}\right)_p.
\end{equation}
The scaling function is normalized to $f_2(0,\Theta)=1$. We follow \cite{Stephanov:2017ghc} and neglect its $\Theta$ dependence, $f_2(Q\xi,\Theta)\to f_2(Q\xi)$, and  assume $f_2$ to have the Ornstein-Zernike form \cite{RevModPhys.49.435}\footnote{%
    In Refs.~\cite{Akamatsu:2018vjr,Rajagopal:2019xwg}, the function is taken as $1/(1+x^{2-\eta})$ where $\eta$ is a critical exponent ($\approx 0.036$ in the 3D Ising model). Here we set it to zero for simplicity as was also done in Refs.~\cite{Stephanov:2017ghc, Akamatsu:2018vjr, Rajagopal:2019xwg}.}
\begin{equation}
    f_2(x) = \frac{1}{1+x^2}\,.
\label{eq:oz}
\end{equation}
With the simplifications made above, the full $Q$ and $\xi$ dependence of the equilibrium value of the slow mode is given by
\begin{equation}
    \bar\phi_Q = \bar\phi_0 f_2(Q\xi) = \left[\frac{\cpz}{n^2}\left(\frac{\xi}{\xi_0}\right)^2\right]\frac{1}{1+(Q\xi)^2}\,;
\label{eq:eqPhiQ_full}
\end{equation}
here $\cpz$ denotes the ``non-critical'' value of the heat capacity in a system with ``non-critical correlation length'' $\xi_0$. Far away from the critical point when $\xi\to\xi_0\ll Q^{-1}$, the mode spectrum (\ref{eq:eqPhiQ_full}) becomes independent of $Q$ and reduces to Eq.~(\ref{eq:phiQeq}). 

Equations (\ref{phi_eq}, \ref{eq:oz}) show that in thermal equilibrium the shape of the spectrum $\phi_Q$ of critical slow modes (i.e. their dependence on the wave number $Q$) is controlled by the evolution of the correlation length $\xi$ (through $f_2(Q\xi)$) as the fluid cell $x$ passes through the critical region of the phase diagram. In addition to the $Q$-dependence, the critical behaviour of $\xi$ also affects the zero mode $\bar\phi_0$ in Eq.~(\ref{eq:phiQeq}) through the critical growth of the heat capacity $\cp\propto\xi^2$.
Note that the critical behavior of $c_{p}$ near the QCD critical point is different from that of the specific heat (at fixed volume) $c_{V}$ in the 3D Ising model. The latter diverges as $\xi^{\alpha}$ where the critical  exponent is $\alpha\approx 0.11$.
In contrast, 
since the order parameter field at the QCD critical point is a linear combination that includes a baryon density component, 
$c_{p}$ diverges more strongly, as $\xi^{2}$
(see for example Sec.~IIB of Ref.~\cite{Akamatsu:2018vjr} for more details). See more discussions on the critical behavior of other transport coefficients in Sec.~\ref{subsec-trans}  below. The off-equilibrium dynamics of $\phi_Q$ shall be studied in Ch.~\ref{ch.fluctuations}.

\section{Transport coefficients}
\label{transcoeff}

The EoS and transport coefficients describe the medium properties of the expanding fluid and as such must be determined microscopically. While for the EoS detailed knowledge is available now from lattice QCD (see Sec.~\ref{subsec-eos}), the same is not true for the transport coefficients. We will here use rough estimates for the transport coefficients that have been obtained from kinetic theory, but have to leave their precise determination to future theoretical or phenomenological work. Specifically, \code\ implements the transport coefficients from Ref.~\cite{Denicol:2012cn, Denicol:2014vaa} which starts from the Boltzmann equation in Relaxation Time Approximation (RTA) and employs the 14-moment approximation \cite{Denicol:2010xn, Denicol:2012cn} for a one-component gas of Boltzmann particles with non-zero but small mass $m\ll T$. In all expressions we only keep the leading non-zero term in powers of $m/T\ll 1$. This is motivated by the approximate masslessness of the microscopic quark-gluon degrees of freedom that make up the fluid described by \code. We shall also briefly discuss critical properties of a few transport coefficients in Sec.~\ref{subsec-trans} which will be used in Ch.~\ref{ch:diffcp}.

\subsection{Shear stress tensor}

For the evolution of the shear stress we take the transport coefficients
\begin{eqnarray}
\frac{\eta}{\tau_\pi} &=& \frac{\ed+\peq}{5}\,, 
\label{etatau}
\\
\frac{\delta_{\pi\pi}}{\tau_\pi} = \frac{4}{3}\,, \qquad
\frac{\tau_{\pi\pi}}{\tau_\pi} &=& \frac{10}{7}\,, \qquad
\frac{\lambda_{\pi\Pi}}{\tau_\pi} = \frac{6}{5}\,.
\label{shearcoeff}
\end{eqnarray}
Following \cite{Liao:2009gb} we express shear viscous effects in terms of the {\it kinematic shear viscosity} 
\begin{equation}
    \bar\eta = \frac{\eta T}{\ed+\peq}\label{etabar}\;.
\end{equation}
For $\mu=0$ the kinematic shear viscosity reduces to the {\it specific shear viscosity} $\eta/\s$, where $\s$ denotes the entropy density, but it differs greatly from $\eta/\s$ at large net baryon densities, and this can lead to significant differences in the hydrodynamic flow patterns \cite{Denicol:2013nua}. Parametrizations of $\bar\eta$ as a function of $T$ and $\mu$ are discussed in \cite{NoronhaHostler:2008ju, Denicol:2015nhu}; in \code\ the default setting for $\bar\eta$ is a constant, $\bar\eta=0.2$. Given $\bar\eta$, Eq.~(\ref{etatau}) is used to calculate $\tau_\pi=5\bar\eta/T$, and Eqs.~(\ref{shearcoeff}) to obtain the remaining transport coefficients. 

Recently, a few more studies on chemical potential dependent $\eta/\s(\mu, T)$ came up, using methods including extrapolation from an $\eta/\s(\mu=0, T)$ \cite{Shen:2020jwv}, HRG model \cite{McLaughlin:2021dph}, and dynamical quasi-particle model (DQPM) \cite{Soloveva:2019xph}, etc. Rapidity-dependent harmonic flows are expected to have constraining power on $\eta/\s(\mu, T)$ \cite{Denicol:2015nhu}, and these parametrization and calculations of $\eta/\s(\mu, T)$ should be tested against experimental data. The parametrization used in Ref.~\cite{Shen:2020jwv} indicates rapid enhancement in $\eta/s$ at lower temperature or higher chemical potential, which is found to give stronger suppression in $v_2(\eta_s)$ at larger pseudo-rapidity, compared to constant or only temperature-dependent $\eta/s$. This is expected, as along the longitudinal direction towards forward- and backward-rapidities, the fireball is cooler and thus the temperature is lower, where a larger $\eta/s$ is expected with such a parametrization of $\eta/\s(\mu, T)$, which will have a stronger suppression over anisotropic flows.

\subsection{Bulk viscous pressure}

To evolve the bulk viscous pressure we take the transport coefficients 
\begin{eqnarray}
\frac{\zeta}{\tau_\Pi}&=&15\left(\frac{1}{3}-c^2_s\right)^{2}(\ed+\peq)\,, 
\label{betaPi}
\\
\frac{\delta_{\Pi\Pi}}{\tau_\Pi}&=&\frac{2}{3}\,,\qquad
\frac{\lambda_{\Pi\pi}}{\tau_\Pi}=\frac{8}{5}\left(\frac{1}{3}-c^2_s\right)\,,
\label{bulkcoeff}
\end{eqnarray}
where $c_s$ is the speed of sound in the medium (see Eq. (\ref{cs2EoS})). Similar to the shear viscosity, we start from a parametrization as a function of $T$ and $\mu$ of the {\it kinematic bulk viscosity}
\begin{equation}
    \bar\zeta = \frac{\zeta T}{\ed+\peq}\label{zetabar}
\end{equation}
and determine from it the bulk relaxation time $\tau_\Pi$ using Eq.~(\ref{betaPi}) and the remaining transport coefficients using Eqs.~(\ref{bulkcoeff}). In the code, we use a parametrization that interpolates between lattice QCD data for the QGP phase and results obtained from the hadron resonance gas model for the hadronic phase, connected quadratically around the pseudocritical temperature $T_{pc}=155$\,MeV \cite{Denicol:2009am}:
\begin{equation}
{\small
  \bar{\zeta} =
  \begin{cases}
  A_0 + A_1 x + A_2 x^2\,, &0.995\,T_{pc} \ge T \ge 1.05\,T_{pc}\,, 
  \\
  \lambda_1 \exp [-(x-1)/\sigma_1] + \lambda_2 \exp [-(x-1)/\sigma_2]+0.001\,,
     &T > 1.05\,T_{pc}\,, 
  \\
  \lambda_3 \exp [(x-1)/\sigma_3] + \lambda_4 \exp [ (x-1)/\sigma_4]+0.03\,,
     &T < 0.995\,T_{pc}\,, 
  \end{cases}
  \label{eq-zetas}
}
\end{equation}
with $x = T/T_{pc}$ and fitted parameters 
\begin{eqnarray*}
  &&A_0=13.45\,,\quad A_1=27.55\,,\quad A_2=-13.77\,, \\
  &&\lambda_1=0.9\,,\quad \lambda_2=0.25\,,\quad\lambda_3=0.9\,,\quad \lambda_4=0.22\,,\\
  &&\sigma_1=0.025\,,\quad \sigma_2=0.13\,, \quad\sigma_3=0.0025\,,\quad \sigma_4=0.022\,.
\end{eqnarray*}

The bulk viscous pressure describes the deviation from the thermal pressure for a non-perfect expanding or contracting fluid. Bulk viscosity has been shown to generate important effects on the slope of the transverse momentum spectra and their azimuthal anisotropy \cite{Ryu:2015vwa}. Bulk viscous effects are expected to be strongest near the quark-hadron phase transition \cite{Paech:2006st, Arnold:2006fz, Kharzeev:2007wb, Karsch:2007jc, Meyer:2007dy, Moore:2008ws, Sasaki:2008fg}, especially near the QCD critical point where critically enhanced contributions associated with critical slowing-down play a key dynamical role
\cite{Berdnikov:1999ph, Song:2009rh, Stephanov:2017ghc}. See also a recent calculation of $\zeta/s(\mu, T)$ within the dynamical quasi-particle model (DQPM) \cite{Soloveva:2019xph}. 

\subsection{Charge diffusion currents}

\subsubsection{Baryon diffusion current}\label{sec:barycoeffcurr}

Compared to the transport coefficients related to the bulk and shear stresses, those controlling baryon diffusion are much less explored (see, e.g., Refs.~\cite{Denicol:2018wdp,Soloveva:2019xph,Soloveva:2020hpr,Rougemont:2015ona}). Following Ref.~\cite{Denicol:2018wdp}, we use the coefficients obtained from the Boltzmann equation for an almost massless ($m/T\ll1$) classical gas (for a calculation of the baryon diffusion coefficient $\kappa_n$ for a massive gas of hadrons see Ref.~\cite{Albright:2015fpa}):
\begin{eqnarray}
\label{kappa}
    \kappa_n &=& \tau_n\n\left[\frac{1}{3}\coth\left(\frac{\mu}{T}\right)-\frac{\n T}{\ed+\peq} \right]\,,
\\
\label{diffcoeff}
    \delta_{nn} &=& \tau_n\;,\qquad \lambda_{nn} = \frac{3}{5}\tau_n\;.
\end{eqnarray}
Here $\tau_n$ is the relaxation time of the baryon diffusion current in Eq.~(\ref{eq-relax-n}) and parametrized as 
\begin{equation}
\label{taun}
    \tau_n = \frac{C_n}{T},
\end{equation}
with a free parameter $C_n$.\footnote{%
        Note that this procedure is different from the bulk and shear viscosity where we parametrized the first-order transport coefficients and computed the relaxation times from them, whereas here we parametrize the relaxation time and use it to compute the diffusion coefficient.}
The expression (\ref{kappa}) for the diffusion coefficient $\kappa_n$ was derived in first-order Chapman-Enskog approximation \cite{Denicol:2018wdp} whereas the second-order transport coefficients (\ref{diffcoeff}) were obtained in the 14-moment approximation \cite{Denicol:2014vaa, Denicol:2010xn, Denicol:2012cn}. In the limit of small net baryon density ($\mu\to0$) the diffusion coefficient $\kappa_n$ reduces to
\begin{equation}
\label{kappa_small_mu}
    \frac{\kappa_n}{\tau_n} = \frac{\n T}{3\mu}\;.
\end{equation}

In RTA, where the collision term in the Boltzmann equation is parametrized with a single relaxation time $\tauR$, the relaxation times for $\Pi$, $\pi^{\mu\nu}$ and $\V^\mu$ are all the same, i.e. $\tau_\Pi=\tau_\pi = \tau_n = \tauR$. Here we allow them to be different. If one does, however, impose the constraint $\tau_n=\tau_\pi=\tauR$, the baryon diffusion and shear viscosity coefficients can be related as follows \cite{Jaiswal:2015mxa}:
\begin{equation}
\label{kappaeta}
    \frac{\kappa_n T}{\eta} = C\left(\frac{\mu}{T}\right)\,
    \left(\frac{\pi T}{\mu}\right)^2
    \left(\frac{\n T}{\ed+\peq}\right)^2\,.
\end{equation}
In contrast to Eqs.~(\ref{kappa},\ref{kappa_small_mu}), this expression takes into account quantum statistics. The function $C(\mu/T)$ exhibits a weak dependence on $\mu/T$, interpolating between 5/3 at large $\mu/T$ and a somewhat smaller value at small $\mu/T$ whose precise magnitude depends on the number of massless degrees of freedom in the gas \cite{Jaiswal:2015mxa}. Note that, since for small $\mu/T$ the net baryon density $\n\propto\mu$, both (\ref{kappa_small_mu}) and (\ref{kappaeta}) yield nonzero baryon diffusion coefficients at zero net baryon density. At large $\mu/T$, the authors of Ref.~\cite{Jaiswal:2015mxa} have shown that the ratio $\kappa_n T/\eta$ approaches zero, i.e., at large net baryon densities and low temperatures baryon diffusion effects can generally be neglected in comparison with shear viscous stresses. 

\begin{figure}[tb!]
\centering\includegraphics[width=0.9\linewidth]{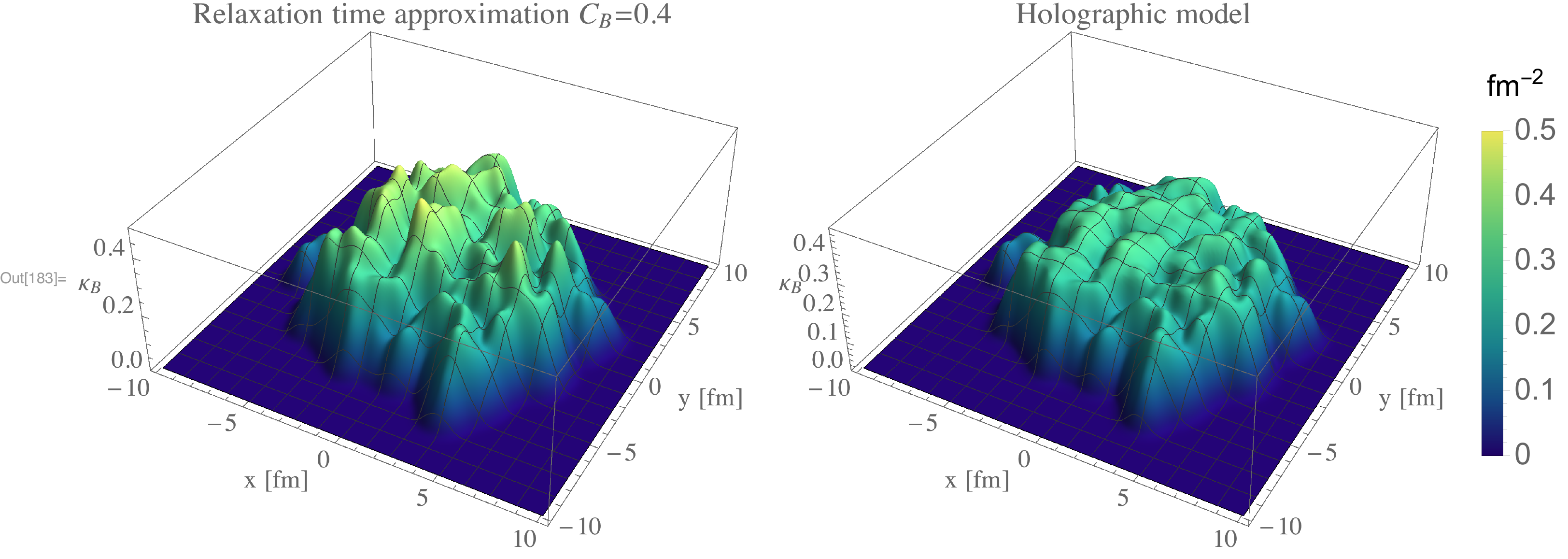}
\caption{Distribution of the baryon diffusion coefficient $\kappa_n$ from kinetic theory in the relaxation time approximation \cite{Denicol:2018wdp} (left panel) and from the holographic model \cite{Rougemont:2015ona} (right panel), for identical MC-Glauber initial energy and baryon density profiles. $C_n=0.4$ lies in the lower half of the typically explored range  \cite{Denicol:2018wdp}.}
\label{fig:F2_kappa}
\end{figure}
%

In addition to kinetic theory, gauge/gravity duality has also been widely used to determine the transport properties of the QGP (see, for example, \cite{Natsuume:2007ty, Rougemont:2015ona, Son:2006em}). With this method it is possible to study the transport properties of strongly coupled gauge theories for which no kinetic theory description exists, including the critical dynamics near a critical point. The latter is expected to provide critical signatures for the experimental identification of the QCD critical point \cite{Berdnikov:1999ph, Stephanov:2017ghc}. Studying and comparing the baryon number evolution with baryon diffusion coefficients corresponding to a weakly \cite{Denicol:2018wdp} or a strongly coupled QGP \cite{Rougemont:2015ona} can be interesting \cite{Du:2018mpf, Li:2018fow, du2021baryon}. Fig.~\ref{fig:F2_kappa} compares the initial distribution of $\kappa_n$ in the transverse plane at midrapidity computed from the initial energy and baryon density profiles with MC-Glauber input. Clearly visible significant differences between the two models, in both magnitude and ``bumpiness'', are expected to affect the diffusion of net baryon number and the final baryon spectra. Such a comparison will be also discussed in Ch.~\ref{ch:diffcp}.

\subsubsection{Multi-charge diffusion matrix}\label{sec:coef_multi}

Once attention is paid to the evolution of net baryon charge and its diffusion current, one quickly realizes that a hydrodynamic description of multiple charges and their couplings, including baryon charge ($B$), electric charge ($Q$) and strangeness ($S$), can be essential for quantitive studies of systems produced in heavy-ion collisions, especially at low beam energies (see Refs.~\cite{Greif:2017byw, Fotakis:2019nbq, Rose:2020sjv, Fotakis:2021diq}). This is because some quarks (or hadrons) carry more than one of those charges, and thus the diffusion of one charge may introduce that of another type of charge as well. In other words, the gradient of one charge density can drive diffusions of all types of charges, and in this sense, the diffusion coefficient of baryon current should be replaced by a  matrix of multi-charge diffusion coefficients \cite{Greif:2017byw} (cf.~Eq.~\ref{eq-ns-n}):
\begin{equation}
\begin{pmatrix}
\begin{tabular}{c}
$j_{B}^{\mu }$ \\ 
$j_{Q}^{\mu }$ \\ 
$j_{S}^{\mu }$%
\end{tabular}%
\end{pmatrix}%
=%
\begin{pmatrix}
\begin{tabular}{ccc}
$\kappa _{BB}$ & $\kappa _{BQ}$ & $\kappa _{BS}$ \\ 
$\kappa _{QB}$ & $\kappa _{QQ}$ & $\kappa _{QS}$ \\ 
$\kappa _{SB}$ & $\kappa _{SQ}$ & $\kappa _{SS}$%
\end{tabular}%
\end{pmatrix}%
\cdot 
\begin{pmatrix}
\begin{tabular}{c}
$\nabla ^{\mu }\alpha _{B}$ \\ 
$\nabla ^{\mu }\alpha _{Q}$ \\ 
$\nabla ^{\mu }\alpha _{S}$%
\end{tabular}%
\end{pmatrix}%
. \label{eq:LinearDiff}
\end{equation}%
Here the diagonal terms are the diffusion coefficients of baryon charge, electric charge and strangeness, respectively, and the off-diagonal terms describe the couplings between different charges.  $\alpha_i\equiv\mu_i/T\; (i=(B, Q, S))$ are the ratios between the chemical potentials and the temperature. The studies \cite{Greif:2017byw, Fotakis:2019nbq, Rose:2020sjv, Fotakis:2021diq} show that the off-diagonal terms can be as large as the diagonal ones, indicating the strong coupling among different diffusion currents and the insufficiency of simulating only one of them, e.g., baryon diffusion. We note that, nevertheless, even the studies on a single charge diffusion, i.e., the baryon diffusion, are still at an early stage, and such simulations when done realistically, are already very computationally costly. A full (3+1)D hydrodynamic framework which includes the full diffusion matrix is still lacking.

\subsection{Critical behavior of transport coefficients}\label{subsec-trans}

Near the critical point, fluctuations at the length scale $\xi$ significantly modify the physical transport coefficients, giving rise to their correlation length dependence. As noted in Sec.~\ref{sec:phase_diagram}, the QCD critical point belongs to the dynamical critical universality class of Model H, and accordingly the critical behavior of many transport coefficients can be obtained. In this section, we focus on the critical scaling of those relevant to baryon diffusion, especially its relaxation time, which will be used in Ch.~\ref{ch:diffcp}.

In Model H, the shear stress tensor and bulk viscous pressure play important roles in critical dynamics through fluid advection \cite{Rajagopal:2019xwg, Du:2020bxp}. We shall only consider critical effects arising from fluctuations in the hydrodynamic regime (i.e., carrying small frequencies and wave numbers, $\omega, k\to0$). Feedback from off-equilibrium fluctuations that are non-analytic in $\omega$ or $k$, commonly referred to as long-time tails, is suppressed by phase space \cite{Rajagopal:2019xwg, Du:2020bxp} and will be neglected. In other words, the scaling parametrizations below derive from equilibrium fluctuations for thermodynamic quantities and from analytic non-equilibrium fluctuations for  transport coefficients. In the presence of a bulk viscous pressure, the critical contribution to bulk viscosity diverges as $\xi^z$ where $z=3$ for the QCD critical point \cite{RevModPhys.49.435}. Besides, the associated relaxation time for the bulk viscous pressure also diverges as $\xi^3$ \cite{Monnai:2016kud}. In this case, the relaxation rate of the fluctuation modes contributing to bulk viscous pressure is much smaller than the typical hydrodynamic frequency, and they can no longer be treated hydrodynamically, requiring instead an extended framework such as {\sc hydro+}/++ \cite{Stephanov:2017ghc,An:2019csj}. 

In this thesis, we focus on effects from baryon diffusion and remark that the Navier-Stokes limit of the baryon diffusion current in Eq.~\eqref{eq-ns-n} can be rewritten in terms of density and temperature gradients,
\begin{subequations}
\label{eq:nmu_NS_decomposition}
\begin{eqnarray}
n^{\mu}_{\rm NS} 
= D_B\nabla^{\mu}n+D_T\nabla^{\mu}T\,,
\end{eqnarray}
where the two coefficients are
\begin{equation}
\label{eq:D_BT}
    D_B=\frac{\kappa_n}{T\chi}, \quad D_T=\frac{\kappa_n}{Tn}\left[\left(\frac{\partial p}{\partial T}\right)_n-\frac{w}{T}\right]\,.
\end{equation}
\end{subequations}
Here $\chi\equiv(\partial n/\partial\mu)_T$ is the isothermal susceptibility, and $w=\ed+p$ is the enthalpy density. We note that the gradient expansion is not unique, and writing it in different ways can be used to explore individual contributions separately (see Ch.~\ref{ch:diffcp}). For later convenience of discussing critical behavior, we also introduce the heat diffusion coefficient,
\begin{equation}\label{eq:D_p}
    D_p=\frac{\lambda_T}{c_p}\,,
\end{equation}
where $c_p\equiv nT(\partial m/\partial T)_p$ is the specific heat, with $m\equiv s/n$, i.e., the entropy per baryon density \cite{Son:2004iv, Stephanov:2017ghc}; $\lambda_T$ is the thermal conductivity, which can be related to the baryon diffusion coefficient $\kappa_n$ by
\begin{equation}\label{eq:kappa-lambda}
    \lambda_T = \left(\frac{w}{nT}\right)^2\kappa_n\,.
\end{equation}
Using Eq.~\eqref{eq:kappa-lambda} one can relate the heat diffusion coefficient $D_p$ to $D_B$ by
\begin{equation}\label{eq:Dp-DB}
    D_p=\left(\frac{w^2\chi}{n^2Tc_p}\right)D_B.
\end{equation}

The following second-order thermodynamic coefficients (isothermal susceptibility $\chi$ and specific heat $c_p$) as well as the first-order transport coefficients (baryon diffusion coefficient $\kappa_n$ and thermal conductivity $\lambda_T$) scale with the correlation length as \cite{RevModPhys.49.435}
    \begin{equation}\label{eq:coeff_scaling}
        \chi\sim c_p\sim \xi^2\,, \quad \kappa_n\sim\lambda_T\sim \xi\,,
    \end{equation}
where the exponents are rounded to their nearest integers for simplicity. Therefore, according to Eqs.~\eqref{eq:D_BT} and \eqref{eq:D_p},
\begin{equation}\label{eq:D_scaling}
    D_T\sim \xi, \quad D_B\sim D_p\sim \xi^{-1}. 
\end{equation}

We now turn to the critical behavior of the relaxation time $\tau_n$. It is worth remembering that the Israel-Stewart-type equations (cf. Eq.~\eqref{eq-n-simple}) provide an ultraviolet completion of the naive (Landau-Lifshitz) hydrodynamic theory. The microscopic relaxation times associated with the new dissipative dynamical degrees of freedom (such as, in our case here, the baryon diffusion current $n^\mu$) play the role of ultraviolet regulators which modify the short-distance (high-frequency) behavior of the theory. For the baryon diffusion current $n^\mu$, $\tau_n$ characterizes the relaxation time to its Navier-Stokes limit $n^\mu_{\text{NS}}$ (which is zero in a homogeneous background). Since $n^\mu$ can only equilibrate as long as {\it all\,} fluctuating degrees of freedom contributing to $n^\mu$ also equilibrate, $\tau_n$ can be considered as the typical equilibration time scale of the slowest fluctuation mode near the critical point. Indeed, in {\sc hydro+}/++, the non-hydrodynamic slow-mode evolution equations for critical fluctuations with typical momenta $q\sim\xi^{-1}$ have the same structure as the Israel-Stewart relaxation equations for the dissipative flows arising from thermal fluctuations with wave numbers $k\sim T$. As already mentioned, Israel-Stewart type equations neglect the non-analytic contributions from long-time tails which we argued above to be negligible (see also Ch.~\ref{ch.fluctuations}).

As mentioned earlier, the {\it slowest} mode contributing to $n^\mu$ is the diffusive-shear two-point correlator between the entropy per baryon density fluctuations $\delta m\equiv\delta (s/n)$ and the flow fluctuations $\delta u_\mu$, i.e., $G_{m\mu}\sim\langle\delta m \delta u_\mu \rangle$ \cite{An:2019csj}. The relaxation rate for this mode with wave number $q$ is given by $\Gamma_G(q)=(\gamma_\eta+D_p)q^2$ where $\gamma_\eta=\eta/w$, with $w=e+p$ being the enthalpy density. The two contributions to this rate stem from the relaxation of the shear stress and of the baryon diffusion, respectively.
Near the critical point $\Gamma_G$ is dominated by  contributions with typical wave numbers $q\sim 1/\xi$. Given $D_p\sim\xi^{-1}$ and approximately $\gamma_\eta\sim\eta\sim\xi^0$ \cite{RevModPhys.49.435}, one finds $\Gamma_G(q)=(\gamma_\eta+D_p)q^2|_{q\sim\xi^{-1}}\sim\xi^{-2}$ and hence $\tau_G = \Gamma_G^{-1} \sim \xi^2$.\footnote{%
     Another mode contributing to the baryon diffusion current is the pressure-shear mode $G_{p\mu}\sim\langle\delta p\delta u_\mu\rangle$ \cite{An:2019csj}. Its relaxation rate at wave number $q$ is $(\gamma_\zeta+\frac{4}{3}\gamma_\eta+\gamma_p)q^2$ where  $\gamma_\zeta=\zeta/w$, $\gamma_\eta=\eta/w$, and $\gamma_p=\kappa_n c_s^2Tw(\partial\alpha/\partial p)_m^2$, with $c_s$ being the speed of sound. In the presence of bulk viscosity (as we assume in order to ensure the correct scaling in Model H), the relaxation rate for $G_{p\mu}$ is dominated by $\gamma_\zeta q^2|_{q\sim\xi^{-1}}\sim\xi$ considering $\gamma_\zeta\sim\zeta\sim\xi^3$, which is much faster than the rate for the diffusive-shear mode which scales like $\Gamma_G\sim\xi^{-2}$. Even in the absence of the viscosities (i.e. for $\eta=\zeta=0$), the relaxation rate $\gamma_p q^2|_{q\sim\xi^{-1}} \sim \xi^{-1}$ is still faster than $\Gamma_G \sim \xi^{-2}$. Therefore, the contribution from the pressure-shear mode, which is not the slowest, can be neglected.}
Thus it is natural to expect $\tau_n\sim\tau_G\sim\xi^2$. 
 
As an aside, let us comment on the consequences, had we tried to ensure the absence of shear stress by demanding that $\eta=0$. In this case $\Gamma_G(q) = D_p q^2|_{q\sim\xi^{-1}}\sim\xi^{-3}$, and therefore $\tau_n\sim\tau_G\sim\xi^3$, which is larger compared to that in the case with shear stress. This arises from the fact that, near the critical point, the shear mode ($\delta u_\mu$) relaxes to equilibrium parametrically faster than the diffusive mode ($\delta m)$. As a result, its dissipation changes the scaling exponent of the relaxation time of the diffusive-shear two-point correlator.

With a proper parametrization of  $\xi(\mu,T)$ and the critical scaling we can arrive at a complete set of relevant thermodynamic quantities and transport coefficients as explicit functions of $T$ and $\mu$ that hold in the entire crossover domain of the QCD phase diagram, both far away from and within the critical region. These will be used for studying critical effects on baryon transport near the critical point in Ch.~\ref{ch:diffcp}, where we shall take the expressions from kinetic theory, Eqs.~(\ref{kappa}) and (\ref{taun}), for calculating the non-critical values of the baryon diffusion coefficient and relaxation time.

\section{Equation of State}\label{subsec-eos}

Another important medium property that crucially affects the dynamical evolution of the fluid is its equation of state 
\begin{equation}
    \peq = \peq(\ed, \n) = \peq(T,\mu)\,.
\label{eq-eos}
\end{equation}
In practice, for the calculation of the transport coefficients and chemical forces we also need the equivalent relations
$T(\ed,\n)$ and $\mu(\ed,\n)$. We use the term EoS generically for any one of these relations.

\subsection{Construction of the Equation of State}
\label{subsec-coneos}

Since the matter produced in nuclear collisions passes through very different physical regimes that differ by orders of magnitude in energy density and must be described with different effective degrees of freedom, we need an EoS that describes the medium properties over a wide range of temperature and length scales \cite{Philipsen:2012nu}. On most scales the degrees of freedom of the evolving system are strongly coupled, rendering  perturbative investigations from first principles unreliable. Over the last decades, lattice QCD (LQCD) has been established as the most precise non-perturbative framework to calculate the EoS of strongly interacting matter at zero baryon chemical potential (see, e.g., \cite{Borsanyi:2010cj}). The method works well at temperatures above $\sim 100$\,MeV; at lower temperatures the lattice signals become weaker and more noisy, necessitating the matching of LQCD data to an analytical hadron resonance gas model.   

Lattice QCD obtains the EoS by calculating the trace of the energy-momentum tensor $T^{\mu\nu}$, $\ed - 3\peq$ (usually referred to as the ``trace anomaly'' or ``interaction measure''), describing deviations from the conformal EoS. Defining the rescaled dimensionless trace anomaly
\begin{equation}
    I(T, \mu) \equiv \frac{\ed(T, \mu) - 3\peq(T, \mu)}{T^4}\;,
\end{equation}
the thermal pressure at zero chemical potential can be written as
\begin{equation}
    \frac{\peq^\mathrm{LAT}(T, 0)}{T^4} = \int^T_0 dT'\frac{I^\mathrm{LAT}(T', 0)}{T'}\;.
\end{equation}

Unfortunately, this method cannot be directly extended to non-zero chemical potential where the evaluation of the QCD path integral for the interaction measure $I(T,\mu)$ suffers from a ``sign problem'' \cite{Philipsen:2012nu}, precluding its direct computation with standard Monte-Carlo methods. This problem can be partially circumvented by using standard LQCD techniques to also compute the $\mu$-derivatives of the pressure $\peq(T,\mu)$ at $\mu=0$ and construct $\peq(T,\mu)$ at non-zero $\mu$ from its Taylor series around $\mu=0$:
\begin{equation}
    \frac{\peq^\mathrm{LAT}(T, \mu)}{T^4} = \frac{\peq^\mathrm{LAT}(T, 0)}{T^4} + \sum_{n=1}^{n_\mathrm{max}} c_{2n}(T)\left(\frac{\mu}{T}\right)^{2n}\,.
\label{Taylor}
\end{equation}
The expansion coefficients are
\begin{equation}
    c_n(T) =  \left.\frac{1}{n!}\frac{\partial^n (\peq/T^4)}{\partial(\mu/T)^n}\right|_{\mu=0}\equiv \frac{1}{n!}\chi_n(T)\,,
\end{equation}
where $\chi_n(T)$ are known as the ``baryon number susceptibilities'' \cite{Borsanyi:2012cr, Parotto:2018pwx, Bazavov:2017dus}.
The computational effort of computing them increases rapidly with their order $n$; at this time, the Taylor expansion (\ref{Taylor}) includes terms up to order $n_\mathrm{max}=3$ and, near $T=T_c$, converges well up to about baryon chemical potentials $\mu/T \lesssim 2$ \cite{Borsanyi:2012cr, Parotto:2018pwx, Bazavov:2017dus}. 

At low temperatures $T\ll T_c\simeq 155$\,MeV, LQCD is increasingly affected by lattice artifacts and the system is more properly described in terms of hadronic degrees of freedom
as a ``hadron resonance gas'' (HRG). In the HRG model, the interactions among different hadronic species are accounted for by including all experimentally identified scattering resonances as additional, non-interacting particle species. In the HRG model the interaction measure $(\ed{-}3\peq)/T^4$ is given as  \cite{Bazavov:2014pvz, Huovinen:2009yb}
\begin{equation}
    I^\mathrm{HRG}(T, 0)=\sum_{m_i\leq m_\mathrm{max}}\frac{g_i}{2\pi^2}\sum^\infty_{k=1}\frac{(-\eta_i)^{k+1}}{k}\left(\frac{m_i}{T}\right)^3K_1\left(\frac{km_i}{T}\right),
\end{equation}
where particle species with spin-isospin degeneracy $g_i$ and mass $m_i$ smaller than some cut-off $m_\mathrm{max}$ can be included, and $\eta_i = -1~(+1)$ for bosons (fermions) describes the effects of quantum statistics. 

In principle, the EoS used for the hydrodynamic evolution should include the same set of hadronic resonances as the hadronic afterburner employed to describe the kinetic final freeze-out stage because otherwise a mismatch of the energy and baryon densities occurs on the conversion surface where we change between these two different dynamical descriptions. In practice these discontinuities tend to be small, and such care is not always taken. In our applications of \bes{} we use different matched equations of state for different hadronic afterburners (e.g., for {\sc UrQMD} \cite{Bass:1998ca, Bleicher:1999xi} and {\sc SMASH} \cite{Weil:2016zrk}); however, the module for matching the lattice QCD data to a HRG with adjustable hadronic mass spectrum is not part of the \bes{} code distribution. 

For the matching procedure between the LQCD and HRG equations of state different methods have been used. For example, in Refs. \cite{Parotto:2018pwx, Denicol:2018wdp} the pressure is interpolated as follows:
\begin{eqnarray}\label{eq:eosinter}
    \frac{\peq(T, \mu)}{T^4} &=& \frac{1}{2}
    \left[1-\tanh\left(\frac{T{-}T'(\mu)}{\Delta T'}\right)\right]
    \frac{\peq^\mathrm{HRG}(T, \mu)}{T^4}
\\
    &+& \frac{1}{2}
    \left[1+\tanh\left(\frac{T{-}T'(\mu)}{\Delta T'}\right)\right]
    \frac{\peq^\mathrm{LAT}(T, \mu)}{T^4}\,.
\end{eqnarray}
Here $\peq^\mathrm{HRG}$ and $\peq^\mathrm{LAT}$ are the equilibrium pressures for the hadron resonance gas and from lattice QCD, respectively, $T'(\mu)$ is the ``switching temperature'' and $\Delta T'$ controls the width of the ``overlap region''. The authors of Ref.~\cite{Marrochio:2013wla}, on the other hand, interpolate the interaction measure at $\mu=0$, $I(T,\mu=0)$ smoothly between $T_1=T_c(\mu{=}0)=155$\,MeV and $T_2=180$\,MeV, using a polynomial interpolation function. 

Once the pressure $\peq(T, \mu)$ is given, other thermodynamic quantities can be calculated from thermodynamic identities:
\begin{eqnarray}
    \frac{\s(T, \mu)}{T^3} &=& \frac{1}{T^3}\left[\frac{\partial \peq(T, \mu)}{\partial T}\right]_{\mu}\;,
\label{sEoS}\\
    \frac{\n(T, \mu)}{T^3} &=& \frac{1}{T^3}\left[\frac{\partial \peq(T, \mu)}{\partial \mu}\right]_T\;,
\label{nEoS}\\
    \frac{\ed(T, \mu)}{T^4} &=& \frac{\s(T, \mu)}{T^3}-\frac{\peq(T, \mu)}{T^4}+\frac{\mu}{T}\frac{\n(T, \mu)}{T^3}\;,
\label{eEoS}\\
    c^2_s(T, \mu) &=& \left[\frac{\partial \peq(\ed, \n)}{\partial \ed}\right]_{\n} + \frac{\n}{\ed+\peq}\left[\frac{\partial \peq(\ed, \n)}{\partial \n}\right]_{\ed}\;.
\label{cs2EoS}
\end{eqnarray}
The last equation requires expressing $\ed$ and $\n$ through $T$ and $\mu$ after taking the derivatives. In practice, the functions $\ed(T,\mu)$ and $\n(T,\mu)$ are numerically inverted, and the quantities $T, \mu, \s, \peq$, and $c^2_s$, as well as the two derivatives on the r.h.s. of Eq.~(\ref{cs2EoS}), are stored in a table on a $(\ed,\n)$ grid which is interpolated by the hydrodynamic code as needed. 

The methods described in this subsection can be readily extended to include non-zero chemical potentials of baryon number, electric charge and strangeness, and to include contribution from a critical point as well. We shall briefly mention some recent progresses in the following two subsections.

\subsection{Equations of State at non-zero charges}

\subsubsection{Equations of State implemented in \code}

In \code\ four different equations of state, EOS1 to EOS4, are implemented, for different purposes: at zero chemical potential, we include a massless (conformal) EoS ($\ed=3\peq$, EOS1) as well as an interpolated LQCD-HRG EoS from the Wuppertal-Budapest collaboration \cite{Borsanyi:2010cj} (EOS2); at non-zero chemical potential, an appropriately generalized conformal EoS (EOS3) and an interpolated LQCD-HRG EoS from Ref.~\cite{Denicol:2018wdp} (EOS4) are used.

EOS1 assumes an ideal gas of massless quarks and gluons:
\begin{equation}
        \ed = 3\peq = 3\left[2(N_c^2-1)+\frac{7}{2}N_cN_f\right]
                      \frac{\pi^2}{90}T^4\;,
\label{EOS1}
\end{equation}
where $N_c = 3$ and $N_f = 2.5$ are the numbers of colors and (approximately) massless quark flavors, respectively.\footnote{%
    Strange quarks, whose mass is of the same order of magnitude as the quark-hadron transition temperature, are (somewhat roughly) counted as 1/2 massless quark flavor.} 
In EOS1, $c_s^2=1/3$ and $\mu = 0$. While the conformal EoS does not properly describe the properties of the matter produced in nuclear collisions, it is, owing to its simplicity, very useful for code testing. Technically, EOS1 can be used in \code\ even when the baryon density and baryon diffusion currents are being evolved; in that case, these currents do not couple to the rest of the hydrodynamic system and evolve purely as background fields.

\begin{figure*}[!t]
\centering
        \includegraphics[width=0.45\textwidth]{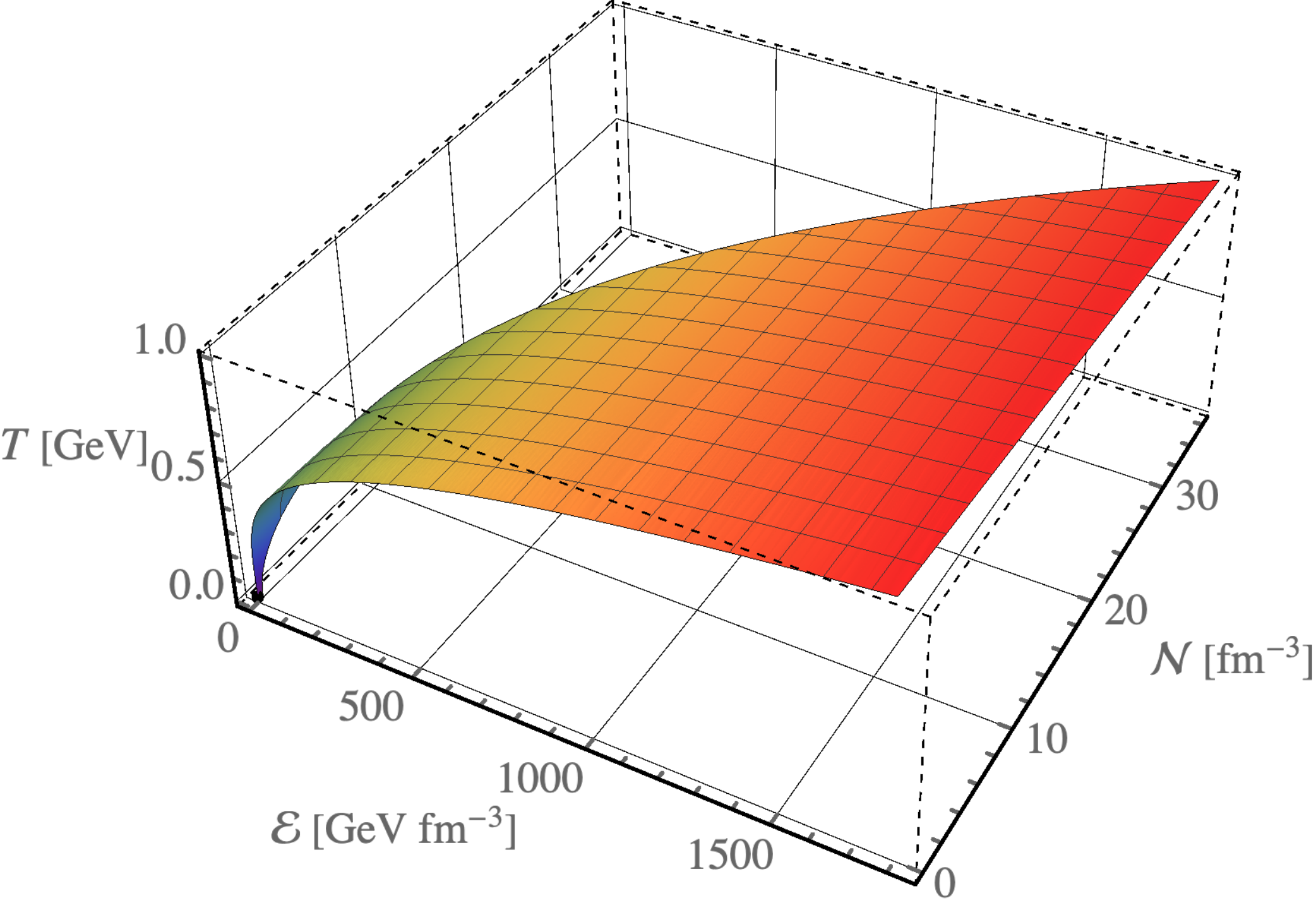}
        \includegraphics[width=0.45\textwidth]{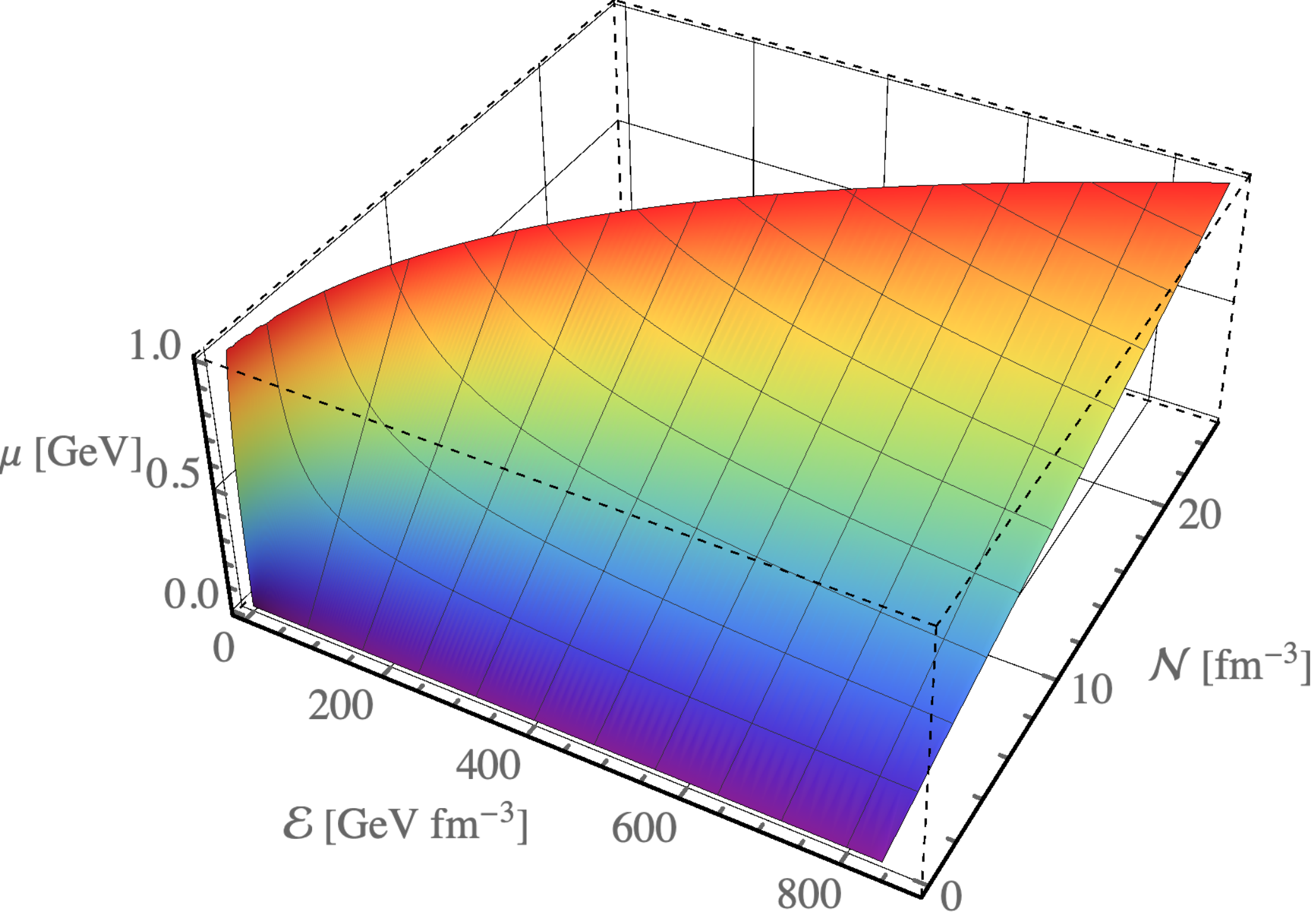}
    \caption{(Color online) Conformal EOS3 with $N_f=2.5$ at finite temperature and chemical potential. {\sl Left:} $T(\ed, \n)$. {\sl Right:} $\mu(\ed, \n)$.    }
    \label{F1}
\end{figure*}

EOS2 from the Wuppertal-Budapest collaboration \cite{Borsanyi:2010cj} can be used for realistic simulations at vanishing $\mu$, i.e. for heavy-ion collisions at ultra-relativistic collision energies with $\sqrt{s_\mathrm{NN}} \gg 100$\,GeV, especially near mid-rapidity. More details about EOS2 can be found in Refs.~\cite{Borsanyi:2010cj, Bazow:2016yra}.

EOS3 is the generalization of EOS1 to non-zero $\mu$. Starting from the ideal massless parton gas expression \cite{Mueller2018}
\begin{equation}
    \frac{\peq}{T^4} = \frac{\pi^2}{90}\left[2(N_c^2-1)+\sum_f 4N_c\left(\frac{7}{8}+\frac{15}{4}\left(\frac{\mu_f}{\pi T}\right)^2+\frac{15}{8}\left(\frac{\mu_f}{\pi T}\right)^4\right)\right],
\label{EOS3}
\end{equation}
where $N_c=3$ and the sum goes over massless quark flavors, we simplify it by setting $\mu_f = \mu/3$ for all flavors (which is appropriate if only baryon number is considered as a conserved charge): 
\begin{eqnarray}
  \frac{\peq(T,\mu)}{T^4}&=&
  p_0+N_f\left[\frac{1}{18}\left(\frac{\mu}{T}\right)^2
     +\frac{1}{324\pi^2}\left(\frac{\mu}{T}\right)^4\right]
     =\frac{\ed(T,\mu)}{3T^4}\;,
\label{EOS3P}
\\
  \frac{\n(T,\mu)}{T^3}&=&
   N_f\left[\frac{1}{9}\left(\frac{\mu}{T}\right)
           +\frac{1}{81\pi^2}\left(\frac{\mu}{T}\right)^3\right]\,,
\label{EOS3N}
\end{eqnarray}
with $p_0=(16+10.5N_f)\pi^2/90$. We again count strange quarks with a factor 1/2, i.e. we set $N_f=2.5$ so that for $\mu=0$ (\ref{EOS3P}) reduces to (\ref{EOS1}). Inverting these functions numerically one obtains the EoS tables used in the hydrodynamic code (see Fig.~\ref{F1}).

%
\begin{figure*}[!t]
\centering
        \includegraphics[width=0.45\textwidth]{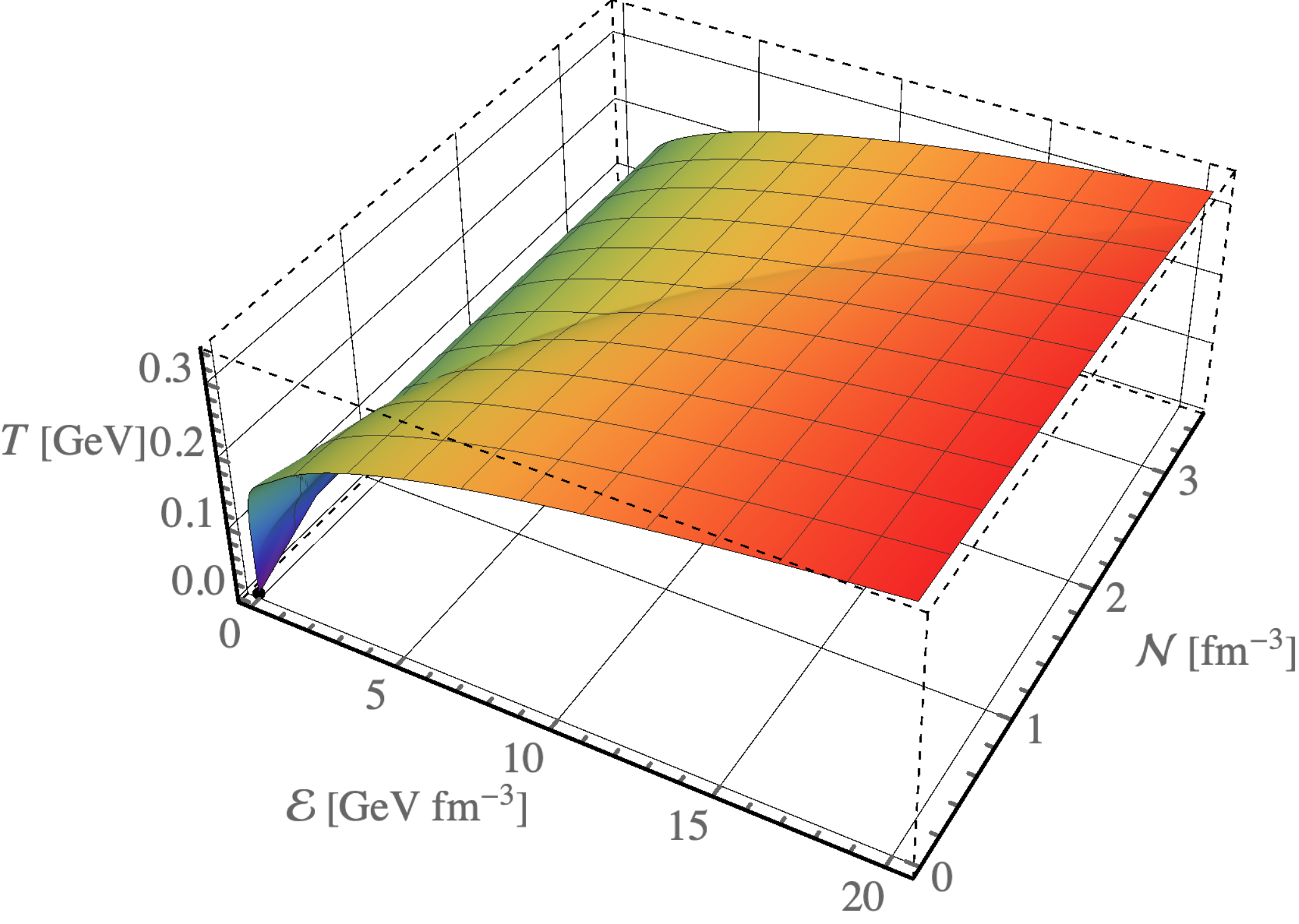}
        \includegraphics[width=0.45\textwidth]{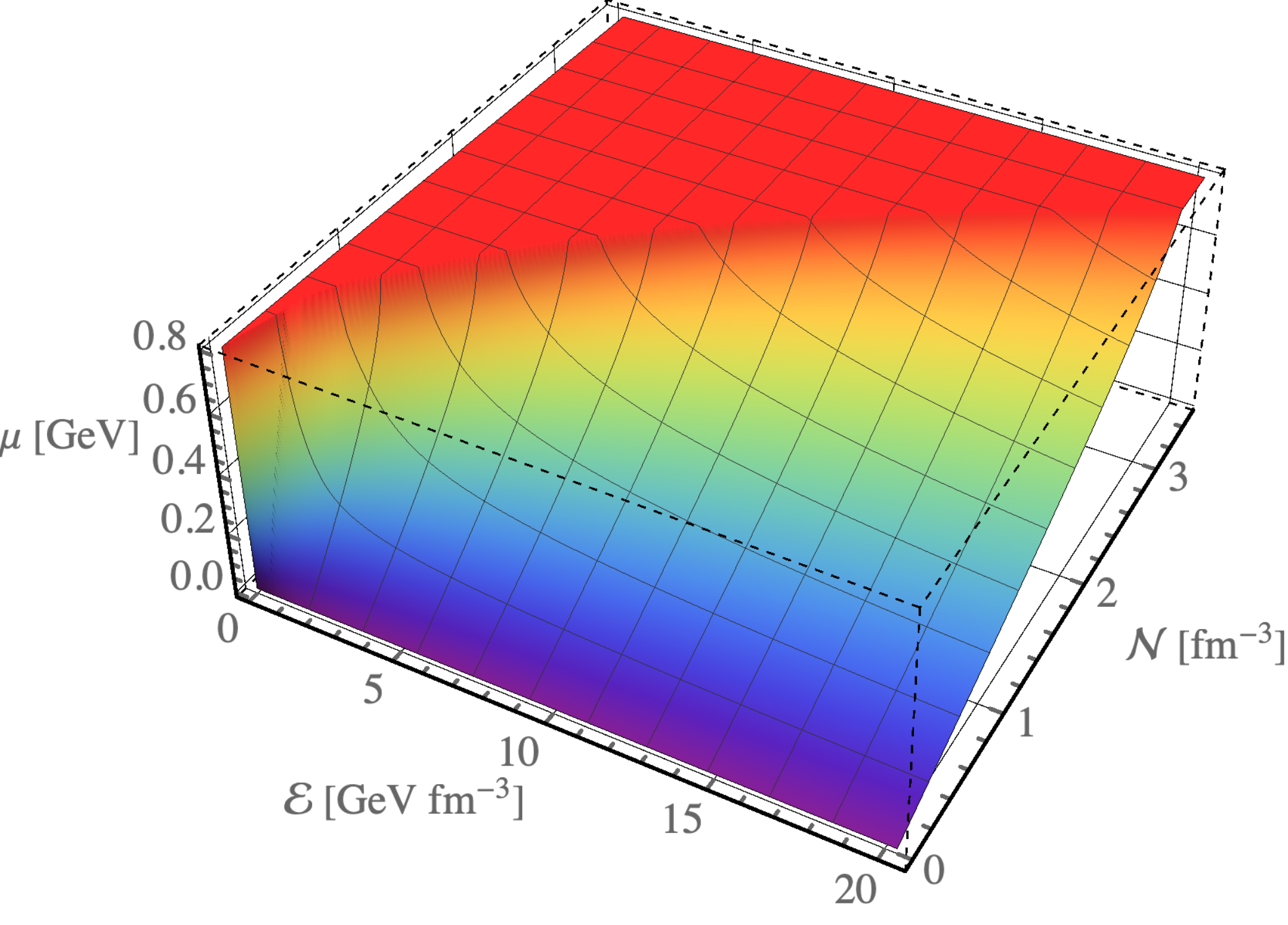}
    \caption{(Color online) EOS4 from Ref.~\cite{Denicol:2018wdp}, restricted to $\mu<800$\,MeV to account for the limited range of validity of the Taylor series extrapolation to non-zero $\mu$ ({\sl left:} $T(\ed, \n)$, {\sl right:} $\mu(\ed, \n)$). In the flat region of the left plot, as $\n$ is increased beyond its edge $\n_\mathrm{edge}(\ed)$, $T(\ed,\n)$, $\mu(\ed,\n)$ and $\peq(\ed,\n)$ are set by hand to remain constant (i.e. $T(\ed,\n>\n_\mathrm{edge}(\ed)) = T(\ed,\n_\mathrm{edge}(\ed))$, etc.). (Note that in the flat region the entropy density $\s$ can go negative if naively calculated from $\s=(\ed+\peq-\mu\n)/T$. Correspondingly, this EOS should not be used for collision systems for which the code makes regular excursions into this region.) The same prescription is then also used in the right plot.
    \label{F2}
    }
\end{figure*}%

EOS4 from Ref.~\cite{Denicol:2018wdp}, extended to finite baryon chemical potential by combining a lattice EoS at high temperature and a HRG EoS at low temperature with a Taylor expansion in $\mu/T$ using techniques discussed in Sec.~\ref{subsec-coneos}, allows to study the evolution of systems with non-zero net baryon density. It is plotted in Fig.~\ref{F2}. In the code, tabulated values for $\peq,\, T$ and $\mu/T$ as functions of $(\ed, \n)$ are included. If the code requires the EoS at $(\ed, \n)$, $\peq, T$ and $\mu/T$ are calculated from nearest neighbors in the table using 2D bilinear interpolation. We note that EOS4 does not include a critical point or first-order phase transition at large $\mu$. A lattice QCD based EoS that includes these features, with adjustable location of the critical point and strength of the first-order transition beyond that point, was constructed by the BEST Collaboration \cite{Parotto:2018pwx} and could be imported into \code\ for future dynamical simulations aiming at helping to locate the QCD critical point.

\subsubsection{Equations of State at finite baryon number, electric charge and strangeness}\label{sec:eos_multi}

As stated above, EoS is determined by the microscopic properties of the medium. In heavy-ion collisions at BES energies, we may assume that the light quarks ($u$, $d$ and $s$) can get thermalized in the QGP, and thus the EoS needs to include non-zero chemical potentials of baryon number, electric charge and strangeness. Recently,  EoS's have been constructed that include these three charges \cite{Noronha-Hostler:2019ayj,Monnai:2019hkn, Monnai:2021kgu}. 

In this case, the chemical potentials of baryon number ($\mu_B$), electric charge ($\mu_Q$) and strangeness ($\mu_S$) are related to those of relevant quarks, in the following way:
\begin{equation}\label{eq:quark_poten}
\mu_u = \frac{1}{3}\mu_B+\frac{2}{3}\mu_Q\,,\quad \mu_d = \frac{1}{3}\mu_B-\frac{1}{3}\mu_Q\,,\quad \mu_s = \frac{1}{3}\mu_B-\frac{1}{3}\mu_Q-\mu_S\,.
\end{equation}
The chemical potential of a hadron species $i$ in HRG model now becomes
\begin{equation}\label{eq:totalchemical}
\mu_i = B_i\mu_B+Q_i\mu_Q+S_i\mu_S\,,
\end{equation}
where $B_i, Q_i$ and $S_i$ are the quantum numbers of baryon number, electric charge and strangeness, respectively. 
The construction of EoS discussed in Sec.~\ref{subsec-coneos} can be extended to non-zero chemical potentials in a straightforward way \cite{Noronha-Hostler:2019ayj,Monnai:2019hkn}. The Taylor series in Eq.~\eqref{Taylor} now becomes
\begin{equation}\label{eq:taylorexp_multi}
    \frac{\peq(T, \mu_B, \mu_Q, \mu_S)}{T^4} = \frac{\peq(T, 0)}{T^4} + \sum_{l,m,n} \frac{\chi^{BQS}_{l,m,n}(T)}{l!m!n!} \left(\frac{\mu_B}{T}\right)^{l}\left(\frac{\mu_Q}{T}\right)^{m}\left(\frac{\mu_S}{T}\right)^{n}\,,
\end{equation}
where the expansion coefficients, i.e., susceptibilities are \cite{Noronha-Hostler:2019ayj,Monnai:2019hkn, Monnai:2021kgu}
\begin{equation}
    \chi^{BQS}_{l,m,n}(T) = \left.\frac{\partial^l\partial^m\partial^n (\peq(T, \mu_B, \mu_Q, \mu_S)/T^4)}{\partial(\mu_B/T)^l\partial(\mu_Q/T)^m\partial(\mu_S/T)^n}\right|_{\mu_{B,Q,S}=0}\,.
\end{equation}
Here matter-antimatter symmetry requires $l+m+n$ to be even. Different methods are used when people interpolate the EoS between HRG and Lattice QCD. Refs.~\cite{Monnai:2019hkn, Monnai:2021kgu} first construct the EoS  at non-zero $\mu_{B,Q,S}$ for the high temperature region using Lattice calculations of $\chi^{BQS}_{l,m,n}(T)$, and then interpolate it with HRG EoS at low temperature using the method in Eq.~\eqref{eq:eosinter}. On the other hand, Ref.~\cite{Noronha-Hostler:2019ayj} instead introduces interpolations for $\chi^{BQS}_{l,m,n}(T)$ between HRG and Lattice QCD calculations for the entire relevant temperature region, and then uses Eq.~\eqref{eq:taylorexp_multi} with the interpolated $\chi^{BQS}_{l,m,n}(T)$ to construct the full EoS at non-zero chemical potentials.

In nuclear collisions, because the colliding nuclei do not have any strange valence quarks and $s$-$\bar s$ pair creation from gluons is a local process, the strangeness density in the produced systems should be initially zero, i.e., $n_S=0$ -- the so-called strangeness neutrality condition. On the other hand, electric density $n_Q$ is related to net baryon density $n_B$, considering the proton-to-nucleon ratio $Z/A$ of the colliding nuclei. For example, in Au and Pb, $Z/A$ is about 0.4, and thus in the systems produced in the collisions of those two nuclei, one can assume $n_Q=0.4n_B$ \cite{Noronha-Hostler:2019ayj,Monnai:2019hkn}. The relations \eqref{eq:quark_poten} between the chemical potentials  can affect the ratios between the final yields of hadron species. As an example, in collisions of neutron-rich nuclei, one may have $\mu_d>\mu_u$, and thus Eq.~\eqref{eq:quark_poten} indicates that in the produced system, $\mu_Q<0$ when $\mu_B>0$. This results in larger yields of $\pi^-$ than $\pi^+$, and Ref.~\cite{Monnai:2019hkn, Monnai:2021kgu} indeed found that using EoS at non-zero chemical potentials can improve the calculated final particle yield ratios compared to experimental data.

\subsection{Equation of State with a critical point}
\label{subsec-coneos-cp}

Near the QCD critical point, the thermodynamic properties of the matter are expected to show singularities, and thus to describe dynamics near the critical point, EoS should include correct singular behaviors. As mentioned in Sec.~\ref{sec:phase_diagram}, the QCD critical point belongs to the same universality class as 3D Ising model \cite{Guida:1996ep, zinn2002quantum, Berges:1998rc, Halasz:1998qr}, from which the universal static critical behavior of QCD critical point can be known. Some authors have constructed EoS's to include critical contribution by mapping the 3D Ising singularities to that of QCD  \cite{Nonaka:2004pg, Parotto:2018pwx, Stafford:2021wik}. Here we briefly summarize the method used in these papers.

Given the pressure of 3D Ising model, $p^\mathrm{Ising}(r, h)$, where $r$ is the reduced temperature and $h$ the magnetic field, one first needs to map it to the pressure of QCD, as a function of $(T, \mu)$. However, the map is not universal and thus different maps would result in different shapes of the critical region where the critical pressure has a significant contribution \cite{Nonaka:2004pg, Parotto:2018pwx, Stafford:2021wik,Pradeep:2019ccv}. Besides, the global scale of the critical pressure is also not known, and larger values for it correspond to larger critical regions. Considering these non-universalities, one can parametrize the critical contribution to the pressure in the critical region, once a map between $(r, h)$ and $(T, \mu)$ is chosen:
\begin{equation}
p^\mathrm{critical}_\mathrm{QCD}(T, \mu) \sim p^\mathrm{Ising}(r(T, \mu), h(T, \mu))\,,
\end{equation}
where a function of $(T, \mu)$ can be added to set the overall scale (see Refs.~\cite{Parotto:2018pwx, Stafford:2021wik}). Then the full pressure is written as the sum of the Ising contribution (i.e., $p^\mathrm{critical}_\mathrm{QCD}(T, \mu)$) and a non-Ising one where the latter is not known {\it a priori}. Refs.~\cite{Parotto:2018pwx, Stafford:2021wik} construct the non-Ising pressure using the same Tylor expansion method described above, and the expansion coefficients at $\mu=0$ are calculated by subtracting the Taylor coefficients of the Ising model (with a factor relevant to the overall function mentioned  above) from the ones calculated from Lattice QCD. We note that large uncertainties and many parameters exist in the construction of an EoS with a critical point. Constraining these parameters from model-to-data comparisons, when the entire framework is already very complicated, is certainly not going to be straightforward.

\section{Particlization and hadronic afterburner}\label{sec:part_hadr}

During the final stage of the evolution, because of the expansion, the system gets dilute and temperature becomes low, and thus the microscopic d.o.f.~are expected to change from quarks and gluons to hadron resonances, through color reconfinement. This process increases the mean free path and thus the Knudsen number, and as a consequence, the hydrodynamic description rapidly breaks down. In practice, hybrid models of heavy-ion collisions switch from a macroscopic hydrodynamic description to a kinetic hadron description on a so-called particlization surface. 

Usually, particlization is carried out on a space-time hyper-surface with a constant switching temperature $T_f$ or energy density $e_f$. Although the hyper-surface on which particlization happens is often called ``freeze-out surface'', the corresponding temperature or energy density is not necessarily the one where chemical freeze-out happens. Particlization is not a dynamic physical process but rather a change of langue, from a macroscopic description to a microscopic one, both of which are expected to be equivalent around the switching temperature or energy but each of which uses different dynamical degrees of freedom and evolution equations. 

\subsection{Converting a fluid to particles}
\label{subsec-particl}

Particlization of a fluid can be carried out via the Cooper-Frye prescription which conserves energy-momentum and charges  \cite{Cooper:1974mv}. In this prescription the particle distributions are given as
\begin{equation}\label{eq:distr1}
    \frac{d^3N_i}{p_Tdp_Td\phi_pdy}=\frac{1}{(2\pi)^3}\int_\Sigma d^3\Sigma_\mu(x) p^\mu f_i(x, p)=\frac{1}{(2\pi)^3}\int_\Sigma d^3\Sigma_\mu(x) p^\mu (f_{0,i}(x, p)+\delta f_i(x, p))\,,
\end{equation}
where $\Sigma$ is the 3-dimensional particlization surface, with surface element at point $x$ encoded by an outward-pointing normal vector $d^3\Sigma_\mu(x)$, $p^\mu$ is the four-momentum of the particle,  and $f_i(x, p)$ is the one-particle distribution function for species $i$. Generally $f_{0,i}(x, p)$ is chosen as the Maxwell-J\"uttner distribution, assuming the fluid cell is in local thermal and chemical equilibrium, while $\delta f_i(x, p)$ describes the off-equilibrium corrections:
\begin{equation}\label{eq:deltf_terms}
\delta f_i(x, p) = \delta f_i^\mathrm{shear}(x, p)+\delta f_i^\mathrm{bulk}(x, p)+\delta f_i^\mathrm{diffusion}(x, p)\,,
\end{equation}
where the three terms represent contributions from the dissipative effects from viscous shear stress, bulk viscous pressure, and baryon diffusion current, respectively \cite{Denicol:2018wdp, McNelis:2019auj, McNelis:2021acu}.

Specifically, for single conserved charge $Q$, the leading contribution $f_{0,i}(x, p)$ reads
\begin{equation}\label{eq:distr_f1}
    f_{0,i}(x, p)= g_i \left[\exp{\left(\frac{u^\mu p_\mu-Q_i\mu_f}{T_f}\right)} +\Theta_i\right]^{-1}\,.
\end{equation}
Here $i$ denotes the particle species, $g_i$ the corresponding spin degeneracy factor,\footnote{%
	We treat different members of an isospin multiplet as different hadron species.
}
$\Theta_i\eq\pm1$ accounts for the quantum statistics of fermions (+) and bosons ($-$), $Q_i$ the baryon charge of species $i$, $u^\mu$ the flow velocity, $T_f$ the freeze-out temperature and $\mu_f$ freeze-out chemical potential associated with the conserved charge $Q$. Note that the hydrodynamic quantities $u^\mu$, $T_f$ and $\mu_f$ describing fluid properties are different from point to point on the freeze-out surface. 

The chemical potential in Eq.~\eqref{eq:distr_f1} can be extended to multiple conserved charges according to Eq.~\eqref{eq:totalchemical}. As mentioned, Refs.~\cite{Monnai:2019hkn, Monnai:2021kgu} showed that using an EoS with non-zero $BQS$-chemical potentials imposing strangeness neutrality and a fixed ratio of electric and baryon charges can help to improve the particle-antiparticle ratios for strange hadrons. These authors also found that, although the electric chemical potential does not make much difference in the particle yield ratios, it is of phenomenological importance to explain experimental measurements, e.g., a small excess of $\pi^+$ over $\pi^-$ among the detected particles.

\subsubsection{Viscous correction models}

Constraints from matching the net baryon current and energy-momentum tensor, i.e., the first and second momentum moments of the distribution function $f_i(x, p)$, can not fully specify the viscous correction $\delta f_i(x, p)$ during particlization, leaving on infinity of higher-order moments unconstrained. Different ansaetze for the momentum dependence of $\delta f_i(x, p)$ resolve this ambiguity by making different assumption, leading to different models for the viscous correction terms. Examples are the linearized viscous corrections (such as the Grad (or 14-moments) and RTA Chapman-Enskog approximations), exponentiated viscous corrections (including PTM and PTB distributions), and the so-called maximum entropy viscous correction (see Refs.~\cite{McNelis:2019auj,McNelis:2021acu,Everett:2021ulz} for more details). Recently, the JETSCAPE collaboration \cite{Putschke:2019yrg} studied the differences between four viscous correction models for $f_i^\mathrm{shear}$ and $\delta f_i^\mathrm{bulk}$ in Eq.~\eqref{eq:deltf_terms}, which provide different prescriptions on how the collective energy-momentum of a hydrodynamic fluid are distributed among different hadron species and across their momenta, for top RHIC and LHC energies \cite{Everett:2020xug, Everett:2020yty}. By including the associated model ambiguities in the particlization process, the collaboration provided the most reliable phenomenological constraints to date on the QGP viscosities using Bayesian inference methods \cite{Everett:2020yty}.

\subsubsection{Conservation laws and statistical ensembles}

We note that in a conventional sampling procedure, given a freeze-out hyper-surface $\Sigma$ and the choice of $\delta f_i(x, p)$, hadron species are sampled according to Eq.~\eqref{eq:distr1} on each surface element independently (see e.g., Refs.~\cite{McNelis:2019auj,Shen:2014vra}). To increase the statistics and thus reduce the statistical uncertainties, one usually samples particles from a given hydrodynamic hypersurface multiple times -- known as the ``over-sampling'' method. The samples obtained in this way are members of a grand canonical ensemble; i.e., during the sampling, only the temperature, chemical potentials and volume are fixed, while the energy, momentum and the quantum numbers are only conserved on average. With over-sampling, the grand conical sampling can provide good approximations for bulk observables and is computationally fast in practice, while on an event-by-event level it can violate the conservation laws at particlization \cite{Huovinen:2012is, Schwarz:2017bdg}. The error caused by this violation can be expected to be larger for lower collision energies, smaller collision systems and rare species \cite{Schwarz:2017bdg}. The fluctuations in final particle multiplicity introduced by the violation may also affect the selection of centrality classes \cite{Huovinen:2012is}. 

Using observables involving fluctuations and correlations to search for the critical point would need a faithful calculation of those observables, and thus the unphysical fluctuations caused by the grand canonical sampling should be avoided and physical fluctuations from the fluid should be preserved properly. For example, in the conventional sampling, particles on different surface elements are sampled independently, and thus the correlations propagated in, say, \bes+~will be lost. Depending on whether only the net charges including net baryon number, strangeness and electric charge, or additionally the total energy  are conserved as well, one should instead employ canonical or micro-canonical sampling, respectively \cite{Schwarz:2017bdg}. Note that the conservation laws can be fulfilled globally or locally on the freeze-out surface, where the former conserves quantities on the entire surface \cite{Schwarz:2017bdg} while the latter on smaller regions \cite{Oliinychenko:2019zfk,Oliinychenko:2020cmr}.

\begin{figure}[!t]
\begin{center}
\includegraphics[width= 0.55\textwidth]{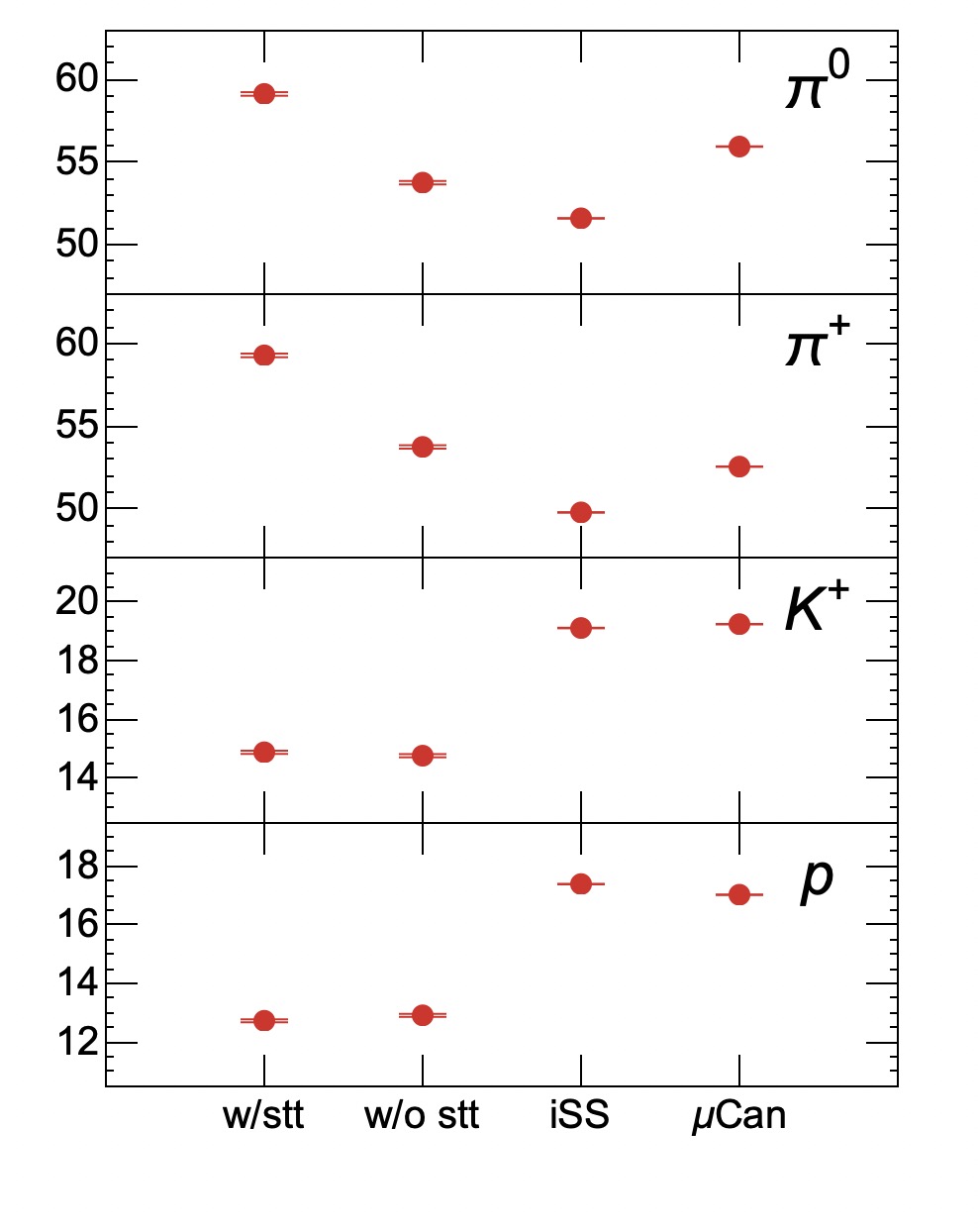}
\caption{Comparison for identified hadron yields among three particle samplers with the same freeze-out surface obtained from Au-Au collisions at $\snn=19.6$ GeV, where for all three samplers viscous corrections are ignored. The left two columns are obtained from the {\sc iS3D} sampler \cite{McNelis:2019auj}, and the right two columns from the {\sc iSS} sampler \cite{Shen:2014vra} and the micro-canonical sampler \cite{Oliinychenko:2019zfk}, respectively. Only the left-most column uses quantum statistics while others use Boltzmann statistics. Figure shows preliminary results (L. Du and S. Shi, unpublished).}
\label{fig:samp_comp}
\end{center}
\end{figure}

The authors of Refs.~\cite{Oliinychenko:2019zfk,Oliinychenko:2020cmr} implemented the local micro-canonical sampling, where the conservation laws are strictly implemented on ``patches'' -- space-time regions smaller than the entire surface but still containing many particles. However, how to split the surface into patches, and thus the number of such patches, is not known and was treated as a physical parameter in the work \cite{Oliinychenko:2019zfk,Oliinychenko:2020cmr}. Because of the conservation laws, the sampled particles of different species are found to be correlated, deviating from the conventional grand canonical sampling method with no such constraints. As mentioned, mapping the hydrodynamically propagated critical correlations in coordinate space onto final-state correlations among the particles  in momentum space is essential where searching for the critical point through model-data comparison, and employing local micro-canonical sampling during particlization will be crucial in such studies.

Fig.~\ref{fig:samp_comp} compares the identified hadron yields obtained from three samplers for the same freeze-out surface. Considerable differences can be seen even for bulk observables such as particle yields, which implies that more investigations are needed to better understand the sampling methods.
We also note that there are attempts to sample particles with interactions, for example, according to the Equation of State of an interacting HRG model \cite{Vovchenko:2020kwg}, or calculating (without sampling) correlations of multiplicity fluctuations at different (pseudo-)rapidities from off-equilibrium critical fluctuations propagated in the hydrodynamic stage within the {\sc hydro+} framework (Maneesha Pradeep {\it et al.}, CPOD2021, unpublished).

\subsection{Kinetic hadron transport}
\label{subsec-hadtrans}

Once the particles are sampled on the hyper-surface, they can be handed over to a kinetic hadron transport description for further scattering, resonance formation and decay until all reactions cease (``kinetic freeze-out''). {\sc UrQMD} \cite{Bass:1998ca,Bleicher:1999xi} and SMASH \cite{Weil:2016zrk} are two popular transport codes used by the heavy-ion community, among a total of  about 15 such codes existing (see Refs.~\cite{Zhang:2017esm,Ono:2019ndq} for systematic comparisons among those codes). Those transport models provide the phase space information of all particles, and solve the following relativistic Boltzmann equation \cite{Weil:2016zrk}:
\begin{equation}
p^\mu\partial_\mu f_i(x,p)+m_iF^\alpha\partial_\alpha^p f_i(x,p) = C^i_\mathrm{coll}\,,
\end{equation}
where $C^i_\mathrm{coll}$ is the collision term describing all scatterings, resonance formation and decays, and $F^\alpha$ are any extermal forces experienced by each particle. The external force $F^\alpha$ is usually taken to be zero for high energy collisions, but at low beam energies, $F^\alpha=-\partial^\alpha U(x)$ with $U(x)$ being the mean-field interaction potential among the hadrons. Within such a transport approach, particles of different species can achieve their chemical and kinetic freeze-out dynamically.

Depending on the observables, the hadronic afterburner stage with scatterings may play roles of various significance, and when scatterings change certain observables only slightly, the afterburner can be replaced by performing only resonance decays. For example, because of large annihilation cross section, yields of baryon and anti-baryon can change significantly by collisions, and thus carrying out scatterings in the afterburner stage is still important for obtaining the correct net proton yields. On the other hand, hadronic rescatterings are less important for computing the total charged hadron's pseudo-rapidity densities $dN_\mathrm{ch}/dy$. For phenomenological studies focusing on direct resonance decays without rescatterings, Monte Carlo simulations of decay processes of resonances are needed. We note that there are studies which bypass calculating intermediate decay processes and directly achieve final particle spectra from hydrodynamic fields on the freeze-out surface \cite{Mazeliauskas:2018irt}.
At BES energies anti-baryons are rare and baryon-antibaryon annihilation plays essentially no role. Therefore full hadronic transport only slightly changes the pseudo-rapidity distributions of charged particles and the rapidity-distributions of net protons, in comparison to carrying out only resonance decays without scatterings \cite{Denicol:2018wdp}. However, the $p_T$-differential elliptic flows of pions, protons and anti-protons are found to be affected considerably by a hadronic afterburner stage, compared to the case with only resonance decays. This is because the hadronic afterburner increases the lifetime of the evolving system, and thus residual spatial anisotropies can continue to be converted into momentum anisotropies \cite{Denicol:2018wdp}.

On the other hand, the hadronic rescattering stage is essential in theoretical calculations of conserved charges' cumulants when searching for the critical point. First, it is needed to understand the effects of efficiency and acceptance corrections in the experimental measurements. Besides, the cumulants of conserved charges may continue to evolve dynamically during the hadronic stage (see, e.g., Refs.~\cite{Steinheimer:2016cir,Asakawa:2019kek}). Existing studies usually  compare the susceptibilities from Lattice QCD and HRG calculations  at the pseudo-critical temperature $T_{pc}$ directly to experimental measurements of cumulants of charges, by assuming they are comparable. The evolution of such cumulants and correlations of conserved charges during the hadronic stage certainly should be investigated more carefully, once we know how to sample correlated particles appropriately.

\section{Summary}

In this chapter, we briefly reviewed some essential ingredients of a multistage framework for heavy-ion collisions at BES energies and recent theoretical developments on various topics. Because of the highly dynamical nature of systems produced in nuclear collisions, such a framework capturing all dynamics of phenomenological importance is the foundation for making experimental discoveries on QCD critical phenomena. Even with a sophisticated framework readily at hand, carrying out realistic simulations is computationally expensive with a set of fixed parameters, not to mention the necessary large parameter space scan for model-data comparison. For this reason, testing the importance of different dynamical ingredients and identifying the ones that are really of essence is a useful exercise.

\chapter{(3+1)D hydrodynamic simulations at non-zero baryon density}
\label{ch:numerics}

In this chapter, we present the (3+1)-dimensional diffusive relativistic hydrodynamic code \code\ which solves the equations of motion of second-order Denicol-Niemi-Molnar-Rischke (DNMR) theory in Sec.~\ref{sec:relat_hydro}, including bulk and shear viscous currents and baryon diffusion currents. {\sc BEShydro} features a modular structure that allows to easily turn on and off baryon evolution and different dissipative effects and thus to study their physical effects on the dynamical evolution individually. An extensive set of test protocols for the code, including several novel tests of the precision of baryon transport that can also be used to test other such codes, is documented here and supplied as a permanent part of the code package. As far as we know, at this point in time the only other open source code that shares all of the main features of \code\ is the latest version of {\sc MUSIC} \cite{Denicol:2018wdp}, while other codes (e.g. \cite{Karpenko:2013wva, Bazow:2016yra, Pang:2018zzo}) so far ignore the evolution of the net baryon diffusion current or, in some cases, even that of the net baryon charge. \code\ has been developed completely independently of {\sc MUSIC}; it can thus serve as a platform for detailed code validations and comparisons, even if in future applications the two codes will likely be applied to different collision systems, using different initialization modules and hadronic afterburners.

This chapter is based on material published in Ref.~\cite{Du:2019obx}. The code is open source and can be freely downloaded from Ref.~\cite{beshydrogit}.

\section{Conservative form of the evolution equations}
\label{appa}

In this section, using the definition of $d \equiv u^\mu \partial_\mu$, we recast Eqs.~(\ref{relEqs_Tmut}-\ref{relEqs_n}) and  (\ref{relEqs_Pi}-\ref{relEqs_pi}), from Sec.~\ref{sec2.1.3}, in the same first-order flux-conserving form
\begin{equation}
    \partial_\tau q + \partial_x (v^x q) + \partial_y (v^y q) + \partial_\eta (v^\eta q) = S_q\;,
\label{numericalEquations}
\end{equation}
where $v^i \equiv u^i/u^\tau$ ($i = x, y, \eta_s$) is the 3-velocity of the fluid, the conserved quantity $q$ can be any component (or linear combination of components) of $T^{\mu\nu}$ and $N^{\mu}$, and $S_q$ is the corresponding source term. This form allows all quantities to be evolved with the same numerical transport algorithm. The procedure follows Refs.~\cite{Bazow:2016yra, Molnar:2009tx}, adding here the equations for baryon evolution. The equations reproduced here are written in a form that facilitates direct comparison with the \code\ code.

With the scaled flow velocities $v^i$, we can write down the following constituent relations for the components of $T^{\mu\nu}$ and $N^\mu$:
\begin{align}
T^{\tau\tau}&=(\ed+P)u^{\tau}u^{\tau}-P+\pi^{\tau\tau}\;,
\label{Ttt} \\
T^{\tau i}&=(\ed+P)u^{\tau}u^{i}+\pi^{\tau i} =v^{i}T^{\tau\tau}+P v^{i}-v^{i}\pi^{\tau\tau}+\pi^{\tau i}\;,
\label{Tti} \\
T^{ij}&=(\ed+P)u^{i}u^{j}-P g^{ij}+\pi^{ij} =v^{i}T^{\tau j}-P g^{ij}-v^{i}\pi^{\tau i}+\pi^{ij}\;,
\label{Tij} \\
N^\tau &= \n u^\tau + n^\tau\;, \\
N^i &=\n u^i + n^i = v^i N^\tau - v^i n^\tau + n^i\;.
\end{align}
Here we introduced $P=\peq+\Pi$. Inserting these into the conservation laws in Eqs.~(\ref{relEqs_Tmut}-\ref{relEqs_n}), one obtains
\begin{align}
\partial _{\tau }T^{\tau \tau }+\partial _{x}(v^{x}T^{\tau \tau})+\partial_{y}(v^{y}T^{\tau \tau })+\partial _{\eta }(v^{\eta}T^{\tau
\tau }) &= I^\tau_2 + I^\tau_x +I^\tau_y +I^\tau_\eta\,, 
\\%
\partial _{\tau }T^{\tau x}+\partial _{x}(v^{x}T^{\tau x})+\partial
_{y}(v^{y}T^{\tau x})+\partial _{\eta }(v^{\eta}T^{\tau x}) &= I^x_2 + I^x_x +I^x_y +I^x_\eta\,,
\\%
\partial _{\tau }T^{\tau y}+\partial _{x}(v^{x}T^{\tau y})+\partial
_{y}(v^{y}T^{\tau y})+\partial _{\eta }(v^{\eta}T^{\tau y}) &= I^y_2 + I^y_x +I^y_y +I^y_\eta \,, 
\\%
\partial _{\tau }T^{\tau \eta }+\partial _{x}(v^{x}T^{\tau \eta
})+\partial _{y}(v^{y}T^{\tau \eta })+\partial _{\eta }(v^{\eta}T^{\tau
\eta }) &= I^\eta_2 + I^\eta_x +I^\eta_y +I^\eta_\eta \,, \label{dtT03}
\\%
\partial _{\tau }N^{\tau}+\partial _{x}(v^{x}N^{\tau})+\partial _{y}(v^{y}N^{\tau})+\partial _{\eta }(v^{\eta}N^{\tau})&= J^\tau_2 + J^\tau_x +J^\tau_y +J^\tau_\eta\;, \label{baryonconservation}
\end{align}
with the following source terms for $T^{\tau\tau }$:
\begin{align}
I^\tau_2=&-\frac{1}{\tau }\left( T^{\tau \tau }+\tau ^{2}T^{\eta \eta}\right)-\left(\peq+\Pi-\pi ^{\tau \tau }\right)\partial _{i}v^i-v^i\partial_i \peq\,, \\
I^\tau_x= &-v^{x}\partial _{x}\left(\Pi-\pi ^{\tau \tau }\right)-\partial _{x}\pi ^{\tau x}\,, \\
I^\tau_y= &-v^{y}\partial _{y}\left(\Pi-\pi ^{\tau \tau }\right)-\partial _{y}\pi ^{\tau y}\,, \\
I^\tau_\eta= &-v^{\eta}\partial _{\eta}\left(\Pi-\pi ^{\tau \tau }\right)-\partial _{\eta}\pi ^{\tau \eta}\,; 
\end{align}
for $T^{\tau x}$:
\begin{align}
I^x_2=&-\frac{1}{\tau }T^{\tau x}-\partial_{x}\peq +\pi ^{\tau x}\partial _{i}v^i\,, \\
I^x_x= &-\partial _{x}\left( \Pi+\pi ^{xx}\right)+v^{x}\partial _{x}\pi ^{\tau x}\,,\\
I^x_y =  & -\partial_{y}\pi ^{xy}+v^{y}\partial_{y}\pi ^{\tau x}\,, \\
I^x_\eta=  &-\partial_{\eta}\pi ^{x\eta}+v^{\eta}\partial_{\eta}\pi ^{\tau x}\,;
\end{align}
for $T^{\tau y}$:
\begin{align}
I^y_2=&-\frac{1}{\tau }T^{\tau y}-\partial_{y}\peq +\pi ^{\tau y}\partial _{i}v^i\,,\\
I^y_x=&-\partial _{x}\pi ^{xy}+v^{x}\partial _{x}\pi ^{\tau y}\,,\\
I^y_y=&-\partial_{y}\left(\Pi+\pi ^{yy}\right)+v^{y}\partial_{y}\pi ^{\tau y}\,, \\
I^y_\eta=&-\partial_{\eta}\pi ^{y\eta}+v^{\eta}\partial_{\eta}\pi ^{\tau y}\,; 
\end{align}
for $T^{\tau \eta}$:
\begin{align}
I^\eta_2=&-\frac{3}{\tau}T^{\tau\eta}
          -\frac{\partial_{\eta}\peq}{\tau^{2}}
          +\pi^{\tau\eta}\partial_{i}v^i\,,
\\
I^\eta_x=&-\partial_{x}\pi^{x\eta}+v^{x}\partial_{x}\pi^{\tau\eta}\,,
\\
I^\eta_y=&-\partial_{y}\pi ^{y\eta}+v^{y}\partial_{y}\pi^{\tau\eta}\,,
\\
I^\eta_\eta=&-\partial_{\eta }\left(\Pi/\tau ^{2}+\pi^{\eta\eta}\right)
+ v^{\eta}\partial_{\eta }\pi^{\tau \eta} \,;
\end{align}
and for $N^\tau$:
\begin{align}
J^\tau_2=&-\frac{1}{\tau } N^{\tau}+n^\tau\partial _{i}v^i \,,
\\
J^\tau_x=&-\partial _{x}n ^{x} +v^{x}\partial _{x}n ^{\tau}\,,
\\
J^\tau_y=& -\partial _{y}n^{y} +v^{y}\partial _{y}n^{\tau}\,,
\\
J^\tau_\eta=& -\partial_{\eta }n^{\eta} 
           +v^{\eta} \partial_{\eta}n^{\tau}\;. 
\end{align}
Note that the above subscripts $x, y, \eta$ of the source terms are chosen based on which components the derivatives are with respect to on the RHS.

Considering $d \equiv u^\mu \partial_\mu$, the relaxation equations (\ref{relEqs_Pi}-\ref{relEqs_pi}) for the dissipative flows can be written as
\begin{eqnarray}
    \partial_{\tau}\Pi+\partial_{x}(v^{x}\Pi)+\partial_{y}(v^{y}\Pi)+\partial_{\eta}(v^{\eta}\Pi)&=&S^\Pi_2 \;,
\label{relEqs_Pia}\\
    \partial_{\tau}\V^\mu+\partial_{x}(v^{x}\V^\mu)+\partial_{y}(v^{y}\V^\mu)+\partial_{\eta}(v^{\eta}\V^\mu) &=&S^\V_2
\label{relEqs_na}\;,\\
    \partial_{\tau}\pi^{\mu\nu} +\partial_{x}(v^{x}\pi^{\mu\nu})+\partial_{y}(v^{y}\pi^{\mu\nu})+\partial_{\eta}(v^{\eta}\pi^{\mu\nu}) &=&S^\pi_2 \;,
\label{relEqs_pia}
\end{eqnarray}
where we used $\partial_{i}v^{i} \equiv \partial_{x}v^{x} + \partial_{y}v^{y} + \partial_{\eta}v^{\eta}$. The source terms in Eqs.~(\ref{relEqs_Pia}-\ref{relEqs_pia}) are given by 
\begin{eqnarray}
    S^\Pi_2&=&\frac{1}{u^\tau}\left(-\frac{\zeta}{\tau_\Pi}\theta-\frac{\Pi}{\tau_\Pi}-I_\Pi\right)+\Pi\partial_i v^i
    \label{relEqs_SPia}\;,\\
    S^\V_2&=&\frac{1}{u^\tau}\left(\frac{\kappa_n}{\tau_\V} \nabla^{\mu }\left(\frac{\mu}{T}\right)-\frac{\V^{\mu}}{\tau _{\V}}-I_\V^\mu-G_\V^\mu\right)+\V^\mu\partial_i v^i 
    \label{relEqs_Sna}\;,\\
    S^\pi_2&=&\frac{1}{u^\tau}\left(\frac{2\eta}{\tau_{\pi}}\sigma^{\mu\nu}-\frac{\pi^{\mu\nu}}{\tau_{\pi}}-I^{\mu\nu}_{\pi}-G^{\mu\nu}_{\pi}\right)+\pi^{\mu\nu}\partial_i v^i\;.
    \label{relEqs_Spia}
\end{eqnarray}
Here the $I$-terms and $G$-terms are defined in Eqs. (\ref{relEqs_Pi})-(\ref{relEqs_pi}). In Milne coordinates the terms $G^\mu_n = u^\alpha \Gamma^\mu_{\alpha\beta} n^\beta$ in Eq.~\eqref{relEqs_Sna} evaluate to 
\begin{equation}
G^\tau_n = \tau u^\eta n^\eta\,,\quad
G^x_n = 0\,,\quad
G^y_n = 0\,,\quad
G^\eta_n = (u^\tau n^\eta + u^\eta n^\tau)/\tau\,.
\end{equation}
To calculate the LRF gradient of $\mu/T$ in Eq.~\eqref{relEqs_Sna} numerically we use
\begin{equation}
 \nabla^{\mu}\left(\frac{\mu}{T}\right) \equiv \Delta^{\mu\nu} d_\nu\left(\frac{\mu}{T}\right) = (g^{\mu\nu}-u^\mu u^\nu)\partial_\nu\left(\frac{\mu}{T}\right)
\end{equation}
and work out the partial derivatives $\partial_\mu (\mu/T)$ in the computational frame numerically from the EoS tables. Finally, the last two terms in Eq.~(\ref{I1-I4-mu}) can be expressed as
\begin{align}
I_3^\mu& = n_{\nu }\omega ^{\nu \mu } = n^{\tau }\omega ^{\tau \mu } - n^x\omega ^{x \mu } - n^y\omega ^{y \mu } - \tau^2 n^{\eta }\omega ^{\eta \mu }\;,\\
I_4^\mu& = n_{\nu }\sigma ^{\nu \mu }= n^\tau\sigma^{\tau \mu } - n^x\sigma^{x \mu } - n^y\sigma^{y \mu }  - \tau^2 n^\eta \sigma^{\eta \mu }\;.
\end{align}

\section{Numerical scheme}
\label{numericalscheme}

We now describe the numerical scheme used in \code\ to solve the coupled set of evolution equations (the conservation laws together with the dissipative relaxation equations and the EoS) discussed in the previous section. Initial values for all components of the baryon charge current and energy-momentum tensor are set on a surface of constant longitudinal proper time $\tau$.\footnote{%
        A dynamical initialization routine with sources for the divergences of the baryon current and energy-momentum tensor that describe the gradual ``hydrodynamization'' of the matter produced in the collision \cite{Shen:2017bsr, Akamatsu:2018olk, Du:2018mpf}
        has been discussed in Sec.~\ref{sec:dyna_init}.}
We focus our attention on aspects of the algorithm related to the evolution and influence of the baryon density and diffusion currents, referring interested readers to Refs. \cite{Schenke:2010nt, Molnar:2009tx, Bazow:2016yra} for additional technical details.

\subsection{The Kurganov-Tadmor algorithm}
\label{KT}

\code\ is designed for flexibility so that different physical limits can be easily studied, by allowing one to switch off some components in Eq.~\eqref{eq:qvector}. A few examples of physical cases that can be studied are listed in Table~\ref{tab:codecase}. Only the propagation of $T^{\tau\tau}$, $T^{\tau x}$, $T^{\tau y}$, and $T^{\tau \eta}$ is always turned on. The evolution of all other variables (dissipation and/or baryon related) can be conveniently turned on and off independently; only the propagation of $\V^\mu$ requires the evolution of $N^{\tau}$ to be turned on.  In the code, the components are arranged as follows:
\begin{eqnarray}\label{eq:qvector}
T^{\tau\tau}, T^{\tau x}, T^{\tau y}, T^{\tau \eta},
\pi^{\tau\tau}, \pi^{\tau x}, \pi^{\tau y}, \pi^{\tau \eta}, \pi^{xx}, \pi^{xy}, \pi^{x\eta}, \pi^{yy}, \pi^{y\eta}, \pi^{\eta\eta},
\Pi, N^\tau, \V^\tau, \V^x, \V^y, \V^\eta\,,
\end{eqnarray}
in $q$, when all of them are turned on.

\begin{table}[htp]
\begin{center}
\begin{tabular}{c|c}
\hline\hline
Components switched on & Physical cases\\
\hline\hline
$T^{\tau\tau}, T^{\tau x}, T^{\tau y}, T^{\tau \eta}$ & ideal hydrodynamics at $\mu=0$\\
\hline
$T^{\tau\tau}, T^{\tau x}, T^{\tau y}, T^{\tau \eta}, N^\tau$ & ideal hydrodynamics at non-zero $\mu$\\
\hline
$T^{\tau\tau}, T^{\tau x}, T^{\tau y}, T^{\tau \eta}, N^\tau, \V^\tau, \V^x, \V^y, \V^\eta$ & diffusive hydrodynamics at non-zero $\mu$\\
\hline
$T^{\tau\tau}, T^{\tau x}, T^{\tau y}, T^{\tau \eta}, \pi^{\tau\tau}, \pi^{\tau x}, \pi^{\tau y}$,\\ $\pi^{\tau \eta}, \pi^{xx}, \pi^{xy}, \pi^{x\eta}, \pi^{yy}, \pi^{y\eta}, \pi^{\eta\eta}, N^\tau$ & viscous hydrodynamics at non-zero $\mu$ (no diffusion)\\
\hline\hline
\end{tabular}
\end{center}
\caption{A few examples of physical cases which can be studied in \code, by switching on the relevant components in Eq.~\eqref{eq:qvector}.}
\label{tab:codecase}
\end{table}%

To solve equation (\ref{numericalEquations}) we use the Kurganov-Tadmor (KT) algorithm \cite{KURGANOV2000241}, with a second-order explicit Runge-Kutta (RK) ordinary differential equation solver \cite{leveque_2002} for the time integration step. This scheme is widely used in relativistic hydrodynamic simulations (see, e.g., \cite{Schenke:2010nt, Pang:2018zzo, Bazow:2016yra}). 

\subsection{Numerical derivatives}

For the source terms $S_q$ we must evaluate spatial and temporal derivatives of the thermodynamic variables and dissipative flows. For the time derivatives the code uses first-order forward differences:
\begin{equation}
    \partial_{\tau}A^{n}_{i,j,k}=\frac{A^{n}_{i,j,k}-A^{n-1}_{i,j,k}}{\Delta\tau}\,. 
\label{eq-time-de}
\end{equation}
Here $A$ is the quantity to be differentiated, $i,j,k$ are  integer labels for the $x, y$, and $\eta_s$ coordinates of the grid point, and $\Delta \tau$ is the temporal grid size (numerical resolution in the $\tau$ coordinate). $n$ and $n-1$ are temporal indices denoting the present and preceding time step. To initialize the temporal evolution code at the first time step $n=1$ we set $A^{0}_{i,j,k} = A^{1}_{i,j,k}$. This is especially important when the initial flow velocity $u^\mu$ is non-zero, for example in the case of the Gubser flow test described in Ch.~\ref{ch.gubser}.

The code provides two methods for calculating spatial derivatives. The first uses second-order central differences, i.e. the derivative of any quantity $A$, say, with respect to $x$ is calculated as
\begin{equation}
    \partial_{x}A^n_{i,j,k} = \frac{A^n_{i+1,j,k}-A^n_{i-1,j,k}}{2\Delta x},\label{eq-cendiff}
\end{equation}
where $\Delta x$ is the numerical resolution (grid size) in $x$ direction. The boundary conditions are taken care of by introducing ghost cells on the boundary as described in \cite{Bazow:2016yra}.

The second method calculates the spatial derivative from a combination of second-order central and first-order backward and forward derivatives, using the generalized minmod flux limiter:
\begin{equation}
    \partial_{x}A^n_{i,j,k} = \mathrm{\texttt{minmod}}\left(
\theta_\mathrm{f}\frac{A^n_{i,j,k} - A^n_{i-1,j,k}}{\Delta x},\,
\frac{A^n_{i+1,j,k} - A^n_{i-1,j,k}}{2\Delta x},\,
\theta_\mathrm{f}\frac{A^n_{i+1,j,k} - A^n_{i,j,k}}{\Delta x}
\right)\;,
\label{eq-approx-derivative}
\end{equation}
where the multivariate minmod function is defined as
\begin{align}
\mathrm{\texttt{minmod}}(x,y,z)\equiv \mathrm{\texttt{minmod}}(x,\mathrm{\texttt{minmod}}(y,z))\,,
\end{align}
with $\mathrm{\texttt{minmod}}(x,y)\equiv[\mathrm{\texttt{sgn}}(x)+\mathrm{\texttt{sgn}}(y)]\cdot
\mathrm{\texttt{min}}(|x|,|y|)/2$ and $\mathrm{\texttt{sgn}}(x)\equiv |x|/x $. In other words, $\mathrm{\texttt{minmod}}(x,y,z)$ always gives the value which is the closest to 0 among $(x,y,z)$.  The parameter $\theta_\mathrm{f}\in[1,2]$; $\theta_\mathrm{f}=1$ ($\theta_\mathrm{f}=2$) corresponds to the most (least) dissipative limiter. In \code\ Eq.~(\ref{eq-approx-derivative}) is used only for the derivatives of $u^\mu$ and $\peq$, and only when selected by the user as an option.

\subsection{Root finding with baryon current}
\label{sec-root}

The code evolves the components of the energy momentum tensor in the global computational frame, but the EoS (which is needed to close the set of evolution equations) and the computation of the source terms on the r.h.s. of Eq.~(\ref{numericalEquations}) require knowledge of fluid velocity $u^\mu$ and the energy and baryon density in the local rest frame of the fluid. Computing the latter from the former is known as the ``root finding'' problem. This must be done as efficiently as possible since this problem must be solved at every point of the computational space-time grid.

At finite baryon density, with nonzero baryon diffusion current, the root finding algorithm becomes more complex than described in Ref. \cite{Bazow:2016yra}. We here describe the most general form of the root finding problem: assuming that $T^{\tau\mu}$, $N^{\tau}$, $\pi^{\tau\mu}$, $\Pi$, and $n^\tau$ are all known from the latest temporal update step, we want to compute $\ed$, $\n$,  and $u^{\mu}$. As will be demonstrated in Sec.~\ref{numericaltests}, the following algorithm \cite{Karpenko:2013wva, Shen:2014vra, Pang:2018zzo} works for both ideal and dissipative fluids, i.e. for both vanishing and non-vanishing dissipative flows. We start by introducing the ``ideal fluid contributions'' $M^\mu$ and $J^\tau$ to the energy-momentum current $T^{\tau\mu}$ and baryon density $N^\tau$ in the computational frame:
\begin{align}
  M^{\tau} &= T^{\tau\tau}-\pi^{\tau\tau} = ( \ed+P) ( u^{\tau} )^{2} -P\;,\label{eq-mtau}\\
  M^{i} &= T^{\tau i}-\pi^{\tau i} = ( \ed+P) u^{\tau} u^{i} \quad (i = x, y, \eta_s)\;,\label{eq-mi}\\
  J^\tau &= N^{\tau}-n^\tau =\n u^{\tau}\,.
\label{eq-jt}
\end{align}
Note that for a viscous fluid $P$ above includes implicitly the bulk viscous pressure (i.e., $\peq + \Pi$). We use the following only when baryon evolution is turned on; otherwise, we use the simpler algorithm described in \cite{Bazow:2016yra} where $\ed$ is found first, using a 1-dimensional zero search. Here we first find the magnitude of the flow velocity, $v$, by solving iteratively \cite{Karpenko:2013wva, Shen:2014vra, Pang:2018zzo}
\begin{equation}
    v \equiv \frac{M}{M^\tau +P} = 
    \frac{M}{M^\tau +\peq\bigl(\ed(v),\n(v)\bigr) + \Pi}\,,
\label{root-v-expression}
\end{equation}
where $M \equiv \sqrt{(M^x)^2+(M^y)^2+\tau^2(M^\eta)^2}$ and $\ed(v)$, $\n(v)$ are obtained from the known quantities $M^\tau$, $M$, and $J^\tau$ as
\begin{align}
  \ed(v) &= M^\tau - vM\,,
\label{root-e}\\
  \n(v) &= J^\tau\sqrt{1-v^2}\,.
\label{root-n1}
\end{align}
Once the flow magnitude $v$ is known, we also know the flow 4-velocity:
\begin{align}
  u^\tau &=\frac{1}{\sqrt{1-v^2}}\,,
\label{eq-utau}\\
  u^i &= u^\tau\frac{M^i}{M^\tau + P}\,,
\end{align}
for $i = x,y,\eta_s$. Note that the algorithm makes active use of all the numerically known components listed above. 

Rewriting Eq.~(\ref{root-v-expression}) in the form
\begin{equation}
    f(v) \equiv v-\frac{M}{M^\tau+\peq(\ed(v),\n(v))+\Pi}
    =0\,,
\label{root-v-function}
\end{equation}
we solve it by the standard Newton-Raphson method, by repeatedly updating the velocity with
\begin{equation}
    v_{i+1} = v_i - \frac{f\bigl(v_i\bigr)}{f'\bigl(v_i\bigr)}\,,
\label{eq-root-vnewton}
\end{equation}
where
\begin{equation}
    f'(v) \equiv \frac{\partial f(v)}{\partial v} = 1+\frac{M}{(M^\tau+P)^2}\frac{d \peq}{d v}\,,
\label{dfdv}
\end{equation}
until a sufficiently accurate value is reached. The last term in (\ref{dfdv}) is evaluated numerically using 
\begin{equation}
    \frac{d \peq}{d v}=-\left[M\frac{\partial \peq}{\partial \ed}+J^\tau \frac{v}{\sqrt{1-v^2}}\frac{\partial \peq}{\partial \n}\right]\;, \label{eq-pvderivative}
\end{equation}
where (for equations of state like EOS4) the derivatives $\partial \peq/\partial \ed$ and $\partial \peq/\partial \n$ must be interpolated from the values stored in the EoS table to the pair $(\ed,\n)$ tried in each step of the iteration. 

When $v$ gets close to the speed of light, the Newton-Raphson iteration must be modified to avoid excursions into the causally forbidden region $v>1$. This can lead to numerical instabilities and/or poor convergence. We therefore follow the recipe proposed in \cite{Shen:2014vra} and use Eq.~(\ref{root-v-function}) to solve for $v$ only if in the previous time step, at the spatial grid point in question, $v \leq 0.563624$ or, equivalently (see (\ref{eq-utau})), $u^\tau \leq 1.21061$. Otherwise we instead solve for $u^\tau$ (which has no upper limit), by employing the Newton-Raphson algorithm to find the zero of
\begin{equation}
    f(u^\tau) \equiv u^\tau - \sqrt{\frac{M^\tau+\peq\bigl(\ed(u^\tau),\n(u^\tau)\bigr)+\Pi}
         {\ed+\peq\bigl(\ed(u^\tau),\n(u^\tau)\bigr)+\Pi}}\,.
\label{root-u-function}
\end{equation}
In this case we need to evaluate in each iteration 
\begin{equation}
    f'(u^\tau) = 1 - \frac{1}{2}\left[\frac{\ed-M^\tau}{(\ed+P)^{3/2}(M^\tau+P)^{1/2}}\right]\frac{d\peq}{du^\tau}\;,
\end{equation}
where
\begin{equation}
    \frac{d\peq}{du^\tau} = -\frac{1}{(u^\tau)^2}\left[\frac{M}{vu^\tau}\frac{\partial \peq}{\partial \ed} + J^\tau\frac{\partial \peq}{\partial \n}\right]\,,
\label{eq-puderivative}
\end{equation}
with $v = \sqrt{1-1/(u^\tau)^2}$. In the Gubser test of Ch.~\ref{ch.gubser} it will be shown that the switch between the two schemes, as implemented in \code, works seamlessly, moving smoothly from $v = 0.563624$ to $u^\tau = 1.21061$ across the switching point. 

When interpolating the EoS table to obtain the derivatives needed on the r.h.s. of Eqs.~(\ref{eq-pvderivative},\ref{eq-puderivative}) one can encounter numerical errors in regions where derivatives of the EoS change discontinuously, e.g. in the recently developed BEST EoS \cite{Parotto:2018pwx} which adds (using a certain prescription) a critical point and first-order phase transition to the LQCD-HRG interpolated EOS4. For such situations, \code\ offers another option for the root finding that avoids calculating these derivatives, at the price of somewhat degraded convergence which can slow the root finding algorithm compared to the Newton-Raphson method. This second regulation scheme may also be preferred when evolving more than one conserved charge, in which case not having to compute the thermodynamic derivatives may overcompensate for the slower intrinsic convergence of the root finding algorithm.

The modified root finder employs the following simple iteration scheme: starting with an initial guess $v_i$ for the velocity (e.g. the solution at this grid point from the preceding time step), we determine $\bigl(\ed(v_i), \n(v_i)\bigr)$ from Eqs.~(\ref{root-e},\ref{root-n1}) and the EoS $\peq(\ed,\n)$, compute an updated value $v_{i+1}$ of the velocity from
\begin{equation}
    v_{i+1} = \frac{M}{M^\tau +\peq(\ed(v_i),\n(v_i)) + \Pi}\,,
\label{root-v-expression-2}
\end{equation}
and iterate these steps until convergence is reached. For $v \geq 0.563624$ or $u^\tau \geq 1.21061$, one instead updates $u^\tau$ using the equation 
\begin{equation}
  u^\tau_{i+1} = 
  \sqrt{\frac{M^\tau+\peq\bigl(\ed(u^\tau_i),\n(u^\tau_i)\bigr)
        +\Pi}
       {\ed +\peq\bigl(\ed(u^\tau_i),\n(u^\tau_i)\bigr)+\Pi}}
\label{root-u-expression}
\end{equation}
until convergence is reached. 

In principle, the two methods are equivalent and should find the same root, within the prescribed numerical precision. In the Gubser test described in Ch.~\ref{ch.gubser} they are indeed shown to yield identical numerical results. In both methods, higher numerical precision of the solution should be demanded when solving for $v$, due to the speed limit $v<1$.

We point out that extra hydrodynamic variables are propagated in the code that are not used in the root finding algorithm, such as $\pi^{xx}$, $\pi^{xy}$, $\pi^{x\eta}$, and $\V^\eta$. In principle, these could be computed from the other components of the shear stress and baryon diffusion current by using the tracelessness of $\pi^{\mu\nu}$ and the orthogonality of $\pi^{\mu\nu}$ and $\V^\mu$ to $u^\mu$. Instead, we propagate all shear stress and baryon diffusion components dynamically and use the tracelessness and orthogonality conditions to check the numerical precision of the code.

\subsection{Regulation scheme}
\label{sec-regul}

The solution of the hydrodynamic equations of motion on discretized grids, the bilinear interpolation of the EoS table at each grid point, the iterative nature of the root finding algorithm, and the need for calculating derivatives numerically all engender unavoidable numerical errors. In addition, the numerical solution for the hydrodynamic variables can make excursions into regions where the approximations under which the evolution equations were derived (such as ignoring higher-order gradient terms) are no longer valid, and the numerical evolution algorithm produces unphysical results. This happens, in particular, because Nature provides us with initial conditions that exhibit unavoidable quantum fluctuations which can lead to local excursions outside the region of validity of dissipative hydrodynamics. Although dissipation usually erases such large fluctuations over short time scales \cite{Shen:2014vra, Bazow:2016yra}, this may not happen quickly enough to avoid numerical instability of the evolution algorithm.\footnote{%
    Note that we are not even talking about stochastic thermal fluctuations during the hydrodynamic evolution discussed in Sec.~\ref{sec:fluct} (our code solves deterministic equations of motion) which add possibly large stochastic fluctuations throughout the evolution history.}

In realistic event-by-event simulations such fluctuations can result in large gradients in both space and time which the code has to be able to deal with. Large gradients of the macroscopic variables can yield locally large Knudsen numbers (for which the fluid dynamic approximation breaks down) or large inverse Reynolds numbers (in which case the applicability of the 14-moment approximation used to simplify the hydrodynamic equations of motion is doubtful) \cite{Shen:2014vra,Bazow:2016yra}. To ensure numerical stability of the code, such excursions must be regulated. To avoid the undesirable consequence that, after regulation, the algorithm no longer solves the underlying evolution equations, the regulation must be local, i.e. it must affect only very localized space-time regions, and its effects must be monitored so that the user is warned when the regulation becomes so strong and the regulated regions become so large that the code no longer correctly simulates the physics encoded in the evolution equations. 

In practice, large gradients can drive large shear stress, bulk viscous pressure and baryon diffusion currents, and these can result in numerical instability or failure of the root finding algorithm. When this happens it is typically during the earliest evolution stage (where both the physical inhomogeneities driven by quantum fluctuation and the longitudinal expansion rate are largest) and/or in the very dilute regions near the transverse and longitudinal edge of the computing grid where the dissipative corrections to the leading thermodynamic quantities become large and the matter can no longer be reasonably treated as a fluid. Since the latter regions are typically far outside the domain where the matter is in the quark-gluon plasma phase (and thus outside the region where we want to apply the hydrodynamic picture), regulating them is innocuous as long as the regulation effects do not have sufficient time to propagate back inwards into the QGP region. Regulating large initial fluctuations is more dangerous because the fluctuations can be large both in- and outside the QGP phase. Both types of regulations must be carefully monitored.

Regulation schemes can be tricky, and a variety of implementations exist.\footnote{%
    For example, CLVisc \cite{Pang:2018zzo} requires $\texttt{max} (|\pi^{\mu\nu}|) < T_0^{\tau\tau}$; when this is violated for some cell in the dilute region, $\pi^{\mu\nu}$ is set to 0 locally. vHLLE \cite{Karpenko:2013wva} requires $\texttt{max}(|\pi^{\mu\nu}|/|T_0^{\mu\nu}|) < C$ and $|\Pi|/\peq < C$, with $C$ being a constant of order but smaller than 1; if one of these conditions is violated, $\Pi$ and/or $\pi^{\mu\nu}$ are rescaled by a factor (which is common for all components of $\pi^{\mu\nu}$) to satisfy this requirement.}
In \code, we follow the lead of {\sc iEBE-VISHNU} \cite{Shen:2014vra}, {\sc GPU-VH} \cite{Bazow:2016yra} and {\sc MUSIC} \cite{Denicol:2018wdp} and implement two types of regulation that build on the schemes suggested in these earlier codes. Both are triggered by large dissipative flows which are then regulated. The trigger criterion compares (in ways defined more precisely below) $\pi^{\mu\nu}$ with $T_0^{\mu\nu}=\ed u^\mu u^\nu - \peq\Delta^{\mu\nu}$, $\Pi$ with $\sqrt{\ed^2+3\peq^2}$, and $\V^\mu$ with $\n u^\mu$. 

\begin{figure}[!hbtp]
    \centering
    \includegraphics[width=0.9\textwidth]{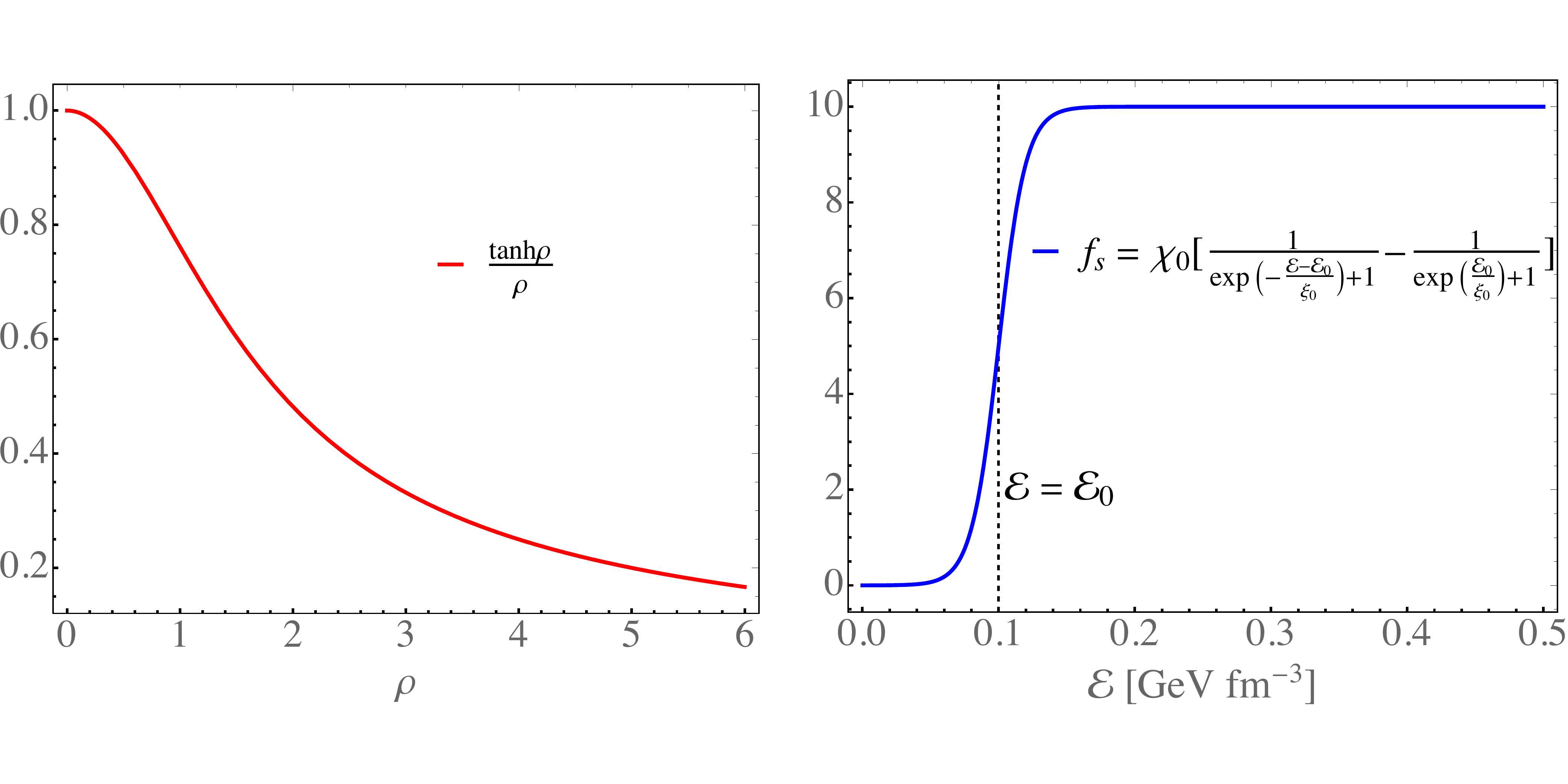}
    \caption{The regulation functions used in our two regulation schemes. {\sl Left:} Regulation function used in the first scheme. Large dissipative components yield large values of $\rho$ where the regulation is stronger. {\sl Right:} Regulation strength function $f_s$ used in the second regulation scheme, for the parameter choice $\chi_0 = 10$, $\ed_0 = 0.1$\,GeV/fm$^3$ and $\xi_0 = 0.01$\,GeV/fm$^3$. $\ed_0$ is the critical energy density below which strong regulation kicks in.
    \label{F3} 
    }
\end{figure}

For the shear stress tensor, {\sc iEBE-VISHNU} \cite{Shen:2014vra} and {\sc GPU-VH} \cite{Bazow:2016yra} require 
\begin{equation}
    \sqrt{\pi^{\mu\nu}\pi_{\mu\nu}} \leq \rho_\mathrm{max}\sqrt{T_0^{\mu\nu}T_{0\mu\nu}} = \rho_\mathrm{max}\sqrt{\ed^2+3\peq^2}\,,
\end{equation}
with $\rho_\mathrm{max}\leq 1$. In addition, the tracelessness of $\pi^{\mu\nu}$ and its orthogonality to $u_\nu$ are required,
\begin{equation}
    \pi^{\mu}_{\mu}\leq\xi_0\sqrt{\pi^{\mu\nu}\pi_{\mu\nu}}\quad
    \textrm{and}\quad \pi^{\mu\nu}u_\nu\leq\xi_0\sqrt{\pi^{\mu\nu}\pi_{\mu\nu}}\;,
\end{equation}
where $\sqrt{\pi^{\mu\nu}\pi_{\mu\nu}}$ sets the scale and $\xi_0\ll 1$ is a small number \cite{Shen:2014vra}. At grid points where these trigger conditions are violated, Refs.~\cite{Shen:2014vra, Bazow:2016yra} regulate the shear stress tensor $\pi^{\mu\nu}$ by (see the left plot of Fig.~\ref{F3})
\begin{equation}
  \pi^{\mu\nu}\to\frac{\tanh\rho_\pi}{\rho_\pi}\pi^{\mu\nu}\;,
\label{regPimunu}
\end{equation}
where
\begin{equation}
  \rho_\pi\equiv\texttt{max}
  \left[
  \frac{\sqrt{\pi^{\mu\nu}\pi_{\mu\nu}}}
       {\rho_\mathrm{max}\sqrt{\ed^2+3\peq^2}}\,,\quad
  \frac{g_{\mu\nu}\pi^{\mu\nu}}
       {\xi_{0}\rho_\mathrm{max}\sqrt{\pi^{\mu\nu}\pi_{\mu\nu}}}\,,\quad
  \frac{\pi^{\lambda\mu}u_{\mu}}
       {\xi_{0}\rho_\mathrm{max}\sqrt{\pi^{\mu\nu}\pi_{\mu\nu}}}\; \forall\;\lambda
  \right]. 
\label{rhoPimunu}
\end{equation}

For the bulk viscous pressure, which can make the root finding process fail when negative and too large, Refs. \cite{Shen:2014vra, Bazow:2016yra} regulate $\Pi$ during the root finding process to ensure existence of at least one non-negative solution for $\ed$, $v$ or $u^\tau$. In \code\ we demand instead that 
\begin{equation}
  \sqrt{3\Pi^2} \leq \rho_\mathrm{max}\sqrt{\ed^2+3\peq^2}\;,
\label{Pilimit}
\end{equation}
and where this trigger condition is violated we regulate $\Pi$ by
\begin{equation}
  \Pi\to\frac{\tanh\rho_\Pi}{\rho_\Pi}\Pi\;,
\label{regPi}
\end{equation}
with
\begin{equation}
\rho_\Pi\equiv \frac{1}{\rho_\mathrm{max}}\sqrt{\frac{3\Pi^2}{\ed^2+3\peq^2}}\;.
\label{rhoPi}
\end{equation}

In the same spirit, we require for the (space-like) baryon diffusion current
\begin{equation}
    -\V^\mu \V_\mu \ll \n^2 \quad\textrm{and}\quad \V^\mu u_\mu = 0\;.
\end{equation}
In the code we replace these conditions by
\begin{equation}
    \sqrt{-\V^\mu \V_\mu} \leq \rho_\mathrm{max}\sqrt{\n^2} \quad\textrm{and}\quad \V^\mu u_\mu \leq \xi_0\sqrt{-\V^\mu \V_\mu}\;.
\end{equation}
When one of these conditions is violated in a cell it triggers the following regulation of the baryon diffusion current:
\begin{equation}
  \V^\mu\to\frac{\tanh\rho_\V}{\rho_\V}\V^\mu\;,
\label{regV}
\end{equation}
with
\begin{equation}
  \rho_\V\equiv\texttt{max}
  \left[
  \frac{\sqrt{-\V^\mu \V_\mu}}{\rho_\mathrm{max}\sqrt{\n^2}}\,,\quad
  \frac{\V^\mu u_\mu}{\xi_0\rho_\mathrm{max}\sqrt{-\V^\mu \V_\mu}}
  \right]\;.
\label{rhoV}
\end{equation}
%

Equations (\ref{regPimunu},\ref{regPi},\ref{regV}) define the first of our two regulation schemes. In contrast to Refs.~\cite{Shen:2014vra, Bazow:2016yra} where $\Pi$ was regulated during the root finding process, our regulation here is performed only after finishing each step of the two-step RK-KT algorithm. After implementing it we find that there is no need for additional regulation of $\Pi$ during the root finding process. As defaults for the regulation parameters we use the same values $\xi_0 = 0.1$ and $\rho_\mathrm{max} = 1$ as proposed in \cite{Shen:2014vra, Bazow:2016yra}. For documentation of the sensitivity studies leading to these default values we refer the reader to Ref.~\cite{Shen:2014vra}.

A second, different regulation scheme is based on the one implemented in the latest version of {\sc MUSIC} \cite{Denicol:2018wdp}. (A flag in the code allows the user to select the preferred regulation scheme before running it.) In this second scheme, the dissipative components are rescaled by
\begin{equation}
    \pi^{\mu\nu}\to\frac{r^\mathrm{max}_\pi}{r_\pi}\pi^{\mu\nu}\;,\quad \Pi\to\frac{r^\mathrm{max}_\Pi}{r_\Pi}\Pi\;,\quad  \V^\mu\to\frac{r^\mathrm{max}_\V}{r_\V}\V^\mu\;,\label{eq-rmaxr}
\end{equation}
where regulation of quantity $i$ is triggered whenever $r_i$ exceeds the corresponding maximally allowed value $r_i^\mathrm{max}$ $(i=\pi, \Pi, \V)$,\footnote{%
    If the regulation for the shear stress or baryon diffusion is triggered, all components of $\pi^{\mu\nu}$ or $n^\mu$ are regulated by a common regulation factor.}
with $r_i$ defined by 
\begin{equation}
    r_\pi = \frac{1}{f_s}\sqrt{\frac{\pi^{\mu\nu}\pi_{\mu\nu}}{\ed^2+3\peq^2}}\;,\quad 
    r_\Pi = \frac{1}{f_s}\sqrt{\frac{3\Pi^2}{\ed^2+3\peq^2}}\;,\quad 
    r_\V = \frac{1}{f_s}\sqrt{\frac{-\V^\mu \V_\mu}{\n^2}}\;.
\label{eq-r}
\end{equation}
By comparing Eq.~(\ref{eq-rmaxr}) to Eqs.~(\ref{regPimunu},\ref{regPi},\ref{regV}) and Eq.~(\ref{eq-r}) to Eqs.~(\ref{rhoPimunu},\ref{rhoPi},\ref{rhoV}) above, we see that the factors $r_i^\mathrm{max}/r_i$ play the same role as $\tanh{\rho_i}/\rho_i$ in the first regulation scheme, causing stronger regulation for larger $r_i$, with $r_i$ playing the role of the quantity $\rho_i$ whereas $f_s$, defined by
\begin{equation}
    f_s = \chi_0
    \left[\frac{1}
               {\exp\bigl(-(\ed{-}\ed_0)/\xi_0\bigr)+1}
         -\frac{1}{\mathrm{exp}(\ed_0/\xi_0)+1}
    \right],
\end{equation}
playing a similar role as $\rho_\mathrm{max}$: as $f_s$ or $\rho_\mathrm{max}$ grows larger, the regulation gets weaker. $f_s$ is designed to approach $\chi_0$ when $\ed\gg\ed_0$ and 0 when $\ed\ll\ed_0$.

The right plot of Fig.~\ref{F3} shows that for $\ed < \ed_0$, i.e. in the dilute region, $f_s$ decreases exponentially and the regulation strength increases accordingly. On the other hand, for large values of the parameter $\chi_0$, the regulation will hardly ever be triggered in the dense region $\ed>\ed_0$. Thus, unlike the first method, which always regulates larger dissipative components more strongly, irrespective of the energy density at the grid point, for the choice $\ed_0=0.1$\,GeV/fm$^3$ \cite{Denicol:2018wdp} the second method causes hardly any regulation at grid points in the dense QGP region but more frequent and stronger regulation in the dilute region far outside the QGP fluid. The authors of Ref.~\cite{Denicol:2018wdp} used $\chi_0=10$, and chose $r^\mathrm{max}_\V = 1$ for regulating the baryon diffusion current; we adopt the same value for $\chi_0$ and identical maximum $r$ values for all dissipative flows: $r^\mathrm{max}_\pi = r^\mathrm{max}_\Pi = r^\mathrm{max}_n = 1$. 

The default set for the regulation parameters is not universal and may need adjustments for different initial conditions, collision systems, and collision energies. The user is encouraged to play with these parameters to achieve maximal code stability with minimal changes to the physics encoded in the evolution equations. The regulation scheme may need to become more involved in future versions of dissipative hydrodynamics that include possibly large thermal and/or critical fluctuations in the dynamics (see e.g. \cite{Singh:2018dpk, Sakai:2018sxp}).

\section{Code validation with semi-analytical solutions}
\label{numericaltests}

Numerical codes solving second-order (``causal'') relativistic dissipative fluid dynamics in 3+1 dimensions have only been developed over the last decade. They solve a problem for which in general no analytic solutions are available. Careful validation of any such code by testing its various components in simplified settings for which analytic or semi-analytic solutions {\it are} available is therefore mandatory. Some of these tests are nowadays standard and are included with this distribution precisely for that reason. The precursor of this code, the CPU-version of {\sc GPU-VH} \cite{Bazow:2016yra} was carefully validated using similar tests, but the entire baryon evolution sector in \code\ is new such that direct comparisons with {\sc GPU-VH} are of limited value. We therefore include here in particular novel semi-analytic tests of the baryon evolution equations. Direct code-to-code comparisons with {\sc MUSIC} (whose latest version \cite{Denicol:2018wdp} also includes baryon evolution) will become possible when that version of {\sc MUSIC} becomes public.

Building on validation protocols described in Refs.~\cite{Karpenko:2013wva, Shen:2014vra, Bazow:2016yra, Pang:2018zzo, Denicol:2018wdp}, we here discuss tests in which we compare, for identical initial conditions, the numerical solutions from \code\ with (semi-)analytic solutions using {\sc Mathematica} \cite{Mathematica} for the Riemann problem for the Euler equations \cite{toro2009riemann, sod1978survey, marti_muller_1994, RISCHKE1995346, RISCHKE1995383}, for Bjorken flow \cite{PhysRevD.27.140}, and for Gubser flow \cite{Gubser:2010ze,Gubser:2010ui} extended to systems with non-zero net baryon density and baryon diffusion current induced by a fluctuation in the initial state. We also include a direct comparison of \code\ with the independently developed numerical algorithm described in Ref.~\cite{Denicol:2010xn} for a system with non-zero net baryon density in a (1+1)-dimensional setting with general longitudinal but vanishing transverse flow. By generalizing previously developed validation protocols to systems with non-zero net baryon density and baryon diffusion currents the work described in this section prepares the ground for code validation of other hydrodynamic codes at finite baryon density that are presently being developed elsewhere for the study of heavy-ion collisions at BES energies.

All tests described in this section are done without code regulation, i.e. all the regulation schemes described in Sec.~\ref{sec-regul} are turned off. In the code, all dimensionful quantities are represented by numbers given in length units, using the appropriate powers of [fm]; when plotting the results we sometimes convert them to physical units by multiplying with the appropriate powers of $\hbar c = 0.197$\,GeV\,fm.

In the tests and all other applications of the code completed to date we have used the following grid spacings: $\Delta x = \Delta y = 0.05$~fm and $\Delta\eta_s = 0.02$; $\Delta\tau$ is adjusted as needed and described in each case below. In realistic simulations the choice of grid spacing has to be a compromise between computational economy and capturing relevant physical information (e.g. large gradients, especially at early times). For smooth, ensemble-averaged initial conditions, larger grid spacings than those listed above can be used, whereas simulating small collision systems (such as proton-proton collisions) may require finer grids.

\subsection{The Riemann problem}
\label{sec4.1}

We start with testing the code against an analytical solution of the Riemann problem for the Euler equations, which historically has played an important role in fluid dynamics research and in the development of hydrodynamic codes \cite{toro2009riemann, marti_muller_1994}. Here a special case of the relativistic Riemann problem, known as Sod's shock tube problem \cite{sod1978survey, marti_muller_1994, RISCHKE1995346, RISCHKE1995383}, is considered describing the 1D evolution of an ideal fluid in the transverse plane during the decay of a discontinuity across a $(y,z)$ surface placed at $x{\,=\,}0$, separating two constant initial states (``left'' ($l$) for $x<0$ and ``right'' ($r$) for $x>0$) at rest:
\begin{equation}
    (\p,\n,u^x) =
    \begin{cases}
         (\p_l, \n_l, u^x_l = 0)\;,\qquad x < 0\;,\\
         (\p_r, \n_r, u^x_r = 0)\;,\qquad x > 0\;.
    \end{cases}
\label{eq-riemann-ic}
\end{equation}
In the longitudinal ($z$) direction the fluid is assumed to expand with a boost-invariant velocity profile $u^z/u^\tau{\,=\,}z/t$ (i.e. $u^\eta{\,=\,}0$), and the system is initialized along a surface of constant longitudinal proper time $\tau_0$.

\begin{figure}[!t]
    \centering \includegraphics[width=\textwidth]{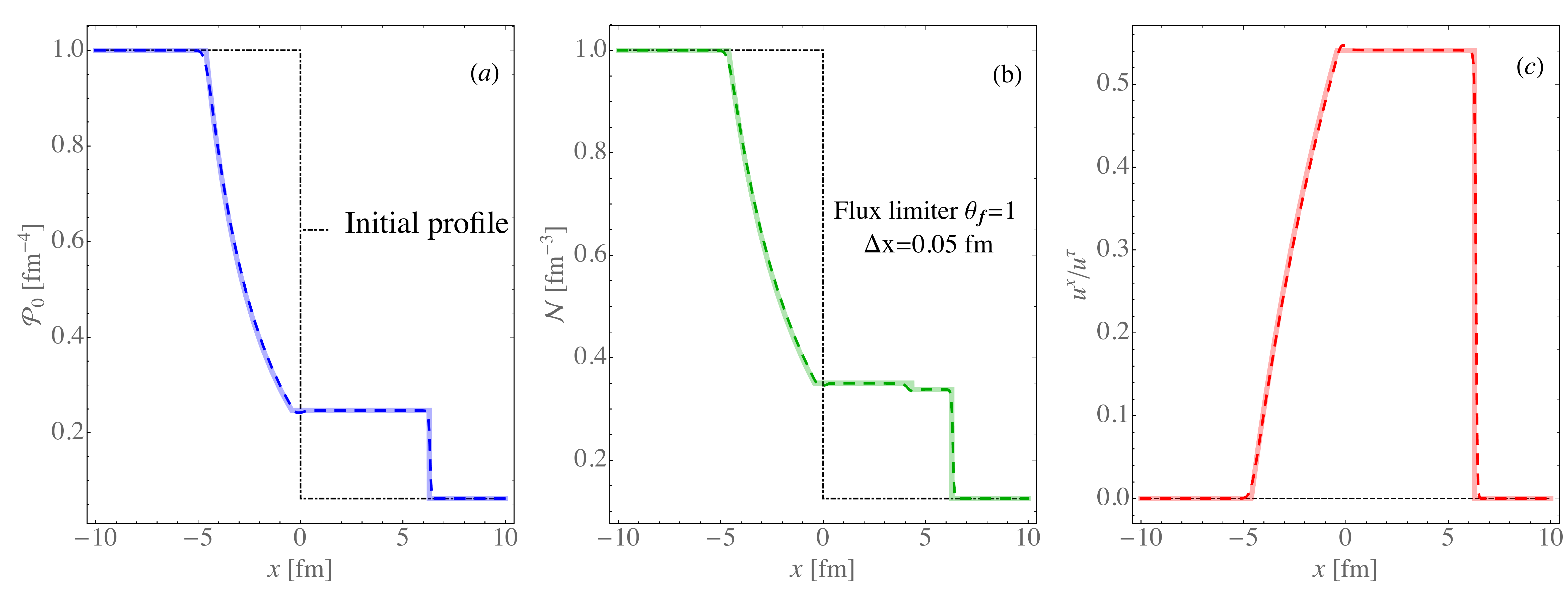}
    \caption{Analytical (Ref.~\cite{Blaga:2016kug}, continuous lines) vs. numerical results (\code, broken lines) for Sod's relativistic shock-tube problem in an ideal fluid with a conformal EoS: (a) pressure; (b) baryon density; (c) scaled velocity $u^x/u^\tau$. The numerical simulation starts at $\tau_0 = 0.5$\,fm/$c$, the grid spacing in the transverse $(x,y)$ plane is $\Delta x=\Delta y = 0.05$\,fm, and the flux limiter is set to $\theta_\mathrm{f} = 1$. The initial conditions to the left ($l$) and right ($r$) of the shock discontinuity are $\p_l{\,=\,} 1$\,fm$^{-4}$,\; $\n_l{\,=\,}1$\,fm$^{-3}$,\; $u^x_l{\,=\,}u^x_r{\,=\,}0$,\; $\p_r{\,=\,}0.0625$\,fm$^{-4}$,\; and $\n_r{\,=\,}0.125$\,fm$^{-3}$. The plots show results at $\tau=8.5$\,fm/$c$ for the numerical and $\tau=8.0$\,fm/$c$ for the analytical solution (see text for explanation).
    \label{F4}
    }
\end{figure}

The decay of this initial discontinuity gives rise to general features of the Riemann problem, characterized by three elementary waves. Two of them are a rarefaction wave and a shock, moving into the two initial state regions of high and low density, respectively. Between them, two additional states emerge, separated by the third wave, which is a contact discontinuity moving with the fluid \cite{toro2009riemann, marti_muller_1994} (see Fig.~\ref{F4}(b)). For a conformal EoS, an analytical solution for this problem can be derived from the conservation laws with the boundary condition that across the contact discontinuity pressure and velocity are constant (Figs.~\ref{F4}(a,c)) while the density has a jump (Fig.~\ref{F4}(b)) \cite{marti_muller_1994, Molnar:2009tx, Blaga:2016kug}. For non-conformal equations of state at non-zero net baryon density no general analytic solution is known, and the numerical solution can give rise to complex features (see, e.g., \cite{marti_muller_1994, SCHNEIDER199392}). Since for EOS1 $\partial \peq/\partial \n$ (which is needed in Eqs.~(\ref{eq-pvderivative},\ref{eq-puderivative}) for the velocity finding algorithm) is zero, the evolution of baryon density decouples from that of the energy density and pressure. 

In the code, the assumption of longitudinal boost-invariance is implemented by setting the number of cells in the longitudinal direction to 1. The initial profiles of $\ed$ and $T$ are obtained from the EoS. For ideal fluid dynamics the equations of motion become
\begin{align}
    D\n =& -\n\, \theta\;, \label{eq-riemann-n}\\
    D\ed =& -(\ed+\p)\theta\;,\label{eq-riemann-p}\\
    (\ed+\p)Du^\mu =& - \Delta^{\mu \nu} \partial_\nu \p\;. \label{eq-riemann-eom}
\end{align}
They are solved with initial conditions (\ref{eq-riemann-ic}) with the default parameters listed in Fig.~\ref{F4}. For details about the analytical solution we refer the reader to Ref.~\cite{Blaga:2016kug} (see also \cite{RISCHKE1995346, RISCHKE1995383}). We point out that in the analytical solution from Ref.~\cite{Blaga:2016kug} the evolution starts at time zero whereas in the code the hydrodynamic evolution is initialized at $\tau_0 = 0.5$\,fm. Since the solution is self-similar and depends only on the variable $x/(\tau-\tau_0)$, we therefore compare in Fig.~\ref{F4} the numerical results at $\tau=8.5$\,fm/$c$ to the analytical solution at $\tau-\tau_0=8.0$\,fm/$c$. For simplicity, the numerical test is done in Cartesian coordinates where all Christoffel symbols vanish. Also, to adequately capture the large discontinuity in the initial state, derivatives should be evaluated using Eq.~(\ref{eq-cendiff}), and not Eq.~(\ref{eq-approx-derivative}) which would yield zero initial gradients and result in no evolution at all.

Figure~\ref{F4} demonstrates very good overall agreement between the analytical and numerical solutions; the shocks and contact discontinuities are well captured. Although the baryon evolution is decoupled, this test still demonstrates excellent performance of the root finding algorithm. 

\subsection{Bjorken flow}
\label{sec4.2}

In this subsection we test \code\ in Milne coordinates for a transversally homogeneous dissipative fluid undergoing longitudinally boost-invariant Bjorken expansion \cite{PhysRevD.27.140}. Boost-invariant systems are characterized by space-time rapidity independent macroscopic observables and a flow profile that looks static (i.e. $u^\mu = (1,0,0,0)$) in Milne coordinates \cite{Jeon:2015dfa}. In spite of experimental evidence for longitudinal density gradients, there are strong phenomenological indications that near mid-rapidity a longitudinally boost-invariant flow profile is a good approximation for relativistic heavy-ion collisions at $\sqrt{s}\gtrsim 100$\,GeV/nucleon (see, e.g., \cite{Jeon:2015dfa}). The additional assumption of transverse homogeneity, however, is clearly unrealistic, given the finite transverse size of the colliding nuclei. Still, it provides a useful test bed because the resulting independence of all macroscopic quantities from all three spatial dimensions simplifies the dissipative hydrodynamic evolution equations to a set of coupled ordinary differential equations which can be solved with {\sc Mathematica}:
\begin{eqnarray}
\dot{\ed} &=&-\frac{\ed+\peq+\Pi -\pi }{\tau }\;,
\label{evolutionEd} 
\\
\tau _{\Pi }\dot{\Pi}+\Pi &=&-\frac{\zeta }{\tau }-\delta _{\Pi \Pi }\frac{%
\Pi }{\tau }+\lambda _{\Pi \pi }\frac{\pi }{\tau }\;,
\label{evolutionPi} 
\\
\tau _{\pi }\dot{\pi}+\pi &=&\frac{4}{3}\frac{\eta}{\tau} -\left( \frac{1}{3}\tau
_{\pi \pi }+\delta _{\pi \pi }\right) \frac{\pi }{\tau }+\frac{2}{3}\lambda _{\pi \Pi }%
 \frac{\Pi}{\tau}\;,
\label{evolutionpi}
\\
%
\dot{\n} &=&-\frac{\n}{\tau }\;,
\label{evolutionn}
\\
\tau_{\V} \dot{\V}^\eta+\V^\eta&=&-\left(\tau_{\V}+\delta_{\V\V}+\frac{2}{3}\lambda_{\V\V}\right)\frac{\V^\eta}{\tau}\;,
\label{evolutionneta}
\end{eqnarray}
where $\pi\equiv-\tau^2\pi^{\eta\eta}$ has been introduced. 

\begin{figure*}[!htb]
    \centering
    \includegraphics[width= 0.95\textwidth]{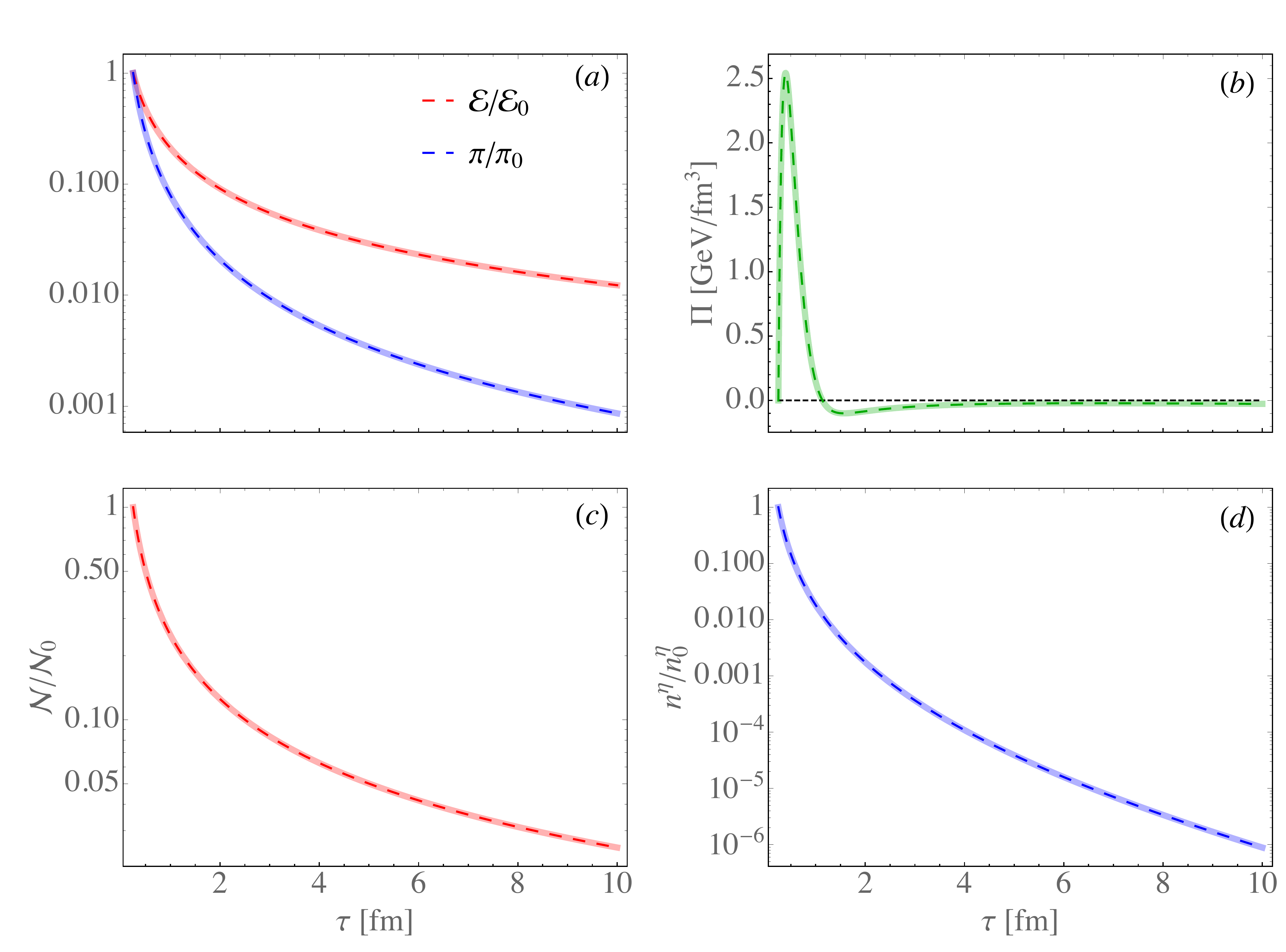}
    \caption{Comparing the semi-analytic results (continuous lines) with numerical output from \code\ (broken lines) for a fluid undergoing Bjorken expansion under the influence of the  Wuppertal-Budapest EoS (EOS2): (a) normalized energy density $\ed$ and shear stress $\pi=-\tau^2\pi^{\eta\eta}$; (b) bulk viscous pressure $\Pi$; (c) normalized baryon density $\n$; (d) normalized baryon diffusion current $n^\eta$. We use $\theta_\mathrm{f} = 1$ for the flux limiter, $\bar\eta=0.2$ for the kinematic shear viscosity, and the parametrization (\ref{eq-zetas}) for the kinematic bulk viscosity $\bar\zeta$. From these the relaxation times $\tau_\pi$, $\tau_\Pi$ and the viscosity related transport coefficients are computed using Eqs.~(\ref{etatau}-\ref{eq-zetas}). For $\tau_n$ we use Eq.~(\ref{taun}) with $C_n=4.0$ when $c\tau_n$ and $1/T$ are measured in fm, and Eq.~(\ref{diffcoeff}) for the baryon related second-order transport coefficients. The expansion is initialized at $\tau_0 = 0.25$\,fm/$c$ with initial conditions $T_0 = 4.5$\,fm$^{-1} = 0.89$\,GeV, 
    $\n_0= 500$ fm$^{-3}$, $\Pi_0=0$, $\pi_0 = \pi_{\mathrm{NS}} = \frac{4}{3} \frac{\eta}{\tau_0}$, and $n_0^\eta=10$ fm$^{-4}$. The initial energy density and pressure are obtained from the EoS.
    \label{F5}
    }
\end{figure*}

In the \code\ simulation the numbers of cells in $(x, y, \eta_s)$ directions are all set to 1.\footnote{%
    Of course, one can also set the number of cells in $(x, y)$ larger than 1 but to ensure transverse spatial homogeneity quantities like $\ed$ and $\peq$ should be the same in all cells.}
With boost-invariant and transversally homogeneous initial conditions, the numerical results from the (3+1)D \code\ code can be tested against a solution of the coupled ODEs (\ref{evolutionEd}-\ref{evolutionneta}) from a separate {\sc Mathematica} code. As for the Riemann problem, the baryon evolution decouples from the rest of the system if an EoS $\peq = \peq(\ed)$ is used; in this case, even with a nonzero baryon diffusion current in longitudinal direction, the Bjorken expansion remains unchanged.
Although longitudinal boost-invariance and transverse homogeneity do not allow any gradients of the chemical potential that could drive a baryon diffusion current, we can then still test the baryonic sector of the code by initiating it with nonzero initial values for the net baryon density and diffusion current. These out-of-equilibrium initial values will then relax according to Eqs.~(\ref{evolutionn},\ref{evolutionneta}), without affecting the Bjorken flow profile.  

By using EOS2 (the Wuppertal-Budapest EoS at $\mu=0$) which features a non-zero interaction measure, we can also test the evolution of the bulk viscous pressure $\Pi$ which is propagated by Eq.~(\ref{evolutionPi}) as an additional dissipative degree of freedom \cite{Bazow:2016yra}. Noting that Eq.~(\ref{evolutionPi})
involves a rather complex parametrization of the transport coefficients (\ref{betaPi})-(\ref{eq-zetas}), we emphasize that this test is indeed non-trivial.

As Fig.~\ref{F5} shows, the agreement of \code\ with the semi-analytic {\sc Mathematica} solution is excellent. The evolution of the baryon diffusion current follows the exact decay very precisely over six orders of magnitude. This would not be possible if the root finding algorithm did not perform with high accuracy.\footnote{%
    As for the Riemann problem, this test again checks this property only for the case $\partial\peq/\partial\n = 0$.}

Still, because of boost-invariance and transverse homogeneity many terms in the full set of evolution equations vanish in this example. The absence of any kind of transverse expansion in this test is particularly worrisome. This question will be addressed in Ch.~\ref{ch.gubser} devoted to the ``Gubser test''.

%
\subsection{Longitudinal evolution}
\label{sec:longevo}
%
In the absence of transverse gradients and flows, the hydrodynamic equations can be solved semi-analytically for the diffusive case without shear stress and bulk viscous pressure, using a setup described in Ref.~\cite{Fotakis:2019nbq} In this situation the 5 conservation laws (\ref{eq:vhydro-N}-\ref{eq:vhydro-u}) in Sec.~\ref{sec:relat_hydro} simplify to
\begin{eqnarray}
 D\n &=& -\n\theta -\nabla_\mu n^\mu\;,
\label{eq:vhydro-N2}
\\
 D\ed &=& -(\ed+p)\theta\;,
\label{eq:vhydro-E2}
\\
  (\ed+p)\, Du^\mu &=& \nabla^\mu \peq\;,
\label{eq:vhydro-u2}
\end{eqnarray}
and the equation of motion for the baryon diffusion is given by
\begin{equation}
    d \V^\mu = \frac{\kappa_n}{\tau_\V} \nabla^{\mu}\alpha - \frac{\V^{\mu}}{\tau_{\V}} - n^\nu u^\mu D u_\nu 
    - u^\alpha \Gamma^\mu_{\alpha\beta} n^\beta\,.
    \label{eq:diff_val2}
\end{equation}
Note that because of the assumed transverse homogeneity all hydrodynamic quantities are only functions of $(\tau, \eta_s)$, and this can be used to simplify the equations above further.

\begin{figure}[!htb]
\begin{center}
    \hspace{-0.5cm}\includegraphics[width=  0.9\textwidth]{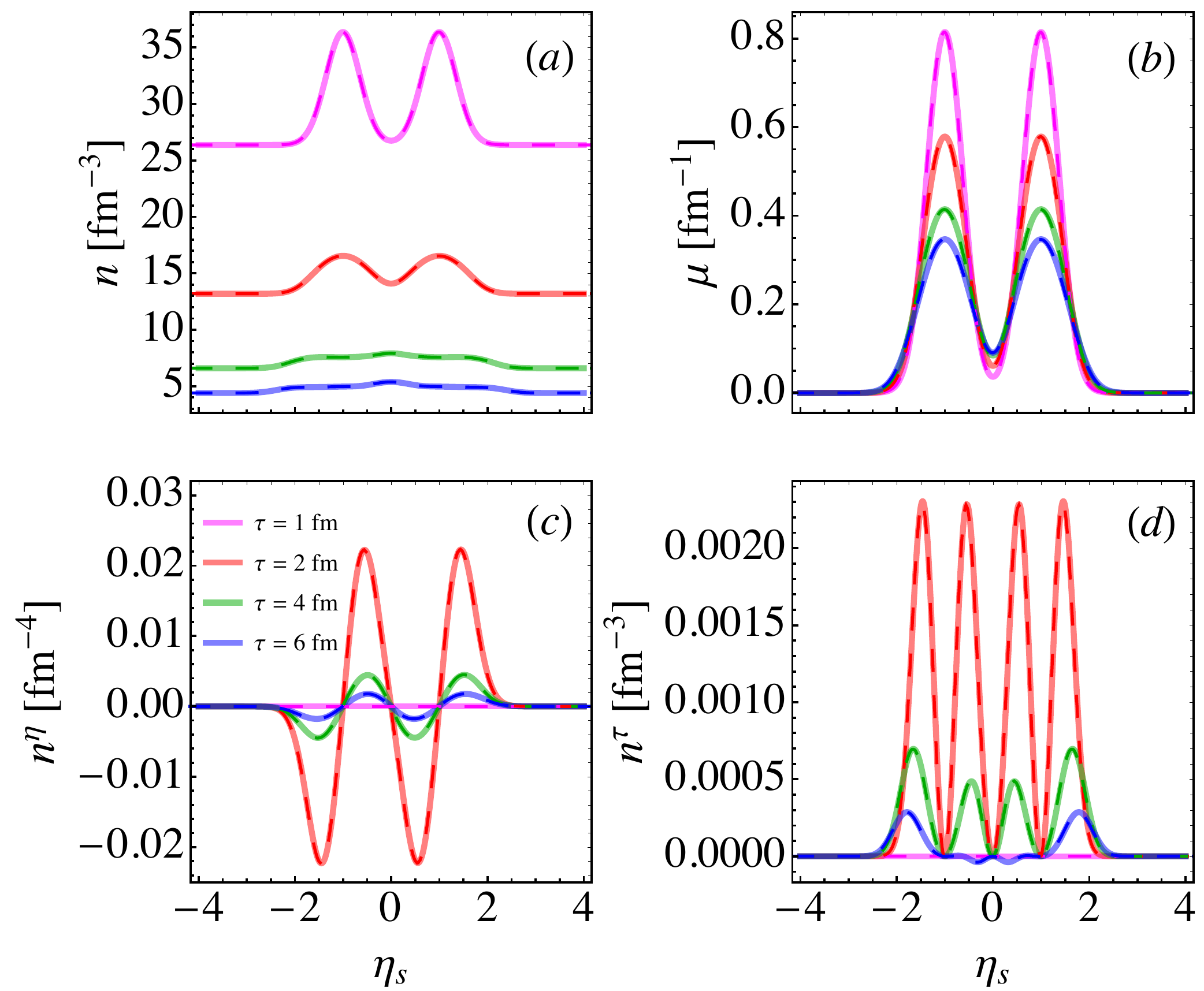}
    \caption{Validation of \bes{} for solving hydrodynamic equations Eqs. \eqref{eq:vhydro-N}-\eqref{eq:simple_eos}, by comparing its numerical results (dashed lines) to semi-analytical solutions (solid lines) at four different times. Here the baryon sector is focused on: (a) baryon density, (b) baryon chemical potential, (c) longitudinal baryon diffusion current and (d) $\tau$-component of baryon diffusion current. Excellent consistency can be observed from the plots.}
    \label{fig:difftest}
\end{center}
\end{figure}

We start the system at initial time $\tau_i= 1$~fm, using the initial conditions where $u^\mu = (1,0,0,0)$ and the initial diffusion current is zero, and  the initial longitudinal distribution of baryon density is given as
\begin{eqnarray}
    n(\tau_i, \eta_s) &=& \frac{g_s}{\pi^2}T_0^3 + n_\mathrm{max} \times  \left[\exp\left(-\frac{(\eta_s+\eta_+)^2}{\sigma_\eta^2}\right)+\exp\left(-\frac{(\eta_s+\eta_-)^2}{\sigma_\eta^2}\right)\right],\nonumber
\end{eqnarray}
where the initial temperature $T(\tauI, \eta_s) = 0.5$~GeV and $n_\mathrm{max} = 10$~fm$^{-3}$, $\sigma_\eta = 0.5$ and $\eta_\pm = \pm 1$. One also needs the two transport coefficients in Eq. \eqref{eq:diff_val2}, and we use \cite{Denicol:2012cn}
\begin{equation}
    \kappa_n = \frac{3}{16\sigma_\mathrm{tot}}\,,\quad \tau_n=\frac{9}{4}\frac{1}{n\sigma_\mathrm{tot}}\,,
\end{equation}
with $\sigma_\mathrm{tot} = 10$ mb = 1 fm$^2$. To close the equations, we use the following EoS
\begin{equation}\label{eq:simple_eos}
    e = 3p = 3nT\,,\quad n = \frac{g_s}{\pi^2}T^3\exp{\left(\frac{\mu}{T}\right)}\,.
\end{equation}
We note that this EoS can be analytically inverted to get $T(e, n)$ and $\mu(e, n)$. With the setup above, the hydrodynamic equations with longitudinal diffusion current can be solved semi-analytically using {\sc Mathematica} and used for validating \bes, and the comparison of them is shown in Fig.~\ref{fig:difftest}, for the baryonic sector. One can see from the figure that \bes\ works perfectly well in the longitudinal evolution.

%
\subsection{Comparison to other codes}
\label{sec4.4}
%

Owing to their shared assumption of longitudinal boost-invariance, none of the tests described in the preceding subsections addresses the performance of the \code\ code in describing the expansion along the rapidity direction. To remedy this we have compared \code\ output for a transversally homogeneous system undergoing arbitrary longitudinal expansion without transverse expansion with the results from an independent (1+1)D hydrodynamic code developed by Monnai \cite{Monnai:2012jc}. To be able to study baryon number transport, EOS4 is used in both codes. A similar comparison was also made in Ref. \cite{Denicol:2018wdp} to test the performance of {\sc MUSIC}. Rather than directly using Monnai's code \cite{Monnai:2012jc} we compare our \code\ results with those reported in the comparison \cite{Denicol:2018wdp} with {\sc MUSIC}. Qualitatively consistent results from an earlier \code\ study were already reported in \cite{Du:2018mpf}.

\begin{figure*}[!b]
\centering
\includegraphics[width=\textwidth]{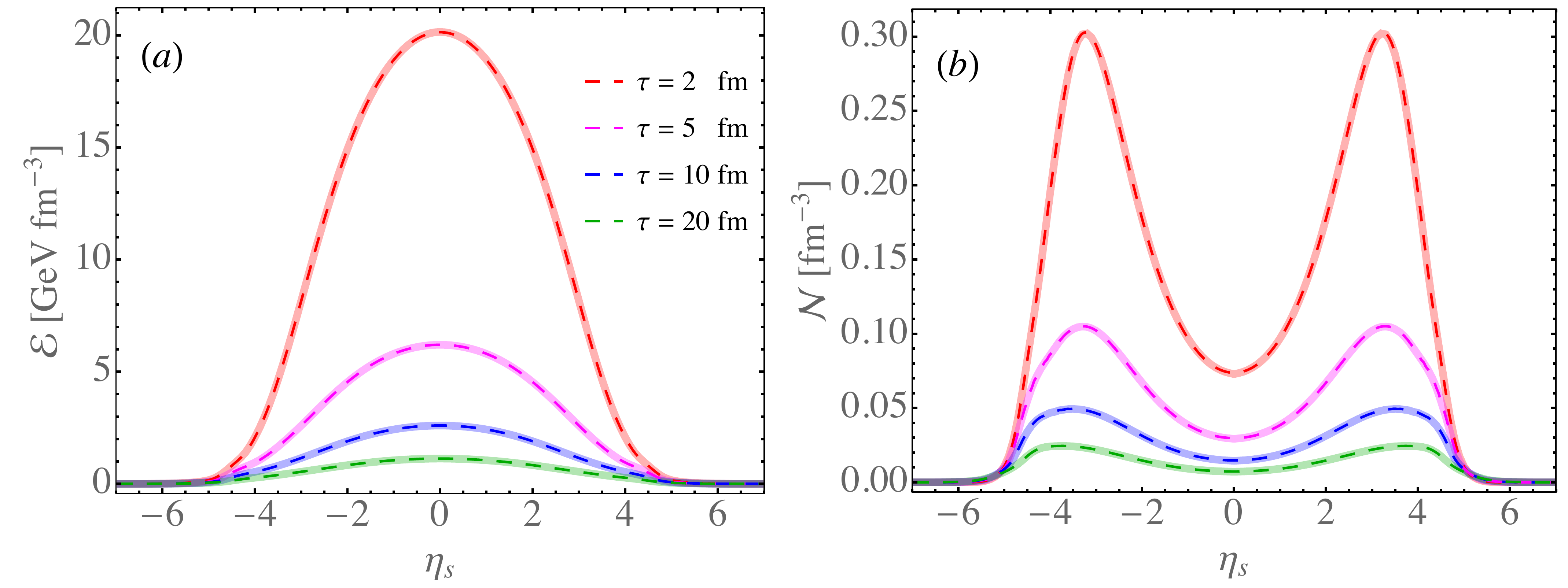}
\caption{Comparison between the numerical results from \code\ (broken lines) and the {\sc MUSIC} simulations \cite{Denicol:2018wdp} of the (1+1)D setup described in Ref.~\cite{Monnai:2012jc} (continuous lines): (a) energy density; (b) net baryon density. The simulation starts at $\tau_0 = 1$\,fm/$c$ and covers the space-time rapidity range $\eta_s\in[-6.94, 6.94]$, with grid spacing $\Delta\eta_s = 0.02$.
\label{F7}
}
\end{figure*}

We here focus on the baryon number evolution in the longitudinal direction, with non-vanishing longitudinal gradients of $\mu/T$, by setting bulk and shear stresses to zero.\footnote{%
    In principle, since shear stress is known to affect the evolution of the energy density it might be expected to also change the baryon number flow; this is an interesting physics question which we postpone for a separate study.}
Specifically, we check how \code\ handles the evolution equation (\ref{relEqs_nmu}) for baryon number diffusion,
\begin{equation}
    d \V^\mu = \frac{\kappa_n}{\tau_\V} \nabla^{\mu} \left(\frac{\mu}{T}\right) - \frac{\V^{\mu}}{\tau_{\V}} - n^\nu u^\mu D u_\nu 
    - u^\alpha \Gamma^\mu_{\alpha\beta} n^\beta\,,
\end{equation}
together with the net baryon conservation law (\ref{relEqs_n}). For the baryon transport coefficients we choose $\kappa_n = 0.2\,\n/\mu$ and $\tau_n = 0.2/T$. 

In Fig. \ref{F7} we show a comparison of the distributions in space-time rapidity $\eta_s$ of energy density $\ed$ and net baryon density $\n$ at four different times. Broken (continuous) lines show the
results from \code\ ({\sc MUSIC} \cite{Denicol:2018wdp}); the agreement between these two codes is very good. In addition to testing the longitudinal dynamics this comparison also shows that the root-finding algorithm works correctly with a realistic EoS\, $\peq(\ed,\n)$ that depends on both energy and net baryon density. 

\subsection{Tests summary}
\label{sec4.5}
We briefly summarize which parts of the \code\ code were tested with the test protocols described in this section. As described in Sec.~\ref{numericalscheme}, the same RK-KT algorithm is applied for solving the equations of motion for all hydrodynamical variables propagating in the system, and the same root-finding algorithm is used for all equations of state, whether they depend on baryon density or not. The observed good performance in propagating all hydrodynamical variables indicates the efficiency of the RK-KT algorithm, and the proper evolution of baryon density and baryon diffusion validates the root-finding method in systems with non-zero baryon currents. 

The Riemann problem in ideal hydrodynamics with EOS1 (Sec.~\ref{sec4.1}) shows the ability of the RK-KT algorithm in \code\ to capture shocks and contact discontinuities. The (0+1)D Bjorken expansion with EOS2 (Sec.~\ref{sec4.2}) tests the programming of the equations of motion, especially for the shear and bulk components, including the non-trivial parametrization of the transport coefficients, and the root-finding algorithm with non-zero but decoupled baryon density and diffusion current. The Gubser flow test with EOS1 (see Ch.~\ref{ch.gubser} below) provides extra validation in situations with strong transverse expansion featuring large temporal and transverse gradients. The validation for the longitudinal dynamics in Sec.~\ref{sec:longevo} and the comparison  with {\sc MUSIC} and Monnai's (1+1)D code (Sec.~\ref{sec4.4}) at finite baryon density with the realistic EOS4 validates the longitudinal dynamics of density and baryon diffusion without the simplification of longitudinal boost-invariance, as well as the root-finding algorithm with non-zero baryon density and baryon diffusion, including the bilinear interpolation of the EoS tables. The figures shown in this section demonstrate that \code\ passes all these tests without struggle.

\section{Baryon diffusion in an expanding QGP}
\label{sec5}

In this section we illustrate the evolution of energy and baryon number in an expanding QGP with realistic ``bumpy'' initial conditions, by visualizing the evolution of the corresponding densities in the transverse plane at $\eta_s=0$.\footnote{%
    For a similar earlier study with smooth, ensemble-averaged initial conditions see \cite{Denicol:2018wdp}.}
This generalizes many similar visualizations made in the past for systems without conserved charges. Since the physics of initial-state fluctuations along the longitudinal direction is still not very well explored (for a few examples see, Refs.~\cite{Denicol:2015nhu, Monnai:2012jc, Denicol:2018wdp}), we here use smooth longitudinal initial conditions and refrain from showing the (mostly uninteresting) evolution along the beam direction.

\begin{figure}[!b]
    \centering
    \includegraphics[width=\textwidth]{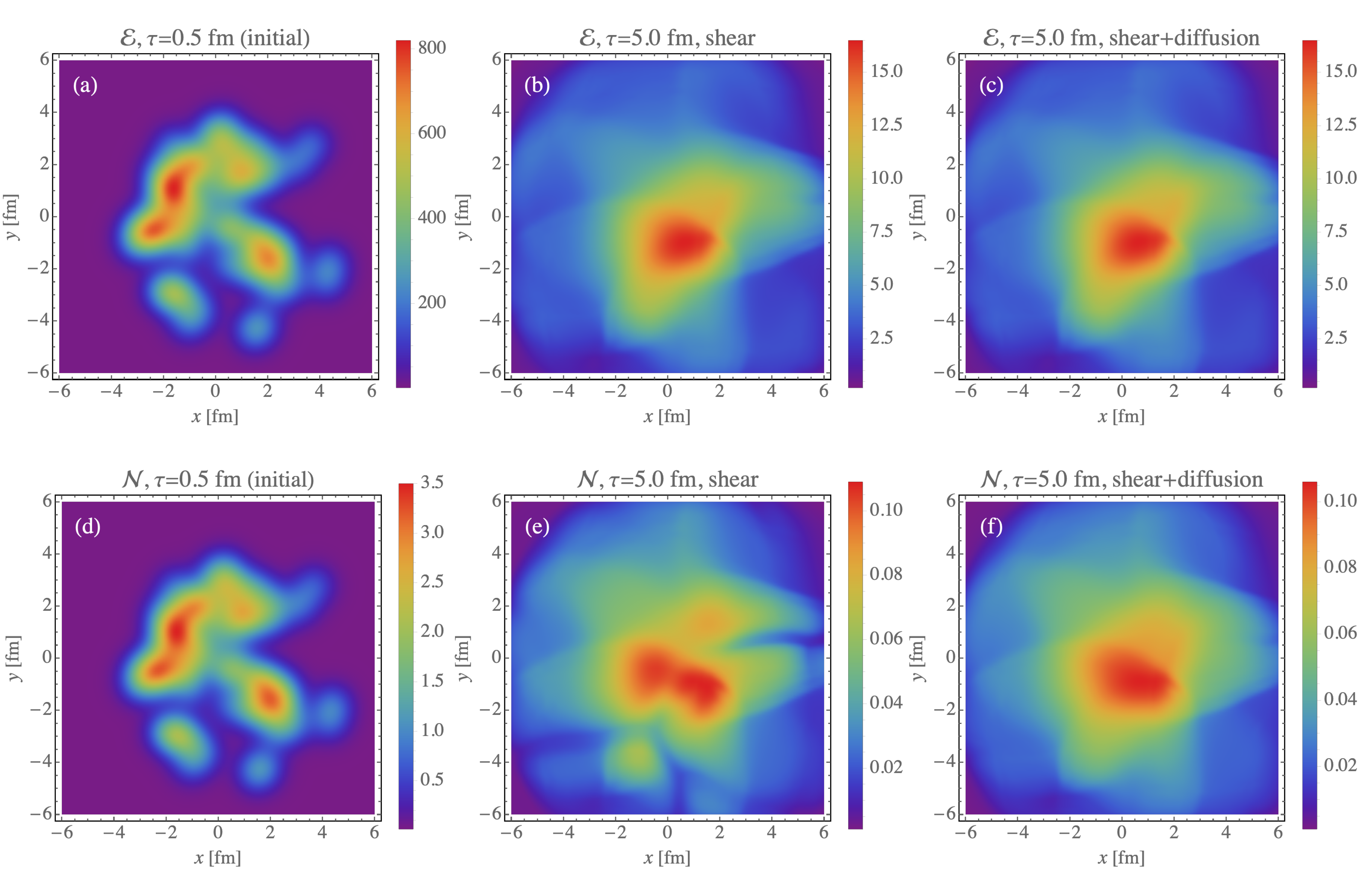}
    \caption{(Color online) Evolution of an expanding QGP with non-zero net baryon density with a bumpy initial condition, for a Cu+Cu collision at $b=4$\,fm. {\sl Top row:} Energy density $\ed$ in fm$^{-4}$ at initial time $\tau_0=0.5$\,fm/$c$ (a) and at time $\tau=5$\,fm/$c$ with kinematic shear viscosity $\bar\eta = 0.2$ and zero (b) or non-zero (c) baryon diffusion. {\sl Bottom row:} Same as top row, but for the net baryon density $\n$ in fm$^{-3}$. EOS4 is used for the equation of state, and $C_n = 0.4$ for evaluating the baryon diffusion coefficient $\kappa_n$ in Eq.~(\ref{kappa}). The bulk viscous pressure is set to zero.}
    \label{fig:visualization}
\end{figure}

Following Refs.~\cite{Shen:2017ruz, Du:2018mpf, Denicol:2018wdp} we use a 3-dimensional initial condition at non-zero baryon density which extends a transverse profile obtained from the Monte Carlo (MC) Glauber model \cite{Miller:2007ri} into the longitudinal direction with the following prescription:
\begin{eqnarray}
    \ed(\tau_0,x,y,\eta_s) &=& \frac{\ed_0}{\tau_0}\,
    \Bigl[T_A(x,y)\ed_A(\eta_s)+T_B(x,y)\ed_B(\eta_s)\Bigr]\;,
\\
    \n(\tau_0,x,y,\eta_s) &=& \frac{1}{\tau_0}\,
    \Bigl[T_A(x,y)\n_A(\eta_s)+T_B(x,y)\n_B(\eta_s)\Bigr]\;,
\end{eqnarray}
where $T_{A/B}(x,y)$ are the transverse profiles of the right- and left-moving nuclei from the MC-Glauber model and $\ed_{A/B}(\eta_s)$, $\n_{A/B}(\eta_s)$ are the corresponding longitudinal profiles for the energy and net baryon density, respectively \cite{Shen:2017ruz, Du:2018mpf, Denicol:2018wdp}. $\ed_0$ is a normalization factor which can be tuned to reproduce the final multiplicity while $\n(\tau_0,x,y,\eta_s)$ is normalized to the total number of participant baryons \cite{Denicol:2018wdp}.

In Fig. \ref{fig:visualization}, we show the transverse distributions at $\eta_s=0$ of the energy density (upper panels) and net baryon density (lower panels), at the hydrodynamic starting time $\tau_0 =0.5$\,fm/$c$ (left column) and later at $\tau=5$\,fm/$c$, for evolution with only shear stress turned on (middle column) and both shear stress and baryon diffusion turned on (right column). By comparing the middle and right columns we note that baryon diffusion leaves no pronounced signatures in the evolution of the energy density but smoothes out gradients in baryon density. The authors of Refs.~\cite{Shen:2017ruz,Du:2018mpf,Denicol:2018wdp} came to a similar conclusion for the evolution in the longitudinal direction.

\section{Summary}

In this chapter we have described the structure and performance of a new code called \code\ describing the (3+1)-dimensional space-time evolution of hot and dense matter created in relativistic heavy-ion collisions using second-order dissipative relativistic fluid dynamics. \code\ differs from most other publicly available algorithms by evolving, together with energy and momentum, a conserved current related to net baryon number, including its dissipative diffusion current, and it evolves the medium with an equation of state that depends on both the energy and net baryon densities. The generalization of the \code\ algorithm to the simultaneous propagation of multiple conserved charge currents \cite{Greif:2017byw} should be a straightforward task for the near future. The code can also be plugged in the JETSCAPE framework \cite{Putschke:2019yrg} as a hydrodynamic evolution module.

After briefly describing the physics to be addressed with \code\ simulations, we presented in detail the equations \code\ solves, and the numerical schemes it employs, including the root-finding algorithm for finding the flow velocity field at each time step and the regulation schemes used to regulate overly large dissipative flows caused by numerical or anomalously large physical fluctuations. The performance of the code was verified with high precision using a series of standard tests involving comparison with analytically or semi-analytically known solutions for problems of reduced dimensionality, characterized by additional symmetries that are usually not respected in real life situations but can be exploited for code verification. The code is distributed together with this suite of verification test protocols, thus enabling the user to check the continued accuracy of the code after changing or generalizing some of its parts. Finally, we presented a simple example illustrating the practical effects of baryon number diffusion on the evolution of energy and net baryon density for a collision between copper (Cu) nuclei, demonstrating the smoothing effects of baryon diffusion on large gradients of the net baryon density in the transverse plane.    

\chapter{Baryon-charged medium undergoing Gubser expansion}
\label{ch.gubser}

Gubser flow \cite{Gubser:2010ze,Gubser:2010ui} describes conformally symmetric systems that, in addition to longitudinally boost-invariant Bjorken flow, undergo at the same time strong azimuthally symmetric (``radial'') transverse flow. Contrary to heavy-ion collisions where transverse flow is initially zero and only generated after the collision in response to transverse pressure gradient, in Gubser flow the transverse flow exists at all times (i.e. even on a hypersurface corresponding to some very early ``initial'' longitudinal proper time $\tau_0$). 

In spite of its simplicity and high symmetry, Gubser flow captures essential features of the expansion pattern exhibited by systems produced in heavy-ion collisions and, more importantly, it has analytical solutions which we can use to validate numerical codes or carry out exploratory studies. For these reasons we summarize a number of key ingredients of Gubser flow in this chapter which will be used elsewhere in this thesis. For example, we used Gubser flow to validate our hydrodynamic code \code{} in Ref.~\cite{Du:2019obx} and the embedded freeze-out surface finder in Ref.~\cite{du2021jet}, and we studied the fluctuation dynamics on top of the Gubser flow in Ref.~\cite{Du:2020bxp}. This chapter is based on materials from these references \cite{Du:2019obx, du2021jet, Du:2020bxp}.

\newpage
%
%

\section{Gubser coordinates}
\label{sec4.3.1}

Gubser flow originates from an ingenious symmetry that, like Bjorken symmetry, makes the flow appear static in an appropriately chosen set of coordinates called ``Gubser coordinates'' which map Minkowski space onto a 3-dimensional de Sitter space times a line, dS$_3\otimes\mathbb{R}$ \cite{Gubser:2010ze,Gubser:2010ui}. As a result of this symmetry, macroscopic quantities do not depend on any of the space-like coordinates but only on the time-like coordinate $\rho\in\mathbb{R}$ in this coordinate system, and the dissipative hydrodynamic equations reduce to a set of coupled ODEs in $\rho$.

To introduce the Gubser coordinates we first rescale the invariant distance measure of Minkowski space in Milne coordinates with a Weyl transformation\footnote{%
    To make our equations readily comparable with those in the by now vast literature on Gubser flow we temporarily switch our metric signature convention to the mostly-plus metric, i.e. $g^{\mu\nu} = \textrm{diag}(-1,1,1,1)$ in Cartesian coordinates.}  
\begin{equation}
    ds^2\to d\hat s^2\equiv ds^2/\tau^2 = (-d\tau^2+dr^2+r^2d\phi^2)/\tau^2+d\eta_s^2\;.
    \label{eq-weyl}
\end{equation}
Next we perform the coordinate transformation $x^\mu = (\tau, r, \phi, \eta_s) \mapsto \hat x^\mu = (\rho,\theta,\phi, \eta_s)$,\footnote{%
    Here $r\equiv\sqrt{x^2+y^2}$ and $\phi\equiv\tan^{-1}(x/y)$. All quantities expressed as functions of Gubser coordinates are made unitless by scaling them with the appropriate powers of the Milne time $\tau$ and labeled with a hat.} 
by introducing \cite{Gubser:2010ze,Gubser:2010ui}
\begin{align}
    \rho(\tau,r) &\equiv - \sinh^{-1}\left(\frac{1-q^2\tau^2+q^2r^2}{2q\tau}\right)\;,\label{eq-rho}\\
    \theta(\tau,r) &\equiv\tanh^{-1}\left(\frac{2qr}{1+q^2\tau^2-q^2r^2}\right)\;,\label{eq-theta}
\end{align}
where $q$ is an arbitrary energy scale that defines the physical size of the system (the solution is invariant under a common rescaling of $q$, $\tau$ and $r$ such that $qr$ and $q\tau$ remain unchanged). In these coordinates the Weyl-rescaled invariant distance measure becomes
\begin{equation}
    d\hat s^2 = -d\rho^2+\cosh^2\rho\,
    \bigl(d\theta^2+\sin^2\theta d\phi^2\bigr)+d\eta_s^2\;,
\label{eq-measure-desitter}
\end{equation}
with the metric 
\begin{equation}
    \hat g_{\mu\nu} = \textrm{diag}(-1,\, \cosh^2\rho,\, \cosh^2\rho\sin^2\theta,\, 1)\;.
\label{eq-metric-desitter}
\end{equation}
A system that appears static in the coordinates $\hat x^\mu = (\rho, \theta, \phi, \eta_s)$, i.e. has flow velocity $\hat u^\mu=(1,0,0,0)$, is said to exhibit Gubser flow in Minkowski space. 

To map quantities expressed in Gubser coordinates back to Milne coordinates in Minkowski space one uses metric rescaling \cite{Gubser:2010ui} and the definitions (\ref{eq-rho},\ref{eq-theta}), for example
\begin{align}
    u_{\mu}(\tau, r) &=\tau\frac{\partial \hat x^\nu}{\partial x^\mu}\hat u_\nu(\rho(\tau, r))\;,
\label{eq-gubser-umu}
\\
    \pi_{\mu\nu}(\tau, r) &=\frac{1}{\tau^2}\frac{\partial \hat x^\alpha}{\partial x^\mu}\frac{\partial \hat x^\beta}{\partial x^\nu}\hat\pi_{\alpha\beta}(\rho(\tau, r))\;,
\\
T(\tau, r) &= \hat T(\rho(\tau, r))/\tau \;,
\\
\ed(\tau, r) &= \hat \ed(\rho(\tau, r))/\tau^4 \;.
\label{eq-gubser-energy}
\end{align}
With these transformation rules the Gubser flow profile can be expressed in Milne coordinates through the components
\begin{align}
u^\tau(\tau, r) &= \cosh\kappa(\tau,r)\;, 
\label{eq-gubser-ut}
\\
u^{x}(\tau, r) &= \frac{x}{r}\sinh\kappa(\tau,r)\;,
\label{eq-gubser-ux}
\\
u^{y}(\tau, r) &= \frac{y}{r}\sinh\kappa(\tau,r)\;,
\label{eq-gubser-uy}
\\
u^\phi(\tau, r) &= u^\eta(\tau, r) = 0\;,
\label{eq-gubser-uphi}
\end{align}
where $\kappa(\tau,r)$ is the transverse flow rapidity, corresponding to the transverse flow velocity 
\begin{equation}
    v_\perp(\tau,r)=\tanh \kappa(\tau,r) \equiv 
    \frac{2q^2\tau r}{1+q^2\tau^2+q^2r^2}\;.
\label{eq-gubser-kappa}
\end{equation}
Note that the transverse flow components are azimuthally symmetric. 

This flow is dictated by symmetry, so it applies to both ideal and dissipative fluids whose thermodynamic functions have Gubser symmetry (i.e. depend only on $\rho$ when expressed in Gubser coordinates). Different initial conditions for the hydrodynamic quantities and different transport coefficients yield different $\rho$ dependencies for their evolution, translating into different characteristics $r(\tau)$ and different $(\tau, r(\tau))$ profiles when expressed in Minkowski space coordinates. 

To ensure invariance of the hydrodynamic equations under the Weyl transformation (\ref{eq-weyl}) the energy momentum tensor must be traceless. This means that a conformal EoS must be used and the  bulk viscosity and bulk viscous pressure must be set identically to zero.
 
\section{Gubser flow with baryon diffusion}
\label{sec4.3.2}

In this section we extend the existing semi-analytical solutions for conformal Israel-Stewart hydrodynamics with Gubser flow \cite{Marrochio:2013wla} to systems with a non-zero baryon diffusion current. One could argue that the longitudinal reflection symmetry under $\eta_s\to-\eta_s$ in the Gubser symmetry indicates that the baryon diffusion current should be zero \cite{Denicol:2018wdp}. However, when the conformal EoS is used, as required by the Weyl invariance of the hydrodynamics, the baryon evolution decouples from the rest of the system, and the baryon density and diffusion current evolve as background fields. This means that a non-zero baryon diffusion current does not modify the Gubser flow profile, and the numerical results for the evolution of the baryon diffusion current can be tested by comparing them to the semi-analytical solutions of the equations of motion obtained from $\hat u^\mu=(1,0,0,0)$ under Gubser symmetry. 

In this section we derive these baryon equations of motion in de Sitter space. We rewrite Eqs.~(\ref{eq:vhydro-N}, \ref{eq:vhydro-E}) and (\ref{eq-n-simple}, \ref{eq-pi-simple}) with the mostly-plus metric tensor and apply the Gubser flow profile $\hat u^\mu=(1,0,0,0)$ and the de Sitter metric (\ref{eq-metric-desitter}) to obtain\footnote{%
    Similar to Bjorken flow, the shear stress for Gubser flow has only one independent component for which we take $\hat\pi^{\eta\eta}$. The other non-vanishing components $\hat\pi^{\theta\theta}$ and $\hat\pi^{\phi\phi}$ are related to $\hat\pi^{\eta\eta}$ by tracelessness (which gives $\hat\pi^\eta_\eta = -\hat\pi^\theta_\theta - \hat\pi^\phi_\phi$) and azimuthal symmetry (which implies $\hat\pi^\eta_\eta = -2\hat\pi^\theta_\theta = -2\hat\pi^\phi_\phi$ for evolution with azimuthally symmetric initial conditions).
    }
\begin{eqnarray}
   \partial_\rho\hat \ed + 2\tanh\rho\,\hat\ed &=& 
   2\tanh\rho\,
   \left(\frac{1}{2}\hat{\pi}^{\eta\eta} - \hat{\peq}\right),
\label{eq-gubser-evolution-e} 
\\
   \tau _{\pi}\partial_\rho\hat \pi^{\eta\eta}+\hat \pi^{\eta\eta} &=& 2\tanh\rho\left(\frac{2}{3}\hat{\eta}-\delta_{\pi\pi}\hat \pi^{\eta\eta}+\frac{1}{6}\tau_{\pi\pi}\hat \pi^{\eta\eta}\right),  
\label{eq-gubser-evolution-pi}
\\
   \partial_\rho\hat \n + 2\tanh\rho\,\hat{\n}&=&0\;,
\label{eq-gubser-evolution-n}
\\
   \tau _{\V}\partial_\rho\hat \V^\eta +\hat \V^\eta &=& -2\tanh\rho\left(\delta_{\V\V} - \frac{1}{3}\lambda_{\V\V}    
             \right) \hat \V^\eta\;.
\label{eq-gubser-evolution-neta}
\end{eqnarray}
(Note that $\hat\theta=2\tanh\rho$ is the scalar expansion rate for Gubser flow.) For the transport coefficients we use the same parametrization as described before in Fig.~\ref{F5} (see also the caption of Fig.~\ref{F6}). The transformation rules for the baryon density and diffusion current are
\begin{eqnarray}
   \n(\tau, r) &=& \hat\n(\rho(\tau, r))/\tau^3 \;,
\label{eq-gubser-n}
\\
   \V^\eta(\tau, r) &=& \hat\V^\eta(\rho(\tau, r))/\tau^4\;.
\label{eq-gubser-nmu}
\end{eqnarray}

Given initial conditions $\hat\ed(\rho_0)$, $\hat\pi^{\eta\eta}(\rho_0)$, $\hat\n(\rho_0)$, and $\hat\V^\eta(\rho_0)$ at, say, $\rho_0 = 0$, obtaining $\hat\peq(\rho_0)$ and $\hat T(\rho_0)$ with the help of EOS1 in Sec.~\ref{subsec-eos}, Eqs.~(\ref{eq-gubser-evolution-e}-\ref{eq-gubser-evolution-neta}) can be solved with {\sc Mathematica} in Gubser coordinates and then translated into Milne coordinates $(\tau,x,y,\eta_s=0)$ using the transformation rules (\ref{eq-gubser-umu}-\ref{eq-gubser-energy}) and (\ref{eq-gubser-n}-\ref{eq-gubser-nmu}). This semi-analytic solution can then be used to obtain initial conditions for \code\ on an initial proper time hypersurface $\tau_0$, which are then further evolved with the (3+1)-dimensional \code\ code. Specifically, \code\ requires initial data at $\tau_0$ for 
    $$\ed, \peq, T, u^x, u^y, \pi^{xx}, \pi^{yy}, \pi^{xy}, \pi^{\tau\tau}, \pi^{\tau x}, \pi^{\tau y}, \pi^{\eta\eta}, \n, n^\eta$$
on the computational $(x, y)$ grid (due to longitudinal boost-invariance these are only required at $\eta_s=0$). All remaining hydrodynamic components are either zero or can be obtained from the above by symmetry. For example, all shear stress components $\pi^{\mu\nu}(\tau_0, x, y, \eta_s=0)$ for \code\ can be obtained from the semi-analytic solution $\hat\pi^{\eta\eta}(\rho)$ by using tracelessness and azimuthal symmetry,
\begin{align}
    \hat\pi_{\theta\theta}(\rho)&=-\frac{1}{2}\cosh^2\rho\;\hat\pi_{\eta\eta}(\rho)\;, \quad
    \hat\pi_{\phi\phi}(\rho)=-\frac{1}{2}\cosh^2\rho\sin^2\theta\;\hat\pi_{\eta\eta}(\rho)
\end{align}
(with all other Gubser components being zero by symmetry), followed by 
\begin{equation}
    \pi_{\mu\nu}(\rho(\tau, r(x, y))) =\frac{1}{\tau^2}\frac{\partial \hat x^\alpha}{\partial x^\mu}\frac{\partial \hat x^\beta}{\partial x^\nu}\hat\pi_{\alpha\beta}(\rho)\;.
\end{equation}
This gives, for example, at $(\tau{=}\tau_0,\eta_s{=}0)$
\begin{equation}
    \pi^{\tau\tau}\Bigl(\rho(\tau_0, r(x, y))\Bigr) = -\frac{q^2\sin^2\theta}{2\tau_0^2}\,\hat\pi_{\eta\eta}(\rho)\;,\label{eq-gubseric}
\end{equation}
where the value of $\rho$ depends on the transverse grid point $(x,y)$.\footnote{%
    Obviously, whenever the transport coefficients are changed in \code, the semi-analytic solution must be recomputed accordingly for comparison, also because the full exact solution (not just its initial conditions in Gubser coordinates) is required to obtain initial conditions for the \code\ code in Milne coordinates.}

%

In the case of ideal Gubser flow, where the dissipative components all vanish, the equations of motion in Eqs.~(\ref{eq-gubser-evolution-e}-\ref{eq-gubser-evolution-neta}) are further simplified as follows:
\begin{subequations}
\label{eq-gubser-evolution2}
\begin{eqnarray}
  \partial_\rho\hat n + 2\tanh\rho\, \hat n &=& 0\;,
\label{eq-gubser-evolution-n2}
\\
  \partial_\rho\hat e + 2\tanh\rho\, \hat e &=& -2\tanh\rho\,\hat p\;,
\label{eq-gubser-evolution-e2}
\end{eqnarray}
\end{subequations}
whose analytical solutions are \cite{Gubser:2010ze}
\begin{equation}
    \hat{n}(\rho) \propto \frac{1}{(\cosh\rho)^{2}}\;,\quad \hat{e}(\rho) \propto \frac{1}{(\cosh\rho)^{8/3}}\;.
\end{equation}
Using the translation rules illustrated above, one can get their profiles in Milne coordinates:
\begin{eqnarray}
    n(\tau, r) &\propto& \frac{1}{\tau^{3}}  \frac{(2q\tau)^{2}}{\bigl[1+2q^2(\tau^2{+}r^2)+q^4(\tau^2{-}r^2)^2\bigr]}\;,\label{idealGn}\\
    e(\tau, r) &\propto& \frac{1}{\tau^{4}}  \frac{(2q\tau)^{8/3}}{\bigl[1+2q^2(\tau^2{+}r^2)+q^4(\tau^2{-}r^2)^2\bigr]^{4/3}}\label{idealGe}\;.
\end{eqnarray}
Here $n$ and $e$ can have arbitrary dimensionless prefactors which fix their scales, and we shall discuss our choice of them motivated by phenomenology in the following section.

%
\section{Phenomenologically motivated parameters}\label{sec:gubser_para}
%

In this section, we will fix the scales for energy and baryon densities in Eqs.~\eqref{idealGn} and \eqref{idealGe}, from the final  particle multiplicities measured in Au-Au collisions at top RHIC energy, assuming ideal Gubser flow. The thus obtained ideal Gubser profile shall be used for study off-equilibrium fluctuation dynamics in Ch.~\ref{ch.fluctuations} and medium response to energetic partons in Ref.~\cite{du2021jet}. 

For consistency with the conformal symmetry, we use the EOS3 in Sec.~\ref{subsec-eos}, which using $\alpha=\mu/T$ can be rewritten as \cite{Hatta:2015era}
\begin{equation}
    e = 3p \equiv f_*(\alpha) T^4\,,\quad n = g_*(\alpha)\mu T^2 \equiv \alpha g_*(\alpha) T^3\,,\label{eq:eos1}
\end{equation}
and thus
\begin{equation}
    s = \frac{1}{T}(e+p-\mu n)=T^3\Bigl(\frac{4}{3}f_*(\alpha)-\alpha^2g_*(\alpha)\Bigr)\equiv h_*(\alpha) T^3\,,
\label{eq:eos2}
\end{equation}
with the unitless coefficients
\begin{subequations}
\label{eq:fgh}
\begin{eqnarray}
  f_*(\alpha)&=&3p_0+\frac{N_f}{6}\alpha^2+\frac{N_f}{108\pi^2}\alpha^4\,,
\label{eq:f}
\\
  g_*(\alpha)&=&\frac{N_f}{9}+\frac{N_f}{81\pi^2}\alpha^2\,,
\label{eq:g}
\\
  h_*(\alpha)&=&4p_0+\frac{N_f}{9}\alpha^2\,.
\label{eq:h}
\end{eqnarray}
\end{subequations}
One easily verifies that Eqs.~(\ref{eq-gubser-evolution2}) together with the conformal EoS (\ref{eq:eos1}) are solved consistently if $\alpha=\mu/T$ is a $\rho$-independent constant.\footnote{%
    This is generally not true for dissipative Gubser flow at non-zero baryon density.}
The corresponding temperature evolution is \cite{Gubser:2010ze}
\begin{equation}
    \hat{T}(\rho) = \frac{C}{(\cosh\rho)^{2/3}}\;,
\end{equation}
or, equivalently, in Milne coordinates
\begin{equation}
    T(\tau,r) = \frac{C}{\tau}\frac{(2q\tau)^{2/3}}{\bigl[1+2q^2(\tau^2{+}r^2)+q^4(\tau^2{-}r^2)^2\bigr]^{1/3}}\,,
\label{eq-gubser_temp}
\end{equation}
where $C$ is a constant of integration. 

We see that the background fluid is defined by providing three constants, the size parameter $q$ (with larger $q$ corresponding to smaller transverse size), the chemical potential in units of the temperature $\alpha=\mu/T$ (with larger $\alpha$ corresponding to lower collision energies which are characterized by larger baryon number stopping), and the normalization constant $C$ (which should increase with collision energy). To simulate central Au+Au collisions we follow \cite{Gubser:2010ze} and set $q^{-1}{\,=\,}4.3$\,fm. For phenomenological guidance how to fix $\alpha{\,=\,}\mu/T$ we follow Ref.~\cite{Hatta:2015era} and note that, since ideal Gubser flow evolves at constant $\alpha$, we can determine the latter from experimental data at chemical freeze-out.  A thermal analysis of hadron yield ratios measured in Au+Au and Pb+Pb collisions at RHIC and LHC leads to the following collision energy dependence of the temperature and baryon chemical potential at chemical freeze-out \cite{Cleymans:2005xv}:
\begin{equation}
    T(\mu) = a - b\mu^2 - c\mu^4\,,\quad \mu(\snn) = \frac{d}{1+e\snn}\,,
\end{equation}
where $a{\,=\,}0.166$\,GeV, $b{\,=\,}0.139$\,GeV$^{-1}$, $c{\,=\,}0.053$\,GeV$^{-3}$, $d{\,=\,}1.308$\,GeV, $e{\,=\,}0.273$\,GeV$^{-1}$, and $\snn$ is the collision energy per nucleon pair in GeV. This curve can be well described by \cite{Cleymans:2005xv, Hatta:2015era}
\begin{equation}
    \alpha(\snn) = \frac{\mu}{T}(\snn) 
    \approx d/(a\,e\snn) \approx 29\,\mathrm{GeV}/\snn\,;
\end{equation}
for $\snn = 200$\,GeV this yields $\alpha \approx 0.145$.
Using $N_f=2.5$ in Eqs.~(\ref{eq:fgh}) this leads to
\begin{equation}
    f_* \approx 13.91\,,\quad g_* \approx 0.28\,,\quad h_* \approx 18.54\,,\label{eq:estfgh}
\end{equation}
again for $\snn = 200$\,GeV.

Finally, we fix $C$ from the total entropy produced in the collision. Again, since ideal Gubser flow conserves entropy, this can be determined from the experimentally measured final charged hadron multiplicity. Following Ref.~\cite{Gubser:2010ze} we estimate the entropy per unit space-time rapidity $\eta_s$ (which is conserved in ideal fluid dynamics with longitudinal boost-invariance) from the measured pseudorapidity density of charged particles as
\begin{equation}
    \frac{dS}{d\eta_s} \approx 7.5 \frac{dN_\mathrm{ch}}{d\eta} \approx 5000\,,
\label{eq:gubser_init_s}
\end{equation}
where we used boost-invariance to identify $\eta_s{\,=\,}\eta$ (which also makes $dS/d\eta_s$ independent of $\eta_s$) and inserted $dN_\mathrm{ch}/d\eta{\,\simeq\,}660$ for the most central Au+Au collisions at $\snn{\,=\,}200$\,GeV. $C$ is now fixed by evaluating $dS/d\eta_s$ with the entropy density (\ref{eq:eos2}) corresponding to the ideal Gubser temperature profile (\ref{eq-gubser_temp}), by integrating the entropy flux $s_\mu d^3\Sigma^\mu$ through a suitably chosen hypersurface $\Sigma$ over the entire fireball:
\begin{equation}
    \frac{dS}{d\eta_s} = \int_\Sigma s_\mu(\tau,r)\, \frac{d^3\Sigma^\mu(\tau,r,\eta_s)}{d\eta_s}, 
\label{eq:gubser_init_s1}
\end{equation}
where $s_\mu{\,=\,}s\,u_\mu$. Using for simplicity a $\tau{\,=\,}$const. surface with $d^3\Sigma^\mu=(\tau\,2\pi r\,dr\,d\eta_s,0,0,0)$, this simplifies to \cite{Gubser:2010ze}
\begin{equation}
    \!\!\!
    \frac{dS}{d\eta_s} = 2\pi h_*
    \int_0^\infty T^3(\tau,r)\, u^\tau(\tau,r)\, \tau\,r\,dr\,
    = 4\pi h_* C^3,\ 
\label{eq:gubser_init_s2}
\end{equation}
where in the last step we inserted Eq.~(\ref{eq-gubser-ut}) in the form
$$
  u^\tau(\tau,r) = \frac{1 + q^2\tau^2 + q^2r^2}      {\bigl[1 + 2q^2(\tau^2{+}r^2) + q^4(\tau^2{-}r^2)^2\bigr]^{1/2}}\,,
$$
as well as Eq.~(\ref{eq-gubser_temp}) and performed the integral.\footnote{%
    Due to scale invariance and entropy conservation the integral is independent of both $q$ and the value of $\tau$ that defines the hypersurface.}
Comparing this expression with Eq.~(\ref{eq:gubser_init_s}) yields $C\simeq 2.8$.

\section{Particle production on a closed hypersurface}
\label{sec:pheno}
%

After evolving the baryon-charged medium undergoing Gubser flow, we can evaluate particle production and corrections to it on an arbitrary hypersurface  (for example an isothermal surface). In this Section we demonstrate the formalism of calculating the correction to total entropy from the off-equilibrium slow modes (whose dynamics shall be discussed in detail in Ch.~\ref{ch.fluctuations}) as well as the final spectra of protons and anti-protons to validate \bes{} + \isd. 

With the off-equilibrium slow modes evolved on top of the ideal Gubser flow (\ref{idealGn}, \ref{idealGe}), we can compute their non-equilibrium correction to the entropy density, $\Delta s(x)$ (see Eq.~\eqref{eq:deltas} in Ch.~\ref{ch.fluctuations}), and we denote the corresponding correction to the entropy current as $\delta s^\mu\equiv u^\mu\Delta s$. When integrated over a hypersurface $\Sigma$ that encloses the entire system this generates a correction to the total entropy 
\begin{equation}
    \delta S = \int_\Sigma \delta s^\mu(x)\,d^3\Sigma_\mu(x) 
    = \int_\Sigma \Delta s(x)\, u(x){\,\cdot\,}d^3\Sigma(x)\,,
\label{eq:fods}
\end{equation}
where $d^3\Sigma_\mu$ is the four-vector of freeze-out surface element.
Similarly, the final particle distribution on the hypersurface can be obtained through the Cooper-Frye formula \cite{Cooper:1974mv} as described in Sec.~\ref{subsec-particl}:
\begin{equation}\label{eq:distr}
    \frac{d^3N_i}{p_Tdp_Td\phi_pdy}=\frac{1}{(2\pi)^3}\int_\Sigma d^3\Sigma_\mu(x) p^\mu f_i(x, p)\,,
\end{equation}
where $p^\mu$ is the four-momentum of the particle and $f_i(x, p)$ is the one-particle distribution function of species $i$, for which we here (for ideal Gubser flow) simply ignore the viscous correction and take the the Maxwell-J\"uttner distribution in Eq.~\eqref{eq:distr_f1}
\begin{equation}\label{eq:distr_f}
    f_i(x, p)= g_i \left[\exp{\left(\frac{u^\mu p_\mu-Q_i\mu_f}{T_f}\right)} +\Theta_i\right]^{-1}\,.
\end{equation}
Taking this distribution allows us to carry out the integral later in analytical form with special functions (see Eq.~\eqref{eq:valid_spec_2} below).

Parametrizing the surface $\Sigma$ in Milne coordinates
$(\tau, x, y, \eta_s)$ as
\begin{eqnarray}
    \Sigma^\mu(x)=(\tau_f(x,y,\eta_s),x,y,\eta_s)\;,
\label{eq:fosurface}
\end{eqnarray}
where $\tau_f(x,y,\eta_s)$ is the longitudinal proper time associated with the surface point located at spatial position $(x,y,\eta_s)$, the surface normal vector at point $x$ is given by
\begin{eqnarray}
    d^3\Sigma_\mu(x) 
    &=& -\epsilon_{\mu\nu\kappa\lambda} \frac{\partial\Sigma^\nu}{\partial x} \frac{\partial\Sigma^\kappa}{\partial y} \frac{\partial\Sigma^\lambda}{\partial\eta_s}\, \sqrt{-g}\,dxdyd\eta_s
\nonumber\\
\label{eq:foel}
    &=& \left(1, -\frac{\partial\tau_f}{\partial x},     -\frac{\partial\tau_f}{\partial y}, -\frac{\partial\tau_f}{\partial\eta_s}\right)
    \tau_f dxdyd\eta_s\;.
\end{eqnarray}
Here $\epsilon_{\mu\nu\kappa\lambda}$ is the Levi-Civita symbol, and $\sqrt{-g}=\tau$ is the metric determinant for Milne coordinates. The expressions (\ref{eq:fosurface}) and (\ref{eq:foel}) are completely general and allow to parametrize any hypersurface $\Sigma$ (although $\tau_f(x,y,\eta_s)$ may be multivalued). For the Gubser flow, which possesses longitudinal boost-invariance and azimuthal symmetry, it is advantageous to use polar coordinates $(r,\phi)$ instead of $(x,y)$ in the transverse plane, such that Eq.~(\ref{eq:foel}) simplifies to 
\begin{equation}
    d^3\Sigma_\mu(x)=
    \left(1, -\frac{\partial\tau_f}{\partial r}, 0, 0 \right)\tau_f rdrd\phi d\eta_s\,.
\label{eq:foe2}
\end{equation}
For an isothermal hypersurface defined by $T(\tau_f(r),r)=T_f$, $dT(\tau_f, r) = (\partial T/\partial r)dr + (\partial T/\partial \tau_f)d\tau_f = 0$, the second term can be evaluated as follows \cite{Gursoy:2014aka}:
\begin{equation}
    \frac{\partial\tau_f}{\partial r} = -\left[\left(\frac{\partial T}{\partial r}\right)\left(\frac{\partial T}{\partial \tau}\right)^{-1}\right]_{T_f}\,.
\label{eq:fode}
\end{equation}
The two factors inside square brackets are easily evaluated using the ideal Gubser temperature profile (\ref{eq-gubser_temp}). 

We shall apply \eqref{eq:fods} to our Gubser background flow without back-reaction from non-equilibrium slow modes, and study the slow-mode correction to the entropy per unit space-time rapidity $d\delta S/d\eta_s$ on an isothermal hypersurface defined by $T(\tau_f(r),r)=T_f$ in Ch.~\ref{ch.fluctuations}. In this situation,  the entropy correction becomes
\begin{equation}
    \frac{d\delta S}{d\eta_s} = 2\pi\int_0^{r_\mathrm{max}} \Delta s
    \left(u^\tau-u^r\frac{\partial\tau_f}{\partial r} \right)\tau_f(r) r dr\;. 
\label{eq:fodsgubser}
\end{equation}
Here the finite upper limit $r_\mathrm{max}$ accounts for the finite radial extent of this isothermal surface --- for points with $r{\,>\,}r_\mathrm{max}$ the temperature never exceeds $T_f$. 

The final particle spectra in Eq.~\eqref{eq:distr} can be simplified as well, since with boost-invariance and radial symmetry, only two components of particle momentum are non-zero:
\begin{equation}\label{eq:four_m}
    p^\tau=m_T\cosh(y-\eta_p)\,,\quad p^r=p_T\cos(\phi_p-\phi)\,.
\end{equation}
When using the Boltzmann distribution and setting $\mu_f\eq0$ in Eq.~\eqref{eq:distr_f}, the integral can be carried out with special functions, and the particle spectra become
\begin{eqnarray}\label{eq:valid_spec_2}
    \frac{d^3N_i}{p_Tdp_Td\phi_pdy}&=&\frac{g_i}{2\pi^2(\hbar c)^3}\int dr\,r\tau_f(r)\left[m_T K_1\left(\frac{m_Tu^\tau}{T_f}\right) I_0\left(\frac{p_Tu^r}{T_f}\right)\right.\\
    &+&\left.\left(-\frac{\partial\tau_f(r)}{\partial r}\right)p_T K_0\left(\frac{m_Tu^\tau}{T_f}\right) I_1\left(\frac{p_Tu^r}{T_f}\right) \right]\,,\nonumber
\end{eqnarray}
where $K_n(z)$ and $I_n(z)$ are the modified Bessel function of the second kind and the first kind, respectively. Note that the $\hbar{}c$ factor is restored here so that the results can be directly compared to what is given by \isd. We shall use Eq.~\eqref{eq:valid_spec_2} to test our freeze-out finder and particle sampler in the next section.

\section{Validation of \bes{} + \isd}

\begin{figure*}[!htbp]
    \centering
    \includegraphics[width= 0.95\textwidth]{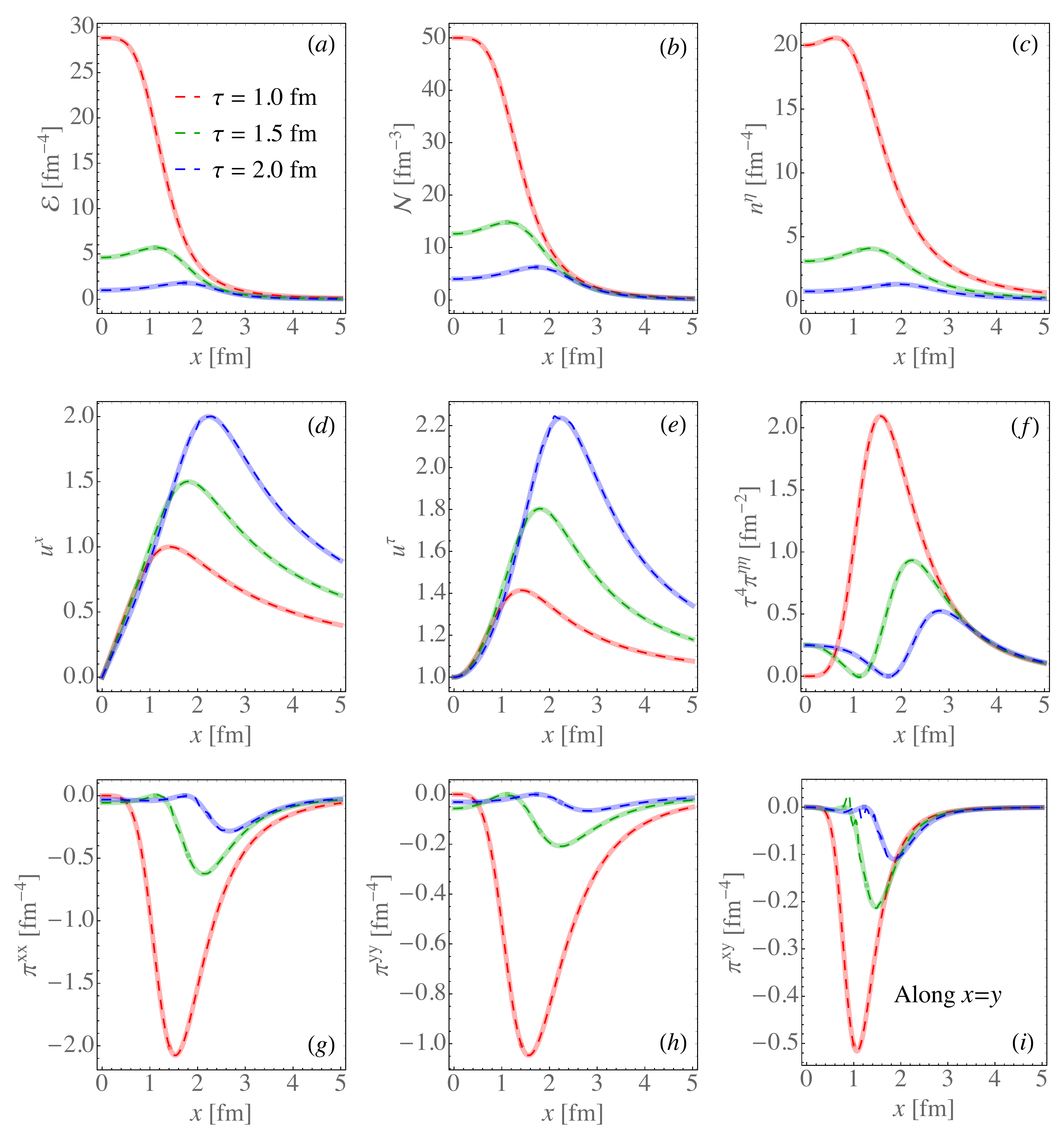}
    \caption{Comparison for Gubser flow between the semi-analytical solutions (continuous lines) and the numerical results from \code\ (using the Newton–Raphson method as root-finder) (broken lines) at $\tau = 1.0,\; 1.5,\; 2.0$ fm$/c$: (a) energy density $\ed$; (b) net baryon density $\n$; (c) baryon diffusion current $n^\eta$; (d),(e) flow velocity components $u^x$ and $u^\tau$; (f)-(i) shear stress tensor components $\tau^4\pi^{\eta\eta}$, $\pi^{xx}$, $\pi^{yy}$, and $\pi^{xy}$. Except for $\pi^{xy}$ in the last panel, which is plotted along the line $x=y$ (or $\phi=\pi/4$), all quantities are shown as functions of $x$ at $\eta_s=\phi=0$. The comparison is made for $q=1$\,fm$^{-1}$ and $\theta_\mathrm{f} = 1.8$, using temporal and spatial grid spacings $\Delta\tau=0.005$\,fm/$c$ and $\Delta x = \Delta y = 0.05$\,fm. For the transport parameters we use $\bar\eta = 0.2$, $C_n=4$, and $\tau_{\pi\pi} = 0$; the remaining transport coefficients are specified in the text. The simulation starts at $\tau_0 = 1$\,fm/$c$ with the following initial conditions: at $\rho=0$ (or equivalently at $(\tau = \tau_0,\; r=0)$) we set $T = 1.2$\,fm$^{-1}$, $\pi^{\eta\eta}=0$, $\n=50$\,fm$^{-3}$, and $n^\eta=20$\,fm$^{-3}$.
    \label{F6}
    }
\end{figure*}

First, we test the performance of our hydrodynamic simulation, by comparing the \code\ output to the semi-analytic Gubser solution with baryon diffusion in Fig.~\ref{F6}, for the setup described in the figure caption. One observes excellent agreement. Owing to the non-trivial transverse expansion of Gubser flow it allows to test additional source terms in the \code\ evolution equations when compared with the Bjorken flow test in Ch.~\ref{ch:numerics}. We can also use it to study the performance of the root-finding algorithm in \code. 

As discussed in Sec.~\ref{sec-root}, when searching for the flow velocity given the energy-momentum tensor two different methods are used in complementary ranges of the flow velocity separated by the critical value $v = 0.563624$ or, equivalently, $u^\tau = 1.21061$. Fig.~\ref{F6}(e) shows that the root-finding algorithm works equally well on both sides of the critical value of $u^\tau$. We also checked the precision and relative speed of convergence of the Newton-Raphson method and modified iteration schemes described in Sec.~\ref{sec-root}. For both methods, the maximum number of iterations is set to 100 (which is never reached), and the root finding stops when the relative error $|v_{i+1}-v_{i}|/v_{i+1} < 10^{-6}$ or $|u^\tau_{i+1}-u^\tau_{i}|/u^\tau_{i+1} < 10^{-4}$. For the conformal EoS used here we found both methods to converge about equally well to the same result (within the specified uncertainty), with the modified iteration scheme being about 15\% faster than the Newton-Raphson method.\footnote{%
    During the early evolution stages the Newton-Raphson method converges somewhat faster but at later times the modified interaction scheme is found to be more efficient.} 


Next, we perform another test to validate the continuous spectra of final particles given by \isd, especially those of protons and anti-protons  (for our study in Ref.~\cite{du2021jet}), where non-zero chemical potential plays an important role. At the same time, the test can validate the freeze-out finder incorporated in \bes{} based on {\sc Cornelius} \cite{Huovinen:2012is} as well, since the freeze-out surface file  used by \isd{} is generated from \bes{}. For this test, we use the ideal Gubser flow which can make the validation straightforward, since analytical solutions are available for both the freeze-out surface and the final particle spectra. To perform the test, we start the hydrodynamic evolution at $\tau_0\eq{}1$~fm, and $q\eq{}1$~fm$^{-1}$, with $C\eq1.2$ (corresponding to an initial temperature $T_0\simeq 237$~MeV at the center of the fireball), and $\alpha\eq\mu/T\eq0.5$ (Note that $\alpha$ remains a constant in the ideal evolution). The initial temperature and corresponding chemical potential give us the initial values for energy and baryon densities according to the Equation of State in Eqs.~\eqref{eq:eos1}, which are used to initialize \bes{}. The freeze-out surface is defined by the freeze-out temperature $T_f\eq148$~MeV, and thus the associated freeze-out chemical potential is $\mu_f\eq\alpha{}T_f\eq74$~MeV.

%
\begin{figure}[!t]
    \centering
    \includegraphics[width=0.8\textwidth]{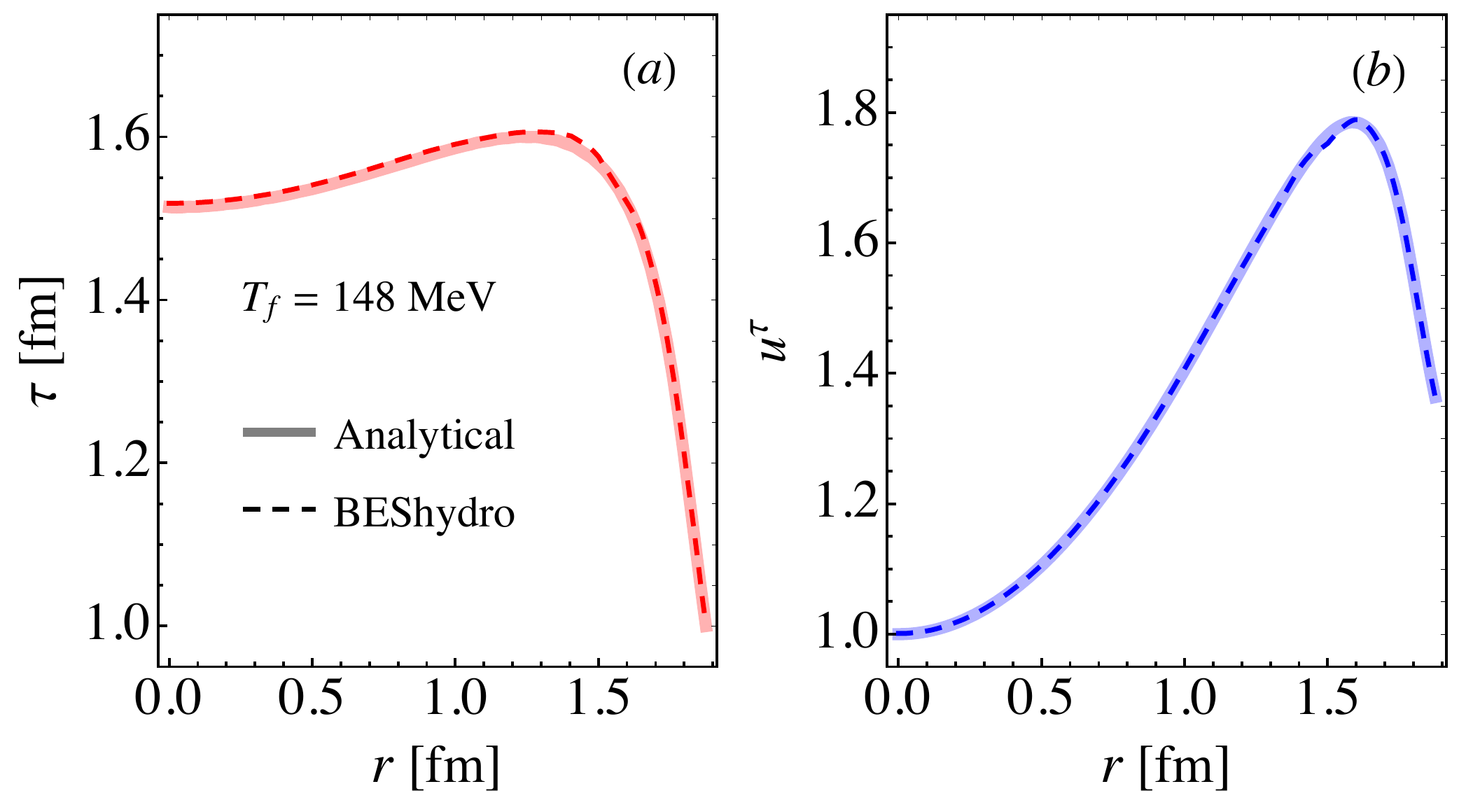}
    \caption{Validation of the freeze-out surface finder of \bes, by comparing its results (dashed lines) to analytical profiles (solid lines), at freeze-out temperature $T_f{\,=\,}148$~MeV. (a) The space-time profile of the freeze-out surface, and (b) time component of flow $u^\tau$ on the surface.}
    \label{fig:fz_validation}
\end{figure}
%

We start with testing the freeze-out finder by comparing numerical results from \bes{} to the analytical profiles on the freeze-out surface, whose space-time coordinates $\tau_f(r)$ can be obtained by solving $T(\tau,\,r)\eq{}T_f$ where $T(\tau,\,r)$ is from Eq.~\eqref{eq-gubser_temp}. Given the space-time profile $\tau_f(r)$, the other quantities on the freeze-out surface are readily available, by plugging $\tau_f(r)$ in their analytical space-time profiles, for example, the temperature distribution in Eq.~\eqref{eq-gubser_temp}. (Of course, one would get a constant distribution for $T(\tau_f(r), r)$.) We show the comparison in Fig.~\ref{fig:fz_validation}, where Fig.~\ref{fig:fz_validation}(a) shows the space-time profile of the freeze-out surface, and Fig.~\ref{fig:fz_validation}(b) shows $u^\tau$ on the surface, i.e., $u^\tau(\tau_f(r), r)$. The consistency between the numerical results from \bes{} (dashed lines) and the analytical profiles (solid lines) shows the good performance of the code, both for the hydrodynamic evolution and the freeze-out finder.

\begin{figure}[!t]
    \centering
    \includegraphics[width=0.8\textwidth]{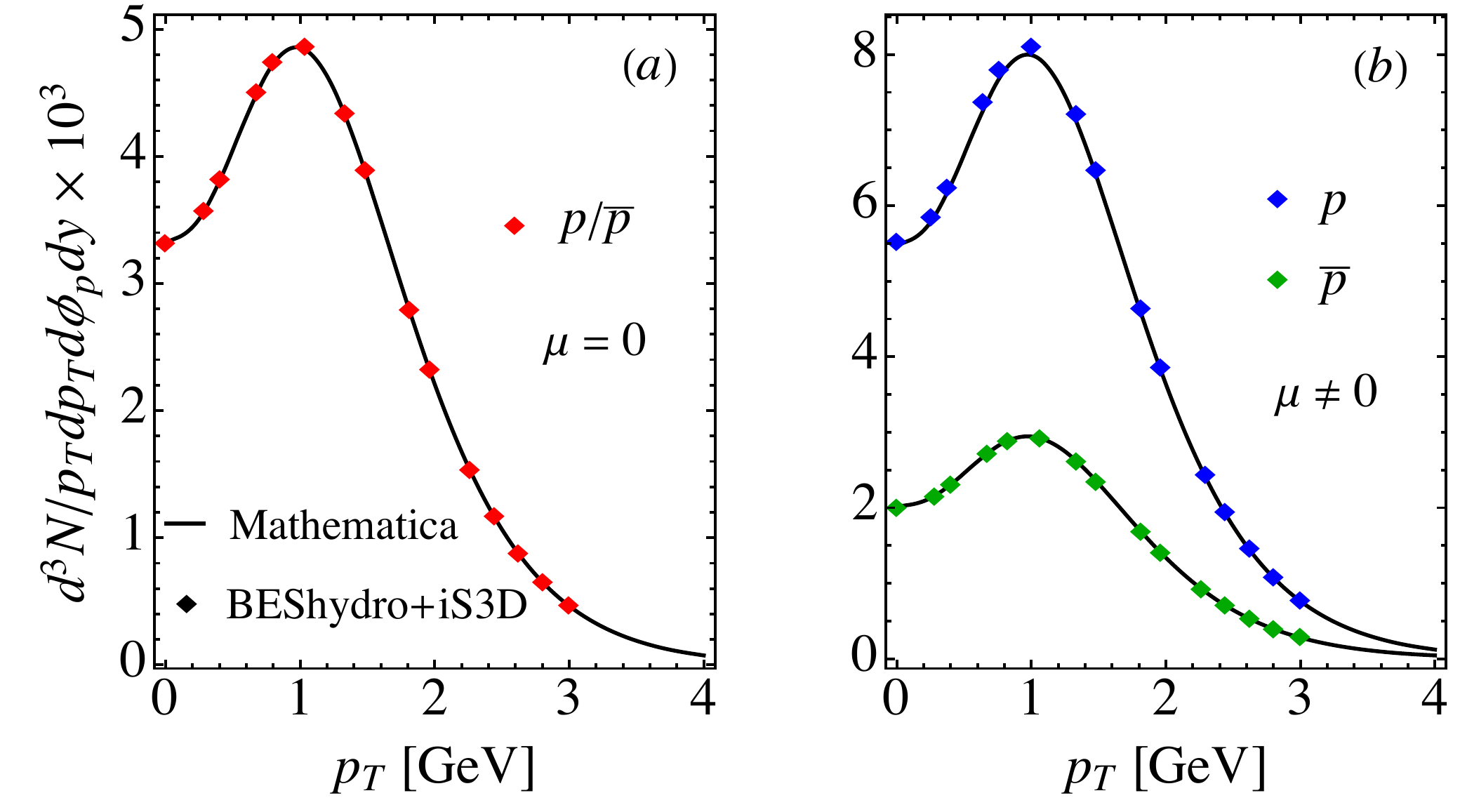}
    \caption{Validation of the combined \bes{} + \isd, by comparing the spectra of proton and anti-proton from the simulation (rhombus) and analytical distribution (solid line), with the freeze-out temperature $T_f{\,=\,}148$~MeV and $\mu_f{\,=\,}74$~MeV, with $\alpha{\,=\,}0.5$. (a) Comparison for the case by setting $\mu_f\eq 0$ in \isd{}, and the analytical distribution is given by Eq.~\eqref{eq:valid_spec_2}. (b) Comparison for the case with non-zero baryon chemical potential $\mu_f$, where the analytical distributions are calculated by {\sc Mathematica}. We note that changing grid size from 0.05 fm to 0.025 fm doesn't change the final spectra or yields by $O(10^{-3})$, showing the convergence with such resolution.}
    \label{fig:spec_valid}
\end{figure}

With the analytical Gubser flow profile, all quantities in Eq.~\eqref{eq:valid_spec_2} are known or calculable once the freeze-out surface is defined, and then the particle spectrum can be calculated using {\sc Mathematica}. We first carry out the integral in Eq.~\eqref{eq:valid_spec_2},  which assumed $\mu_f\eq0$, with $m_p\eq938$~MeV and $g_i=2$ for protons and anti-protons, which have the same spectra in this case. On the other hand, using \isd{} with the $T_f=\mathrm{const.}${} freeze-out surface found by \bes{} and setting $\mu_f\eq0$ by hand, we can get the corresponding spectra from our (3+1)D code. The comparison between results from {\sc Mathematica} (solid line) and \bes{} + \isd{} (rhombus) is shown in Fig.~\ref{fig:spec_valid}(a). We note excellent agreement. We point out that in \isd, the Fermi-Dirac distribution is used instead of the Boltzmann distribution used in Eq.~\eqref{eq:valid_spec_2}, but nevertheless a reasonable comparison can be made, since $m_p/T_f\gg1$ and the Fermi-Dirac distribution is thus very well approximated by the Boltzmann distribution on the freeze-out surface for protons and anti-protons. Finally, to test the performance of \isd{} at non-zero chemical potential, we also compare its results to those from {\sc Mathematica} by solving Eqs.~(\ref{eq:distr},\ref{eq:distr_f}), Eqs.~(\ref{eq:foe2},\ref{eq:fode}) together with Eq.~\eqref{eq:four_m}. This comparison is shown in Fig. \ref{fig:spec_valid}(b) for both protons and anti-protons, again validating the good performance of \bes{} + \isd{}.

\chapter{Baryon transport and the QCD critical point}
\label{ch:diffcp}

Baryon diffusion is an important dissipative effect in the hydrodynamic evolution of systems carrying a conserved net baryon charge. It smoothes out chemical inhomogeneities by transporting baryon charge relative to the momentum flow from regions of large to smaller net baryon density. 
Critical effects on the bulk viscous pressure was shown to have non-negligible phenomenological consequences for the rapidity distributions of hadronic particle yields, implying that critical effects might  play an important role in the calibration of the bulk medium \cite{Monnai:2016kud}. In this chapter we study critical effects on the bulk evolution in the baryon sector, by including the critical scaling of the relaxation time for the baryon diffusion current and of the baryon diffusion coefficient, as well as (in a simplified treatment) the critical contribution to the Equation of State (EoS) \cite{Nonaka:2004pg, Parotto:2018pwx, Stafford:2021wik}. We study the phenomenological consequences of baryon diffusion in a system both away from and close to the QCD critical point; the former is essential for modeling heavy-ion collisions at the high end of the BES collision energy range \cite{Shen:2017ruz, Denicol:2018wdp, Monnai:2019hkn, Du:2019obx, Greif:2017byw, Fotakis:2019nbq, Shen:2020jwv}. Including only the baryon diffusion while neglecting other dissipative effects helps us to study its hydrodynamical consequences in isolation. A more comprehensive study including all dissipative effects simultaneously is left for future work. 

This chapter is based on material published in Ref.~\cite{du2021baryon}.

\newpage
\section{Baryon diffusion and critical behavior}\label{sec:hydro}

As discussed in Ch.~\ref{ch:multistage}, in heavy-ion collisions the conserved quantities are energy, momentum, as well as net baryon charge, electric charge and strangeness, among which we only study the net baryon charge in this thesis. The baryon diffusion current $n^\mu$ is the dissipative part of the net baryon current which in the Landau matching scheme manifests itself as a non-zero spatial vector in the local rest frame of the fluid. Its evolution equation is obtained from the Denicol-Niemi-Molnar-Rischke (DNMR) theory \cite{Denicol:2010xn,Denicol:2012cn} (see Sec.~\ref{sec:relat_hydro} for more details). In this chapter, to isolate the effects arising from the baryon diffusion current $n^\mu$, we shall ignore other dissipative effects due to $\pi^{\mu\nu}$ and $\Pi$.

With this approximation, the equation of motion for $n^\mu$ from DNMR theory reads
\begin{equation}\label{eq:IS_nmu}
\tau_{n}\dot{n}^{\left\langle \mu \right\rangle}+n^{\mu } = \kappa_n
 \nabla^{\mu}\alpha+ {\cal J}^\mu\,,
\end{equation}
where $\dot{n}^{\left\langle \mu \right\rangle} \equiv \Delta^\mu_\nu\dot{n}^\nu=\Delta^\mu_\nu{}D{n}^\nu$, $\kappa_n$ is the baryon diffusion coefficient, $\alpha\equiv\mu/T$ is the chemical potential in the units of temperature,  and $\tau_{n}$ is the relaxation time, over which the baryon diffusion current relaxes towards its Navier-Stokes limit
\begin{equation}\label{eq:nslimit}
    n^{\mu}_{\rm NS}\equiv\kappa_n \nabla^{\mu}\alpha\,.
\end{equation}
Here $\nabla^\mu \equiv \Delta^{\mu\nu}d_\nu$ is the spatial gradient in the local rest frame. The term ${\cal J}^\mu$ contains higher order gradient contributions \cite{Denicol:2010xn,Denicol:2012cn}. Rewriting Eq.~\eqref{eq:IS_nmu} as a relaxation equation
\begin{equation}\label{eq:IS_nmu2}
\dot{n}^{\left\langle \mu \right\rangle} = -\frac{1}{\tau_{n}}(n^{\mu }-\kappa_n
 \nabla^{\mu}\alpha)+ \frac{1}{\tau_{n}}{\cal J}^\mu\,
\end{equation}
shows that $\nabla^{\mu}\alpha$ is the driving force for baryon diffusion while $\kappa_n$ controls the strength of the baryon diffusion flux arising in response to this force. $\tau_n$ characterizes the response time scale. As mentioned in Sec.~\ref{transcoeff} on transport coefficients, both $\tau_{n}$ and $\kappa_n$ depend on the microscopic properties of the medium, which have been calculated in various theoretical frameworks, including kinetic theory \cite{Denicol:2018wdp} and holographic models \cite{Son:2006em,Rougemont:2015ona}. In principle, they can also be constrained phenomenologically by data-driven model inference, but as of today such studies are still very limited for baryon evolution. 

Eq.~\eqref{eq:IS_nmu} can be recast into
\begin{equation}\label{eq:IS_nmu3}
    u^\nu\partial_\nu \V^\mu = \frac{\kappa_n}{\tau_\V} \nabla^{\mu}\alpha - \frac{\V^{\mu}}{\tau_{\V}}  -\frac{\delta_{nn}}{\tau_n}n^\mu\theta - n^\nu u^\mu D u_\nu
    - u^\alpha \Gamma^\mu_{\alpha\beta} n^\beta\,,
\end{equation}
where the $n^\mu\theta$-term arises from ${\cal J}^\mu$ (the only term we keep from ${\cal J}^\mu$ as given in \cite{Denicol:2012cn}), $\delta_{nn}$ is the associated transport coefficient, and the last two terms come from rewriting $\dot{n}^{\left\langle \mu \right\rangle}$ explicitly, with $\Gamma^\mu_{\alpha\beta}$ being the Christoffel symbols. Eq.~(\ref{eq:IS_nmu3}) is the equation we use to evolve the baryon diffusion current in this chapter. As illustrated in Sec.~\ref{subsec-trans}, critical dynamics will manifest itself in the Navier-Stokes limit of the baryon diffusion current as follows: After rewriting it in terms of density and temperature gradients
\begin{subequations}
\label{eq:nmu_NS_decomposition2}
\begin{eqnarray}
\label{eq:D_BD_T}
n^{\mu}_{\rm NS} 
= D_B\nabla^{\mu}n+D_T\nabla^{\mu}T\,
\end{eqnarray}
with the coefficients
\begin{equation}
\label{eq:D_BT2}
    D_B=\frac{\kappa_n}{T\chi}, \quad D_T=\frac{\kappa_n}{Tn}\left[\left(\frac{\partial p}{\partial T}\right)_n-\frac{w}{T}\right]\,,
\end{equation}
\end{subequations}
where $\chi\equiv(\partial n/\partial\mu)_T$ is the isothermal susceptibility, $\kappa_n$ is the baryon diffusion coefficient, and $w=\ed+p$ is the enthalpy density. Critical scaling affects $\chi$ and $\kappa_n$  as (see Sec.~\ref{subsec-trans})
    \begin{equation}\label{eq:coeff_scaling2}
        \chi\sim \xi^2\,, \quad \kappa_n\sim \xi\,,
    \end{equation}
where the exponents have been rounded to their nearest integers for simplicity. Thus according to Eqs.~\eqref{eq:D_BT2},
\begin{equation}\label{eq:D_scaling2}
    D_B \sim \xi^{-1}\,, \quad D_T\sim \xi\,. 
\end{equation}
In Sec.~\ref{subsec-trans} we also identified the critical scaling of the relaxation time for the diffusion current as
\begin{equation}\label{eq:taun_scaling2}
\tau_n\sim\xi^2\,.
\end{equation}

\begin{figure}[!tb]
\begin{center}
    \includegraphics[width= 0.55\textwidth]{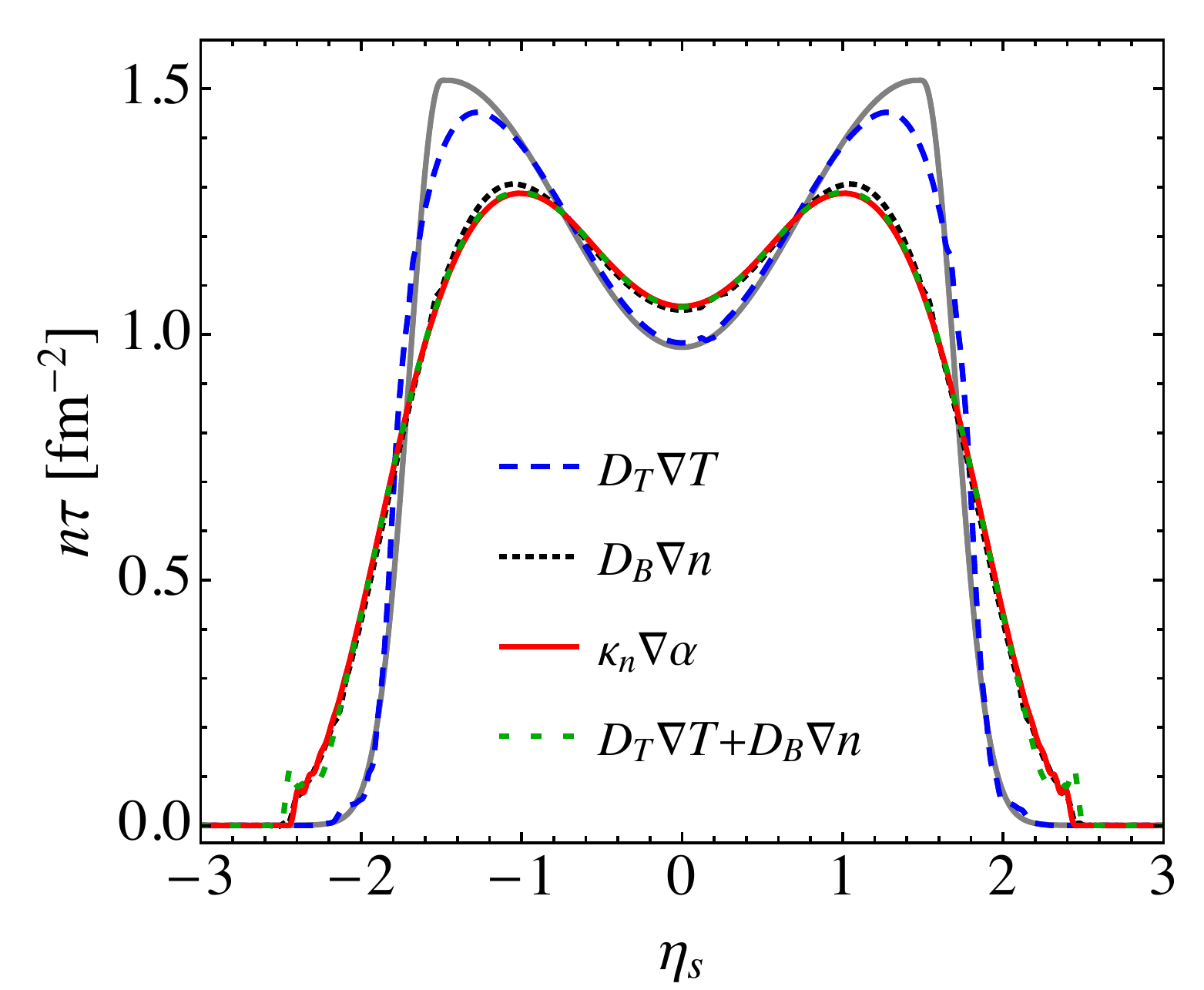}
    \caption{Validation of the equivalence of Eqs.~\eqref{eq:nslimit} and \eqref{eq:nmu_NS_decomposition2}, and comparisons among contributions from different gradient terms in these two expressions to baryon diffusion current. The gray curve shows the initial distribution of baryon density at the initial time $\tauI\eq1.5$\,fm/$c$; colored curves show the evolved distributions at $\tau\eq5.5$\,fm/$c$ for diffusive evolution driven by various gradient terms, respectively.}
    \label{fig:gradtest}
\end{center}
\end{figure}

We start by verifying numerically in Fig.~\ref{fig:gradtest} the equivalence of Eqs.~\eqref{eq:nslimit} and \eqref{eq:nmu_NS_decomposition2}, using the Equation of State {\sc neos} \cite{Monnai:2019hkn}. Note that the consistency shown in Fig.~\ref{fig:gradtest} between the results from $\kappa_n\nabla{}\alpha$ (red solid line) and $D_T\nabla{}T+D_B\nabla{}n$ (green dashed line) is a highly non-trivial test of our numeric methods, since the latter involves calculating $(\partial p/\partial T)_n$ and interpolating a table of values for $\chi=(\partial n/\partial\mu)_T$ provided by Ref.~\cite{Monnai:2019hkn}. Though the overall agreement is very good, small wiggles can be seen in the green dashed lines near the edge, indicating numerical issues when the matter becomes cold and dilute. From the figure, we also see that the baryon transport driven by $D_B\nabla^{\mu}n$ is very close to that driven by $D_B\nabla^{\mu}n+D_T\nabla^{\mu}T$, indicating that density gradients dominates over temperature gradients in the Navier-Stokes limit of the baryon diffusion current.

The above system of hydrodynamic equations is closed by the EoS, either in the form $p(\ed, n)$ or, equivalently, through the pair of relations $\bigl(\mu(\ed, n),\,T(\ed, n)\bigr)$.  For the EoS at non-zero net baryon density we use {\sc neos} \cite{Monnai:2019hkn} which was constructed by smoothly joining Lattice QCD data \cite{Bazavov:2014pvz, Bazavov:2012jq, Ding:2015fca, Bazavov:2017dus} with the hadron resonance gas model.

\section{Setup of the framework}\label{sec:setup_cp}

In this section we set up the framework for simulating the evolution of a fireball close to the QCD critical point. The core of our framework is the hydrodynamic equations discussed in Sec.~\ref{sec:hydro}. In addition to the  EoS ({\sc neos}), we must specify the transport coefficients and the  initial and final conditions. We also discuss some specific aspects of the  particlization process that arise in this setting.

\subsection{Initial conditions}

We start with the initial conditions which, from a physics perspective, describe the initial state of the systems while mathematically providing the initial data for solving the initial value problem associated with our coupled set of partial differential equations.  In this chapter, we try to establish a basic understanding of baryon diffusion dynamics for Au-Au collisions at $\snn\eq19.6$\,GeV in which we focus entirely on the longitudinal dynamics, modeling a (1+1)-dimensional system without transverse gradients initiated instantaneously at a constant proper time $\tauI$ (see also Refs.~\cite{Monnai:2016kud, Li:2018fow}). More specifically, we evolve the system hydrodynamically using the longitudinal initial profiles $e(\tauI, \eta_s),\, n(\tauI, \eta_s)$ provided in Ref.~\cite{Denicol:2018wdp}, starting at $\tauI\eq1.5$\,fm/$c$.\footnote{%
    Ref.~\cite{Denicol:2018wdp} provides longitudinal initial distributions for the entropy and baryon densities. We here adopt the functional form of their initial entropy profile as our energy profile, after appropriate normalization.}
The initial hydrodynamic profiles are shown as gray curves in Fig.~\ref{fig:long_evolution} below. The initial energy density has a plateau covering the space-time rapidity $\eta_s\in[-3.0,\,3.0]$ whereas the initial net baryon density features a double peak structure and covers a narrower region  $\eta_s\in[-2.0,\,2.0]$, reflecting baryon stopping.\footnote{%
    We note that baryon stopping affects the initial momentum rapidity $y$ of the baryon number carrying degrees of freedom and is typically modelled by a rapidity shift $\Delta y\sim 1-1.5$, depending on system size and collision energy. To translate this rapidity shift into a shift in space-time rapidity $\eta_s$ (as done in Fig.~\ref{fig:long_evolution}) requires a dynamical initialization model. Different such models yield different initial density and flow profiles \cite{Shen:2017ruz, Shen:2017bsr, Akamatsu:2018olk, Du:2018mpf, Shen:2020gef, Shen:2020mgh}.}
For the initial longitudinal momentum flow we take the ``static'' flow profile $u^\mu = (1,0,0,0)$ in Milne coordinates (corresponding to Bjorken expansion \cite{Bjorken:1982qr} in Cartesian coordinates), and the initial baryon diffusion current is assumed to vanish, $n^\mu = (0,0,0,0)$. 

\subsection{Transport coefficients}

Given these initial conditions, the hydrodynamic equations are solved by \beshydro\ \cite{Du:2019obx} (see Ch.~\ref{ch:numerics}). As already mentioned, we here focus on baryon diffusion dynamics by ignoring shear and bulk viscous stresses. For the transport coefficients related to baryon diffusion far away from the critical point we rely on the theoretical work in Refs.~\cite{Son:2006em, Rougemont:2015ona, Denicol:2018wdp, Soloveva:2019xph} since phenomenological constraints are still lacking. Specifically, we here use the coefficients obtained from the Boltzmann equation for an almost massless classical gas in the relaxation time approximation (RTA) \cite{Denicol:2018wdp}, which gives for the baryon diffusion coefficient
\begin{equation}
\label{eq:kappa_kinetic}
    \kappa_{n,0} = C_n\frac{n}{T}\,\left(\frac{1}{3}\coth\alpha-\frac{n T}{w} \right)\,,
\end{equation}
and for the relaxation time
\begin{equation}
\label{eq:taun_kinetic}
    \tau_{n,0}= \frac{C_n}{T}\,,
\end{equation}
where $C_n$ is a free unitless parameter (see also Sec.~\ref{sec:barycoeffcurr}).\footnote{%
    In Israel-Stewart theory \cite{ISRAEL1976310}  $\tau_{n,0}=\lambda_{n,0} T \beta_n$ where $\beta_n$ is a second-order transport coefficient and $\lambda_{n,0}$ is the non-critical thermal conductivity associated with $\kappa_{n,0}$ through Eq.~\eqref{eq:kappa-lambda}.} 
Throughout this paper, we set $C_n\eq0.4$; in  Ref.~\cite{Denicol:2018wdp} this value was shown to yield good agreement with selected experimental data. Following the kinetic theory approach \cite{Denicol:2018wdp} we also set $\delta_{nn,0}=\tau_{n,0}$ in Eq.~(\ref{eq:IS_nmu3}) as its non-critical value. In the limit of zero net baryon density,
$\kappa_{n,0}$ remains non-zero at non-zero temperature, $\lim_{\mu\to0} \kappa_{n,0}/\tau_{n,0} = nT/(3\mu)$ \cite{Denicol:2018wdp} --- a feature also seen in  holographic models; for example, using the AdS/CFT correspondence, the (baryon) charge conductivity of $r$-charged black holes translates into \cite{Son:2006em,Li:2018fow}
\begin{equation}\label{eq:kappa_holo}
    \kappa_{n,0} = 2\pi\frac{Ts}{\mu^2}\left(\frac{nT}{w}\right)^2\,.
\end{equation}%
%
\begin{figure}[!tb]
\begin{center}
\hspace{-1cm}
\includegraphics[width=0.55\textwidth]{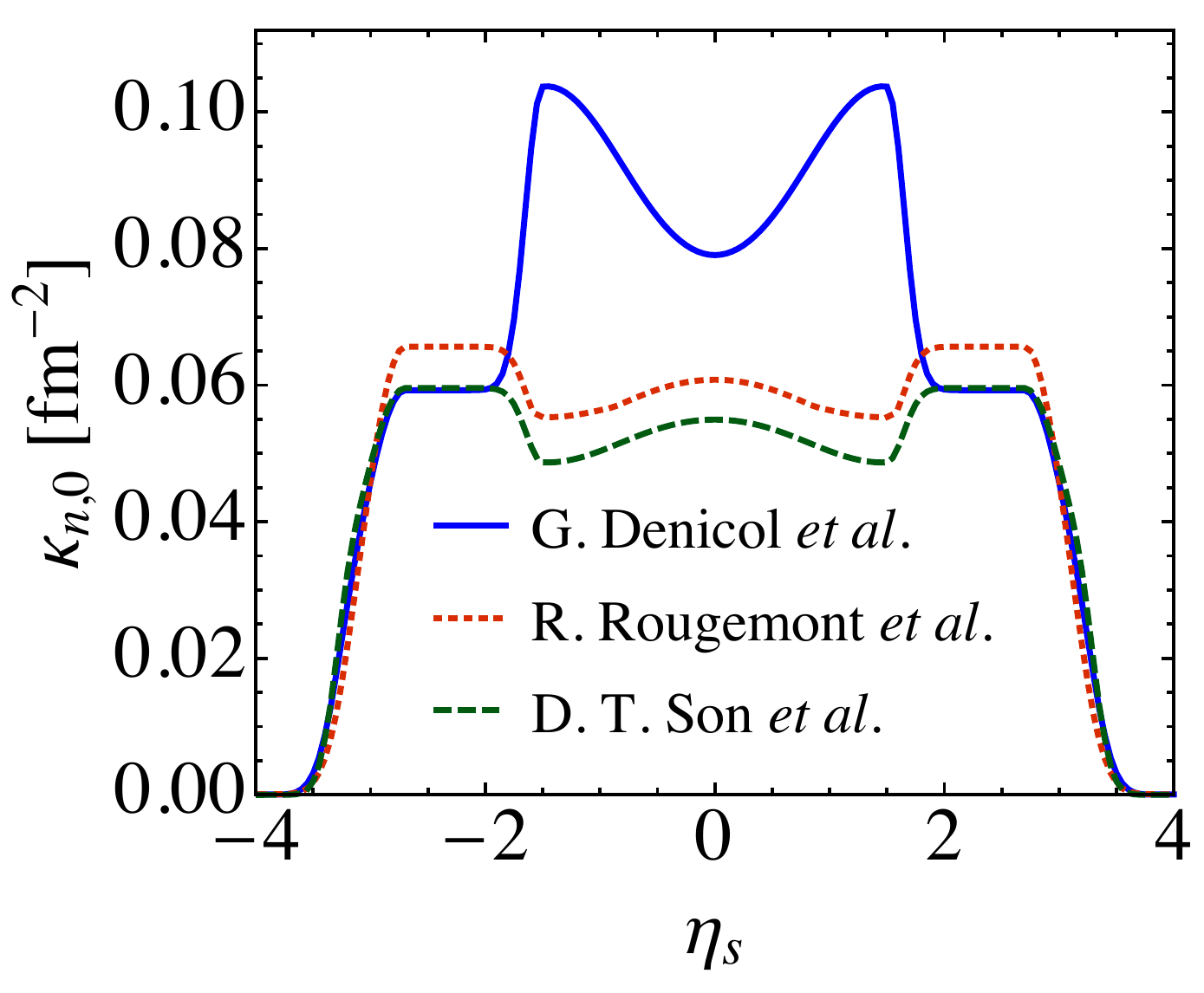}
\caption{%
    Initial longitudinal distributions of $\kappa_{n,0}$, corresponding to the initial profiles in Fig.~\ref{fig:long_evolution} below. $\kappa_{n,0}$ is calculated using different methods, including kinetic theory \cite{Denicol:2018wdp} (Eq.~\eqref{eq:kappa_kinetic}, blue solid line) and two holographic models, using Eq.~\eqref{eq:kappa_holo} (Ref.~\cite{Son:2006em}, green dashed line) and tabulated values found in Refs.~\cite{Rougemont:2015ona, Rougemont:2015wca} (red dotted line).
    \label{fig:initial_kappa_tau}}
\end{center}
\end{figure}%
%

Since $\kappa_{n,0}$ is such an important parameter in our study, we offer some intuition about its key characteristics in Fig.~\ref{fig:initial_kappa_tau}, where its initial space-time rapidity profile is plotted for three of the theoretical approaches referenced above.\footnote{%
    Ref.~\cite{Li:2018fow} compared the expressions of $\kappa_{n,0}$ in Eqs.~(\ref{eq:kappa_kinetic}) and (\ref{eq:kappa_holo}) while a comparison of Eq.~(\ref{eq:kappa_kinetic}) with a different holographic approach \cite{Rougemont:2015ona, Rougemont:2015wca} was previously presented in Ref.~\cite{Du:2018mpf}. See also Sec.~\ref{transcoeff}.
} 
The figure shows that, at large rapidity $|\eta_s|\gtrsim2.5$ where the baryon density approaches zero (cf. Fig.~\ref{fig:long_evolution}d below), the different models for $\kappa_{n,0}$ yield similar distributions which all decrease to zero with decreasing temperature (cf. Fig.~\ref{fig:long_evolution}b). This implies that when the fireball expands and cools down, baryon transport is expected to weaken (see discussion below in Sec.~\ref{sec:tevol}). In the region $|\eta_s|\lesssim2$ where the net baryon density is non-zero, the weakly and strongly coupled approaches shown in Fig.~\ref{fig:initial_kappa_tau} lead to very different $\kappa_{n,0}(\eta_s)$ profiles. In particular, we observe that in the holographic approaches $\kappa_{n,0}$ is {\em suppressed} by baryon density while in the kinetic approach it is {\em enhanced}. Thus, with transport coefficients from the kinetic approach, $\kappa_{n,0}$ peaks near the maxima of the baryon density, causing large baryon diffusion currents on either side of the maxima where the gradient of $\alpha=\mu/T$ is also large (see discussion below in Sec.~\ref{sec:diffnocp}).

When simulating the evolution near the critical point, we shall apply the following parametrizations using Eqs.~(\ref{eq:coeff_scaling2},\ref{eq:taun_scaling2}):
\begin{equation}\label{eq:chi_kappa_tau_scaling}
    \chi=\chi_0\left(\frac{\xi}{\xi_0}\right)^2, \quad \kappa_n=\kappa_{n,0}\left(\frac{\xi}{\xi_0}\right),\quad \tau_n=\tau_{n,0}\left(\frac{\xi}{\xi_0}\right)^2.
\end{equation}
Here $\xi_0$ is the non-critical correlation length, $\kappa_{n,0}$ is the non-critical value of baryon diffusion coefficient, $\chi_0$ is the isothermal susceptibility evaluated in the non-critical region, $\chi_0\equiv(\partial n_0/\partial\mu)_T$ where $n_0$ is the non-critical baryon density,\footnote{%
    While the notational distinction between $\chi$ and $\chi_0$ is needed here for clarity, we generally drop the subscript ``0'' for thermodynamic quantities away from the critical region elsewhere to avoid clutter.} and $\tau_{n,0}$ is the non-critical relaxation time. We use the kinetic expressions (\ref{eq:kappa_kinetic},\ref{eq:taun_kinetic}) to calculate the non-critical values of $\kappa_n$ and $\tau_n$.
With Eq.~\eqref{eq:chi_kappa_tau_scaling} the parametrizations with critical scaling for $D_B$ and $D_T$ are readily obtained from Eqs.~\eqref{eq:D_BT2}.

\subsection{Particlization}

After completion of the hydrodynamic evolution (results of which will be discussed in the next Section) we compute the particle distributions corresponding to the hydrodynamic fields on the freeze-out surface $\Sigma$, using the Cooper-Frye formula \eqref{eq:distr1} with $f_i =f_{0,i} + \delta f_i^{\mathrm{diffusion}}$ where $f_{0,i}$ is the equilibrium distribution and $\delta f_i^{\mathrm{diffusion}}$ the off-equilibrium correction resulting from net baryon diffusion. At first order in the Chapman-Enskog expansion of the RTA Boltzmann equation, this dissipative correction is given by \cite{Denicol:2018wdp, McNelis:2019auj, McNelis:2021acu}
\begin{eqnarray}
    \delta f_i^{\mathrm{diffusion}} = f_{0,i} (1 -\Theta_i f_{0,i}) \left(\frac{n}{w} - \frac{Q_i}{E_p} \right) \frac{p^{\langle \mu \rangle} n_\mu}{\hat{\kappa}}\,,
\label{eq:diffusion_deltaf}
\end{eqnarray}
where $\Theta_i\eq1$ $(-1)$ for fermions (bosons), $Q_i$ is the baryon number of particle species $i$, $p^{\langle \mu \rangle}\equiv\Delta^{\mu\nu}p_\nu$, and $\hat{\kappa}=\kappa_n/\tau_n$. We will discuss how the critical correction is included, as well as its effects on the final particle distributions, in Sec.~\ref{sec:diffcp_spectra}. We evaluate the continuous momentum distribution (\ref{eq:distr1}) numerically using the {\sc iS3D} particlization module \cite{McNelis:2019auj}, ignoring in this exploratory study rescattering among the particles and resonance decays after particlization.

\section{Results and discussion}
\label{sec:results}

In this section we discuss the dynamics of the fireball at a fixed collision energy of $\snn\eq19.6$\,GeV. First we study in Sec.~\ref{sec:diffnocp} baryon diffusion effects on its $T$-$\mu$ trajectories through the phase diagram for cells located at different space-time rapidities \etas, and in Sec.~\ref{sec:diffnocp_particles} its effects on freeze-out surface and final particle distributions, both in the absence of critical dynamics. Then, in Sec.~\ref{sec:diffcp}, we discuss how the critical behavior described in Sec.~\ref{sec:hydro} modifies this dynamics for cells whose trajectories pass close to the critical point, and how this affects particlization. In Sec.~\ref{sec:tevol} we point out some generic features of the time evolution of baryon diffusion.

\subsection{Longitudinal dynamics of baryon evolution}
\label{sec:diffnocp}

\begin{figure*}[!t]
\begin{center}
\hspace{-1cm}
\includegraphics[width=1.05\textwidth]{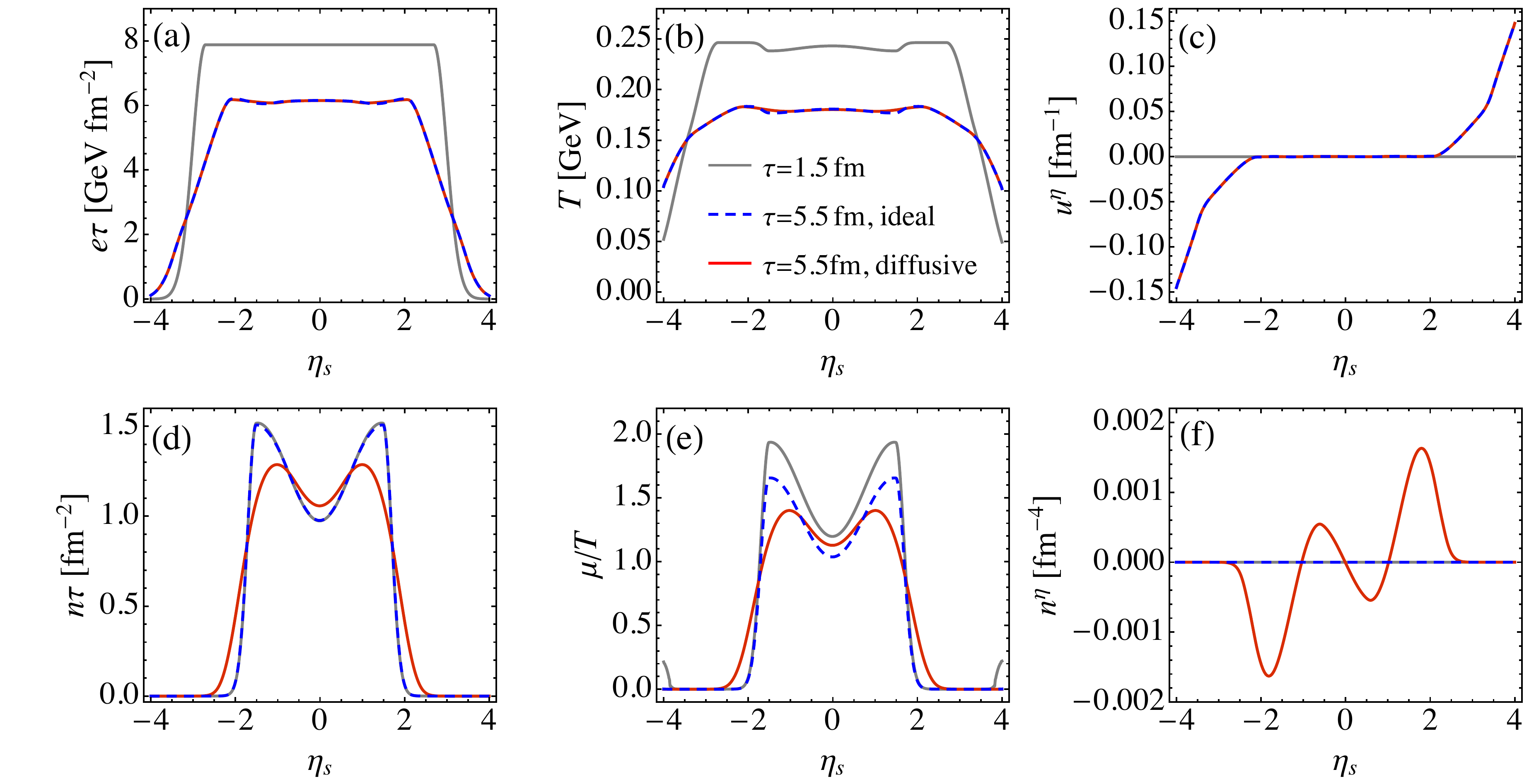}
\caption{%
    Longitudinal evolution for the cases without (``ideal'', blue dashed line) and with baryon diffusion (``diffusive'', red solid line). The gray curves show the initial distributions at the initial time $\tauI\eq1.5$\,fm/$c$; colored curves show the evolved distributions at $\tau\eq5.5$\,fm/$c$ for ideal and diffusive evolution, respectively. Note that both energy and baryon densities are scaled by the proper time $\tau$.}
    \label{fig:long_evolution}
\end{center}
\end{figure*}

In Figure~\ref{fig:long_evolution}, we show snapshots of the longitudinal distributions of the hydrodynamic quantities at two times, the initial time 1.5\,fm/$c$ (gray solid lines) and later at $\tau\eq5.5$\,fm/$c$, with and without baryon diffusion (red solid and blue dashed lines, respectively). The gray curves in panels (a) and (d)  show the initial energy and baryon density distributions from Ref.~\cite{Denicol:2018wdp}. The gray lines in panels (b) and (e) show the corresponding temperature and chemical potential profiles, extracted with the {\sc neos} equation of state \cite{Monnai:2019hkn, Bazavov:2014pvz, Bazavov:2012jq, Ding:2015fca, Bazavov:2017dus}. The gray horizontal lines in panels (c) and (f) show the zero initial conditions for the longitudinal flow and baryon diffusion current.

We see that the temperature profile shares the plateau with the energy density, up to small structures caused by the double-peak structure of the baryon density and baryon chemical potential profiles. Similar small structures are also seen in the pressure (not shown). The chemical potential in panel (e) inherits the double-peak structure from the baryon density in panel (d).\footnote{%
    In the very dilute forward and backward rapidity regions 
    one observes a steep rise of the initial $\mu/T$. This feature is sensitive to the rates at which $e$ and $n$ approach zero as $|\eta_s|\to\infty$, and it is easily affected by numerical inaccuracies. Since both $T$ and $\mu$ are close to zero there, the baryon diffusion coefficient $\kappa_{n,0}$ also vanishes, and (as seen in Fig.~\ref{fig:long_evolution}f) the apparently large but numerically unstable gradient of $\mu/T$ at large $\eta_s$ does not generate a measurable baryon diffusion current.}

We next discuss the blue dashed lines in Fig.~\ref{fig:long_evolution} showing the results of ideal hydrodynamic evolution. Work done by the longitudinal pressure converts thermal energy into collective flow kinetic energy such that the thermal energy density $e$ decreases faster than $1/\tau$ (panel (a)). Small pressure variations along the plateau of the distribution caused by the rapidity dependence of $\mu/T$ lead to slight distortions of the rapidity plateau of the energy density as its magnitude decreases. Longitudinal pressure gradients at the forward and backward edges of the initial rapidity plateau accelerate the fluid longitudinally, generating a non-zero $\eta_s$-component of the hydrodynamic flow at large rapidities (panel (c)). As seen in panels (a) and (c), the resulting longitudinal rarefaction wave travels inward slowly, leaving the initial Bjorken flow profile $u^\eta\eq0$ untouched for $|\eta_s|<2.5$ up to $\tau\eq5.5$\,fm/$c$. For Bjorken flow without transverse dynamics baryon number conservation implies that $n\tau$ remains constant. Panel (d) shows this to be the case up to $\tau\eq5.5$\,fm/$c$ because, up to that time, the initial Bjorken flow has not yet been affected by longitudinal acceleration over the entire $\eta_s$-interval in which the net baryon density $n$ is non-zero. Panel (e) shows, however, that in spite of $n\tau$ remaining constant within that $\eta_s$ range, the baryon chemical potential $\mu/T$ decreases with time, as required by the {\sc neos} equation of state.

The nontrivial evolution effects of turning on baryon diffusion, Eq.~(\ref{eq:IS_nmu3}), are shown by the red solid lines in Fig.~\ref{fig:long_evolution}. Panels (a)-(c) show that baryon diffusion has almost no effect at all on the energy density  (and, by implication, on the pressure), the temperature, and the hydrodynamic flow generated by the pressure gradients. It does, however, significantly modify the rapidity profiles of the net baryon density (d), chemical potential $\mu/T$ (e) and baryon diffusion current $n^\eta$ (f). Generated by the negative gradient of $\mu/T$, the baryon diffusion current moves baryon number from high- to low-density regions, causing an overall broadening of the baryon density rapidity profile in (d) while simultaneously filling in the dip at midrapidity \cite{Shen:2017ruz, Denicol:2018wdp, Li:2018fow, Du:2018mpf, Du:2019obx, Fotakis:2019nbq}. Panel (e) shows how the chemical potential $\mu/T$ tracks these changes in the baryon density profile, and panel (f) shows the baryon diffusion current responsible for this transport of baryon density, with its alternating sign and magnitude tracing the sign and magnitude changes of $-\nabla(\mu/T)$. As we shall see in Sec.~\ref{sec:tevol}, the smoothing of the gradients of baryon density and chemical potential contributes to a fast decay of baryon diffusion effects. 

\begin{figure}[!t]
\begin{center}
\hspace{-.5cm}
\includegraphics[width=0.55\textwidth]{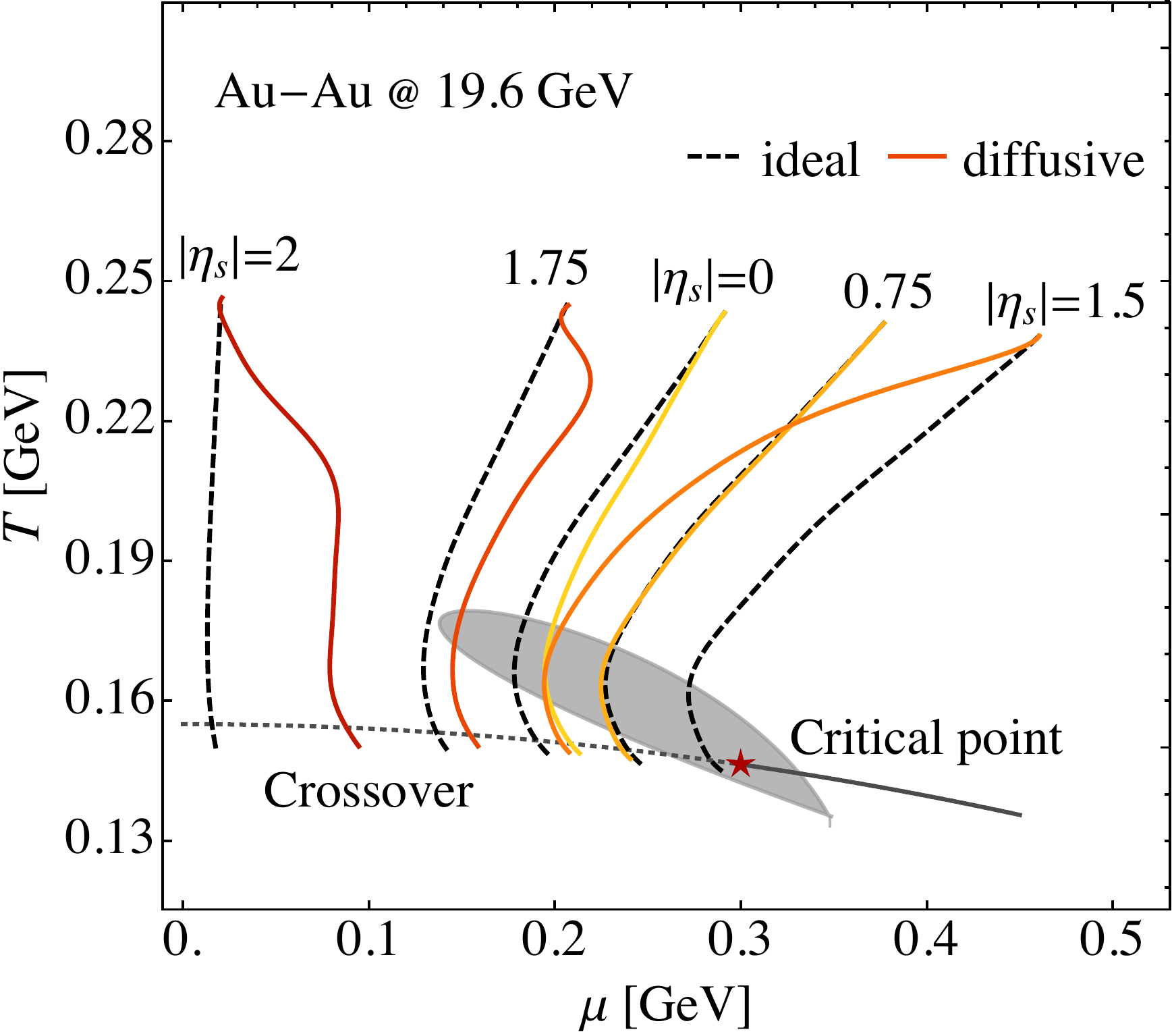}    
\caption{%
    Phase diagram trajectories of fluid cells at different $|\eta_s|$ for the Au+Au collision fireball discussed here.  Black dashed lines indicate ideal evolution while colored solid lines include the effects of baryon diffusion. All fluid cells evolve from high to low temperature. The phase transition line with critical point and critical region are included only to guide the eye -- critical effects on the EoS and transport coefficients are not included in this figure.}
    \label{fig:phase_dia_traj}
\end{center}
\end{figure}

Fig.~\ref{fig:long_evolution} indicates non-trivial thermal and chemical evolution at different rapidities. Fluid cells at different \etas{} pass through different regions of the QCD phase diagram and may therefore be affected differently by the QCD critical point \cite{Monnai:2016kud, Shen:2018pty, Dore:2020jye}. This has led to the suggestion \cite{Brewer:2018abr} of using rapidity-binned cumulants of the final net proton multiplicity distributions as possibly sensitive observables of the critical point.\footnote{%
    We caution that at BES energies the mapping between space-time rapidity \etas{} of the fluid cells and rapidity $y$ of the emitted hadrons is highly nontrivial and requires dynamical modelling.}
To illustrate the point we show in Fig.~\ref{fig:phase_dia_traj} the phase diagram trajectories of fluid cells at several selected $|\eta_s|$ values,\footnote{%
    Cells at opposite but equal space-time rapidities are equivalent because of $\eta_s\to-\eta_s$ reflection symmetry in this chapter.}
both with and without baryon diffusion. As we move from mid-rapidity to $|\eta_s|\eq2.0$, the starting point of these trajectories first moves from $\mu\simeq0.28$\,GeV at $\eta_s\eq0$ to the larger value $\mu\simeq0.45$\,GeV at $\eta_s\eq1.5$, but then turns back to $\mu\simeq0.2$\,GeV at $\eta_s\eq1.75$, and finally to $\mu\simeq0$ at $\eta_s\eq2.0$, without much variation of the initial temperature $T_i\simeq0.25$\,GeV (see Figs.~\ref{fig:long_evolution}b,e). The difference between the dashed (ideal) and solid (diffusive) trajectories exhibits a remarkable dependence on \etas{}: Both the sign and the magnitude of the diffusion-induced shift in baryon chemical potential depend strongly on space-time rapidity. In most cases, we note that the solid (diffusive) trajectories move initially rapidly away from the corresponding ideal ones, but then quickly settle on a roughly parallel ideal trajectory. A glaring exception is the trajectory of the cell at $\eta_s\eq1.5$, which starts at the maximal initial baryon chemical potential and keeps moving away from its initial ideal $T$-$\mu$ trajectory for a long period, settling on a new ideal trajectory only shortly before it reaches the hadronization phase transition. The reason for this behavior can be found in Fig.~\ref{fig:long_evolution}e, which shows that at $\eta_s\eq1.5$ the gradient of $\mu/T$ remains large throughout the fireball evolution. But almost everywhere else baryon diffusion effects die out quickly.

\begin{figure}[!t]
\begin{center}
\hspace{-0.5cm}
\includegraphics[width= 0.55\textwidth]{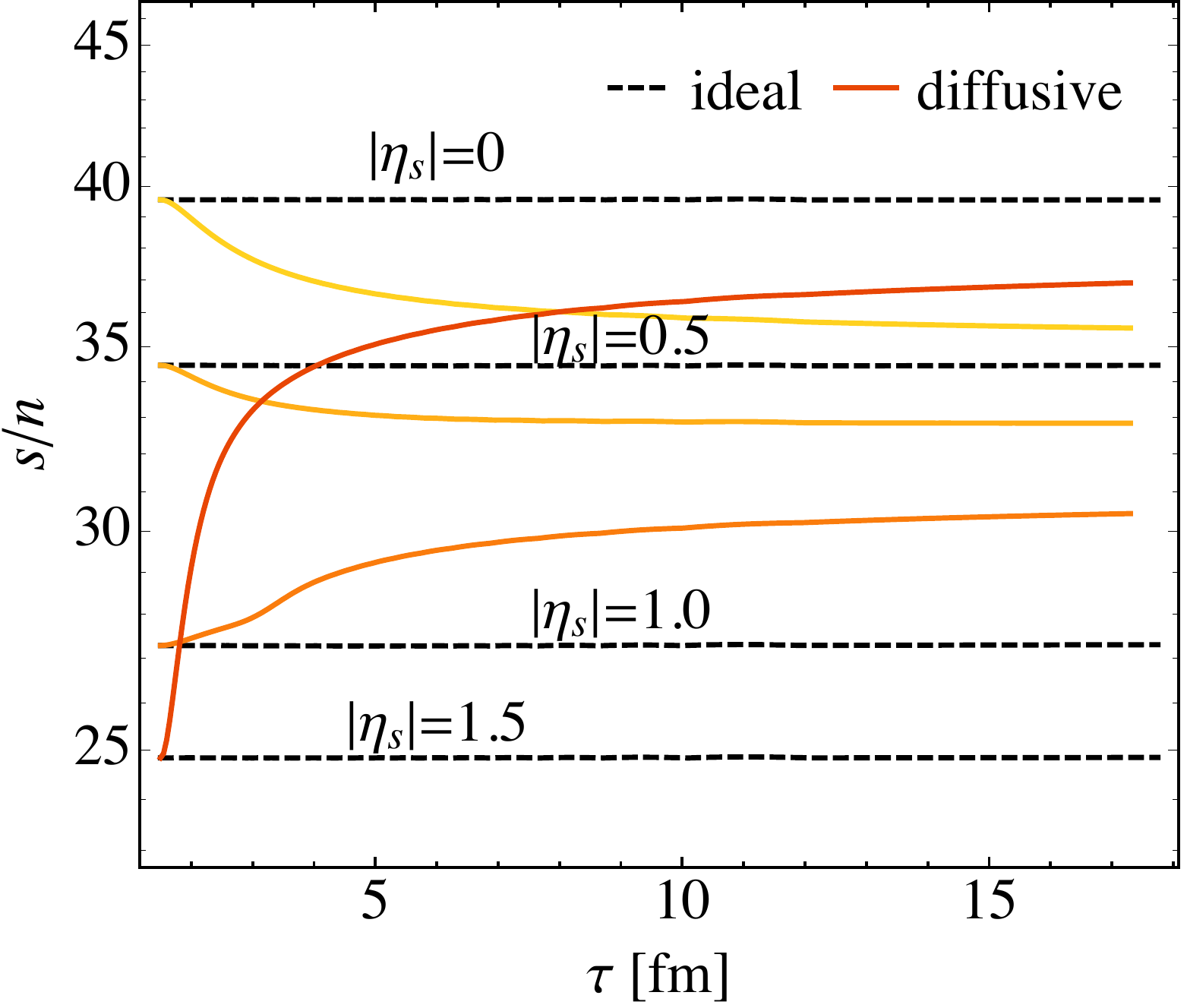}
\caption{%
    Time evolution of entropy per baryon at selected \etas{} values in ideal and diffusive fluid dynamics (dashed and solid lines, respectively).
    }
    \label{fig:sn_evolution}
\end{center}
\end{figure}

Since ideal fluid dynamics conserves both baryon number and entropy, the dashed trajectories are lines of constant entropy per baryon. This is shown by the dashed lines in Fig.~\ref{fig:sn_evolution}. Baryon diffusion leads to a net baryon current in the local momentum rest frame and thereby changes the baryon number per unit entropy. This is illustrated by the solid lines in Fig.~\ref{fig:sn_evolution}. Depending on the direction of the $\mu/T$ gradients, baryon diffusion can increase or decrease the entropy per baryon. 

We close this discussion by commenting on the turning of the dashed $s/n\eq$const.{}~trajectories in Fig.~\ref{fig:phase_dia_traj} from initially pointing towards the lower left to later pointing towards the lower right. This is a well known feature of isentropic expansion trajectories in the QCD phase diagram \cite{Cho:1993mv, Hung:1997du, Monnai:2019hkn} that reflects the change in the underlying degrees of freedom, from quarks and gluons to a hadron resonance gas, at the point of hadronization as embedded in the construction of the EoS.

Figure~\ref{fig:phase_dia_traj} is reminiscent of the QCD phase diagram often shown to motivate the study of heavy-ion collisions at different collision energies in order to explore QCD matter at different baryon doping (see, for example, the 2015 DOE-NSF NSAC Long Range Plan for Nuclear Physics \cite{osti_1296778}). What is shown there are (isentropic) expansion trajectories for matter created {\em at midrapidity in heavy-ion collisions with different beam energies}, whereas Fig.~\ref{fig:phase_dia_traj} shows similar expansion trajectories {\em for different parts of the fireball in a collision with a fixed beam energy}. Fig.~\ref{fig:phase_dia_traj} thus makes the point that in general the matter created in heavy ion collisions can never be characterized by a single fixed value of $\mu/T$. At high collision energies space-time and momentum rapidities are tightly correlated, $\eta_s\simeq y$, and different $\eta_s$ regions with different baryon doping $\mu/T$ can thus be more or less separated in experiment by binning the data in momentum rapidity $y$. This motivates the strategy of scanning the changing baryonic composition in the $T$-$\mu$ diagram by performing a rapidity scan at fixed collision energy rather than a beam energy scan at fixed rapidity \cite{Monnai:2016kud, Shen:2018pty, Brewer:2018abr}. This strategy fails, however, at lower collision energies where particles of fixed momentum rapidity can be emitted from essentially every part of the fireball and thus receive contributions from regions with wildly different chemical compositions, with non-monotonic rapidity dependences that are non-trivially and non-monotonically affected by baryon diffusion.   

\subsection{Freeze-out surface and final particle distributions}
\label{sec:diffnocp_particles}

The expansion trajectories shown in the previous subsection all end at the same constant proper time (see Fig.~\ref{fig:sn_evolution}). In phenomenological applications it is usually assumed that the hydrodynamic stage ends and the fluid falls apart into particles when all fluid cells reach a certain ``freeze-out energy density'', here taken as $e_f\eq0.3$\,GeV/fm$^3$.\footnote{%
    This is lower than the value of 0.4\,GeV/fm$^3$ used in Ref.~\cite{Denicol:2018wdp}, in order to ensure that the expansion trajectories reach into the hadronic phase below the crossover line from Ref.~\cite{Bellwied:2015rza}.}
With such a freeze-out criterion, fluid cells at different \etas{} freeze out at different times $\tau_f(\eta_s)$. In this subsection we discuss this freeze-out surface and the distributions of particles emitted from it. 

\begin{figure}[!htb]
\begin{center}
\hspace{-0.5cm}
\includegraphics[width= 0.8\textwidth]{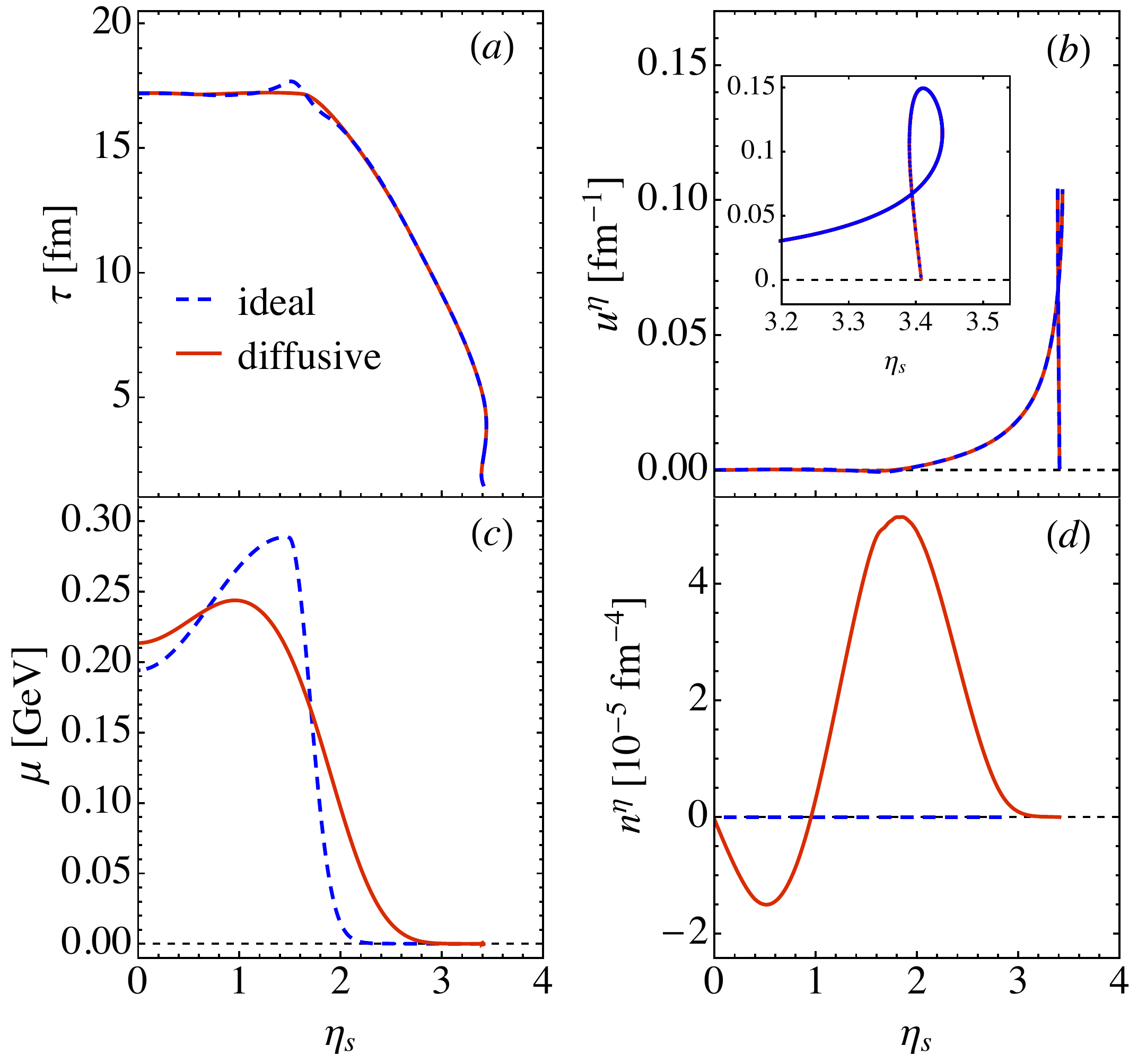}
\caption{%
    Distributions on the freeze-out surface, defined by a constant freeze-out energy density $e_f\eq0.3\,$GeV/fm$^3$. Shown are as functions of \etas{} (a) the space-time profile of the freeze-out surface, (b) the longitinal flow, (c) the baryon chemical potential, and (d) the longitudinal component of the diffusion current.
    Blue dashed and solid red lines correspond to ideal and diffusive fluid dynamics, respectively.}
    \label{fig:fz_surf}
\end{center}
\end{figure}

Fig.~\ref{fig:fz_surf} shows the freeze-out surface $\tau_f(\eta_s)$ in panel (a) as well as the longitudinal flow, baryon chemical potential, and longitudinal component of the baryon diffusion current in panels (b)-(d).\footnote{%
    The freeze-out finder implemented in \beshydro{} is based on {\sc Cornelius} \cite{Huovinen:2012is}, which was previously tested within \beshydro{} at non-zero baryon density in the transverse plane in Ch.~\ref{ch.gubser}.}
Ideal and diffusive hydrodynamics are distinguished by blue dashed and red solid lines. Panel (a) shows that initially the longitudinal pressure gradient causes the fluid to grow in \etas{} direction before it starts to shrink after $\tau\gtrsim4$\,fm/$c$ due to cooling and surface evaporation. As seen in Fig.~\ref{fig:long_evolution}a, the core of the fireball remains approximately boost invariant while cooling by performing longitudinal work, until the longitudinal rarefaction wave reaches it. Once the energy density in this boost-invariant core drops below $e_f$, it freezes out simultaneously, as seen in the flat top of the freezeout surface shown in panel (a). Slight deviations from boost invariance are caused by the effects of the boost-non-invariant net baryon density profile and its (small) effect on the pressure whose gradient drives the hydrodynamic expansion. Baryon diffusion has practically no effect on the freeze-out surface, nor on the longitudinal flow along this surface shown in panel (b), owing to the weak dependence of the EoS on baryon doping. The distributions of the baryon chemical potential and baryon diffusion current across this surface, on the other hand, are significantly affected by baryon diffusion, as seen in panels (c) and (d). It bears pointing out, however, that the magnitude of the baryon diffusion current in panel (d) is very small.

\begin{figure}[!t]
\begin{center}
\includegraphics[width= 0.48\textwidth]{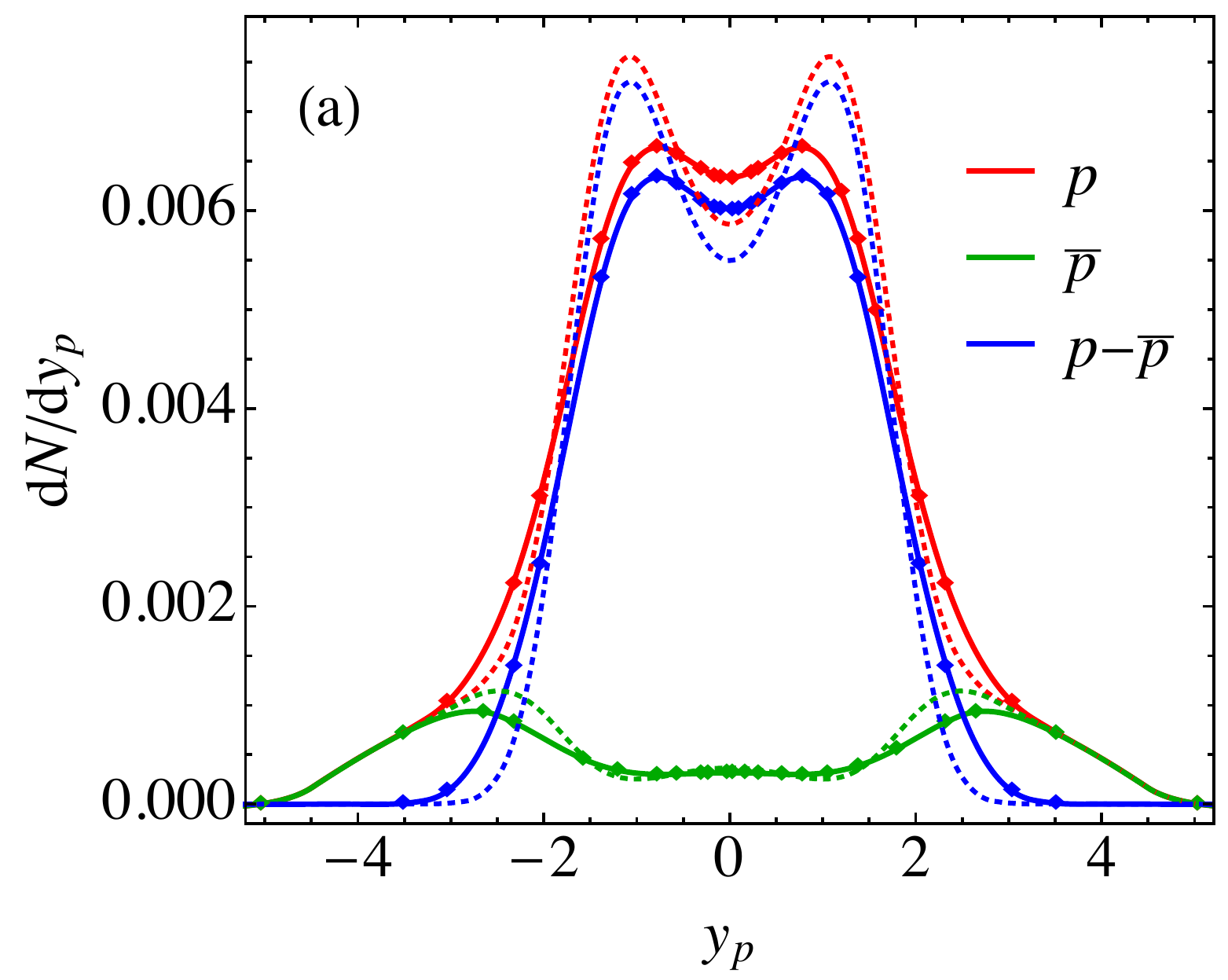}\hspace{0.6cm}\includegraphics[width=0.48\textwidth]{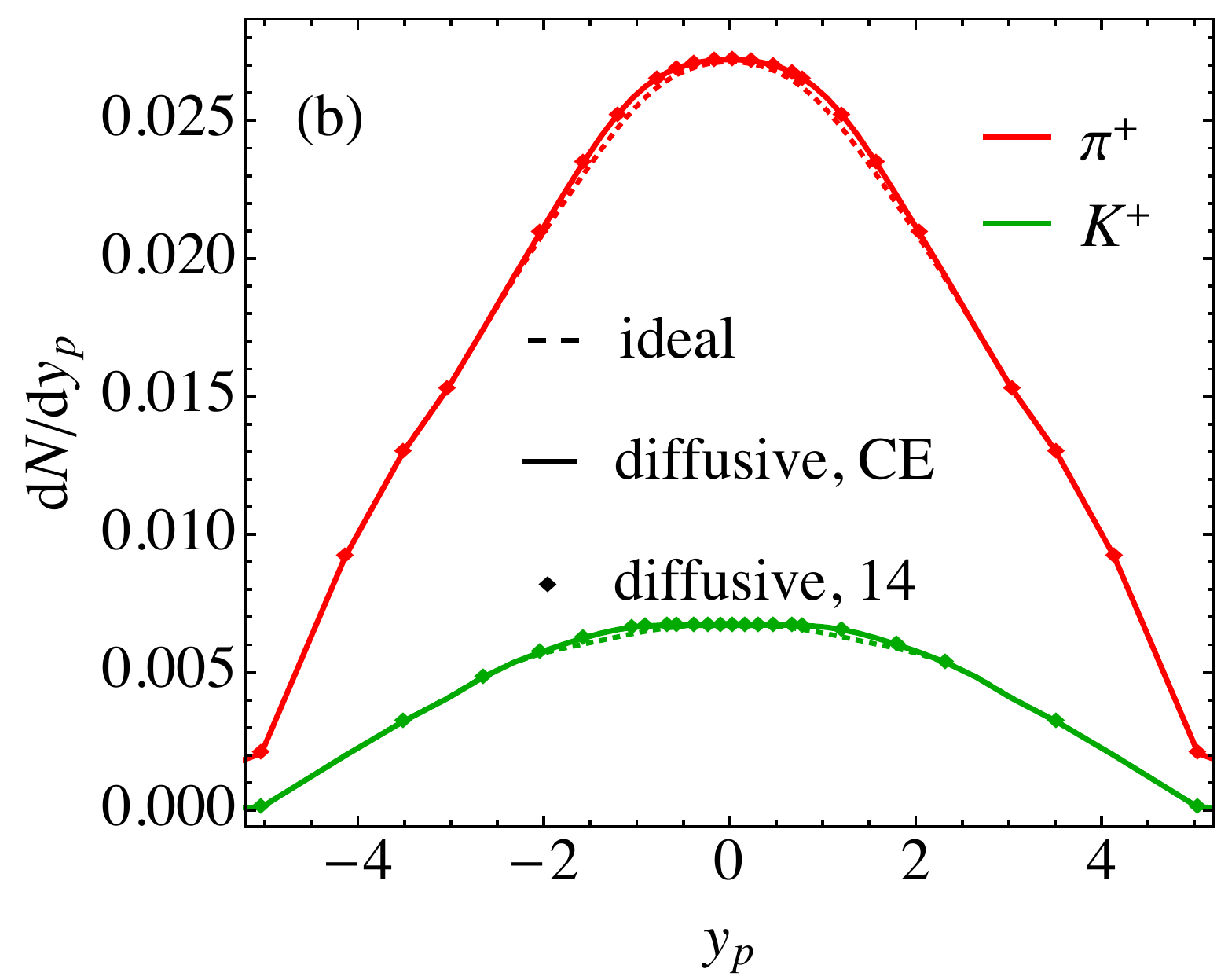}
\caption{%
    Final particle rapidity distributions for (a) baryons (red: protons; green: antiprotons; blue: net protons) and (b) mesons (red: pions; green: kaons). Thin dashed and thick solid lines show results for ideal and diffusive evolution, respectively. For the diffusive case, the solid lines use    the Chapman-Enskog approximation for the diffusive correction, while colored markers show the values obtained from the 14-moment approximation.}
    \label{fig:1D_spec_comp}
\end{center}
\end{figure}

Given these quantities on the freeze-out surface, we use the \isd{} module \cite{McNelis:2019auj} to evaluate the Cooper-Frye integral (\ref{eq:distr1},\ref{eq:diffusion_deltaf}) for the rapidity distributions of hadrons emitted from the freeze-out surface. Results are shown in Fig.~\ref{fig:1D_spec_comp}.
Panel (b) indicates that baryon diffusion has negligible effects on meson distributions. It affects only baryon distributions. Panel (a) shows that baryon diffusion significantly increases the proton and net-proton yields at mid-rapidity and also broadens their rapidity distributions at large rapidity. Both effects were previously shown to increase with the magnitude of the baryon diffusion coefficient $\kappa_n$ \cite{Shen:2017ruz, Du:2018mpf, Denicol:2018wdp, Li:2018fow}. The approximate boost-invariance of the longitudinal flow over a wide range of \etas{} on the freeze-out surface (see Fig.~\ref{fig:fz_surf}b) maps the baryon diffusion effects seen in Figs.~\ref{fig:long_evolution}d,e and \ref{fig:fz_surf}c as functions of space-time rapidity $\eta_s$ onto momentum rapidity $y_p$ in Fig.~\ref{fig:1D_spec_comp}a \cite{Shen:2017ruz, Du:2018mpf, Denicol:2018wdp, Li:2018fow}. Differences between the Chapman-Enskog and 14-moment approximations for the dissipative correction (\ref{eq:diffusion_deltaf}) are negligible, and even ignoring $\delta f_{\mathrm{diff},i}$ in Eq.~(\ref{eq:distr1}) entirely does not make much of a difference (not shown). This reflects the tiny magnitude of the baryon diffusion current on the freeze-out surface seen in Fig. \ref{fig:fz_surf}d.\footnote{%
    Ref.~\cite{Denicol:2018wdp}, with transverse expansion, shows that baryonic observables in the transverse plane, such as the $p_T$-differential proton elliptic flow $v_2^p(p_T)$, are sensitive to the dissipative correction $\delta f_\mathrm{diff}$ from baryon diffusion.}
    
We emphasize that the mapping of baryon diffusion effects seen as a function of spacetime rapidity \etas{} in Figs.~\ref{fig:long_evolution}d,e and \ref{fig:fz_surf}c onto momentum rapidity $y_p$ is expected to be model dependent, and may not work for initial conditions in which the initial velocity profile is not boost-invariant or the initial \etas{}-distribution of the net baryon density looks different. This initial-state modeling uncertainty has so far prohibited a meaningful extraction of the baryon diffusion coefficient from experimental data (see, however, Ref.~\cite{Denicol:2018wdp} for a valiant effort). Additional uncertainties from possible critical effects associated with QCD critical point on the bulk dynamics, especially through baryon diffusion, may further complicate the picture, in particular as long as the location of the critical point is still unknown. In the following subsection we address some of these effects arising from critical dynamics.

\subsection{Critical effects on baryon diffusion}
\label{sec:diffcp}

In this section, we explore whether the QCD critical point can have significant effects on the bulk dynamics, through the baryon diffusion current. For this purpose, we include critical effects as described in Eqs.~\eqref{eq:chi_kappa_tau_scaling}, and explore effects from critical slowing down on the hydrodynamic transport, as well as critical corrections to final particle distributions through the Cooper-Frye formula.

\subsubsection{Critical slowing down of baryon transport}
\label{sec:diffcp_hydro}

In the critical region
baryon transport is affected by critical slowing down \cite{RevModPhys.49.435}. Outside the critical region all thermodynamic and transport properties approach their non-critical baseline described in Sec.~\ref{sec:diffnocp}, but as the system approaches the critical point its dynamics is affected by critical modifications of the transport coefficients involving various powers of $\xi/\xi_0>1$. We study this by incorporating the critical scaling of $\chi$, $\kappa_n$ and $\tau_n$ in Eqs.~\eqref{eq:chi_kappa_tau_scaling}, with the correlation length $\xi(\mu,T)$ parametrized by Eq.~\eqref{eq:xi(mu,T)}. 

Before doing any simulations we briefly discuss qualitative expectations. Eq.~\eqref{eq:D_scaling2} indicates that, as the correlation length grows, $\xi/\xi_0>1$, the coefficient $D_B$ is suppressed while $D_T$ is enhanced. According to  Eqs.~\eqref{eq:nmu_NS_decomposition2} a suppression of $D_B$ reduces the contribution from baryon density inhomogeneities while an enhancement of $D_T$ increases the contribution from  temperature inhomogeneities to the Navier-Stokes limit $n^\nu_\mathrm{NS}=\kappa_n\nabla^\nu(\mu/T)$.\footnote{%
    We note that in the literature sometimes only the baryon density gradient term $D_B\nabla n$ is included in the diffusion current (see, e.g., Refs.~\cite{RevModPhys.49.435, Son:2004iv}) which then leads to its generic suppression close to the critical point.}
In addition to thus moving its Navier-Stokes target value, proximity of the critical point also increases the time $\tau_n$ (see Eqs.~\eqref{eq:chi_kappa_tau_scaling}) over which the baryon diffusion current relaxes to its Navier-Stokes limit -- its response to the driving force is critically slowed down. 

\begin{figure}[!t]
\begin{center}
\hspace{-0.5cm}
\includegraphics[width= 0.55\textwidth]%
    {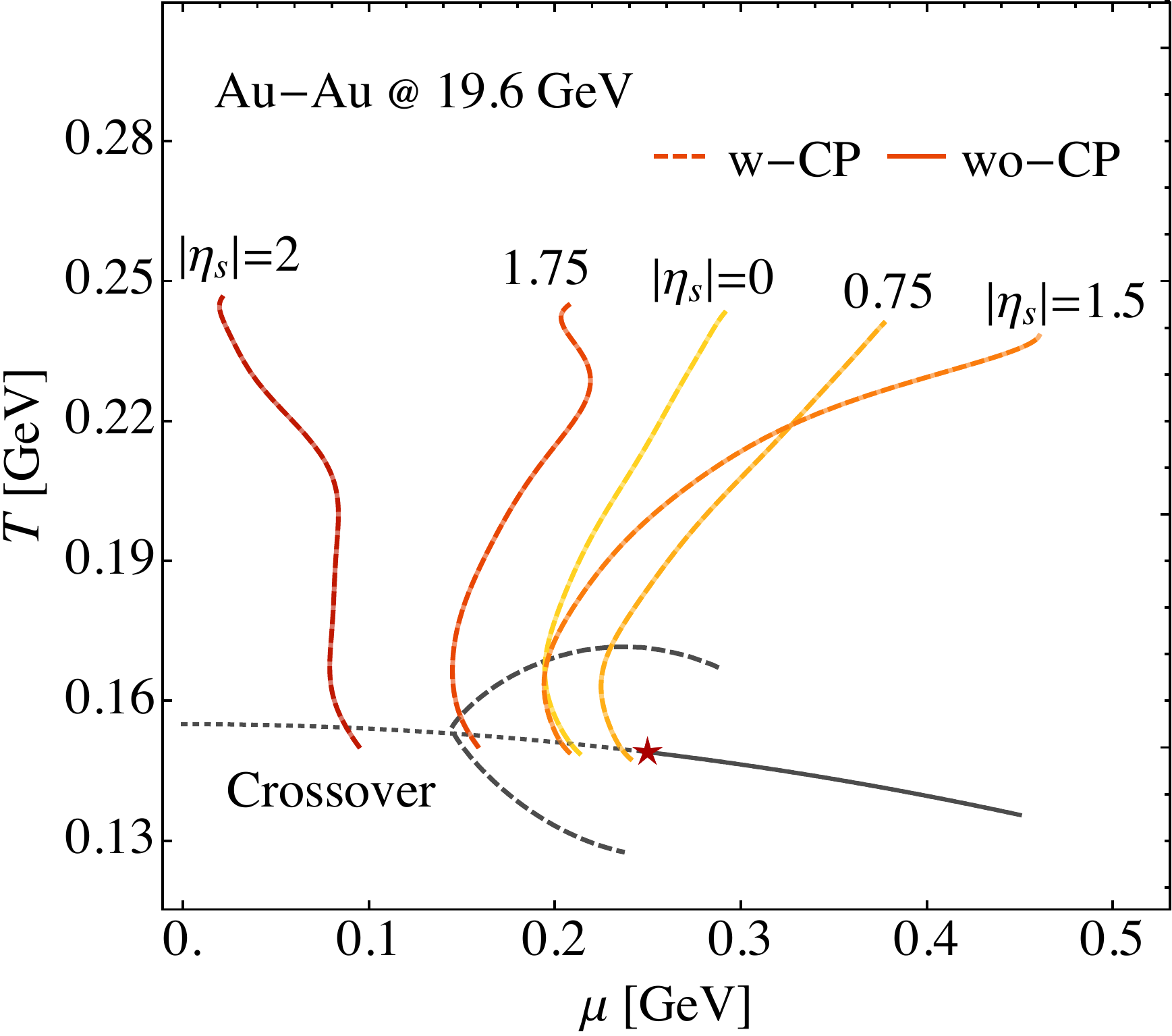}
\caption{%
    Phase diagram trajectories at different space-time rapidities, with (w-CP, dashed lines) and without (wo-CP, solid lines) inclusion of critical effects. The solid lines are taken from Fig.~\ref{fig:phase_dia_traj}. The dashed lines account for the critical scaling of all parameters controlling the evolution. 
    The black dashed line encloses the region on the crossover side where $\xi(\mu,T)>\xi_0\eq1$\,fm.}
    \label{fig:trajectory_CP_noCP}
\end{center}
\end{figure}

Repeating the simulations with the same setup as in Sec.~\ref{sec:diffnocp}, except for the inclusion of critical scaling, yields the results shown in Fig.~\ref{fig:trajectory_CP_noCP}. For the parametrization of the correlation length $\xi(\mu,T)$ we assumed a critical point located at ($T_c\eq149$\,MeV,\,$\mu_c\eq250$\,MeV). This is very close to the right-most trajectory shown in Fig.~\ref{fig:trajectory_CP_noCP} which should therefore be most strongly affected by it.\footnote{%
    Since we do not have the tools here to handle passage through a first-order phase transition, we do not consider any expansion trajectories cutting the first-order transition line to the right of QCD critical point.}
Surprisingly, none of the trajectories, not even the one passing the critical point in close proximity, are visibly affected by critical scaling of transport coefficients.

\begin{figure}[!t]
\begin{center}
\hspace{-0.5cm}
\includegraphics[width= 0.48\textwidth]{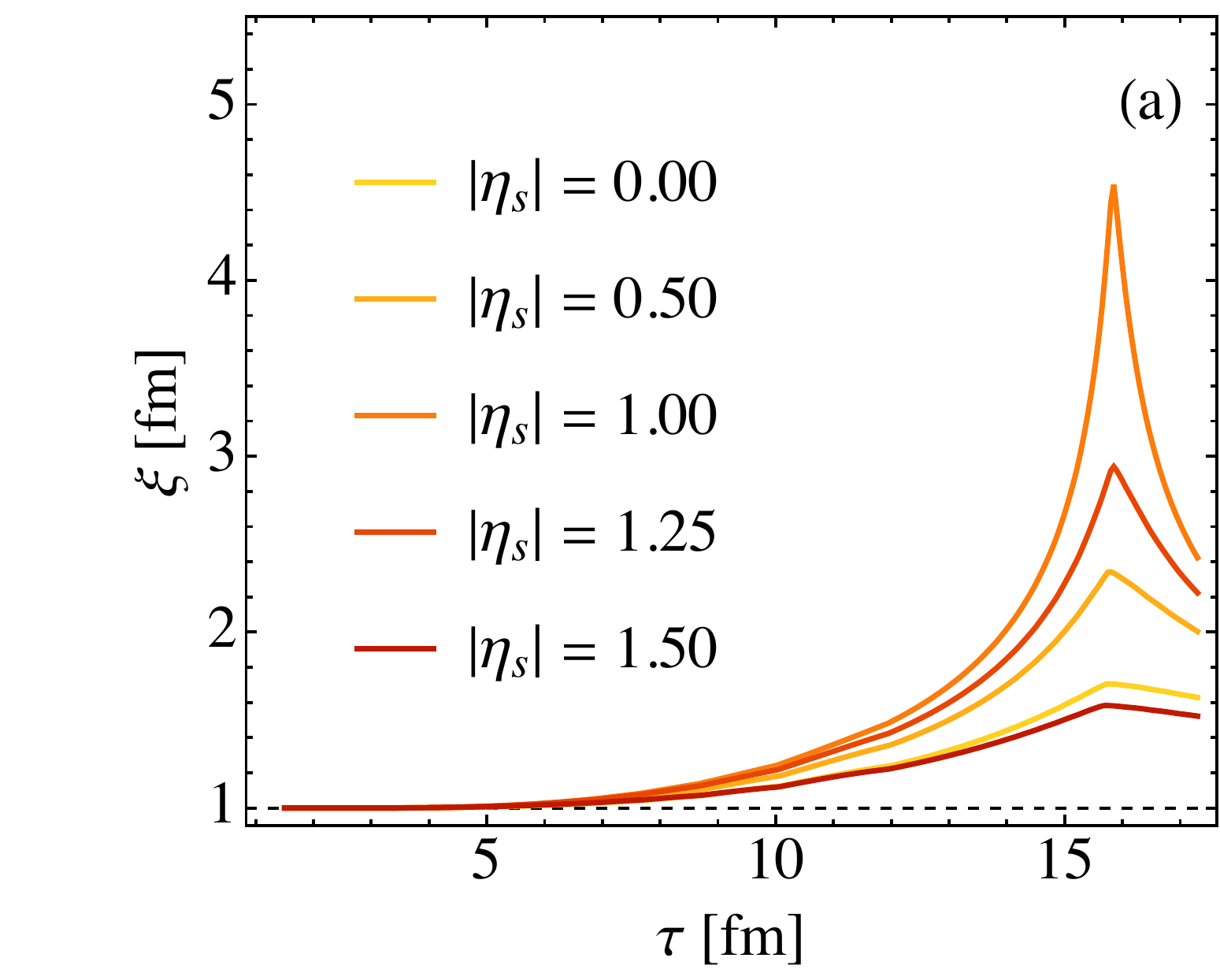}\hspace{0.5cm}\includegraphics[width=0.48\textwidth]{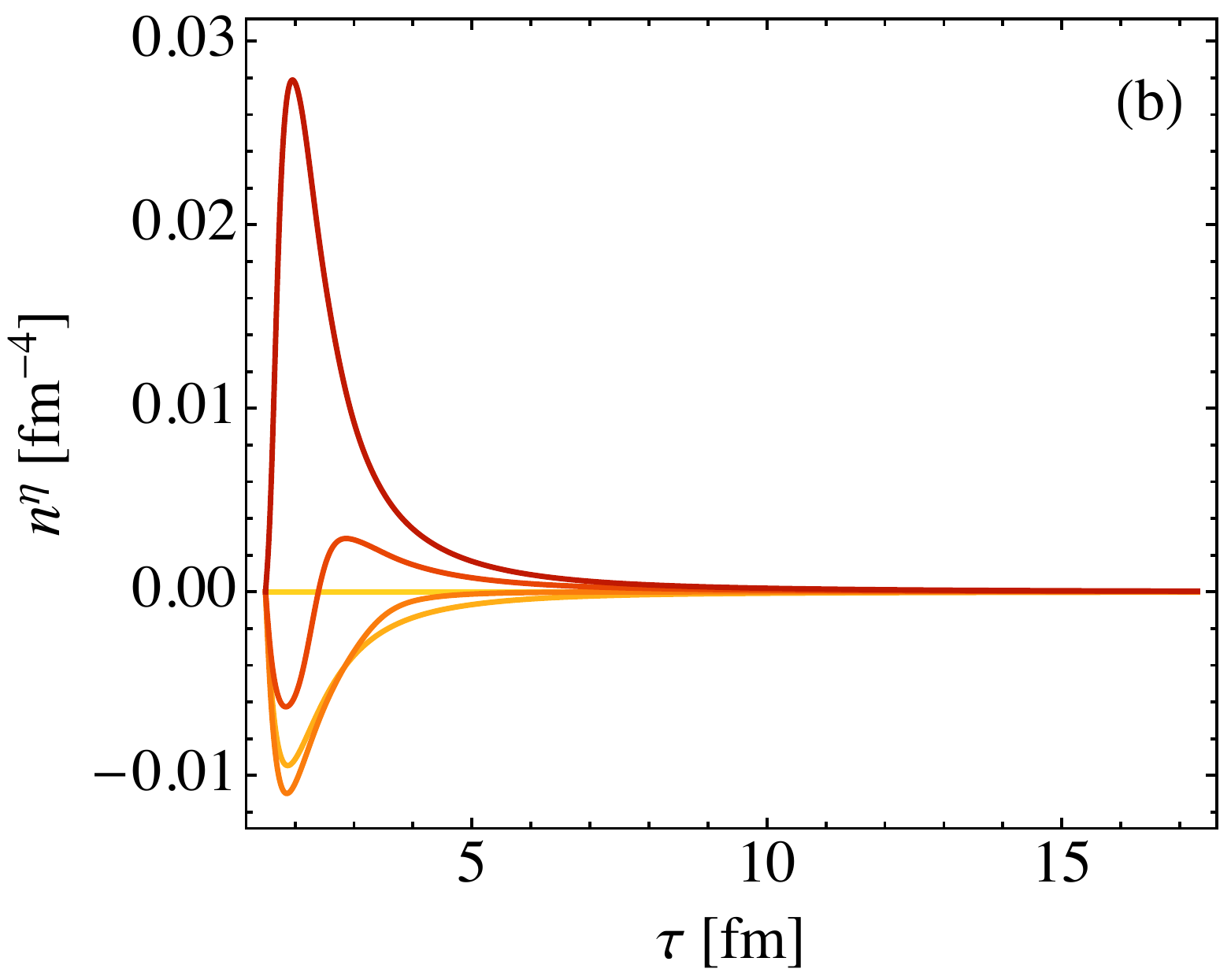}
\caption{%
    Time evolution of (a) correlation length and (b) longitudinal baryon diffusion at selected space-time rapidities. Note that $n^\eta$ at $\eta_s=1.25$ is shown intentionally as its sign changes during the evolution. See text for discussion.
    }
    \label{fig:xi_neta_evolution}
\end{center}
\end{figure}

To better understand this we plot in Fig.~\ref{fig:xi_neta_evolution} the history of the correlation length and baryon diffusion current at different $\eta_s$. In panel (a) we see that $\xi$ does show the expected critical enhancement, by up to a factor $\sim 4.5$ at $\eta_s\eq1$. This maximal enhancement corresponds to $\tau_n\simeq20\,\tau_{n,0}$ and $D_B\simeq0.22\,D_{B,0}$, naively suggesting significant effects on the dynamical evolution. However, the critical enhancement of the correlation length does not begin in earnest before the fireball has cooled down to a low temperature $T\lesssim T_c+\Delta T$. 
%
%
Fig.~\ref{fig:xi_neta_evolution}b shows at at this late time the baryon diffusion current has already decayed to a tiny value. In other words, the largest baryon diffusion currents are created at early times when the temporal gradients are highest but the system is far from the critical point; by the time the system gets close to the critical point, thermal and chemical gradients have decayed to such an extent that even a critical enhancement of the correlation length by a factor 5 can no longer revive the baryon diffusion current to a noticeable level. 

This two-stage feature, with a first stage characterized by large baryon diffusion effects without critical modifications and a second stage characterized by large critical  fluctuations \cite{Rajagopal:2019xwg, Du:2020bxp} with negligible effects on the bulk evolution through baryon diffusion, is an important observation. For a deeper understanding we devote Sec.~\ref{sec:tevol} to a more systematic investigation of the time evolution of the diffusion current, but not before a brief exploration in the following subsection of critical effects on the final single-particle distributions.

\subsubsection{Critical corrections to final single-particle distributions}
\label{sec:diffcp_spectra}

How to consistently include critical fluctuation effects on the finally emitted single-particle distributions, at the ensemble-averaged level, is still a subject of active research. A solid framework may require a microscopic picture involving interactions between the underlying degrees of freedom and the fluctuating critical modes during hadronization. In this work we employ a simple ansatz where critical corrections to the final particle distributions are included only via the diffusive correction $\delta f^\mathrm{diffusion}$ from Eq.~\eqref{eq:diffusion_deltaf} appearing in the Cooper-Frye formula \eqref{eq:distr1}. In this subsection this dissipative correction is computed from the simulations described in the preceding subsection which include critical correlation effects through critically modified transport coefficients, specifically a normalized baryon diffusion coefficient $\hat{\kappa}\equiv\kappa_n/\tau_n$ with critical scaling
\begin{equation}
\label{eq:critical_deltaf}
    \hat{\kappa}=\hat{\kappa}_0\left(\frac{\xi}{\xi_0}\right)^{-1}\,,
\end{equation}
obtained from Eqs.~\eqref{eq:chi_kappa_tau_scaling} using $\hat\kappa_0 = \kappa_{n,0}/ \tau_{n,0}$.\footnote{%
    We note that this ansatz is more straightforwardly implemented in the Chapman-Enskog method used here than in the 14-moment approximation \cite{Monnai:2018rgs,McNelis:2019auj} whose  transport coefficients do not have such obvious critical scaling.}
Since we saw in the preceding subsection that the hydrodynamic quantities on the particlization surface are hardly affected by the inclusion of critical scaling effects during the preceding dynamical evolution, the main critical scaling effects on the emitted particle spectra arise from any critical modification that $\hat{\kappa}$ might experience on the particlization surface.

\begin{figure}[!htb]
\begin{center}
\hspace{-0.5cm}
\includegraphics[width= 0.483\textwidth]{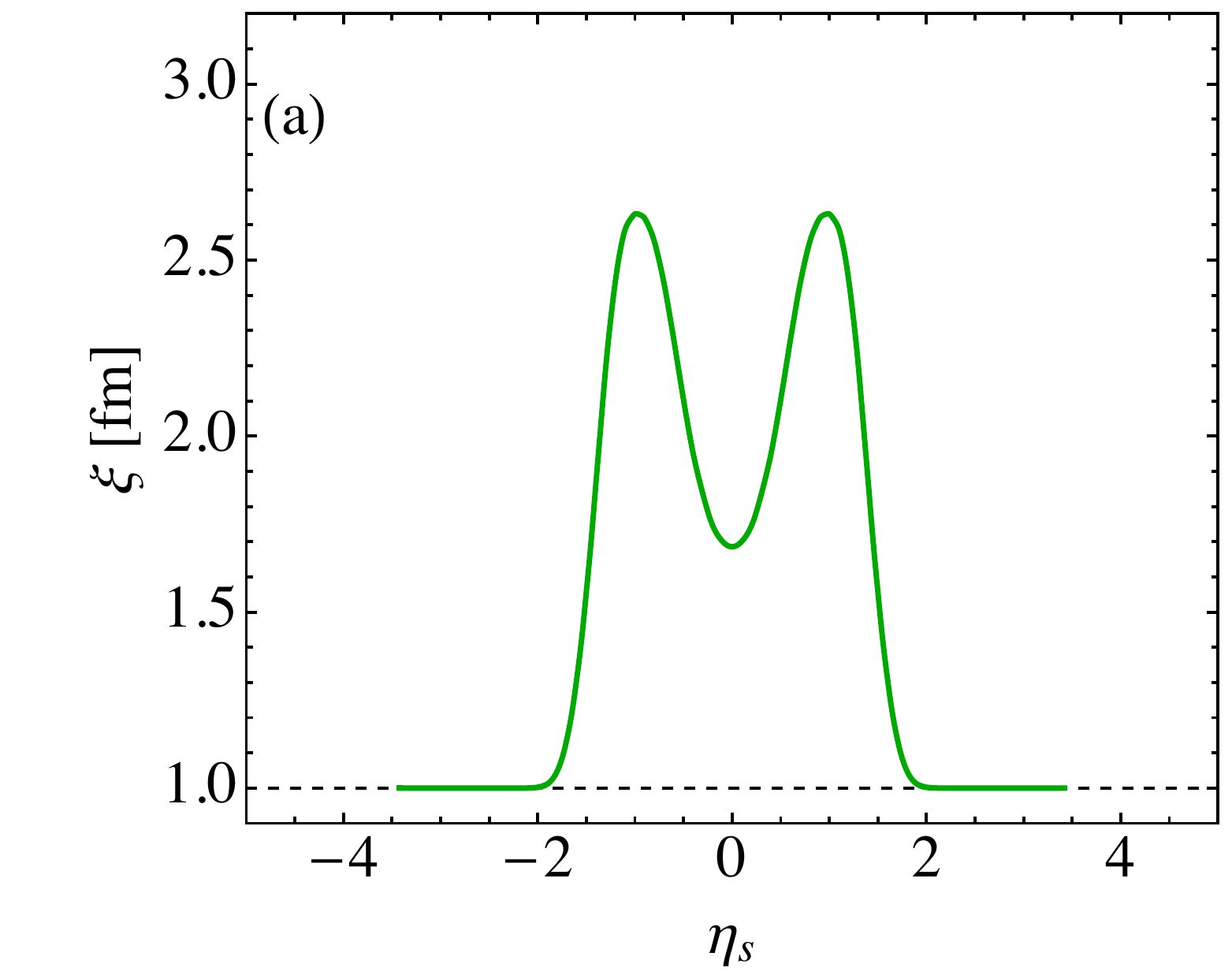}\hspace{0.5cm}\includegraphics[width= 0.48\textwidth]{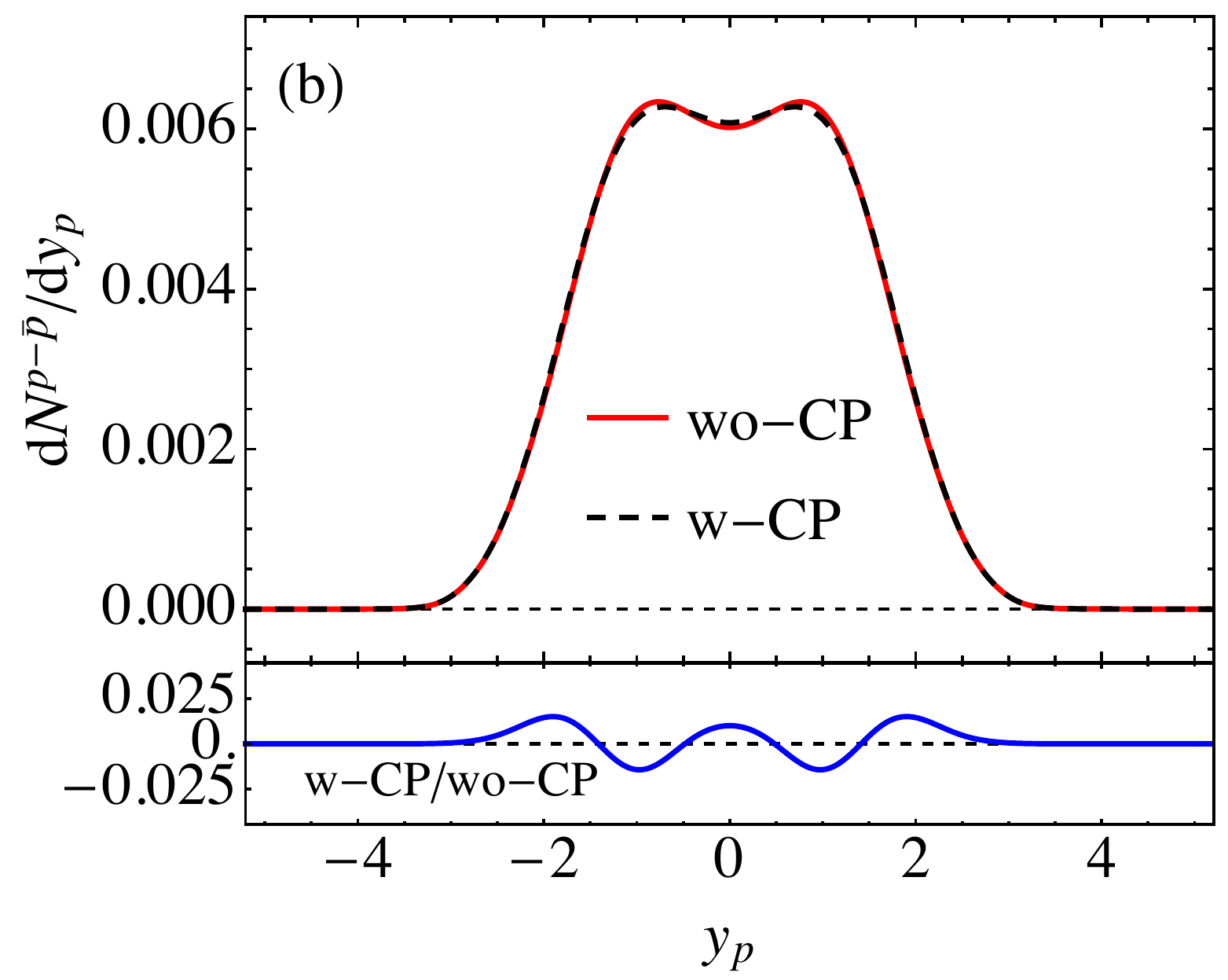}
\caption{%
    (a) Space-time rapidity distribution of the correlation length along the freeze-out surface. (b) {\sl Upper panel:} Net proton rapidity distributions with and without critical scaling effects (w-CP and wo-CP), calculated from the Cooper-Frye formula \eqref{eq:distr1} with diffusive correction \eqref{eq:diffusion_deltaf}. For the w-CP case critical scaling effects are included in both the hydrodynamic evolution and the diffusive correction at particlization. {\sl Lower panel:} Deviation from 1 of the ratio of the rapidity spectra shown in the upper panel.}
    \label{fig:xi_spec_cp}
\end{center}
\end{figure}

The space-time rapidity distribution of the correlation length $\xi$ along the freeze-out surface, as well as the net proton rapidity distributions with and without critical scaling effects, are shown in Fig.~\ref{fig:xi_spec_cp}. Panel (a) shows that $\xi$ peaks near $|\eta_s|\simeq1.0$ on the freeze-out surface, consistent with Fig.~\ref{fig:xi_neta_evolution}a. Note that, although fluid cells at different $\eta_s$ generally freeze out at different times, the freeze-out surface in Fig.~\ref{fig:fz_surf}a shows that within $\eta_s\in[-1.5, 1.5]$ all fluid cells freeze out at basically the same time $\tau_f\sim17$\,fm/$c$. Therefore Fig.~\ref{fig:xi_spec_cp}a indeed corresponds to the $\xi$ values at different \etas{} at the end of the evolution in Fig.~\ref{fig:xi_neta_evolution}a.

Even though Fig.~\ref{fig:xi_spec_cp}a shows a critical enhancement of $\xi/\xi_0{\,\simeq\,}2.7$ near $\eta_s{\,\simeq\,}1.0$, corresponding to $\hat{\kappa}/\hat{\kappa}_0{\,\simeq\,}0.37$, we see in Fig.~\ref{fig:xi_spec_cp}b that the net proton distribution is modified by at most a few percent. The lower panel in Fig.~\ref{fig:xi_spec_cp}b indicates that the largest critical corrections indeed correspond to regions of large $\xi/\xi_0$, sign-modulated by the direction of the baryon diffusion current (cf. Fig.~\ref{fig:fz_surf}d). We also notice a thermal smearing when mapping the distribution of $\xi$ in \etas{} to the modification of net proton distribution in $y_p$. The critical modification of the net proton spectra arising from the diffusive correction to the distribution function is very small also because $\delta f^\mathrm{diffusion}$ in \eqref{eq:diffusion_deltaf} is roughly proportional to the magnitude of $n^\mu$ which is tiny. Such small modifications are certainly unresolvable with current or expected future measurement.

In conclusion, critical scaling effects on both the hydrodynamic evolution of the bulk medium and the finally emitted single-particle momentum distributions are small, mostly because by the time the system passes the critical point and freezes out the baryon diffusion current has decayed to negligible levels.

\subsection{Time evolution of baryon diffusion}
\label{sec:tevol}

In this subsection we further analyze the baryon diffusion dynamics and the origins of its rapid decay. We define Knudsen and inverse Reynolds numbers for baryon diffusion and display their space-time dynamics. The resulting insights are relevant for model building and for the future quantitative calibration of the bulk fireball dynamics at non-zero chemical potential. 

\subsubsection{Fast decay of baryon diffusion}
\label{sec5d1}

As discussed in Sec.~\ref{sec:hydro}, the diffusion current relaxes to its Navier-Stokes limit $n^\nu_\mathrm{NS}\eq\kappa_n\nabla^\nu(\mu/T)$ on a time scale given by $\tau_n$. General features of baryon diffusion evolution can thus be understood by following the time evolution of $n^\mu_\mathrm{NS}$, $\kappa_n$ and $\tau_n$. Here we focus on their evolution without inclusion of critical scaling since we established that the latter has negligible effect on the bulk evolution and therefore the non-critical values of $n^\mu_\mathrm{NS}$, $\kappa_n$ and $\tau_n$ evolve almost identically with and without inclusion of critical effects.

\begin{figure}[!b]
\begin{center}
\hspace{-0.5cm}
\includegraphics[width= 0.55\textwidth]{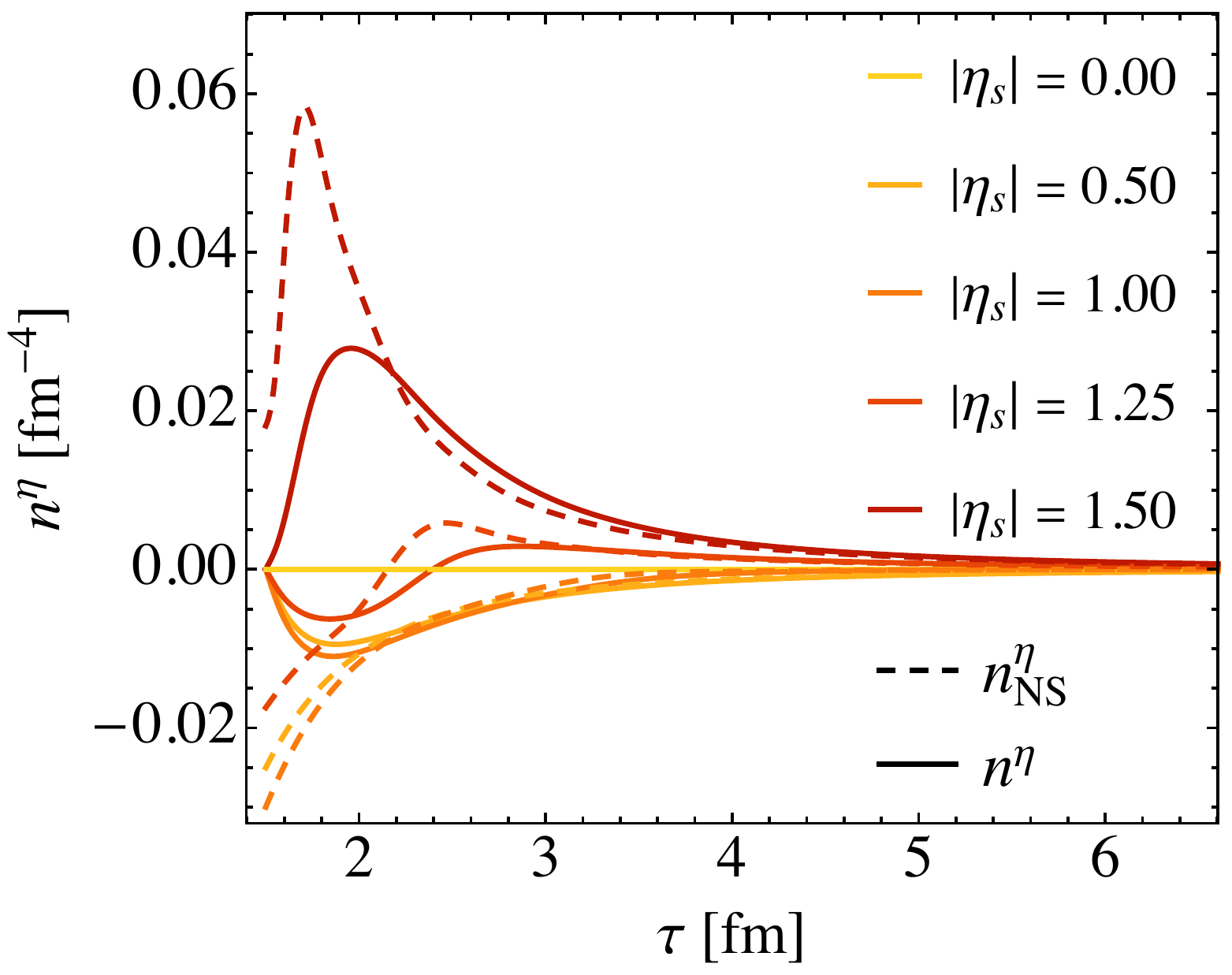}
\caption{%
    Same as Fig.~\ref{fig:xi_neta_evolution}b, but zoomed in onto the early evolution stage and including for comparison as dashed lines the corresponding Navier-Stokes limit $n^\eta_\mathrm{NS}$ of the longitudinal diffusion current.}
    \label{fig:NSlimit_evolution}
\end{center}
\end{figure}

\begin{figure}[!t]
\begin{center}
   \hspace{-0.5cm}\includegraphics[width= 0.48\textwidth]{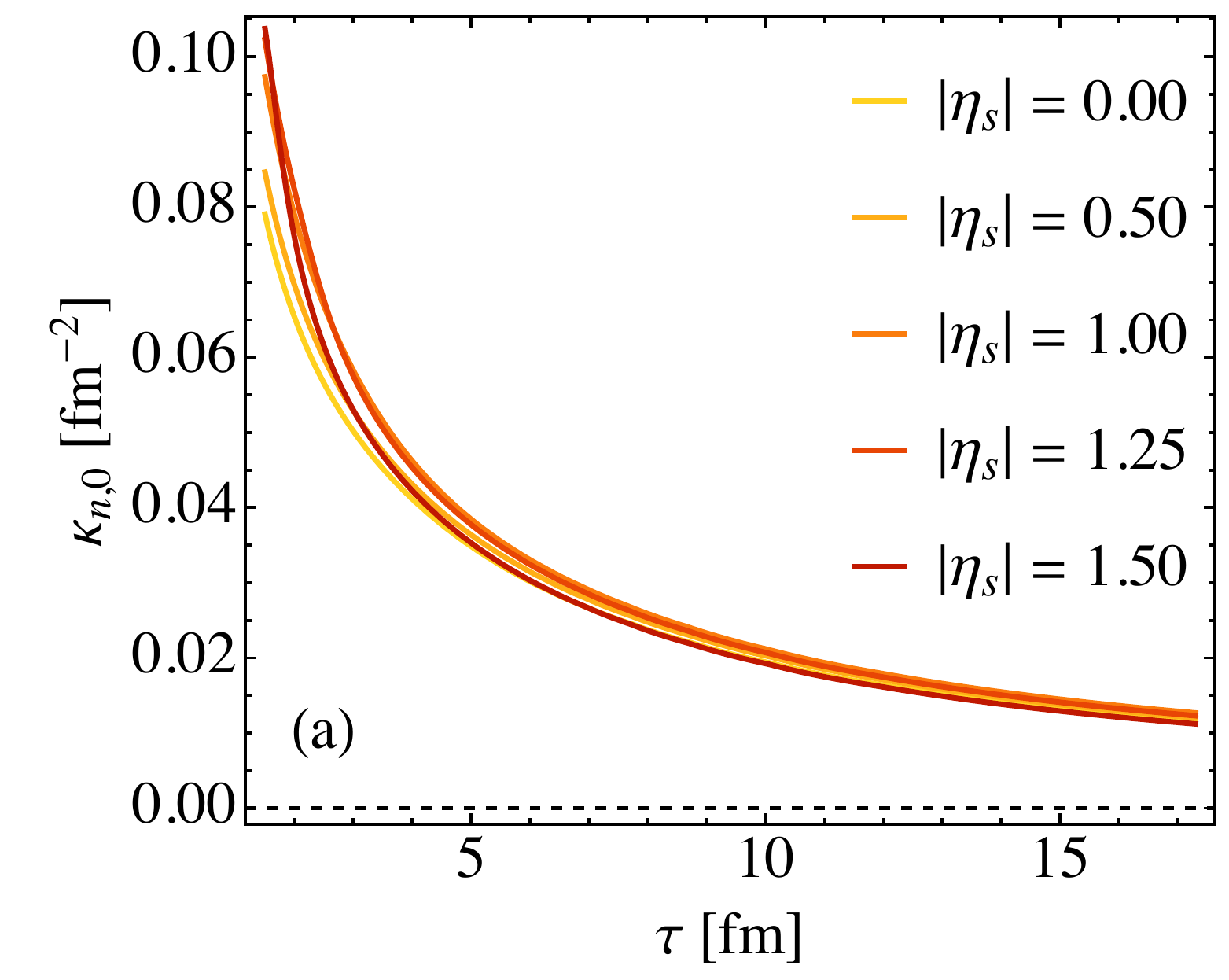}\hspace{0.5cm}\includegraphics[width=0.48\textwidth]{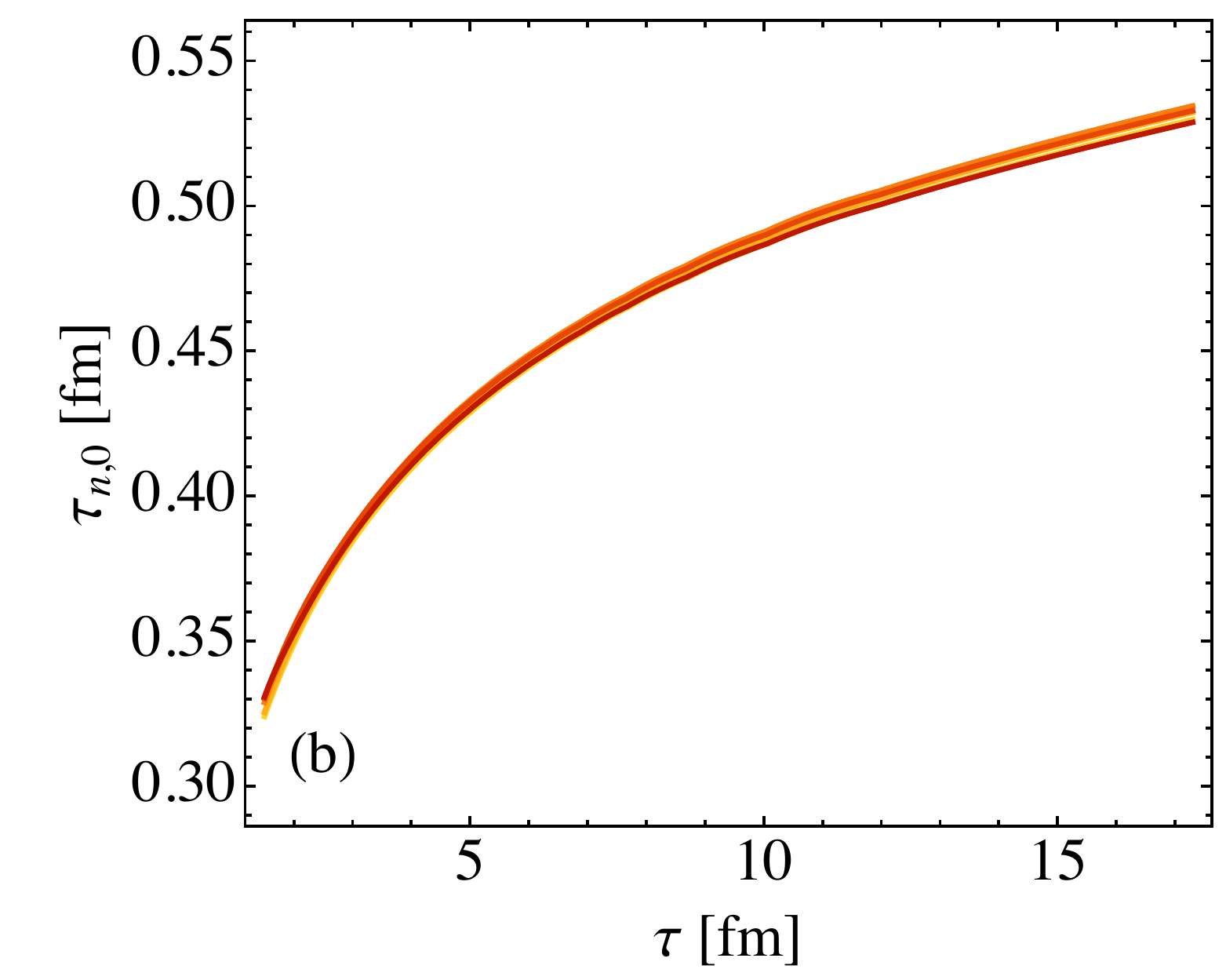}
    \caption{Time evolution of (a) $\kappa_{n,0}$ and (b) $\tau_{n,0}$ at some space-time rapidities, corresponding to Figs.~\ref{fig:xi_neta_evolution} and \ref{fig:NSlimit_evolution}.}
    \label{fig:kappa_tau_evolution}
\end{center}
\end{figure}

Fig.~\ref{fig:NSlimit_evolution} shows a comparison of the longitudinal baryon diffusion current (solid lines) with its Navier-Stokes limit (dashed lines) at different space-time rapidities. One sees that the relaxation equation for $n^\eta$ tries to align the diffusion current with its Navier-Stokes value (which is controlled by the longitudinal gradient $\nabla^\eta(\mu/T)$) but the finite relaxation time delays the response, causing $n^\eta$ to perform damped oscillations around $n^\eta_\mathrm{NS}$. This is most clearly illustrated in Fig.~\ref{fig:NSlimit_evolution} by following the cell located at $\eta_s\eq1.5$ (uppermost): Initialized at zero, $n^\eta$ initially rises steeply, trying to adjust to its positive and rapidly increasing Navier-Stokes value, but at $\tau{\,\simeq\,}1.7$\,fm/$c$ the longitudinal gradient of $\mu/T$ starts to decrease and so does $n^\eta_\mathrm{NS}$. The hydrodynamically evolving $n^\eta$ follows suit, turning downward with a delay of about 0.3\,fm/$c$ (which, according to Fig.~\ref{fig:kappa_tau_evolution}b below, is the approximate value of the relaxation time $\tau_{n,0}$ at $\tau\eq2$\,fm/$c$), but soon finds itself overshooting its Navier-Stokes value. For the cell located at $\eta_s\eq1.25$, $n^\eta$ crosses its Navier-Stokes value even twice. 

As long as the relaxation time $\tau_n$ is short and not dramatically increased by critical slowing down, the rapid decrease of the dynamically evolving diffusion current is seen to be a generic consequence of a corresponding rapid decrease of its Navier-Stokes value: Fig.~\ref{fig:NSlimit_evolution} shows that after $\tau{\,\sim\,}3.5$\,fm/$c$, $n^\eta$ basically agrees with its Navier-Stokes limit $n^\eta_\mathrm{NS}$. Fig.~\ref{fig:kappa_tau_evolution}b shows that in the absence of critical effects the relaxation time $\tau_{n,0}\eq C_n/T$ increases by less than a factor of 2 over the entire fireball lifetime. The rapid decrease of $n^\eta_\mathrm{NS}$ is a consequence of two factors: (i) the gradients of $\mu/T$ decrease with time, owing to both the overall expansion of the system and the diffusive transport of baryon charge from dense to dilute regions of net baryon density, and (ii) the baryon diffusion coefficient $\kappa_{n,0}$ decreases dramatically (by almost an order of magnitude over the lifetime of the fireball as seen in Fig.~\ref{fig:kappa_tau_evolution}a), as a result of the fireball's decreasing temperature.

In summary, three factors contribute to the negligible influence of the QCD critical point on baryon diffusion: First, baryon diffusion is largest at very early times when its relaxation time is shortest and it quickly relaxes to its Navier-Stokes value; the latter decays quickly, due to decreasing chemical gradients and a rapidly decreasing baryon diffusion coefficient. Second, the relaxation time for baryon diffusion increases at late times, generically as a result of cooling but possibly further enhanced by critical slowing down if the system passes close to the critical point. This makes it difficult for the baryon diffusion current to grow again. Third, critical effects that would modify \footnote{%
    According to Eq.~\eqref{eq:nmu_NS_decomposition2} and Eq.~(\ref{eq:D_scaling2}), critical effects can increase or decrease the Navier-Stokes value of the baryon diffusion, depending on the relative sign and magnitude of the density and temperature gradients.}
the Navier-Stokes limit for the baryon diffusion current become effective only at very late times when $n^\eta_\mathrm{NS}$ has already decayed to non-detectable levels. The baryon diffusion current thus remains small even if its Navier-Stokes value were significantly enhanced by critical scaling effects.

\subsubsection{Knudsen and inverse Reynolds numbers}

\begin{figure}[!b]
\begin{center}
\hspace{-.8cm}
\includegraphics[width= 0.8\textwidth]{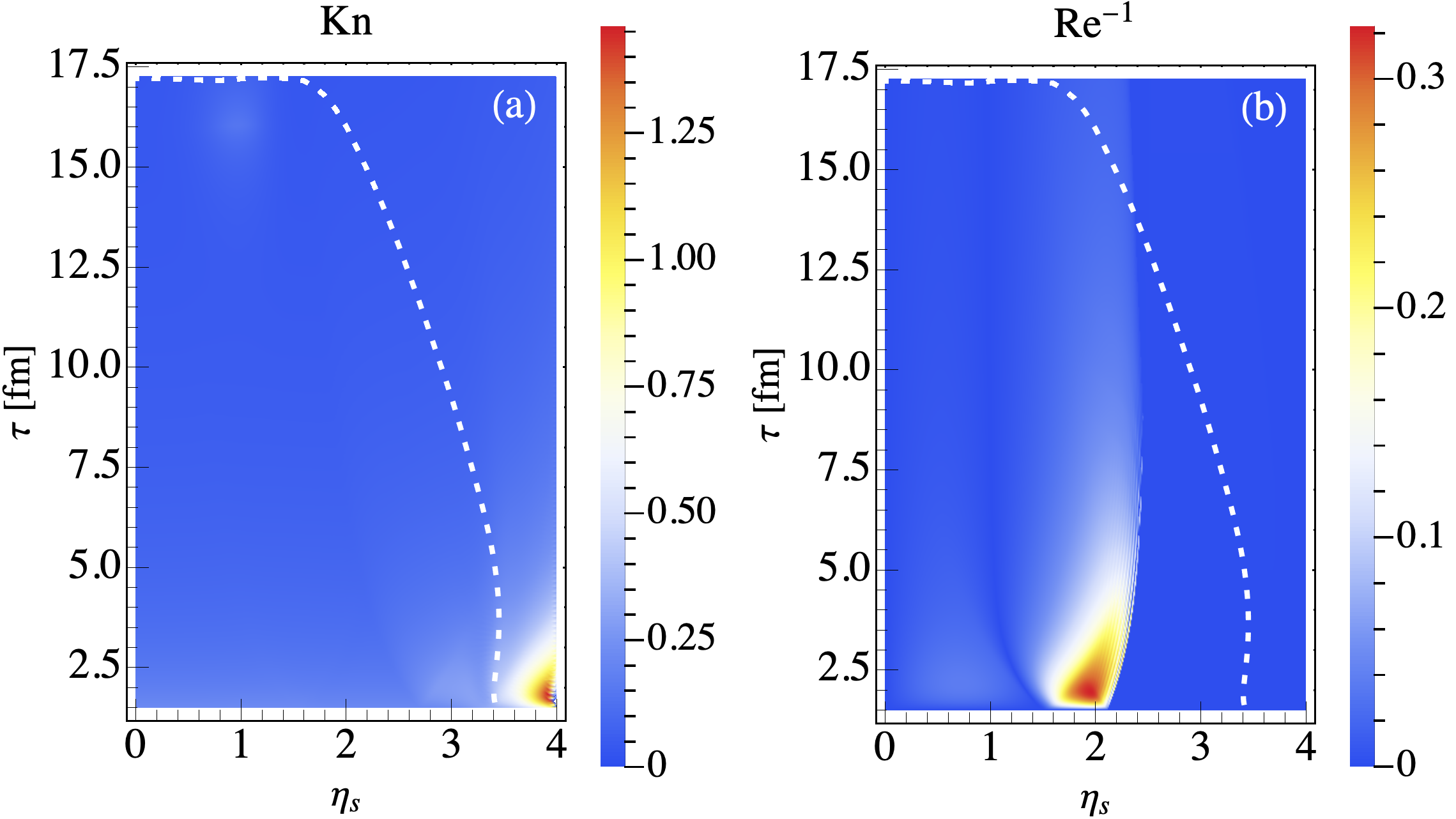}
\caption{%
    Space-time evolution of (a) the Knudsen number Kn and (b) the inverse Reynolds number $\mathrm{Re}^{-1}$ for the baryon diffusion current. The white dashed line shows the same freeze-out surface as Fig.~\ref{fig:fz_surf}a.
    \label{fig:kn_re_evolution}}
\end{center}
\end{figure}

We close this section by investigating the Knudsen and inverse Reynolds numbers associated with baryon diffusion. These are typically taken as quantitative measures to assess the applicability of second order viscous hydrodynamics such as the \beshydro{} framework employed in this work. Copying their standard definitions for shear and bulk viscous effects \cite{Shen:2014vra, Denicol:2018wdp, Du:2019obx}, we here set
\begin{equation}
    \mathrm{Kn}\equiv \tau_n\theta\,,\quad\mathrm{Re}^{-1}\equiv\frac{\sqrt{|n^\mu n_\mu|}}{n}\,,
\end{equation}
for baryon diffusion, where $\theta$ is the scalar expansion rate. Kn is the ratio between time scales for microscopic diffusive relaxation ($\tau_n$) and macrosopic expansion ($\tau_\mathrm{exp}=1/\theta$); the relaxation time $\tau_n$ includes the effects of critical slowing down in the neighborhood of the QCD critical point. $\mathrm{Re}^{-1}$ is the ratio between the magnitude of the off-equilibrium baryon diffusion current and the equilibrium net baryon density in ideal fluid dynamics. Their space-time evolutions are shown in Fig.~\ref{fig:kn_re_evolution}, together with the freeze-out (particlization) surface at $e_f\eq0.3\,$GeV/fm$^3$.

Fig.~\ref{fig:kn_re_evolution}a tells us that $\mathrm{Kn}{\,\gtrsim\,}1$ happens only outside the freeze-out surface, in the fireball's corona where the fluid has already broken up into particles even at the earliest stage of the expansion. The short-lived peak in Kn near $\tau\eq\tau_i=1.5$\,fm/$c$ and $\eta_s{\,\sim\,}4$ is caused by the rapid increase of $\tau_n$ in the dilute and very cold corona of the fireball (note that $\tau_{n,0}=C_n/T$). Critical slowing down near the QCD critical point causes the Knudsen number to increase somewhat around $\eta_s\eq1$ close to the freeze-out surface; this critical enhancement is barely visible as a light cloud on a blue background, indicating critical Knudsen numbers in the range Kn${\,\sim\,}0.5-0.7$. Fig.~\ref{fig:kn_re_evolution}b, on the other hand,  indicates that $\mathrm{Re}^{-1}{\,\lesssim\,}0.3$ during the entire evolution, even close to the places where the Navier-Stokes value of the baryon diffusion current peaks at early times (see Fig.~\ref{fig:NSlimit_evolution}). After $\tau{\,\sim\,}5$\,fm/$c$ (including the entire critical region around the QCD critical point) its maximum value drops below 0.1, reflecting the rapid decay of the baryon diffusion current. 

The maximal $\mathrm{Re}^{-1}$ occurs around $\eta_s\simeq2$ shortly after the hydrodynamic evolution starts at $\tau_i=1.5\,$fm/$c$. From it emerges a region of sizeable inverse Reynolds number which ends at two moving boundaries where $\mathrm{Re}^{-1}\eq0$ (dark blue). The left boundary, moving towards smaller \etas{} values, reflects a sign change of the baryon diffusion current (see Fig.~\ref{fig:long_evolution}f where at $\tau\eq5.5$\,fm/$c$ $n^\eta$ flips sign at $\eta_s\simeq1$). The right boundary, on the other hand, corresponds to where $n^\mu$ decays to zero (which, according to Fig.~\ref{fig:long_evolution}f, happens at $\eta_s{\,\simeq2.5\,}$ when $\tau\eq5.5$\,fm/$c$). The initial outward movement of the right boundary is a result of baryon transport to larger space-time rapidity. It stops moving after the diffusion current has decayed and no longer transports any baryon charge longitudinally.

The small values of $\mathrm{Kn}$ and $\mathrm{Re}^{-1}$ during the entire fluid dynamical evolution validate the applicability of second order viscous hydrodynamics, \beshydro, for describing flow and diffusive transport of baryon charge in the collision system studied here.

\section{Summary and conclusions}
\label{sec:summary}

In this work we studied central Au-Au collisions at $\sqrt{s_\mathrm{NN}}\sim20$\,GeV in which the fireball covers about 7 units of space-time rapidity along the beam direction (Fig.~\ref{fig:long_evolution}a). Assuming that baryon stopping leads to a space-time rapidity shift of about 1.5 units for the incoming projectile and target baryons, the initial net baryon distribution had a width of about 4 units. It was modeled by a double-humped function with two well-separated peaks located at $\eta_s\eq\pm1.5$ (Fig.~\ref{fig:long_evolution}d). After accounting for ideal hydrodynamic evolution and thermal smearing this resulted in a double-humped net proton rapidity distribution whose peaks were still relatively cleanly separated by about 2 units of rapidity, but after including baryon diffusion they almost (though not quite) merged into a single broad hump around midrapidity (Fig.~\ref{fig:1D_spec_comp}a). In our calculation, the QCD critical point was positioned at a baryon chemical potential $\mu=250$\,MeV. Recent Lattice QCD results put the likely location of this critical point at $\mu>400$\,MeV \cite{Bazavov:2017dus, Mukherjee:2019eou, Giordano:2019gev}, which requires lower collision energies for its experimental exploration. At lower collision energies the width of the initial space-time rapidity interval of non-zero net baryon density will be narrower, and the final net-proton distribution will eventually become single-peaked in central collisions \cite{Bearden:2003hx}.   

In the work presented here we focused on the questions how diffusive baryon transport manifests itself along the beam direction in hydrodynamic simulations, what traces it leaves in the finally measured rapidity distributions, and how it is affected by critical scaling of transport coefficients in proximity of the QCD critical point. To address these questions we systematically discussed the static and dynamic critical behavior of thermodynamic properties (especially those associated with baryon transport), introduced an analytical parametrization of the correlation length that correctly reproduces the critical exponents of the 3D Ising universality class, and (based on the Hydro++ framework \cite{An:2019csj}) identified the critical scaling (``critical slowing down'') of the relaxation time for the baryon diffusion current: $\tau_n\sim\xi^2$. We did not discuss the out-of-equilibrium evolution of slow critical fluctuation modes themselves which is studied in Ch.~\ref{ch.fluctuations} using the \beshydro+ framework \cite{Du:2020bxp} where it was found that their feedback to the hydrodynamic bulk evolution was negligible.

The baryon diffusion flows seen in the calculations presented in this paper are characterized by an important feature: They show almost no sensitivity to critical effects even for cells passing close to the critical point. Taken at face value, this implies that the hydrodynamic evolution of baryon diffusion leading to the finally emitted ensemble-averaged single-particle momentum spectra does not carry useful information for locating the QCD critical point. The absence of critical effects on baryon diffusion in this work contrasts starkly with the strong critical effects on the evolution of the bulk viscous pressure found in Ref.~\cite{Monnai:2016kud} which led to significant distortions of the rapidity distributions for all emitted hadron species.

As we noted in Sec.~\ref{sec:theochall} and Ch.~\ref{ch:multistage}, critical effects of significant phenomenological consequences on bulk properties should be identified and then added to the bulk dynamics when one calibrates such a multistage framework for heavy-ion collisions at BES energies. This study for baryon diffusion and related work in Ref.~\cite{Monnai:2016kud} for bulk pressure took the first steps in this direction. Because a more precise evaluation of their significance can only be made once a well-constrained framework is available, additional studies of critical effects on bulk dynamics are needed that simultaneously include {\it all} viscous effects, especially shear stress whose effects on baryon flow has not yet been investigated at all.

\chapter{Fluctuation dynamics near the QCD critical point}
\label{ch.fluctuations}

Reliable predictions of observable fluctuation signatures require complex dynamical simulations of the non-equilibrium dynamics of the critical fluctuation modes, coupled with a comprehensive dynamical evolution package for the medium in which these fluctuations arise \cite{Nahrgang:2011mv, Nahrgang:2011mg,Mukherjee:2015swa,Akamatsu:2016llw,Murase:2016rhl,Herold:2016uvv,Stephanov:2017ghc,Hirano:2018diu,Singh:2018dpk,Herold:2018ptm,Nahrgang:2018afz,Yin:2018ejt}. In this chapter, we use the newly developed {\sc hydro+} framework \cite{Stephanov:2017ghc} to study the coupled dynamics of out-of-equilibrium fluctuations and the bulk hydrodynamic evolution. We will use an analytically solvable model, the ideal Gubser flow described in Ch.~\ref{ch.gubser}, as our background for the non-equilibrium fluctuation dynamics. As already noted in Ref.~\cite{Rajagopal:2019xwg}, the non-equilibrium evolution of critical fluctuations is affected by several different physical mechanisms: (i) the space-time evolution of the thermodynamic properties of the background fluid; (ii) the advection of critical fluctuations from the inside to the outside of the fireball by radial hydrodynamic flow; and (iii) the critical dynamics of the correlation length for these fluctuations in fluid cells that pass close to the critical point. Using the analytically known ideal Gubser flow for the hydrodynamic background allows us to surgically isolate and study these effects, both separately and in combination, without giving up on the simultaneous existence of both longitudinal {\it and} transverse flow in the expanding medium, by applying appropriate analytic manipulations to the background.

This chapter is based on material published in Ref.~\cite{Du:2020bxp}.

%
\section{Non-equilibrium fluctuations}
\label{sec:nonequi}
%
In Sec.~\ref{sec:fluct}, we discussed the dependence of the equilibrium mode spectrum $\bar\phi_Q$ on the local hydrodynamic variables which evolve dynamically. Expansion of the fluid on the macroscopic scale drives both the background fluid and the slow mode spectrum out of thermal equilibrium while microscopic interactions push the system back towards (a modified) local equilibrium state. The evolution of the dissipative non-equilibrium corrections to the hydrodynamic variables is handled by viscous fluid dynamics as discussed in Sec.~\ref{sec:relat_hydro}. In this chapter, we discuss the out-of-equilibrium evolution of the critical slow modes.

The authors of Ref.~\cite{Stephanov:2017ghc} first introduced a set of relaxation equations for these modes (which they called ``fluctuation kinetic equations'' or briefly ``kinetic equations'', see also \cite{An:2019osr}) using an educated ansatz that was later refined in the form of so-called ``hydro-kinetic equations'',  derived within the ``hydro-kinetic approach'' to fluctuating hydrodynamics \cite{Akamatsu:2018vjr,Martinez:2018wia,An:2019csj}, first in the hydrodynamic regime ($Q\ll \xi^{-1}$) and then generalized to the scaling regime near a critical point ($\xi^{-1}\ll Q \ll \ell_0^{-1}$, where $\ell_0$ is a microscopic length scale, e.g. $\ell_0\sim1/T$ in a conformal system). By interpolating between these two regimes they could also cover the region $Q{\,\sim\,}\xi^{-1}$. For completeness we here briefly summarize the results of \cite{Akamatsu:2018vjr}, calling the slow-mode evolution equations simply their ``equations of motion''.

%
\subsection{Dynamics in the hydrodynamic regime}
\label{sec:awaycp}
%

In the hydrodynamic regime ($Q\ll\xi^{-1}$) slow-mode dynamics is governed by the relaxation equations \cite{Stephanov:2017ghc,Akamatsu:2018vjr}
\begin{equation}
\label{eq:phi_hydro}
    D\phi_Q = -\Gamma_Q\,(\phi_Q - \bar\phi_0)\,,
\end{equation}
where $D\equiv u^\mu d_\mu$ (with $d_\mu$ denoting the covariant derivative which reduces to a simple partial derivative for scalar fields such as $\phi_Q$) is the LRF time derivative\footnote{%
    Note that $D$ is the LRF time derivative {\it at fixed wave number $\bm Q$ in the LRF}. In fully covariant notation this constraint requires replacing the covariant derivative by the ``confluent derivative" \cite{An:2019osr,An:2019csj} (see also foonote~\ref{fn1}).}
and $\bar\phi_0$ (i.e. $\bar\phi_Q$ in the hydrodynamic limit) is given by Eq.~(\ref{eq:phiQeq}). The $Q$-dependent relaxation rate $\Gamma_Q$ takes the form \cite{Stephanov:2017ghc, Akamatsu:2018vjr}
\begin{equation}
    \Gamma_Q \equiv 2 D_p Q^2 = 2\left(\frac{\lambda_T}{\cp}\right)Q^2\,,
\label{eq:gammaQ_hydro}
\end{equation}
where the factor 2 arises from the fact that this describes the relaxation of a two-point function 
and $D_p$ is a diffusion coefficient, related to the thermal conductivity $\lambda_T$ though the Wiedemann-Franz type relation $D_p = \lambda_T/\cp$ (see also Eq.~\eqref{eq:D_p}),
consistent with the expectation that the relaxation rate is proportional to the transport coefficient ($\lambda_T$) over the susceptibility ($\cp$). Note that Eq.~(\ref{eq:gammaQ_hydro}) is the equilibrium value of the relaxation rate \cite{Rajagopal:2019xwg} and ignores off-equilibrium corrections to the relaxation process \cite{Akamatsu:2018vjr}.

A few comments are in order here. First, when the system evolves, the relaxation of the slow modes (originally from the fluctuations of entropy per baryon, both conserved quantities in an ideal fluid) requires transport of conserved quantities through diffusion \cite{An:2019osr}. This explains why the equilibration rate for $\phi_Q$ is proportional to the diffusion coefficient $D_p$. Second, since the relaxation is through diffusive processes, the relaxation rate is proportional to $Q^2$. Third, in the hydrodynamic regime $Q\ll\xi^{-1}$, $\phi_Q$ relaxes for all wave numbers $Q$ to the {\it static} ($Q = 0$) equilibrium value $\bar\phi_0$. 

As observed in Ref.~\cite{Akamatsu:2018vjr}, since the equilibrium value $\bar\phi_0$ evolves as Eq.~(\ref{eq:phiQeq}) when the system evolves, its fractional change per unit time is controlled by the scalar expansion rate $\theta$ of the system,
\begin{equation}
\label{eq:expansion}
    \theta \equiv d_\mu u^\mu = \partial_\mu u^\mu + \Gamma^\lambda_{\lambda\mu}u^\mu\,,
\end{equation}
where $\Gamma^\lambda_{\lambda\mu}$ are the Christoffel symbols \cite{carroll2004spacetime}. Deviations from the equilibrium value, on the other hand, decay with the relaxation rate $\Gamma_Q \propto Q^2$, i.e. short wave length modes equilibrate faster than long wave length modes. For any given wave number $Q$, the dynamics of $\phi_Q$ thus results from the competition between the expansion of the system (controlled by the expansion rate $\theta$ of the background fluid) and the relaxation of the slow modes (controlled by the relaxation rate $\Gamma_Q$), and there should exist a (dynamical) wave number scale $Q_\mathrm{neq}$ which separates approximately thermalized fluctuation modes from those that strongly deviate from equilibrium \cite{Akamatsu:2018vjr}. 

This competition between expansion and relaxation towards equilibrium is generic and persists in the critical scaling regime near the critical point, as will be seen formally in the following subsection and studied numerically in Sec.~\ref{sec:results}. To characterize this competition quantitatively it is convenient to introduce the ``critical Knudsen number" for the slow modes as the ratio between the scalar expansion and relaxation rates:
\begin{equation}
    \textrm{Kn}(Q)\equiv\theta/\Gamma_Q\,.
\label{eq:kn}
\end{equation}
Slow modes with large critical Knudsen numbers will lag behind, not being able to follow the hydrodynamic evolution of the equilibrium value $\bar\phi_Q(x)$. Very roughly, the scale $Q_\mathrm{neq}$ defined in the preceding paragraph should correspond to critical Knudsen numbers of $O(1)$.

%
\subsection{Dynamics in the scaling regime}
\label{sec:nearcp}
%

To extend the slow-mode evolution equation (\ref{eq:phi_hydro}) from the hydrodynamic regime $Q\ll\xi^{-1}$ to the scaling regime $\xi^{-1}\ll Q \ll \ell_0^{-1}$ (including the transition region $Q \sim \xi^{-1}$) we generalize it  \cite{Akamatsu:2018vjr} by replacing on the right hand side the equilibrium value $\bar\phi_0(x)$ for the zero mode without critical scaling by the $Q$-dependent equilibrium value $\bar\phi_Q(x)$ from Eq.~(\ref{eq:eqPhiQ_full}), 
\begin{equation}
\label{eq:phi_scaling}
    D\phi_Q = -\Gamma_Q\,(\phi_Q - \bar\phi_Q)\,,
\end{equation}
and the relaxation rate $\Gamma_Q$ by the $\xi$-dependent expression
\begin{equation}
    \Gamma_Q = \Gamma_\xi f_\Gamma(Q\xi) 
\end{equation}
where 
\begin{eqnarray}
\label{eq:fgamma}
    &&f_\Gamma(Q\xi) \equiv (Q\xi)^2 \Bigl(1+(Q\xi)^2\Bigr)\,,
\\
\label{eq:Gamma_xi}
    &&\Gamma_\xi \equiv 2\left(\frac{\lambda_T}{\cpz\xi^2}\right)
    \left(\frac{\xi_0}{\xi}\right)^2\,.
\end{eqnarray}
The factor $(\xi_0/\xi)^2$ in (\ref{eq:Gamma_xi}) accounts for the critical scaling of $\cp$; the extra factor $1/\xi^2$ in (\ref{eq:Gamma_xi}) compensates with a factor $\xi^2$ in $f_\Gamma$ such that $f_\Gamma/\xi^2 \to Q^2$ in the hydrodynamic limit $Q\ll\xi^{-1}$. Equations (\ref{eq:phi_scaling})-(\ref{eq:Gamma_xi}) reduce to Eqs.~(\ref{eq:phi_hydro},\ref{eq:gammaQ_hydro}) in the hydrodynamic limit $\xi\to\xi_0\ll Q^{-1}$. Note that the factor $Q^2$ in Eqs.~(\ref{eq:gammaQ_hydro},\ref{eq:fgamma}) strongly reduces the thermalization rate for slow modes with small wave numbers $Q{\,\ll\,}\xi^{-1}$, even without critical enhancement of the correlation length. In the critical region, where $\xi$ becomes large, the $\xi$-dependence of $\Gamma_\xi$ in (\ref{eq:Gamma_xi}) causes additional ``critical slowing down'' of relaxation processes even for modes with more typical wave numbers $Q\sim\xi^{-1}$: $\Gamma_\xi \propto \xi^{-z}$ with critical exponent $z=4$.

This model for the evolution of the critical slow modes agrees with the one derived in Refs.~\cite{Stephanov:2017ghc, Akamatsu:2018vjr} and recently studied in \cite{Rajagopal:2019xwg} up to the following differences: First, the authors of Ref. \cite{Stephanov:2017ghc} formulate the slow-mode evolution equation (\ref{eq:phi_scaling}) in terms of a relaxation rate $\Gamma_Q$ that differs from the one used here and in \cite{Akamatsu:2018vjr, Rajagopal:2019xwg} by a factor $\phi_Q/\bar\phi_Q$ (which approaches unity in equilibrium). Second, the assumed scaling behavior with the correlation length $\xi$ depends on the assumed universality class of the critical point: following the classification of Ref. \cite{RevModPhys.49.435}, Ref.~\cite{Rajagopal:2019xwg} uses ``model A", Ref.~\cite{Akamatsu:2018vjr} and the present work use ``model B", while Ref.~\cite{Stephanov:2017ghc} uses ``model H". Let us briefly summarize the differences in the dynamical critical exponent $z$ introduced above and in the shape of the scaling functions $f_\Gamma(Q\xi)$ resulting from these model choices.

Ref.~\cite{Rajagopal:2019xwg} studies dynamics at $\mu \approx0$ where the order parameter is the non-conserved chiral condensate ({\it model A}). Ref.~\cite{Akamatsu:2018vjr} and this chapter place the critical point at large non-zero $\mu$ but ignore the critical behavior of $\lambda_T$ ({\it model B}). {\it Model H}, used in Ref.~\cite{Stephanov:2017ghc}, correctly describes the dynamical universality class of the QCD critical point where the order parameter is a combination of the chiral condensate and baryon density (see also Sec.~\ref{sec:phase_diagram}).
In this exploratory study we follow Ref.~\cite{Akamatsu:2018vjr} and use model B, but it would be straightforward to implement model H instead. In contrast to model A, these two models both feature a $Q$-dependent slow-mode relaxation rate $\Gamma_Q\propto Q^2$ which delays the relaxation of low-$Q$ modes.


To summarize, the dynamics of the slow modes in our work is given by equations (\ref{eq:eqPhiQ_full}) and (\ref{eq:phi_scaling})-(\ref{eq:Gamma_xi}). Our slow mode is the diffusion of fluctuations in entropy per baryon at constant pressure, $\delta(s/n)$; this agrees with Ref.~\cite{Stephanov:2017ghc} and, up to a normalization factor, with Ref.~\cite{Akamatsu:2018vjr} (whose authors studied $n\,\delta(s/n)$), whereas the authors of Ref.~\cite{Rajagopal:2019xwg} studied diffusion of fluctuations in the chiral condensate. Our work differs from Ref.~\cite{Stephanov:2017ghc} by using model B instead of model H for the relaxation rate $\Gamma_Q$ of specific entropy fluctuations, while Ref.~\cite{Rajagopal:2019xwg} uses model A for the relaxation rate of chiral fluctuations. Model differences exist also outside the critical region where Ref. \cite{Rajagopal:2019xwg} uses a constant relaxation rate $\Gamma_0$ while we follow Ref.~\cite{Akamatsu:2018vjr} and set it proportional to $\lambda_T/\cp$; Ref.~\cite{Stephanov:2017ghc} uses the simplified prescription $D_p=\lambda_T/\cp=T/(6\pi\eta\xi_0)$. 

A final difference arises from the fact that Ref.~\cite{Akamatsu:2018vjr} ignores advection by setting the fluid velocity to zero while we and Ref.~\cite{Rajagopal:2019xwg} include advection effects arising from the fluid's expansion in the dynamics of the slow modes. The competition between growth of critical fluctuations near the critical point and advection from hydrodynamic expansion of the fluid will be studied in Sec.~\ref{sec:results}.

%
\subsection{Partial-equilibrium equation of state}
\label{sec:quasieos}
%

We now discuss the influence (``back-reaction") of the critical slow modes, whose evolution was discussed in the two preceding subsections, on the  background fluid, through the equation of state (EoS). The slow modes carry energy and entropy and thus contribute to the pressure of the system by adding to the thermal equilibrium values of these quantities in the background fluid. The resulting  ``partial-equilibrium EoS" or ``quasi-equilibrium EoS" is a central ingredient of {\sc hydro+} \cite{Stephanov:2017ghc}.

The slow modes add new, non-equilibrated degrees of freedom to the quantum states of the system. Denoting the complete-equilibrium entropy density of the d.o.f. describing the hydrodynamic background fluid by $s(e, n)$ and additional entropy density contributed by the additional non-equilibrated d.o.f. as $\D s$, the partial-equilibrium entropy density can be written as
\begin{equation}
  s_{\plus}(e, n, \phi)\equiv s(e,n)+\D s(e, n, \phi)=\log\Omega(e, n, \phi)\,,
\end{equation}
where $\Omega(e, n, \phi)$ is the number of quantum states of the system with $(e, n, \phi)$.\footnote{%
    Here $\phi$ is short for the mode spectrum $\phi_Q$.}
$\D s$ is always negative and describes how much entropy the state with non-equilibrium fluctuations or correlations is missing compared to a state in which the fluctuations are completely equilibrated \cite{Stephanov:2017ghc}. When $\phi$ relaxes to its equilibrium value $\bar\phi(e, n)$, the entropy $s_{\plus}$ should also approach its maximum value $s(e, n)$:
\begin{equation}
  \textrm{max}\; s_{\plus}(e, n, \phi) = s_{\plus}(e, n, \bar\phi) = s(e, n)\,.
\end{equation}
%
%
%
In principle the equilibrium entropy $s(e, n)$ should include the thermodynamic behavior near the critical point (see Sec.~\ref{subsec-coneos-cp}). In past work, however, which mostly ignored the back-reaction of the off-equilibrium fluctuations on the partial-equilibrium EoS, the critical scaling properties near the critical point 
were also ignored in the complete-equilibrium EoS.

The slow-mode contribution to the  partial-equilibrium entropy density\footnote{%
    In natural units $\hbar{\,=\,}c{\,=\,}1$, $\D s$ has units of [fm$^{-3}$] while $\D s_Q$ has units of [fm$^{-2}$].}
is given explicitly by \cite{Stephanov:2017ghc}
\begin{eqnarray}
\D s(e,n,\phi) \equiv \int dQ\, \Delta s_Q  = \int dQ \frac{Q^2}{(2\pi)^2} \left[\log\frac{\phi_Q}{\bar\phi_Q(e, n)}-\frac{\phi_Q}{\bar\phi_Q(e, n)}+1\right],
\label{eq:deltas}
\end{eqnarray}
where the local equilibrium value $\bar\phi_Q$ of the slow mode $\phi_Q$ is determined by the local values of $e$ and $n$ (see Eq. (\ref{eq:eqPhiQ_full})) and we used local isotropy to simplify the integration measure (the factor $\frac{1}{2}$ arises from $\phi$ being related to the width of the fluctuations \cite{Stephanov:2017ghc})
\begin{equation}
    \frac{1}{2}\int \frac{d^3Q}{(2\pi)^3} = \frac{1}{2}\int \frac{4\pi Q^2 d Q}{(2\pi)^3} = \int \frac{Q^2 d Q}{(2\pi)^2}.
\end{equation}
Eqs.~(\ref{eq:phi_scaling}) and (\ref{eq:deltas}) show that (for purpose of calculating the back-reaction) the normalization of $\bar\phi_Q$ is irrelevant since only the ratio $\phi_Q/\bar\phi_Q$ appears. The normalization of $\bar\phi_Q$ controls the magnitude of the contribution of the critical fluctuations to any fluctuation observable, but we shall not compute such observables here. The function in square brackets, $\log(x)-x+1$, and hence the non-equilibrium entropy correction (\ref{eq:deltas}) is negative semi-definite. In the derivation of Eq.~\eqref{eq:deltas}, $|\Delta s|$ is assumed to be much smaller than $s$. Therefore, $s_{(+)}$ remains positive definite within the domain of the applicability of Eq.~\eqref{eq:deltas}. Note also that in deriving this expression \cite{Stephanov:2017ghc} the separation of scales $\ell^{-1}\ll Q$ was used and only the contribution of the slowest mode to $\Delta s$ was included.

The off-equilibrium contribution from the slow mode to the entropy density $s_{(+)}$ modifies the pressure to $p_{(+)}$, given by the generalized thermodynamic relation \cite{Stephanov:2017ghc}
\begin{equation}
\label{eq:splus}
    s_{\plus} = \beta_\plus p_\plus+\beta_\plus e - \alpha_\plus n\,,
\end{equation}
with modified inverse temperature and chemical potential defined by
\begin{eqnarray}
\beta_{\plus} = \left(\frac{\partial s_\plus}{\partial e}\right)_{\!\!n\phi} \equiv \frac{1}{T} + \D \beta\,,
\quad
\alpha_{\plus} = - \left(\frac{\partial s_\plus}{\partial n}\right)_{\!\!e\phi} \equiv \frac{\mu}{T} + \D \alpha\,,
\label{eq:alphabeta}
\end{eqnarray}
where the corrections are
\begin{equation}
  \D \beta \equiv \left(\frac{\partial \D s}{\partial e}\right)_{\!\!n\phi}\,,\quad
\D \alpha \equiv - \left(\frac{\partial \D s}{\partial n}\right)_{\!\!e\phi}\,.\label{eq:dba}
\end{equation}
Solving Eq. (\ref{eq:splus}) for $p_\plus{\,\equiv\,}p+\D p$ (where $s=\beta(e{+}p)-\alpha n$ with $\beta\equiv1/T$ and $\alpha\equiv\mu/T$) one finds
\begin{equation}
    \D p = \bigl[-(e{+}p)\D\beta + n \D\alpha + \D s\bigr]\big/\beta_{\plus}.
\label{eq:dp}
\end{equation}

In this chapter, we focus on the non-equilibrium slow-mode correction $\Delta s$ (\ref{eq:deltas}) to the entropy density as a proxy for estimating the expected size of back-reaction effects on the bulk hydrodynamic evolution. More precisely, such hydrodynamical effects would be driven by the gradients of the modified pressure $p_{(+)} = \bigl(s_{(+)}{\,+\,}\alpha_{(+)}n\bigr)/\beta_{(+)}-e$ (see Eqs.~(\ref{eq:splus},\ref{eq:alphabeta})). We expect the fractional non-equilibrium slow mode corrections to the entropy density and pressure gradients to be of similar orders of magnitude.  Note that $\Delta s$ here is the one we mentioned in Ch.~\ref{ch.gubser} for calculating correction to entropy on a closed hypersurface (see Eq.~\eqref{eq:fods}).

%
\section{Setup of the framework}
\label{sec:frame}
%

We now describe our {\sc hydro+} setup as executed in this chapter. Similar to \cite{Rajagopal:2019xwg} we explore a simplified expansion geometry (boost-invariant longitudinal coupled to azimuthally symmetric transverse expansion), but with an analytic solution for the hydrodynamic expansion of the hydrodynamic background (ideal Gubser flow) rather than the numerical solutions studied in \cite{Rajagopal:2019xwg}. The analytic background flow facilitates the study of the slow-mode evolution, for which we explore a different scenario than in Ref.~\cite{Rajagopal:2019xwg} by moving the critical point away from the temperature axis to a region of non-negligible net baryon density. We thus explore a situation with a different critical scaling behavior than the one studied in \cite{Rajagopal:2019xwg}, and we include finite net baryon density effects in the computation of the contribution to the entropy density caused by off-equilibrium critical fluctuations. Unlike Ref.~\cite{Rajagopal:2019xwg} we here ignore the back-reaction of the slow modes on the dynamical evolution of the background fluid. For the specific expansion geometry studied in \cite{Rajagopal:2019xwg}, the authors found very small back-reaction corrections to the background evolution. In this chapter we confirm that also for the (ideal) Gubser flow studied here the off-equilibrium slow-mode contribution $\D s$ to the entropy density is generically small ($\D s/s{\,\sim\,}\mathcal{O}(10^{-5}{-}10^{-4})$), so we expect similarly small modifications to the background flow caused by them.

%
%

\begin{figure*}[!tp]
    \centering
    \includegraphics[width=0.95 \textwidth]{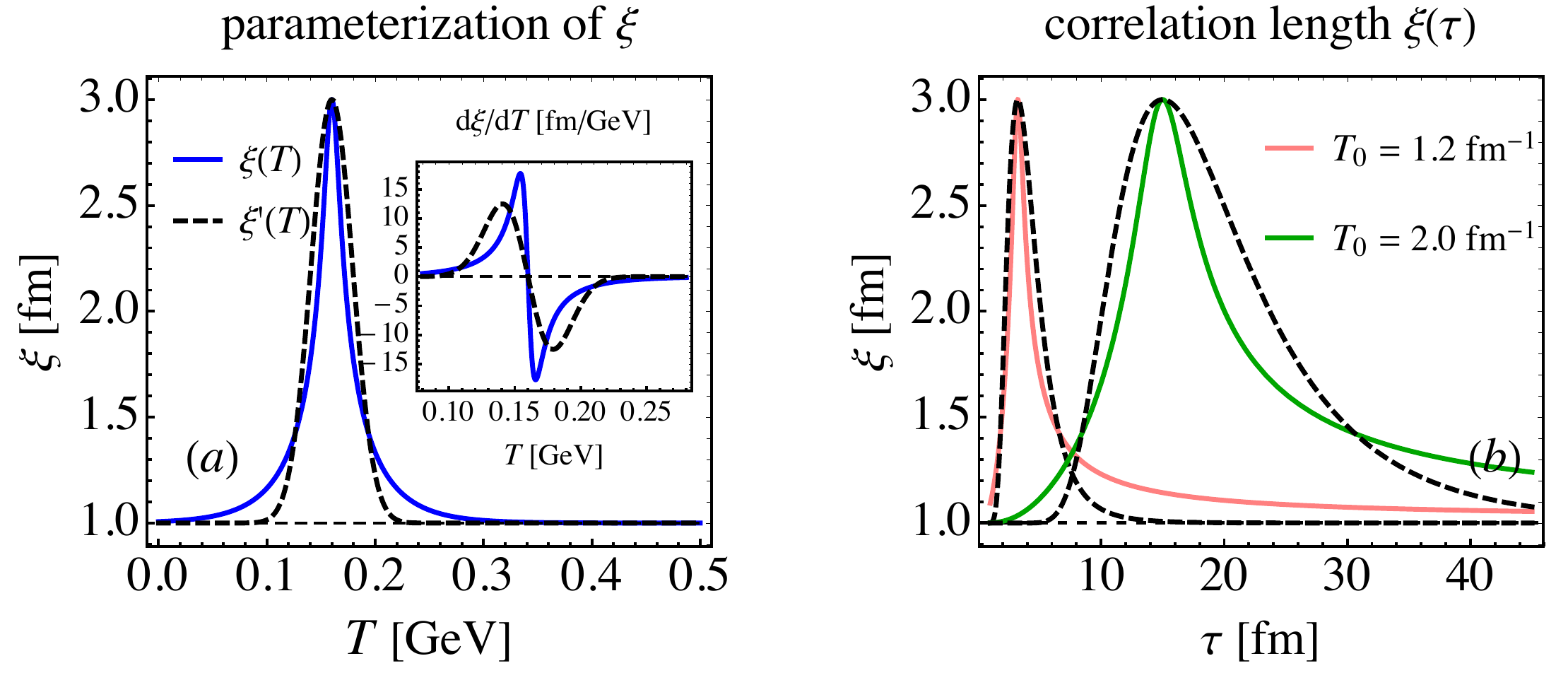}
    \vspace*{-3mm}
    \caption{ 
    (a) Parametrization of the equilibrium correlation length $\xi(T)$ according to Eqs.~(\ref{eq:xipara1}) (solid blue \cite{Rajagopal:2019xwg}) and (\ref{eq:xipara2}) (dashed black). The horizontal dashed line indicates the background correlation length $\xi_0{\,=\,}1$\,fm far away from the critical point. The inset plot shows the derivatives of $\xi(T)$ with respect to $T$. (b) Time evolution of the equilibrium correlation length $\xi\bigl(T(\tau)\bigr)$ for ideal Bjorken flow with $T(\tau) = T_0 (\tau_0/\tau)^{1/3}$ where $\tau_0=1$\,fm, for two choices of $T_0$ as indicated in the legend. Solid and dashed lines refer to the corresponding equilibrium parametrizations in panel (a).}
    \label{fig:param_xi}
    \vspace*{-4mm}
\end{figure*}

We start by supplying the necessary parametrizations of the correlation length $\xi$, heat conductivity $\lambda_T$, and specific heat capacity $\cpz$, which control the relaxation dynamics of the critical slow modes. At finite net baryon density one should expect the equilibrium correlation length to be a function of both $e$ and $n$, $\xi(e,n)$. For our exploratory study we take it as a function of temperature only (i.e. we neglect its dependence on the baryon chemical potential $\mu$).\footnote{%
    Note that in the context of the present study this does not imply that we cannot simulate the effects of a critical {\it point}, situated at a unique location $(T_c,\mu_c)$ in the temperature-chemical potential plane: the specific background flow assumed in this chapter (ideal Gubser flow) evolves at constant $\mu/T$, and we will simply assume that the critical point lies on that trajectory, i.e. $\mu/T=\mu_c/T_c$.}   
We explore two parametrizations: first, we consider the form used in \cite{Rajagopal:2019xwg},
\begin{equation}
    \xi(T) = \frac{\xi_0}{\Bigl[\tanh^2\left(\frac{T-T_c}{\Delta T}\right)\Bigl(1-\left(\frac{\xi_0}{\xi_\mathrm{max}}\right)^4\Bigr)+\left(\frac{\xi_0}{\xi_\mathrm{max}}\right)^4\Bigr]^{1/4}}\,,
\label{eq:xipara1}
\end{equation}
with parameters
\begin{equation}
    T_c = 160\,\textrm{MeV},\ 
    \Delta T = 0.4\,T_c,\ 
    \xi_0 = 1\,\textrm{fm},\ 
    \xi_\mathrm{max} = 3\,\textrm{fm}.
\end{equation}
Without the cutoff $\xi_\mathrm{max}{\,\gg\,}\xi_0$, this parametrization gives $\xi\propto|T-T_c|^{-1/2}$ for small $|T-T_c|$ as motivated by mean field theory \cite{Rajagopal:2019xwg}. We here cut off this singular growth at $\xi_\mathrm{max}=3$\,fm but in such a way that the first derivative of $\xi(T)$ still changes very rapidly near $T=T_c$. (see inset in Fig.~\ref{fig:param_xi}a). Note that our $\Delta T$ is twice that used in Ref.~\cite{Rajagopal:2019xwg}, to avoid the special treatment found necessary in \cite{Rajagopal:2019xwg} to deal with very sharp peaks in the correlation length. While general arguments say that the equilibrium value of $\xi$ should diverge at $T_c$, in an expanding system the actual correlation length will alaways remain finite due to critical slowing down \cite{Berdnikov:1999ph}. Regulating the critical divergence of the equilibrium correlation length at $T_c$ as done in Eq.~(\ref{eq:xipara1}) should therefore not make much of a difference in practice. However, Eq.~(\ref{eq:xipara1}) leads to large temporal gradients $\partial_t \bar\phi_Q/ \bar\phi_Q$, resulting from large derivatives $d\xi/dT$, near $T=T_c$.  
%
%

To explore their importance for the dynamics we also study a second parametrization $\xi'(T)$ which does not share this feature,
\begin{equation}
    \xi'(T) = \xi_0+\Delta\xi\exp\left[-\frac{\left(T-T_c\right)^2}{2\sigma_T^2}\right]\,,
\label{eq:xipara2}
\end{equation}
with $\sigma_T=0.3\,\Delta T=0.12\,T_c$ and   $\Delta\xi=\xi_\mathrm{max}{\,-\,}\xi_0$, whose temperature derivative $d\xi'/dT$ now changes much more slowly near $T_c$ (see inset in Fig.~\ref{fig:param_xi}a). The two parametrizations (\ref{eq:xipara1},\ref{eq:xipara2}) are compared in  Fig.~\ref{fig:param_xi}a; one sees that for the given choice of the Gaussian width $\sigma_T$ Eqs.~(\ref{eq:xipara1}) and (\ref{eq:xipara2}) have very similar overall shapes, but Eq.~(\ref{eq:xipara2}) avoids the sharpness of the peak at $T_c$. Fig.~\ref{fig:param_xi}b shows the resulting evolution of the correlation length $\xi(\tau)$ for the two parametrizations shown in panel (a), for a fluid undergoing Bjorken expansion with two different initial temperature values $T_0$ at time $\tau_0=1$\,fm as shown in the figure. While qualitatively very similar, parametrization (\ref{eq:xipara2}) leads to smoother time dependence as the systems pass through $T_c$, but also to a faster return to the background value $\xi_0$ as the system moves away from the critical point. We will study the evolution of the slow mode with these two parametrizations in Sec.~\ref{sec:expansion}.

The non-critical heat capacity $\cpz\, [\textrm{fm}^{-3}]$ and heat conductivity $\lambda_T\, [\textrm{fm}^{-2}]$ are parametrized as follows \cite{Akamatsu:2018vjr}:
\begin{equation}
    \lambda_T = C_{\lambda} T^2\,,\quad 
    \cpz = \frac{s^2}{\alpha n}\,.
\label{eq:param}
\end{equation}
Here $C_{\lambda}$ is a unitless free parameter and $\alpha=\mu/T$. We choose $C_{\lambda}$ such that the relaxation rate $\Gamma_Q \sim {\cal O}$(1/fm) \cite{Rajagopal:2019xwg}. In principle $\cpz$ should be derived from the EoS of the medium using $\cpz=nT\bigl[\partial(s/n)/\partial T\bigr]_p$. However, in the conformally symmetric background we will be using (the ideal Gubser profiles from Ch.~\ref{ch.gubser}), any parametrization with the correct units should give the same (Gubser- or Milne-) time dependence, up to an overall normalization,\footnote{%
    This holds only as long as back-reaction is neglected because the slow-mode dynamics breaks conformal symmetry.}
and for not too large values of $\alpha$ the parametrization (\ref{eq:param}) of $\cpz$ is numerically quite accurate for the ideal massless gas EOS3 in Sec.~\ref{sec:gubser_para} used in this chapter.

We shall ignore the back-reaction from the slow modes to the conformal background fluid, we can solve Eq.~(\ref{eq:phi_scaling}) in Milne coordinates (which are also the coordinates in which \bes+ is formulated), using the Gubser profile (Ch.~\ref{ch.gubser}) for the Milne components of $u^\mu$ as externally prescribed. We shall use the phenomenologically motivated parameters in Sec.~\ref{sec:gubser_para}, where we got $\alpha=\mu/T\approx0.145$. This value of $\alpha$ is not large, implying that at this collision energy the system will not pass close to the critical point if the latter is at $\mu_c\gtrsim 400$\,MeV \cite{Bazavov:2017dus,Mukherjee:2019eou}. The large-$\alpha$ regime is studied at the lower end of the range of collision energies explored in the RHIC BES program \cite{Adamczyk:2017iwn}. However, since our intent is not to do BES phenomenology but to explore the mechanisms that drive critical fluctuation dynamics, we select $\snn = 200$\,GeV which facilitates comparison with previous work \cite{Gubser:2010ze,Hatta:2015era}. Purely for this convenience we are therefore imagining (similar to Ref.~\cite{Rajagopal:2019xwg}) a critical point at a small value of $\mu_c/T_c$; however, where Ref.~\cite{Rajagopal:2019xwg} focused on the influence of such a critical point on dynamics at $\mu=0$, we here explore its influence on fluctuation dynamics for a system passing close to the critical point on a trajectory with non-zero baryon chemical potential, $\mu\ne0$.

Before moving to numerical studies let us  quickly estimate the expected slow-mode contribution to the partial-equilibrium entropy density with the setup described in this section. As argued, the typical $Q$ contributing to $\Delta s$ in Eq.~(\ref{eq:deltas}) is of the order $Q_\mathrm{neq}$. Therefore $\Delta s \simeq \frac{1}{3}\frac{1}{(2\pi)^2}\,Q^{3}_\mathrm{neq}$ or, as a fraction of the equilibrium entropy density $s$ from Eq.~(\ref{eq:eos2}), %
\begin{equation}
\label{eq:estdeltas}
    \frac{\Delta s}{s} \sim \frac{1}{3}\frac{1}{(2\pi)^2} \frac{1}{h_{*}} \left(\frac{Q_\mathrm{neq}}{T}\right)^3. 
\end{equation}
Using $h_{*}$ from Eq.~(\ref{eq:estfgh}) and  $(Q_\mathrm{neq}/T) \lesssim 1$ we arrived at $(\Delta s/s) \sim {\cal O}(10^{-4})$.
%

%
\section{Results and discussion}\label{sec:results}
%

We now exploit this framework to study the dynamics of the slow modes near the QCD critical point. Here, the availability of analytic expressions for the ideal Gubser flow hydrodynamic background turns out to be helpful: by taking different limits of the background flow, we can easily separate critical dynamics from flow-induced effects. For example, we can turn off transverse flow by taking the limit $q\to0$, corresponding to an infinite transverse radius of the fireball, and we can turn off critical effects by replacing the correlation length $\xi$ by a constant $\xi_0{\,=\,}1$\,fm, the assumed correlation length far away from the critical point. We focus on collective expansion effects on slow-mode evolution in Sec.~\ref{sec:expansion}, on advection effects in Sec.~\ref{sec:advection}, and on critical dynamics due to the growth of the correlation length near the critical point in Sec.~\ref{sec:corrlen}. In the last subsection \ref{sec:phenomenology} we give the reader a feeling for the expected magnitude of phenomenological effects resulting from non-equilibrium slow mode dynamics, by studying the space-time evolution of non-equilibrium corrections to the entropy of the fireball and their imprint on the freeze-out hypersurface.

To simplify the discussion let us introduce some notation. Throughout this section we use a background medium whose hydrodynamic evolution starts at $\tau_0{\,=\,}1$\,fm, with initial velocity and temperature profiles (\ref{eq-gubser-ut})-(\ref{eq-gubser-uphi}) and (\ref{eq-gubser_temp}). We introduce the shorthand $\Phi{\,\equiv\,}\cpz/n^2{\,=\,}\bar{\phi}_{0,\xi_0}$ for the equilibrium value of the static ($Q{\,=\,}0$) slow mode (which depends on space-time position $x$ through the medium properties) and denote by $\Phi_0{\,\equiv\,}\Phi(\tau_0,r)$ its initial $r$-profile at $\tau_0$. We further introduce  $\Gamma{\,\equiv\,}\Gamma_{\xi_0}{\,=\,}2\Bigl(\lambda_T/(\cpz\xi_0^2)\Bigr)$ ({\it cf.} Eq.~(\ref{eq:Gamma_xi})) with $\xi_0{=}1$\,fm and denote by $\Gamma_0\equiv \Gamma_{\xi_0}(\tau_0,r)$ its initial profile at $\tau_0$. To simplify the discussion of critical effects induced by the growth of the correlation length $\xi$ near the critical point, we introduce the ``non-critical reference value" $\bar\phi_{Q,\xi_0}{\,\equiv\,}(\cpz/n^2)/(1+(Q\xi_0)^2)$, i.e. the equilibrium value for the mode with wave number $Q$ in a system with constant correlation length $\xi_0$. When focusing on effects from critical dynamics we therefore plot ratios such as 
\begin{equation}
  \frac{\bar\phi_Q}{\bar\phi_{Q,\xi_0}} = \left(\frac{\xi}{\xi_0}\right)^2 \frac{f_2(Q\xi)}{f_2(Q\xi_0)} = \left(\frac{\xi}{\xi_0}\right)^2\left(\frac{1+(Q\xi_0)^2}{1+(Q\xi)^2}\right)\,.
\end{equation}
When simultaneously looking at the $Q$-dependence we plot
\begin{equation}
  \frac{\bar\phi_Q}{\bar\phi_{0,\xi_0}} = \left(\frac{\xi}{\xi_0}\right)^2 f_2(Q\xi)  =\left(\frac{\xi}{\xi_0}\right)^2
   \left(\frac{1}{1+(Q\xi)^2}\right)\,.
\end{equation}
To facilitate comparison we follow Ref.~\cite{Rajagopal:2019xwg} and initialize the slow modes $\phi_Q$ at their equilibrium values $\bar\phi_Q$.

\subsection{Medium expansion}
\label{sec:expansion}
\subsubsection{Constant correlation length}
\label{constant_xi}
%

In this subsection we focus on effects on the fluctuation dynamics arising from the space-time dependence of the hydrodynamic fields caused by the expansion of the background fluid. We do so by tracing the non-equilibrium evolution of the slow mode $\phi_Q$, as well as its equilibrium value $\bar\phi_Q$, in a noncritical expanding medium with constant correlation length $\xi=\xi_0$. To focus on the dilution and cooling effects caused by the expansion, rather than the collective flow that accompanies it, we remove the spatial gradients in the medium (i.e. advection affects), by letting $q\to 0$. According to Eqs.~(\ref{eq-gubser-ut})-(\ref{eq-gubser-uphi}) this results in $u^\tau \approx 1$ and $u^r \approx 0$, i.e. 1-dimensional Bjorken expansion along the longitudinal direction with the temperature profile
(see Eq.~(\ref{eq-gubser_temp}))
\begin{equation}
    T \approx \frac{C(2q)^{2/3}}{\tau^{1/3}} \equiv T_0\left(\frac{\tau_0}{\tau}\right)^{1/3}\,,
\label{eq:bjorken_temp}
\end{equation}
where $C{\,\propto\,}q^{-2/3}{\,\to\,}\infty$ as $q{\,\to\,}0$ so that $T_0\equiv C(4q^2/\tau_0)^{1/3}$ remains finite. With our conformal EoS this implies the well-known Bjorken scaling laws
\begin{equation}
    n\propto \frac{1}{\tau}\,,\quad 
    s\propto \frac{1}{\tau}\,,\quad 
    e\propto \frac{1}{\tau^{4/3}}
\end{equation}
as $q\to 0$, and the parametrization (\ref{eq:param}) gives
\begin{equation}
  \bar\phi_Q \propto \frac{\cpz}{n^2} \propto \frac{(s^2/n)}{n^2} \propto \tau\,,\quad
  \Gamma_Q \propto \frac{\lambda_T}{\cpz} \propto \frac{T^2}{(s^2/n)} \propto \tau^{1/3}\,.
\end{equation}
One sees that in this geometry, and without critical correlations, both $\Gamma_Q$ and $\bar\phi_Q$ increase monotonically with $\tau$ as the system expands:
\begin{eqnarray}
    {\bar\phi}_Q(\tau) &=& \frac{\tau}{\tau_0}\,{\bar\phi}_Q(\tau_0) = \frac{\tau}{\tau_0}\,\Phi_0\, f_2(Q\xi_0)\,,
\label{eq:bjorken_phi_tau}
\\
    \Gamma_Q(\tau) &=& \left(\frac{\tau}{\tau_0}\right)^{1/3} \!\! \Gamma_Q(\tau_0)
    = \left(\frac{\tau}{\tau_0}\right)^{1/3}\!\!\Gamma_0\, 
    f_\Gamma(Q\xi_0)\,.\quad
\label{eq:bjorken_gamma_tau}
\end{eqnarray}
With this Bjorken flow profile the equation of motion for $\phi_Q$ turns into an ODE,
\begin{equation}
    \partial_\tau \phi_Q(\tau) = - \Gamma_Q(\tau) \Bigl(\phi_Q(\tau) - \bar\phi_Q(\tau)\Bigr)\,,
\label{eq:bjorken_eom}
\end{equation}
which we solve numerically. As the normalization of $\phi_Q$ is arbitrary, and the heat conductivity $\lambda_T$ contains a free parameter $C_\lambda$, we simply set $\Phi_0{\,=\,}1$\,fm$^3$ and $\Gamma_0{\,=\,}1$\,fm$^{-1}$ for the initial conditions in Eqs.~(\ref{eq:bjorken_phi_tau}) and (\ref{eq:bjorken_gamma_tau}) \cite{Rajagopal:2019xwg}.

%
\begin{figure*}[!tp]
    \centering
    \includegraphics[width= \textwidth]{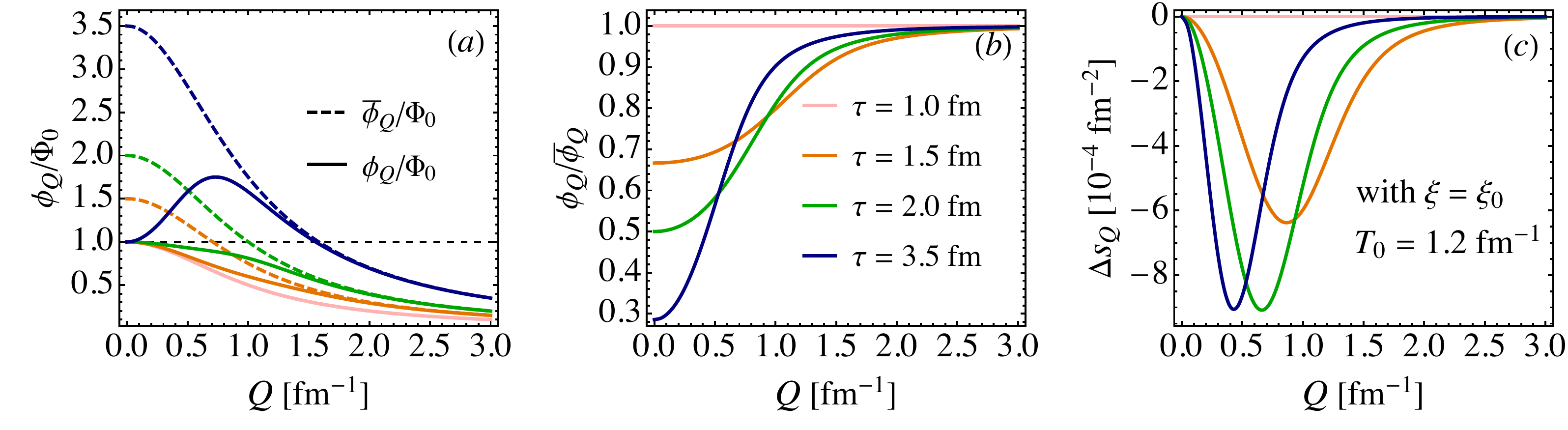}
    \caption{$Q$-dependence of the dynamics of the slow modes in a system undergoing Bjorken expansion (i.e. Gubser expansion in the limit $q\to0$) with constant $\xi=\xi_0$. Lines with different colors correspond to different times as identified in the legend. (a) Comparison of the $Q$-dependences of the scaled equilibrium ($\bar\phi_Q/\Phi_0$, dashed) and non-equilibrium ($\phi_Q/\Phi_0$, solid) values of the correlator $\phi_Q$, as functions of wave number $Q$. (b) $Q$-dependence of the non-equilibrium/equilibrium ratio $\phi_Q/\bar\phi_Q$. (c) $Q$-dependence of the non-equilibrium entropy density correction arising from the slow mode with wave number $Q$, $\Delta s_Q$. Initial conditions are $T_0{\,=\,}1.2$\,\fm, $\Gamma_0{\,=\,}1.0$\,\fm, and $\Phi_0{\,=\,}1.0$\,fm$^{-3}$ at $\tau_0{\,=\,}1$\,fm.
    \label{fig:bjorken_q}}
\end{figure*}
%

Figure~\ref{fig:bjorken_q} shows several snapshots of the $Q$-dependence of $\phi_Q$ (panels (a,b)) and of the corresponding non-equilibrium entropy correction $\Delta s_Q$ (panel (c)) that illustrate their time evolution. Panel (a) shows that the equilibrium value $\bar\phi_Q$ (dashed lines) increases with time but decreases with growing wave number $Q$, reflecting the scaling function $f_2(Q\xi_0)$ (\ref{eq:bjorken_phi_tau}). The solid lines showing the non-equilibrium value $\phi_Q$ exhibit an interesting feature: while they approach their corresponding equilibrium values at large $Q$, they stay close to their initial value $\Phi_0\, f_2(Q\xi_0)$ at small wave number. This reflects the $Q$ dependence of the relaxation rate, $\Gamma_Q\propto Q^2$ for $Q\ll\xi^{-1}$ and $\Gamma_Q\propto Q^4$ for $Q\gg\xi^{-1}$. This feature sets a scale $Q_\textrm{neq}(\tau)$ which decreases with time that separates modes that can equilibrate within time $\tau$ from those that cannot. For small $Q$, $\bar\phi_Q$ increases with time but $\phi_Q$ remains basically frozen,  causing the ratio $\phi_Q/\bar\phi_Q$ to decrease with time, as shown in Fig.~\ref{fig:bjorken_q}b. On the other hand the same panel also shows that, since the relaxation rate $\Gamma_Q$ increases and the expansion rate $\theta$ decreases with time, at sufficiently late times even the low-$Q$ modes equilibrate (i.e. $Q_\textrm{neq}$ decreases with times). As $Q$ decreases from large to smaller values, the decreasing ratio $\phi_Q/\bar\phi_Q$ implies a larger (negative) contribution $\Delta s_Q$ to the entropy density, but as $Q$ approaches zero these are cut off by the phase space factor $(Q/2\pi)^2$ (see Eq.~(\ref{eq:deltas})). The largest contribution to $|\Delta s|$ thus arises from modes with $Q{\,\sim\,}Q_\mathrm{max}{\,\sim\,}\mathcal{O}(Q_\textrm{neq})$ (which decreases with time), as shown in Fig.~\ref{fig:bjorken_q}c (see also Ref. \cite{Rajagopal:2019xwg}).  

\begin{figure}[!tp]
    \centering
    \includegraphics[width= 0.4\linewidth]{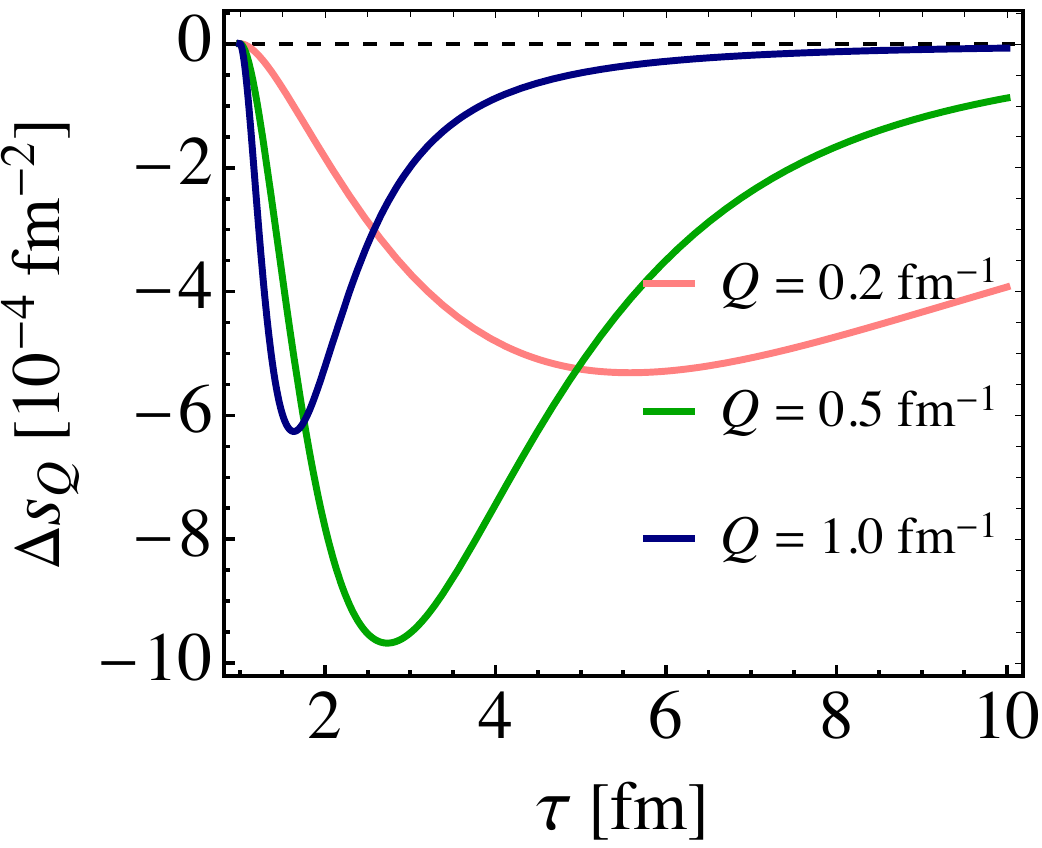}
    \caption{Time evolution of the correction $\Delta s_Q$ to the entropy density for three wave numbers $Q=0.2,\,0.5,\,1.0$\,\fm, for the same setup as in Fig.~\ref{fig:bjorken_q}.
    \label{fig:bjorken_stau}}
\end{figure}

Another way to illustrate this is shown in Fig.~\ref{fig:bjorken_stau} where we plot the time evolution of $\Delta s_Q$ for three typical $Q$ values: $Q{\,=\,}0.5$\,\fm (which is close to $Q_\textrm{max}$) 
as well as $Q{\,=\,}0.2$ and 1.0\,\fm (which are below and above $Q_\textrm{max}$). As time evolves, the modes first drop out of equilibrium (and thus begin to contribute to $\Delta s$) but later re-equilibrate (with their contribution to $\Delta s$ peaking and later decreasing). We note that the entropy contributed by the non-equilibrium slow modes, both for a fixed wave number $Q$ and integrated over $Q$, initially decreases as the slow modes are driven out of equilibrium by the large initial longitudinal expansion rate. As already explained, the negative sign of $\Delta s$ simply reflects the fact that, as long as the slow modes are out of equilibrium, the system has not yet reached a state of maximum entropy. For large $Q$, $|\Delta s_Q|$ peaks early and ceases rather quickly, and for small $Q$ the entropy contribution peaks and decays later. At first sight it looks as if the time integral of $\Delta s_Q$ might diverge as $Q\to0$, but we checked numerically that infrared convergence is ensured by the phase-space factor $(Q/2\pi)^2$.  

\vspace*{-3mm}
\subsubsection{Critical correlations in an expanding medium}
\label{critical_xi}
\vspace*{-2mm}

\begin{figure*}[!tb]
    \centering
    \includegraphics[width= \textwidth]{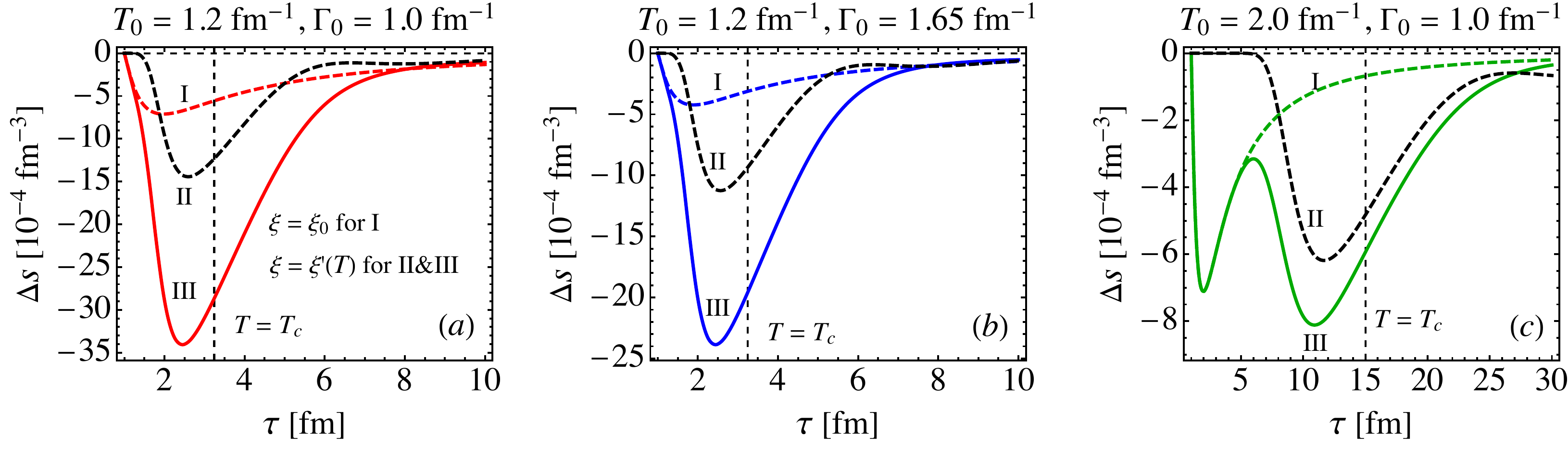}
    \caption{Comparison of the time evolution of the $Q$-integrated non-equilibrium entropy density correction $\Delta s$ for different dynamic models. I: constant $\xi = \xi_0$ (colored dashed lines); II: 
    $\xi=\xi'(T)$ (Eq.~(\ref{eq:xipara2})) with time-independent $\Phi=\Phi_0$ and $\Gamma=\Gamma_0$ (black dashed lines); III: full dynamics with evolving $\Phi$, $\Gamma$ and $\xi=\xi'(T)$ (colored solid lines). The three panels correspond to different initial conditions: (a) $T_0 = 1.2$\,\fm\ and $\Gamma_0=1.0$\,\fm; (b) $T_0 = 1.2$\,\fm\ and $\Gamma_0=1.65$\,\fm; (c) $T_0 = 2.0$\,\fm\ and $\Gamma_0=1.0$\,\fm. The vertical black dashed lines show the time $\tau{\,=\,}3.24$\,fm in (a,b), and $\tau{\,=\,}15$\,fm in (c) when the temperature passes through $T_c$ where $\xi$ peaks.
    \label{fig:bjorken_dynamics_comparison}}
\end{figure*}

\begin{figure*}[!tb]
    \centering
    \includegraphics[width= \textwidth]{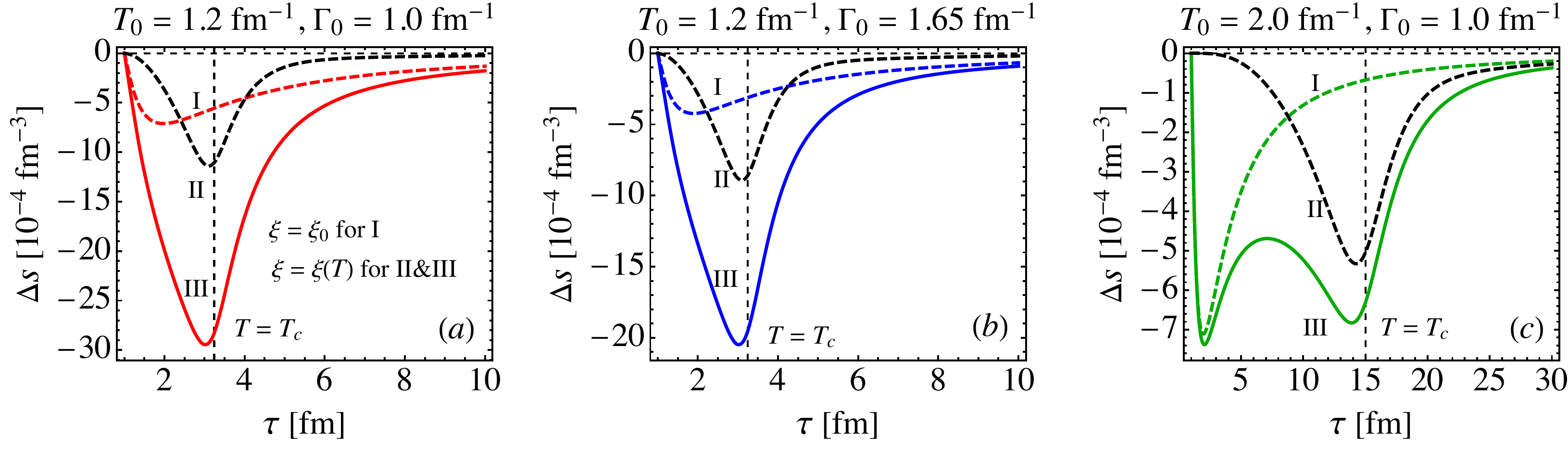}
    \caption{Same as Fig.~\ref{fig:bjorken_dynamics_comparison}, but with the correlation length $\xi(T)$ parametrized by Eq.~(\ref{eq:xipara1}).
    \label{fig:bjorken_dynamics_comparison2}}
\end{figure*}

In this subsection we add critical effects, by generalizing the results from the preceding subsection for Bjorken expansion to include a temperature-dependent correlation length which peaks at a critical temperature $T_c$. We start with the Gaussian parametrization (\ref{eq:xipara2}), showing the corresponding $Q$-integrated non-equilibrium entropy density corrections $\Delta s(\tau)$ in Fig.~\ref{fig:bjorken_dynamics_comparison}, and then compare with the parametrization (\ref{eq:xipara1}) in Fig.~\ref{fig:bjorken_dynamics_comparison2}, to get a feeling for how strongly different parametrizations of $\xi$ might affect the evolution of the critical fluctuations. In each panel we compare three dynamical scenarios for the slow modes (the hydrodynamic background remains always the same): a constant correlation length $\xi{\,=\,}\xi_0$ as in the preceding subsection (I); a correlation length $\xi(T)$ which changes with the time evolving temperature $T$ while the temperature- and resulting time-dependence of $\cpz/n^2$ and $\lambda_T/\cpz$ is ignored, i.e. $\Phi{\,=\,}\Phi_0$ and $\Gamma{\,=\,}\Gamma_0$ are frozen at their initial values \cite{Rajagopal:2019xwg} (II); and (III) a fully dynamical scenario where not only $\xi$, but also the ratios $\cpz/n^2$ and $\lambda_T/\cpz$ (i.e. $\Phi$ and $\Gamma$) change with the evolving temperature. The three panels in each figure correspond to three different initial conditions: a lower initial temperature $T_0{\,=\,}1.2$\,\fm\ in panels (a,b) and a 60\% higher initial $T_0{\,=\,}2$\,\fm\ in panel (c), combined with a somewhat slower relaxation controlled by $\Gamma_0=1.0$\,\fm\ in panels (a,c) and a faster relaxation $\Gamma_0=1.65$\,\fm\ in panel (b). 

%

Comparison of the colored dashed and solid lines for scenarios I and III shows that at times corresponding to temperatures far above or below $T_c$ (where the time corresponding to $T=T_c$ is identified by a thin vertical black dashed line (see Fig. \ref{fig:param_xi}b)) the non-equilibrium entropy density corrections agree --- they differ only around $T_c$ where the correlation length $\xi$ is critically enhanced, leading to larger entropy corrections $|\Delta s|$. Panel (c) with the higher initial temperature $T_0{\,=\,}2.5\,T_c$ is interesting: without critical slowing down (scenario I) the non-equilibrium entropy density correction peaks early and has largely decayed (i.e. the slow modes have largely equilibrated) by the time the system passes through $T_c$; in scenario III, on the other hand, critical slowing down near $T_c$ allows the slow modes to fall out of equilibrium for a second time (starting when $T\lesssim T_c{\,+\,}\Delta T$), leading to a secondary peak of $|\Delta s|$ near $T_c$. In scenario II both the relaxation rates $\Gamma_Q$ and equilibrium values $\phi_Q$ change only because $\xi$ evolves with temperature, and therefore $|\Delta s|$ closely tracks the evolution of $\xi$, with a single peak near $T_c$ \cite{Rajagopal:2019xwg}. The shift of the $\xi$-driven peaks in $|\Delta s|$ towards temperatures $T>T_c$ is the result of a competition between relaxation towards equilibrium of the slow modes and the rate of expansion of the hydrodynamic medium which drives the slow modes away from equilibrium. Since the expansion rate falls like $1/\tau$, the balance is shifted away from equilibrium at $T>T_c$ (earlier times) and towards equilibrium at $T<T_c$ (later times), giving rise to the observed asymmetry and shift of the peak in $|\Delta s|$. 

Comparison of Figs.~\ref{fig:bjorken_dynamics_comparison} and \ref{fig:bjorken_dynamics_comparison2} shows that this asymmetry and shift is less pronounced for the parametrization (\ref{eq:xipara1}) for $\xi(T)$ which peaks more sharply at $T_c$, making the peak in $|\Delta s|$ less sensitive to the changing hydrodynamic expansion rate between $T\gtrsim T_c$ and $T\lesssim T_c$.\footnote{%
    The sharper peak of $\xi(T)$ causes $d\xi/dT$ to get large near $T_c$, causing disequilibrating expansion effects on the slow mode $\sim \partial_\tau \bar\phi_Q/\bar\phi_Q$ to dominate over equilibrating relaxation dynamics, especially close to $T_c$, thereby causing the peak in 
    $|\Delta s|$ near $T_c$.}$^,$\footnote{%
    Note that the colored dashed lines describing scenario I (with a constant $\xi{\,=\,}\xi_0$) are, of course, identical in Figs.~\ref{fig:bjorken_dynamics_comparison}
    and \ref{fig:bjorken_dynamics_comparison2}.}
At a sufficiently detailed level, the choice of parametrization for $\xi$ is thus seen to have a noticeable effect on the evolution of the off-equilibrium modes, especially close to the critical point. Since Eq.~(\ref{eq:xipara1}) is the more realistic parametrization we will from now on use it as our default.

%
\subsection{Transverse flow effects on slow-mode dynamics}
\label{sec:advection}
%

We now turn our attention to transverse flow effects on slow-mode dynamics. As already mentioned in the introduction, advection by transverse flow can affect the evolution of the non-equilibrium fluctuations by carrying them outward from the middle to the edge of the fireball \cite{Rajagopal:2019xwg}. The analytically known structure of the ideal Gubser solution makes it possible to study this effect semi-analytically, without having to solve a (3+1)-dimensional set of coupled differential equations. Things become particularly simple and clear in the early time regime $\tau \ll 1/q$ \cite{Hatta:2014upa, Hatta:2014jva, Hatta:2015era}. In this limit, the flow velocity can be approximated by
\begin{equation}
\label{eq:flow_approx}
    u^\tau \approx 1+\mathcal{O}(\tau^2)\,,\quad u^r\approx\frac{2q^2\tau r}{1+q^2r^2}\,.
\end{equation}
Since we will ignore the $\mathcal{O}(\tau^2)$ corrections this approximation breaks down when $\tau \sim 1/q$ \cite{Hatta:2014upa, Hatta:2014jva, Hatta:2015era}. The expansion rate is approximated by $\theta \approx \partial_r u^r + 1/\tau + u^r/r$, and the equations of motion become $(\partial_\tau + u^r\partial_r)\phi_Q = - \Gamma_Q (\phi_Q{-}\bar\phi_Q)$, or equivalently,
\begin{equation}
    \partial_\tau \phi_Q = - \Gamma_Q (\phi_Q - \bar\phi_Q) - u^r\partial_r\phi_Q \,.
\label{eq:gubser_smallt_eom}
\end{equation}
The last term, driven by the transverse radial flow $u^r$, modifies the Bjorken dynamics studied in the previous subsection, by contributing with a negative sign to the time derivative of the slow mode $\phi_Q$ if its gradient $\partial_r\phi_Q$ points along the flow direction. The approximate temperature profile is
\begin{equation}
    \!\!\!
    T(\tau,r)\approx\frac{C}{\tau^{1/3}}\frac{(2q)^{2/3}}{(1+q^2r^2)^{2/3}}\equiv T_0\left(\frac{\tau_0}{\tau}\right)^{1/3}
    \!\!\mathcal{F}^{2/3}(r)\,,\ 
\label{eq:temp_smallt}
\end{equation}
with $\mathcal{F}(r)\equiv1/(1+q^2r^2)$ and $T_0\equiv C(4q^2/\tau_0)^{1/3}$. The function $\mathcal{F}(r)$ encodes the $r$-dependence of the temperature profile. From the temperature and the condition $\alpha{\,=\,}\mu/T{\,=\,}$const. the other thermodynamic quantities can be derived using the EoS. Comparing Eqs. (\ref{eq:temp_smallt}) and (\ref{eq:bjorken_temp}) we see that at the early times the $\tau$-dependence of the profile agrees with the one for Bjorken flow while the radial profile is modified by the factor $\mathcal{F}(r)$. This is expected since for $\tau \ll 1/q$ the expansion is dominantly along the longitudinal direction \cite{Gubser:2010ze, Gubser:2010ui}. In this limit, the background fluid can be considered as a superposition of fluid cells undergoing Bjorken expansion with different initial conditions. Still, the evolution equation (\ref{eq:gubser_smallt_eom}) for the critical fluctuations now has the additional source term $- u^r\partial_r\phi_Q$ on its right hand side, and  comparing the following results to those presented in Sec. \ref{sec:expansion} therefore allows us to isolate advection effects caused by transverse expansion flow through this term. To gain intuition about transverse flow effects, we will in this subsection use this picture to late times ($\approx 15$ fm) as a simplified background, even though the approximation breaks down for $\tau > 1/q$, i.e. it no longer represents Gubser flow at such late times.

We simplify the dynamics further by setting $\Phi{\,=\,}\Phi_0$ and $\Gamma{\,=\,}\Gamma_0$, i.e. by ignoring the time dependence of $\cpz/n^2$ and $\lambda_T/\cpz$ \cite{Rajagopal:2019xwg}, in order to focus on flow and suppress effects simply caused by cooling through expansion. The time dependence of $\bar\phi_Q$ and $\Gamma_Q$ will then arise solely from the temperature- and resulting time-dependence of $\xi$:
\begin{eqnarray}
    \bar\phi_Q &=& \Phi_0 \mathcal{F}^{-2}(r)\left(\frac{\xi}{\xi_0}\right)^2f_2(Q\xi)\,,
\label{eq:adv_initial1}\\
    \Gamma_Q &=& \Gamma_0\,\mathcal{F}^{-2/3}(r)\left(\frac{\xi_0}{\xi}\right)^4 f_\Gamma(Q\xi)\,.
\label{eq:adv_initial2}
\end{eqnarray}
Even if $\xi$ were a constant, $\bar\phi_Q$ and $\Gamma_Q$ still depend on $r$ because they depend on the temperature of the medium whose spatial variation is described by the profile $\mathcal{F}(r)$. In the general case these quantities acquire additional $r$-dependence through the $T(\tau,r)$-dependence of $\xi(T)$. Initial conditions for $\bar\phi_Q$ and $\Gamma_Q$ are computed with $\Phi_0{\,=\,}1.0$\,fm$^{-3}$ and $\Gamma_0{\,=\,}0.9$\,\fm, using an initial temperature $T_0{\,=\,}2.2$\,\fm\ at $\tau_0{\,=\,}1$\,fm. The corresponding initial temperatures for cells at transverse positions $r{\,=\,}1.69$ and 5.64\,fm are $T(r{=}1.69\,\mathrm{fm},\tau_0){\,=\,}2.0$\,\fm\ and $T(r{=}5.64\,\mathrm{fm},\tau_0)=1.2$\,\fm, with initial relaxation rates $\Gamma(r{=}1.69\,\mathrm{fm}, \tau_0){\,=\,}1.0$\,\fm\ and $\Gamma(r{=}5.64\,\mathrm{fm}, \tau_0)=1.65$\,\fm. The latter agree with the initial conditions studied in Fig.~\ref{fig:bjorken_dynamics_comparison2}b,c in the preceding subsection, to facilitate comparison.

\begin{figure*}[!tp]
    \centering
    \includegraphics[width=\textwidth]{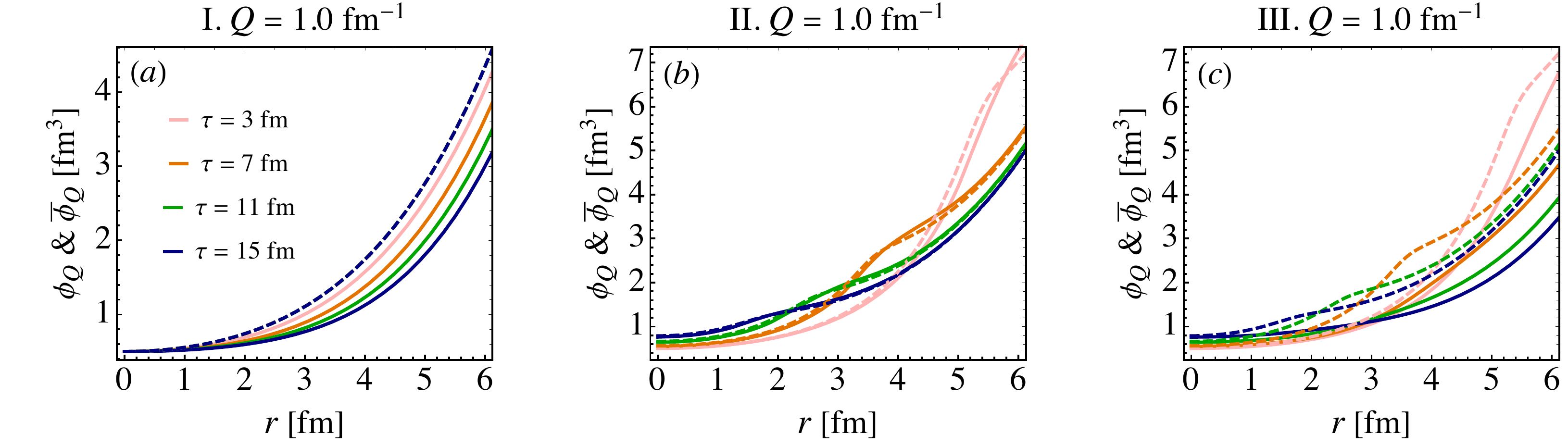}
    \caption{Evolution of the mode with $Q = 1.0$\,fm$^{-1}$ in three dynamic models: (I) with transverse flow $u^r$ and constant $\xi=\xi_0$; (II) without transverse flow (i.e. setting $u^r = 0$) but $\xi=\xi(T)$;  (III) with transverse flow $u^r$ and $\xi=\xi(T)$. Dashed lines show the equilibrium values $\bar\phi_Q$, solid lines the non-equilibrium values $\phi_Q$. Different colors indicate different times, $\tau = 3,\, 7,\, 11,\, 15$\,fm. The initial temperature and relaxation rate at $r=0$ are $T_0=2.2$\,\fm\ and $\Gamma_0=0.9$\,\fm, respectively.
    \label{fig:gubser_phievo_r_0}}
\end{figure*}

We shall see that the additional $r$-dependence introduced by the profile factor $\mathcal{F}(r)$ in Eqs.~(\ref{eq:adv_initial1},\ref{eq:adv_initial2}) results in qualitatively different evolution of the slow modes in this chapter compared to Ref.~\cite{Rajagopal:2019xwg}. In Fig.~\ref{fig:gubser_phievo_r_0} we compare, for a single slow mode with wave number $Q{\,=\,}1$\,\fm, three dynamic models: I (panel a), constant correlation length $\xi{\,=\,}\xi_0$ with non-zero radial flow $u^r$; II (panel b), temperature dependent correlation length $\xi{\,=\,}\xi(T)$ without radial flow, $u^r{\,=\,}0$; and III (panel c), temperature dependent $\xi{\,=\,}\xi(T)$ combined with non-zero radial flow $u^r$.\footnote{%
    Note that we do not change the flow velocity of the background fluid --- we only turn on or off the $u^r$ term in the evolution equation for the slow modes.}
Note that by setting $u^r{\,=\,}0$ in (II), fluid cells at different transverse positions $r$ do not affect each other and, with the profile (\ref{eq:temp_smallt}), evolve independently from each other following Bjorken dynamics; in other words, transverse flow is turned off in scenario (II). For each of these three scenarios, the plots show snapshots at 4 different times of the radial profiles of the equilibrium (dashed lines) and non-equilibrium (solid lines) values of the slow mode, $\bar\phi_{Q}$ and $\phi_Q$, respectively.

%
%

In Fig.~\ref{fig:gubser_phievo_r_0} we first draw the reader's attention to the generic upward-sweeping behavior at large $r$ of both equilibrium and non-equilibrium values of the slow mode. This dependence arises directly from the profile factor $\mathcal{F}(r)$ in Eq.~(\ref{eq:adv_initial1}) and reflects the fact that in the expression $\bar\phi_0=\cpz/n^2$ the square of the baryon density $n^2$ decreases faster with decreasing temperature and hence increasing $r$ than the heat capacity $\cpz$. From Eq.~(\ref{eq:adv_initial2}) it is clear that $\Gamma_Q$ shares with $\bar\phi_Q$ this monotonic rise with $r$, at a somewhat slower rate. Since $\cpz \propto T^3 \propto n$, meaning that $\bar\phi_0\propto 1/n$, the upward rise of this measure of fluctuations at large $r$ corresponds to the increase in fluctuations in a region where there are few particles.  In any future phenomenological analysis, it will contribute little to observables. In our model study, however, we shall see that this $r$-dependence is useful as a device that will enable us to visualize important physical effects.

In scenario I (Fig.~\ref{fig:gubser_phievo_r_0}a) the equilibrium value $\bar\phi_Q$ (dashed line) remains frozen at its initial value because $\xi{\,=\,}\xi_0$ is constant and independent of temperature. However, the radial gradient of $\phi_Q$ couples to the non-zero radial flow $u^r$ and causes $\phi_Q$ to evolve differently at different radial positions $r$. Since the gradient $\partial_r\phi_Q$ points along $u^r$, the time derivative of the slow mode $\phi_Q$ gets a negative contribution in Eq.~(\ref{eq:gubser_smallt_eom}) from $-u^r\partial_r\phi_Q$, and thus $\phi_Q$ is pushed downward (or rather outward) further and further as time progresses. This outward transport of $\phi_Q$ by radial flow is known as ``advection'' \cite{Rajagopal:2019xwg} although (due to the different profile functions $\mathcal{F}$) it manifests itself differently here than in Ref.~\cite{Rajagopal:2019xwg}. As expected, at $r=0$ the transverse flow vanishes, $u^r(0){\,=\,}0$, and the slow mode does not evolve. 

In scenario II without transverse flow (Fig.~\ref{fig:gubser_phievo_r_0}b), cells at different $r$ evolve independently. In this scenario the temperature dependence is included for $\xi(T)$. Since the temperature (\ref{eq:temp_smallt}) is highest at $r{\,=\,}0$, the location of $T_c$ moves inward (due to cooling by longitudinal expansion) as time proceeds. In Figs.~\ref{fig:gubser_phievo_r_0}b,c this is reflected by the leftward movement of the 
``bumps" on the dashed curves which correspond to the location where the temperature of the medium is in the range where $\xi(T)$ has its peak. When plotted in this way, the upward sweep of the curves is more apparent to the eye than the bumps. Recalling, though, that this upward sweep occurs by definition in regions with small $n$ that would contribute little to observable consequences, we plot in the following Fig.~\ref{fig:gubser_phievo_r} a unitless ratio which serves to eliminate this $r$-dependence, making the effects of the critical fluctuations --- which arise in regions with larger $n$ --- more apparent. Figs.~\ref{fig:gubser_phievo_r}b,c are in this sense the better way to visualize the consequences of our analysis, but the dynamics in the equations that govern $\phi$ are more easily understood from Figs.~\ref{fig:gubser_phievo_r_0}b,c so we shall inspect these first. We see that the dashed lines move around with time, as the temperature of the plasma changes in space and time and as the region where $\xi$ peaks moves inward. We then see that in Fig.~\ref{fig:gubser_phievo_r_0}b the solid lines follow the dashed lines and become closer and closer to the dashed lines as time proceeds. This clearly demonstrates relaxation towards equilibrium in the absence of any effects of transverse flow.

%
\begin{figure*}[!tp]
    \centering
    \includegraphics[width=\textwidth]{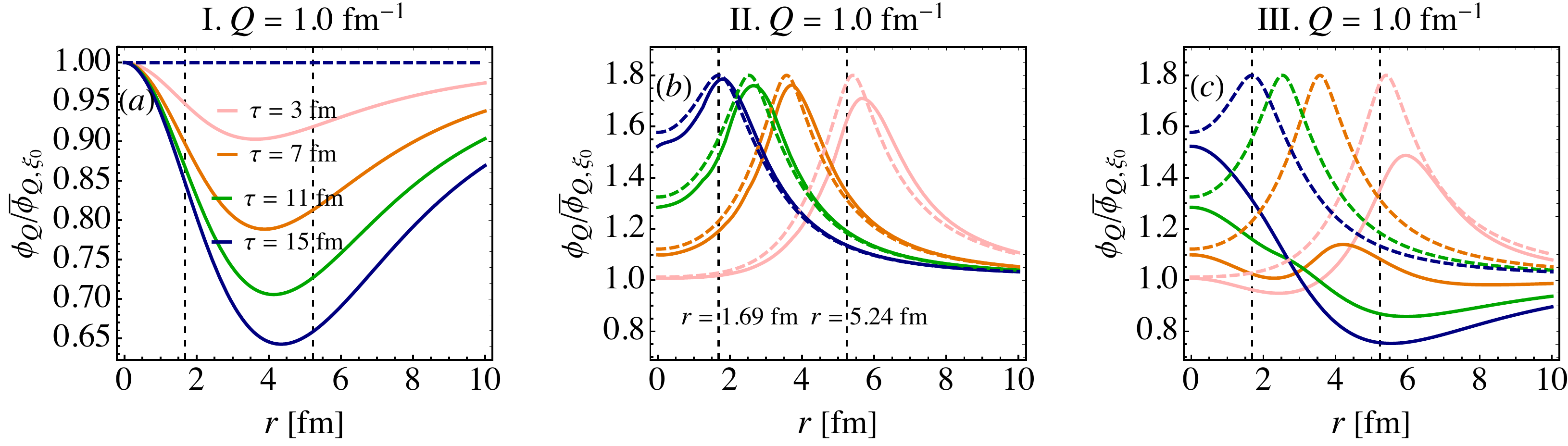}
    \caption{Same as Fig.~\ref{fig:gubser_phievo_r_0}, but using unitless ratios for the vertical axes and an extended range up to $r{\,=\,}10$\,fm for the horizontal axes. Dashed lines show the rescaled equilibrium values $\bar\phi_Q/\bar\phi_{Q, \xi_0}$, solid lines the corresponding non-equilibrium values $\phi_Q/\bar\phi_{Q, \xi_0}$. Here $\bar\phi_{Q, \xi_0}$ denotes $\bar\phi_Q$ for $\xi=\xi_0$ for each respective $r$ at $\tau_0$ (i.e. the dashed line in Fig.~\ref{fig:gubser_phievo_r_0}a). The two vertical dashed lines show the two radial distances $r{\,=\,}5.64$ and 1.69\,fm where the initial temperatures are 1.2 and 2.0 fm$^{-1}$ and the initial relaxation rates are 1.65 and 1.0\,\fm, respectively, corresponding to the cases studied in Fig.~\ref{fig:bjorken_dynamics_comparison2}b,c. 
    \label{fig:gubser_phievo_r}}
\end{figure*}%
%

Scenario III, shown in Fig.~\ref{fig:gubser_phievo_r_0}c, combines the dynamical effects included in scenarios I and II. Transverse flow effects can be uniquely identified by comparing panels b and c: the radial profiles showing snapshots of the non-equilibrium evolution of $\phi_Q$ (solid lines) are pushed outward by advection and thereby away from the corresponding equilibrium profiles $\bar\phi_Q$ (dashed lines) --- dis-equilibration caused by radial flow gradients wins over equilibration by relaxation. 
As we noted above, the growth of $\phi_Q$ towards the dilute periphery of the fireball can be intuitively understood by remembering the increasing relative importance of density fluctuations when the average density gets small. A unitless fluctuation measure that correctly absorbs this trivial Poisson-statistical effect would be the product $n\,\phi_Q$. With our parametrization of $\cpz=s^2/(\alpha n)$, removing the profile function ${\cal F}(r)$ by dividing $\phi_Q$ by the factor $\bar\phi_{Q, \xi_0}$ denoting the initial value at $\tau_0$ at each position $r$ of $\bar\phi_Q$ for fixed $\xi=\xi_0$ (shown as the dashed line in Fig.~\ref{fig:gubser_phievo_r_0}a) achieves the same end. This unitless ratio is shown, for both equilibrium (dashed) and non-equilibrium (solid) slow modes, in Fig.~\ref{fig:gubser_phievo_r}. This way of plotting $\phi_Q$ de-emphasizes the peripheral, low-density regions and brings out more clearly those features that will be phenomenologically relevant in future computations of experimental fluctuation signals.

With this in mind we now discuss Fig.~\ref{fig:gubser_phievo_r}. In scenario I (Fig.~\ref{fig:gubser_phievo_r}a) the normalized equilibrium value $\bar\phi_Q/\bar\phi_{Q,\xi_0}$ (dashed horizontal line) remains frozen at 1. Furthermore, as expected from Eq.~(\ref{eq:gubser_smallt_eom}), the minimum value of the scaled $\phi_Q$ matches the maximum of the transverse flow $u^r$, resulting from a maximum (negative) contribution from the advection term $-u^r\partial_r\phi_Q$ in (\ref{eq:gubser_smallt_eom}).  In scenarios II and III, shown in Figs.~\ref{fig:gubser_phievo_r}b,c, the $r$-dependence from ${\cal F}(r)$ drops out from the normalized equilibrium ratio $\bar\phi_Q/\bar\phi_{Q, \xi_0}$. In consequence, the peaks of the dashed lines in Figs.~\ref{fig:gubser_phievo_r}b,c unambiguously reflect the peak at $T_c$ of the correlation length $\xi(T)$ which enters in the numerator of that ratio. Following the location of the critical point at $T_c$, these peaks in Fig.~\ref{fig:gubser_phievo_r}b move inward as time proceeds and the system cools by expansion.

The difference between panels b and c of Fig.~\ref{fig:gubser_phievo_r} is in the dynamical evolution of the (normalized) non-equilibrium slow mode $\phi_Q/\bar\phi_{Q,\xi_0}$ (solid lines), {\em with} (panel c) and {\em without} (panel b) advection by transverse radial flow. Panel b isolates and nicely illustrates the effects of critical slowing-down: for example, at $r{\,=\,}1.69$\,fm, where $\xi(T)$ increases monotonically from the red (bottom left) to the blue (top left) lines, we see that $\phi_Q$ trails behind the evolution of $\bar\phi_Q$ and always below the equilibrium value; at $r{\,=\,}5.64$\,fm, on the other hand, which for $\tau>3$\,fm sits on the falling side of the $\xi(T)$ curve, we see that $\phi_Q$ first trails below $\bar\phi_Q$ when $\xi$ is still growing but moves above $\bar\phi_Q$ (again trailing behind the equilibrium value) once $T$ drops below $T_c$ and $\xi(T)$ begins to decrease again at this radial position. Generically, the transverse expansion rate is smaller and the slow-mode relaxation rate is larger at large $r$, so that at large $r$ the slow modes relax faster to their equilibrium value than near the center; this, too, is clearly visible in this panel.

In scenario III (Fig.~\ref{fig:gubser_phievo_r}c), we see that when transverse flow is included, the non-equilibrium value $\phi_Q$ falls behind $\bar\phi_Q$ further and further as time proceeds, except at $r=0$ where $u^r=0$ and thus the evolution is the same as in Fig.~\ref{fig:gubser_phievo_r}b. At larger $r\gtrsim 8$ fm, on the other hand, the advection effects seen and discussed in Fig.~\ref{fig:gubser_phievo_r_0} cause the $r$-dependence of the normalized ratio $\phi_Q/\bar\phi_{Q,\xi_0}$ in panel c to develop similarities with what is seen in panel a, especially at late times when the system has passed the critical region and $\xi$ goes back to $\xi_0$. In the intermediate region, $r\approx4$\,fm where the flow is the largest, the $r$-dependence of the solid lines is complicated by the fact relaxation effects are gradually being overshadowed by the transverse flow and advection effects which grow with time.

Note that, unlike Ref.~\cite{Rajagopal:2019xwg}, we do not see clear signs of a second peak in Fig.~\ref{fig:gubser_phievo_r}c, i.e. the critical fluctuation peak is not transported outward to larger $r$ by the transverse flow. The main model feature responsible for this difference is the profile function $\mathcal{F}(r)$ in our expression (\ref{eq:adv_initial2}) for the relaxation rate $\Gamma_Q$ of the slow mode which accelerates relaxation of $\phi_Q$ at large $r$. We confirmed that when the $r$ dependence is removed in Eqs.~(\ref{eq:adv_initial1},\ref{eq:adv_initial2}) we observe a second outgoing peak as shown in Fig.~7 of Ref.~\cite{Rajagopal:2019xwg}. We note that, even if such an outward-moving (advected) second peak in $\phi_Q$ were to show up in our model, it would be dwarfed at large $r$ by the upward-sweeping noncritical fluctuations and would become essentially invisible after normalization with $\bar\phi_{Q,\xi_0}$ as done in Fig.~\ref{fig:gubser_phievo_r_0}. Furthermore, in our model the profile function ${\cal F}(r)$ increases the relaxation rate in the dilute periphery.
 
\begin{figure*}[!t]
    \centering
    \includegraphics[width=0.8\textwidth]{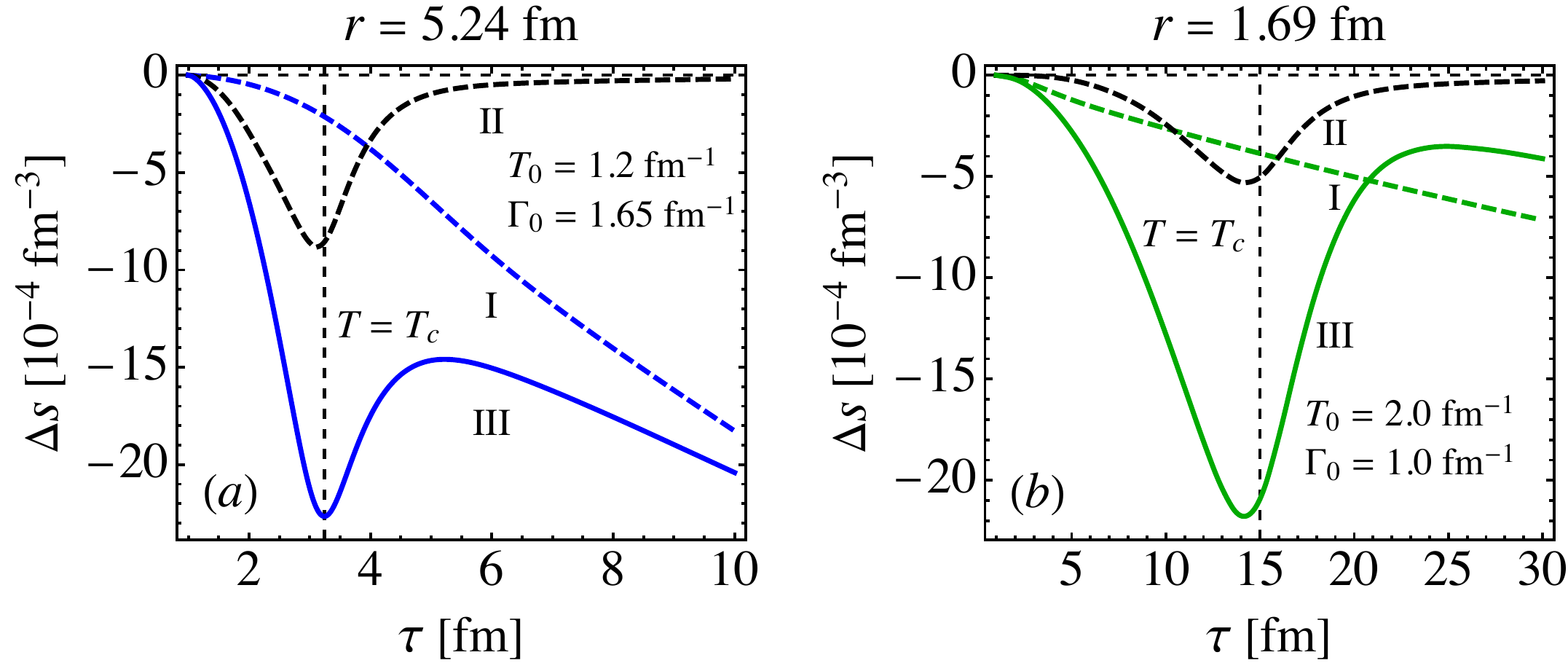}
    \caption{Evolution of the $Q$-integrated non-equilibrium entropy density correction $\Delta s$ at two radial distances, $r = 5.64$ fm (left) and $r = 1.69$ fm (right), for the scenarios I, II, III shown in Fig.~\ref{fig:gubser_phievo_r}a,b,c, respectively. The dashed vertical lines indicate the times when the fluid passes through $T = T_c$. Lines labeled by II correspond to the same evolution as those with the same label shown in Fig.~\ref{fig:bjorken_dynamics_comparison2}b,c.
    \label{fig:gubser_sevo_twors}}
\end{figure*}

The back-reaction of this non-equilibrium slow-mode dynamics on the entropy density of the medium (i.e. the non-equilibrium entropy correction $\Delta s$) is shown as a function of time in Fig.~\ref{fig:gubser_sevo_twors}. The two panels show how this plays out at two different transverse distances from the fireball center, a larger one at $r{\,=\,}5.24$\,fm (Fig.~\ref{fig:gubser_sevo_twors}a) which passes through $T_c$ first at $\tau\approx 3.24$\,fm, and a smaller one at $r{\,=\,}1.69$\,fm (Fig.~\ref{fig:gubser_sevo_twors}b) which passes through $T_c$ later at $\tau\approx 15$\,fm (vertical dashed lines). The black dashed lines describe scenario II without transverse flow and reproduce the identically labeled lines from Figs.~\ref{fig:bjorken_dynamics_comparison2}b,c. The colored dashed lines for scenario I show that transverse flow can induce large $|\Delta s|$ even for a constant correlation length $\xi = \xi_0$, i.e., without critical slowing-down. When critical behavior of $\xi(T)$ is added in scenario III (solid colored lines), the magnitude of $\Delta s$ increases strongly in the critical region around $T_c$. The crossing of the dashed and solid lines at $\tau \approx 20$\,fm in Fig.~\ref{fig:gubser_sevo_twors}b must be attributed to critical slowing-down in scenario III which keeps $\phi_Q$ from reacting to the time-increasing transverse flow effects as quickly as it can when $\xi{\,=\,}\xi_0$ is a (small) constant.

%
%

A qualitative feature of the entropy density evolution shown in Fig.~\ref{fig:gubser_sevo_twors} is that (for the background flow pattern assumed in this subsection) transverse flow appears to cause a non-equilibrium entropy correction from slow-mode dynamics that increases approximately linearly with time at late times.\footnote{%
    Physically this is, of course, an artefact because the flow pattern studied in this subsection results from an approximation that should not be used at large times
    $\tau > 1/q$.}
This can be understood from Fig.~\ref{fig:gubser_phievo_r}, panels a and c, by following the ratio $\phi_Q/\bar\phi_Q$ (i.e. the ratio between the solid and dashed colored lines) in time along the two vertical dashed lines indicating the $r$ positions studied in Fig.~\ref{fig:gubser_sevo_twors}: One sees that, for both constant (a) and temperature dependent (c) correlation length $\xi$, $\phi_Q/\bar\phi_Q$ decreases monotonically with time, explaining the growing magnitude of the entropy correction $\Delta s$ at late times seen in Fig.~\ref{fig:gubser_sevo_twors}.

We close this subsection with a discussion of transverse flow effects on anisotropic perturbations of the transverse profile. In the early time limit, $\tau\ll 1/q$, the perturbed solution is available analytically \cite{Gubser:2010ui,Hatta:2014jva, Hatta:2015era}:\footnote{%
    Here and below the undeformed (``isotropic") profiles (e.g. those in Eqs.~(\ref{eq:flow_approx},\ref{eq:temp_smallt}))
    are labeled by a subscript ``iso".}
\begin{eqnarray}
\label{eq:gubser_aniso}
    &&T=T_\mathrm{iso} (1 - \epsilon_n{\cal A}_n\delta)\,,
\nonumber\\
    &&u^r = u^r_\mathrm{iso} - \epsilon_n\nu_s (\delta u^r)\,,
\\\nonumber
    &&u^\phi = -\epsilon_n\nu_s (\delta u^\phi)\,.
\end{eqnarray}
Here $\epsilon_n$ is related to the eccentricity, $\delta$ and $\nu_s$ are parameters controlling the fluctuations of temperature and flow velocity, and the deformation profile is
\begin{equation}
\label{eq:A}
    {\cal A}_n(r,\phi) \equiv \left(\frac{2qr}{1+(qr)^2}\right)^{\!\!n}\cos(n\phi)\,,
\end{equation}
which is $\propto Y_{n,n}(\vartheta,\phi)+Y_{n,-n}(\vartheta,\phi)$ (here $(\vartheta,\phi)$ are polar coordinates in de Sitter space \cite{Gubser:2010ze}). The flow profile deformations are related to ${\cal A}_n$ by \cite{Hatta:2014jva, Hatta:2015era}
\begin{eqnarray}
    \delta u^r &=& \frac{2q\tau}{1+(qr)^2}\partial_\vartheta{\cal A}_n =n \,\frac{(2qr)^{n-1}(2q\tau)}{\bigl(1{+}(qr)^2\bigr)^n}
        \left(\frac{1{-}(qr)^2}{1{+}(qr)^2}\right)\cos(n\phi)\,,
\nonumber\\
    \delta u^\phi &=& \tau\partial_\phi{\cal A}_n = -n\tau\left(\frac{2qr}{1{+}(qr)^2}\right)^{\!\!n}\sin(n\phi)\,.
\end{eqnarray}

We assume $\epsilon_n\ll1$ so that we can linearize in $\epsilon_n$, e.g. $e=e_\mathrm{iso}(1 - \epsilon_n{\cal A}_n\delta)^4\approx e_\mathrm{iso}(1-4\epsilon_n{\cal A}_n\delta)$, and similarly for the slow modes (remembering $\Phi_0 \propto T^{-3}$ and $\Gamma_0 \propto T^{-1}$):
\begin{equation}
    \Phi_0 \approx \Phi_{0,\mathrm{iso}}(1+3\epsilon_n{\cal A}_n\delta)\,,\quad \Gamma_0 \approx \Gamma_{0,\mathrm{iso}}(1+\epsilon_n{\cal A}_n\delta)\,.
\label{eq:gubser_aniso_phi}
\end{equation}
Using these linearized expressions in Eqs.~(\ref{eq:adv_initial1},\ref{eq:adv_initial2}) we obtain  deformed profiles for $\bar\phi_Q$ and $\Gamma_Q$. Here we only consider elliptic deformations ($n=2$) and follow \cite{Hatta:2014jva, Hatta:2015era} by setting $\delta{\,=\,}1$, $\nu_s{\,=\,}-3/2$ and $\epsilon_2{\,=\,}0.15$. Using Eq.~(\ref{eq:gubser_aniso_phi}) together with the temperature and flow profiles in Eqs.~(\ref{eq:gubser_aniso}), solving the equations of motion for the slow modes as before, one can explore the anisotropic evolution.

%
\begin{figure*}[!tp]
    \centering
    \includegraphics[width= \textwidth]{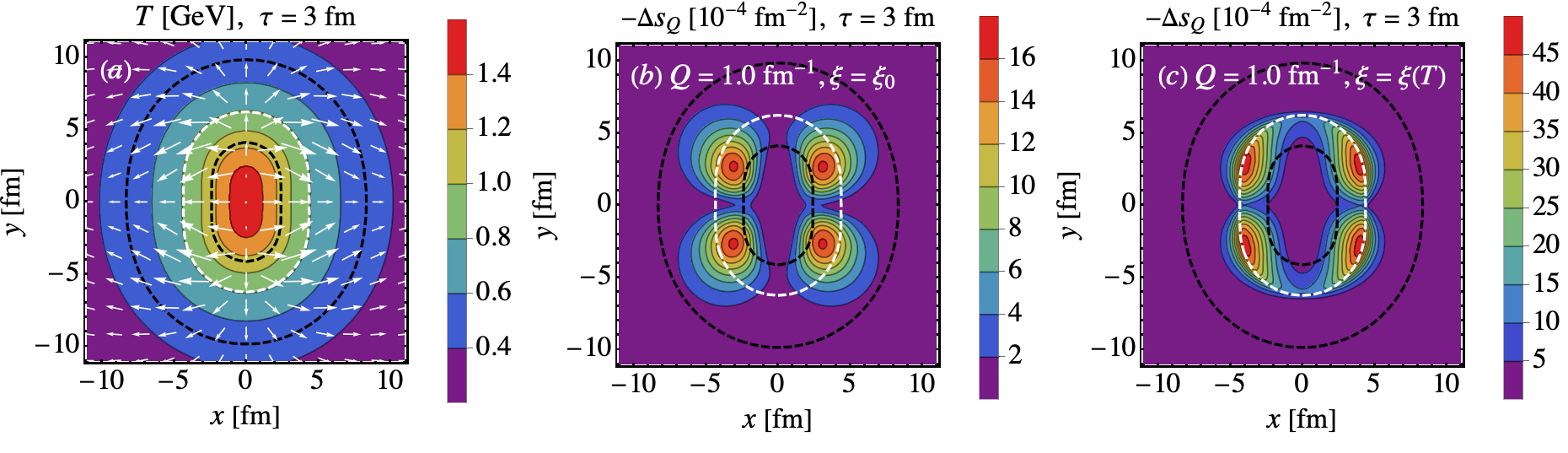}
    \caption{Transverse flow effects in an anisotropic profile at $\tau=3$\,fm. (a) Temperature contours in the transverse plane, elongated along the $y$ direction. Arrows indicate the anisotropic transverse flow. (b) and (c) show the correction $\Delta s_Q$ to the entropy density arising from the mode with wave number $Q=1.0$\,\fm; (b) assumes constant $\xi=\xi_0$ (scenario I) while (c) assumes temperature dependent $\xi(T)$ (scenario III). In all three panels the white dashed contour indicates $T{\,=\,}T_c$ while the two black dashed contours show $T_H=T_c+\Delta T=224$\,MeV and $T_L=T_c-\Delta T=96$\,MeV (where $\Delta T{\,=\,}64$\,MeV), respectively.
    \label{fig:gubser_advection_aniso}}
\end{figure*}
%

The results are shown in Fig.~\ref{fig:gubser_advection_aniso}. Panel (a) shows the temperature distribution in the transverse plane (which is clearly elongated in $y$ direction) and the anisotropic transverse flow. In panels (b,c) we compare, for the same dynamical scenarios I and III studied above, the entropy density modification $\Delta s_Q$ arising from the non-equilibrium evolution of the slow mode with wave number $Q{\,=\,}1$\,\fm. In scenario I with constant correlation length $\xi{\,=\,}\xi_0$, shown in panel (b), $\Delta s_Q$ does not feel the critical temperature and is sensitive only to effects arising from the anisotropic transverse flow; the location of the maximum of $|\Delta s_Q|$ basically coincides with that of the maximum flow velocity which can be identified in panel (a) by scanning for the longest flow arrows. In scenario III with a temperature dependent correlation length $\xi(T)$, shown in panel (c), the $|\Delta s_Q|$ maxima are clearly shifted closer to the $T=T_c$ critical contour. As time procedes, this contour moves inward due to cooling by longitudinal expansion. This means that the peak of $\bar\phi_Q$ associated with the critical peak of $\xi(T)$ moves inward, too, and that $\phi_Q$ tries to catch up with it. At $\tau = 3$\,fm, the flow is not yet very strong, so $\phi_Q$ does not lag too far behind its equilibrium value, and the  non-equilibrium effects causing $|\Delta s_Q|$ are tightly constrained to the critical contour. However, since the relaxation rate $\Gamma_Q$ increases with $r$ through ${\cal F}(r)$ in Eq.~(\ref{eq:adv_initial2}), the deviations from equilibrium tend to be smaller outside than inside the $T=T_c$ contour, explaining the slight inward shift of the maxima of $|\Delta s_Q|$ from the critical contour.\footnote{%
    This reasoning is supported by studying the $\tau{\,=\,}3$\,fm radial profiles of $\phi_Q$ and $\bar\phi_Q$ in Fig.~\ref{fig:gubser_phievo_r}c.} 
We see that, similar to panel (b), the azimuthal variation of the flow velocity causes the appearance of four ``hot spots''\footnote{%
    More accurately these should be called ``cold spots'' because $\Delta s_Q$ is negative and thus reduces the effective temperature.
    \label{fn20}}
of $|\Delta s_Q|$ at angles corresponding to flow maxima, but that the radial position of these maxima is strongly biased towards $T_c$ by the critical peaking of the correlation length.

The anisotropic entropy density correction $\Delta s$ from the slow modes ``reacts back'' on the expanding medium and affects its geometric eccentricity. We can define a slow-mode induced change of ellipticity, $\Delta \epsilon_2$, by using the definition of $\epsilon_2$ in terms of the expectation value of $\cos(2\phi)$, with a conformally weighted entropy density \cite{Hatta:2014jva,Hatta:2015era} as weight function:
\begin{equation}
    \epsilon_2 +\Delta\epsilon_2 = 
    -\frac{\int rdr d\phi\, (s{+}\Delta s)\,         
           \left[(qr)^2/\bigl(1{+}(qr)^2\bigr)\right]
           \cos(2\phi)}
          {\int rdr d\phi\, (s{+}\Delta s)\, \left[(qr)^2/\bigl(1{+}(qr)^2\bigr)\right]}\,.
\end{equation}
The uncorrected background gives $\epsilon_2{\,=\,}0.182284$. Including only the contribution to the correction $\Delta s$ from the slow mode with $Q=1.0$\,\fm\ and weighting it by a $Q$-bin width $\Delta Q{\,=\,}1$\,\fm, we find that at $\tau{\,=\,}3$\,fm the entropy correction changes the ellipticity to $\epsilon_2 + \Delta\epsilon_2 = 0.182293$ for scenario I in panel (b) and to $\epsilon_2 + \Delta\epsilon_2 = 0.182302$ for scenario III in panel (c). Transverse flow thus leads to a slight increase of the ellipticity, resulting from the slightly reduced particle emission (lowered entropy density) from the ``hot spots''$^{\ref{fn20}}$ indicated Fig.~\ref{fig:gubser_advection_aniso}b,c; the ellipticity correction is larger for scenario III which includes critical behavior of the correlation length $\xi$. [Note that altogether the ellipticity correction is tiny, of relative order $\lesssim 10^{-4}$ (consistent with the estimate (\ref{eq:estdeltas})), reflecting the smallness of the non-equilibrium entropy correction on the scale of the overall entropy density of the background fluid.]

%
\subsection{Combined non-equilibrium dynamical effects in full Gubser flow}
\label{sec:corrlen}
%

In the last two subsections we focused on the effects of background medium expansion and advection on slow-mode dynamics, and these studies were facilitated by taking certain limits of the background flow ((0+1)-dimensional Bjorken flow in Sec.~\ref{sec:expansion}, early-time limit for ideal Gubser flow in Sec.~\ref{sec:advection}). While some discussion of the specific effects caused by critical growths of the correlation length $\xi$ near the critical point was already included in these subsections, we will now extend the discussion of correlation length effects to the full (unapproximated) ideal Gubser flow.  

The exact temperature profile for ideal Gubser flow was given in Eq.~(\ref{eq-gubser_temp}). For the discussion in this subsection we take $C=2.78$ from Sec. \ref{sec:gubser_para} for the normalization. For the equilibrium values and the damping rates of the slow modes we use
\begin{eqnarray}
    \bar\phi_Q &=& \frac{\Phi_0}{\mathcal{F}^3(\tau,r)} \left(\frac{\xi}{\xi_0}\right)^{\!\!2} f_2(Q\xi)\,,\\
    \Gamma_Q &=& \frac{\Gamma_0}{\mathcal{F}(\tau,r)} \left(\frac{\xi_0}{\xi}\right)^{\!\!4} f_\Gamma(Q\xi)\,,
\label{eq:general_initial}
\end{eqnarray}
where $\mathcal{F}(\tau,r){\,\equiv\,}T(\tau,r)/T(\tau_0,0)$, $\Phi_0{\,=\,}1.0$\,fm$^{-3}$, and $\Gamma_0{\,=\,}0.9$\,\fm.

%
\begin{figure*}[!tb]
    \centering
    \includegraphics[width= \textwidth]{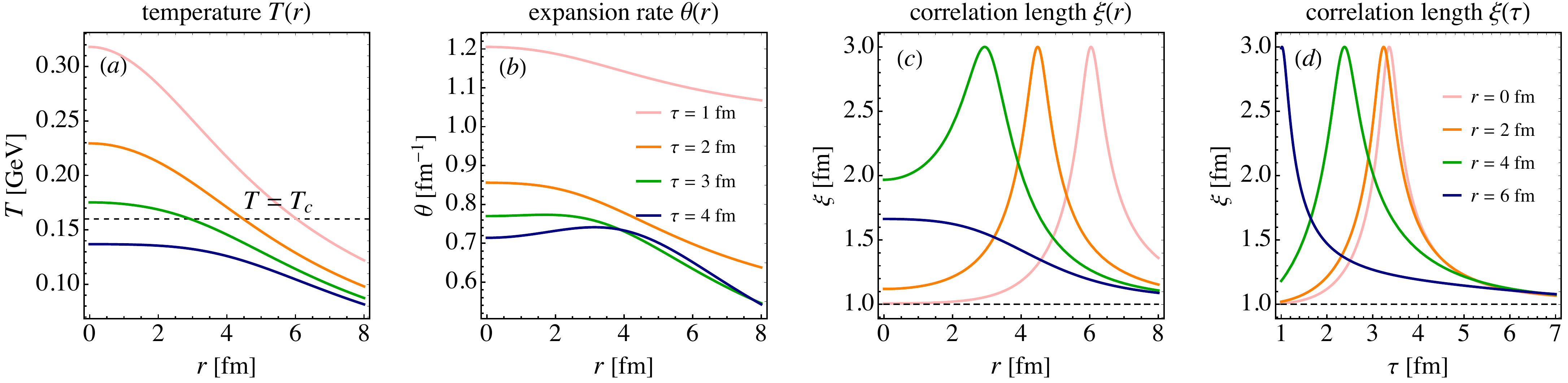}
    \caption{Radial profiles of the temperature $T(r)$ (a), scalar expansion rate $\theta(r)$ (b), and correlation $\xi(r)$ (c) at four different times $\tau{\,=\,}1,\,2,\,3,\,4$\,fm, as well as the time evolution $\xi(\tau)$ of the correlation length (d) at four different transverse distances $r{\,=\,}0,\,2,\,4,\,6$\,fm, for ideal Gubser flow.
    The black dashed lines in (c,d) indicate the constant value $\xi_0{\,=\,}1$\,fm.
    \label{fig:gubser-setup}}
\end{figure*}
%

Figure~\ref{fig:gubser-setup} shows snapshots of the key temperature (a), expansion rate (b) and correlation length (c) profiles for different times, as well as the time evolution of correlation length in panel (d) for different transverse positions. Panel (a) shows that fireball remains hottest at $r{\,=\,}0$ until all of it has cooled below $T_c$, and that (due to cooling by a combination of longitudinal and transverse expansion) the critical surface $T(\tau,r){\,=\,}T_c$ moves inward with time. Panel (b) shows that, generically, the expansion rate {\it decreases} with time, driven by the slowing rate of longitudinal expansion, $\theta_\parallel\sim1/\tau$; this facilitates equilibration of the slow modes at later times. However, the bump at $r\approx3.5$\,fm of the expansion rate at $\tau{\,=\,}4$\,fm in panel (b) also demonstrates the increasing contribution $\theta_\perp$ from transverse flow as time increases, causing the expansion rate to {\it increase} with time for $\tau>3$\,fm in the periphery $r>4$\,fm. For our initial conditions transverse expansion does not, however, dominate the expansion rate until the entire fireball has cooled below $T_c$. Panel (c) illustrates that the peak of the correlation length moves inward together with $T_c$ as time proceeds, and panel (d) shows that fluid cells pass through the critical point earlier at large $r$ than at smaller $r$, with cells at $r>6$\,fm starting out and remaining subcritical.

\begin{figure*}[!htb]
    \centering
    \includegraphics[width= \textwidth]{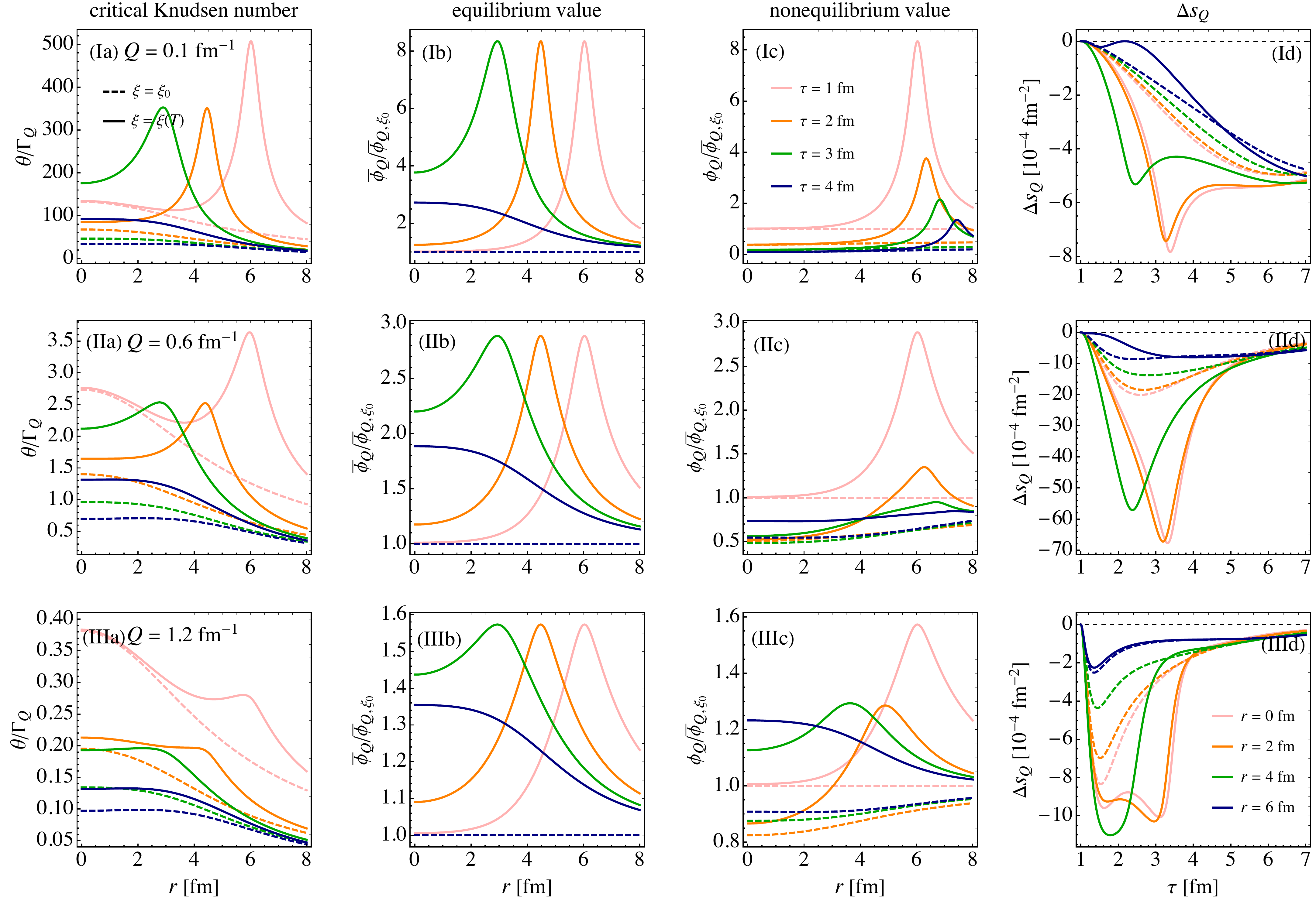}
    \caption{Comparison for the dynamics of three slow modes with widely different wave numbers: $Q = 0.1$\,\fm\ (I, top row), $Q = 0.6$\,\fm\ (II, middle row) and $Q = 1.2$\,\fm\ (III, bottom row). Dashed colored lines indicate evolution with constant correlation length $\xi = \xi_0$, solid colored lines use a temperature-dependent correlation length $\xi(T)$ that peaks at $T_c$. Similar to Fig.~\ref{fig:gubser-setup}, the three left columns (a-c) show radial profiles at different times, for the critical Knudsen number (a) and the (normalized) slow-mode equilibrium (b) and actual non-equilibrium values (c). The right column (panels (d)) shows the time evolution at four different transverse distances of the contribution $\Delta s_Q$ to the non-equilibrium entropy density correction arising from these three slow modes. 
    \label{fig:gubser-dynamics}}
\end{figure*}

In Fig.~\ref{fig:gubser-dynamics} we study the dynamical evolution of the actual and equilibrium values (columns (b,c)) of slow modes with  three different wave numbers, $Q{\,=\,}0.1\,\mathrm{fm}^{-1}<\xi_\mathrm{max}^{-1}$ (top row, I), $\xi_\mathrm{max}^{-1}<Q{\,=\,}0.6\,\mathrm{fm}^{-1}<\xi_0^{-1}$ (middle row, II) and $Q{\,=\,}1.2\,\mathrm{fm}^{-1}\gtrsim\xi_0^{-1}$ (bottom row, III), as well as of two related $Q$-dependent quantities, the critical Knudsen number $\theta/\Gamma_Q$ (column (a)) reflecting the competition between disequilibrating collective expansion and equilibrating mode relaxation, and the non-equilibrium slow-mode correction $\Delta s_Q$ to the entropy density in column (d). Throughout, dashed lines reflect non-critical dynamics with constant correlation length $\xi{\,=\,}\xi_0$ while solid lines show the results for critical dynamics with a correlation length $\xi(T)$ that peaks at $T_c$. 

For noncritical dynamics ($\xi{\,=\,}\xi_0$, dashed lines) the critical Knudsen number $\theta/\Gamma_Q$ in column (a) is seen to decrease with time throughout the fireball, basically following the scalar expansion rate $\theta$ shown in Fig.~\ref{fig:gubser-setup}b. Due to the factor $Q^2$ in $\Gamma_Q$, the critical Knudsen numbers are roughly 100 times larger in the top row than in the bottom row, severely hindering relaxation of the slow mode towards equilibrium. In comparison with the dashed lines, the solid lines show the additional effect of critical slowing down caused by the critical enhancement of the correlation length $\xi(T)$ near $T_c$. This effect can be up to a factor ${\sim\,}10$ (mostly due to the factor $\xi_0^2/\xi^2$ in $\Gamma_\xi$) in the top panel (small $Q$) but is seen to be significantly smaller for more typical $Q$ values like the one shown in the bottom panel (due to the compensating factor $1{+}Q^2\xi^2$ in $f_\Gamma$).  

Due to the small $Q$ value in the top row, the critical Knudsen numbers shown in panel (Ia) of Fig.~\ref{fig:gubser-dynamics} are huge (ranging from the tens to the hundreds). This implies that relaxation towards equilibrium plays practically no role in the dynamical evolution of this particular slow mode, and that the evolution patterns seen in panels (Ib) and (Ic) are entirely due to advection. For non-critical dynamics (dashed lines) the normalized equilibrium value shown in panel (Ib) remains frozen at 1 by definition while the normalized non-equilibrium value in panel (Ic) decreases with time, similar to Fig.~\ref{fig:bjorken_q}b at small $Q$ and Fig.~\ref{fig:gubser_phievo_r}a. For critical dynamics (solid lines) comparison of panels (Ib) and (Ic) shows how advection pushes the critical peak of $\phi_Q$ outward while the peak of its equilibrium value would follow the critical temperature inward as time proceeds. For the modes with the larger wave numbers shown in the two lower rows, the equilibrium value of the slow mode in panels (IIb) and (IIIb)  follows qualitatively the same pattern as for the small-$Q$ mode shown in (Ib), although the critical effects are significantly reduced at the higher $Q$ values. Due to the much smaller critical Knudsen numbers for the mode with $Q{\,=\,}1.2$/fm in the bottom row, panels (IIIb) and (IIIc) show quite different evolution patterns, reflecting the competition between relaxation (thermalization) and advection: the critical peak of $\phi_Q$ is now no longer pushed outward by advection, but moves inward via relaxation towards equilibrium. For the mode with an intermediate wave number $Q{\,=\,}0.6\,$\fm (row II), on the other hand, the dynamics shows a mixture of the characteristics seen in rows I and III --- one can still recognize in panel IIc at $r\gtrsim6$\,fm the initial peak being advected outward while being damped by relaxation (this is the dominant feature in panel Ic) while at the same time the relaxation dynamics that dominates in panel IIIc causes the solid lines in panel IIc at intermediate $r$ values to rise above the dashed lines as time proceeds.

Taken together, the three rows of Fig.~\ref{fig:gubser-dynamics} illustrate that there are two effects at play: (i) the {\it initial} peak in the fluctuations is carried outwards by advection while it is at the same time damped by relaxation; and (ii) as the location where critical fluctuations would occur in equilibrium (indicated by the peaks in the curves shown in column b) moves inward toward smaller $r$ values, the actual out-of-equilibrium fluctuations at these smaller values of $r$ increase, with the solid curves in column c relaxing upward toward the same-colored curves in column b, but more slowly due to critical slowing down. In principle both effects are present in all three rows, but the first effect is invisible in row III because at larger values of $Q$ the initial peak dissipates more rapidly and in addition it is rapidly dwarfed by the increase in the noncritical fluctuations $\phi_Q \propto 1/n$ at larger $r$, and the second effect is invisible in row I because relaxation is very slow at such a small value of Q.

The right column (d) of Fig.~\ref{fig:gubser-dynamics} (which should be compared with Figs.~\ref{fig:bjorken_dynamics_comparison2} and \ref{fig:gubser_sevo_twors}) shows the time evolution of the non-equilibrium entropy density correction $\Delta s_Q$ arising from the three modes studied in the three rows, at four different radial positions. Again, dashed (solid) lines reflect noncritical (critical) dynamics. The differences in (Id) are qualitatively similar to those observed between scenarios I and III in Fig.~\ref{fig:gubser_sevo_twors}, while the differences between dashed and solid lines in panel (IIId) are more similar to those observed between scenarios I and III in Fig.~\ref{fig:bjorken_dynamics_comparison2}. This is because low-$Q$ modes are more affected by advection (which was included in Fig.~\ref{fig:gubser_sevo_twors}) than high-$Q$ modes which can successfully fight advection effects (which Fig.~\ref{fig:bjorken_dynamics_comparison2} did not include). Depending on when the system enters the critical regime, the evolution of $|\Delta s_Q(\tau)|$ can feature two peaks, one due to expansion before reaching $T_c$ and another arising from critical slowing-down when entering the critical region. While both peaks are seen in panel (IIId) at $r{\,=\,}0$ and 2\,fm ({\it cf.} Fig.~\ref{fig:bjorken_dynamics_comparison2}c), only a single peak is observed for $r{\,\geq\,}4$\,fm ({\it cf.} Fig.~\ref{fig:bjorken_dynamics_comparison2}a,b). Consistent with the dynamics plotted in panel IIc we see in panel IId that the time evolution of $|\Delta s_Q|$ for the intermediate-$Q$ mode interpolates smoothly between panels Id and IIId. A notable feature, however, is the much larger magnitude of $|\Delta s_Q|$ in (IId) compared to both (Id) and (IIId): It is explained by referring to Fig.~\ref{fig:bjorken_stau} where we noted that, while off-equilibrium dynamical effects are stronger at small $Q$, their contribution to $|\Delta s(\tau)|$ peaks at an intermediate wave number $Q_\mathrm{max}{\,\sim\,}\mathcal{O}(Q_\textrm{neq})$, due to phase-space suppression by the factor $(Q/2\pi)^2$ at small $Q$.

\begin{figure*}[!tb]
    \centering
    \hspace*{-8mm}
    \includegraphics[width= 1.03\textwidth]{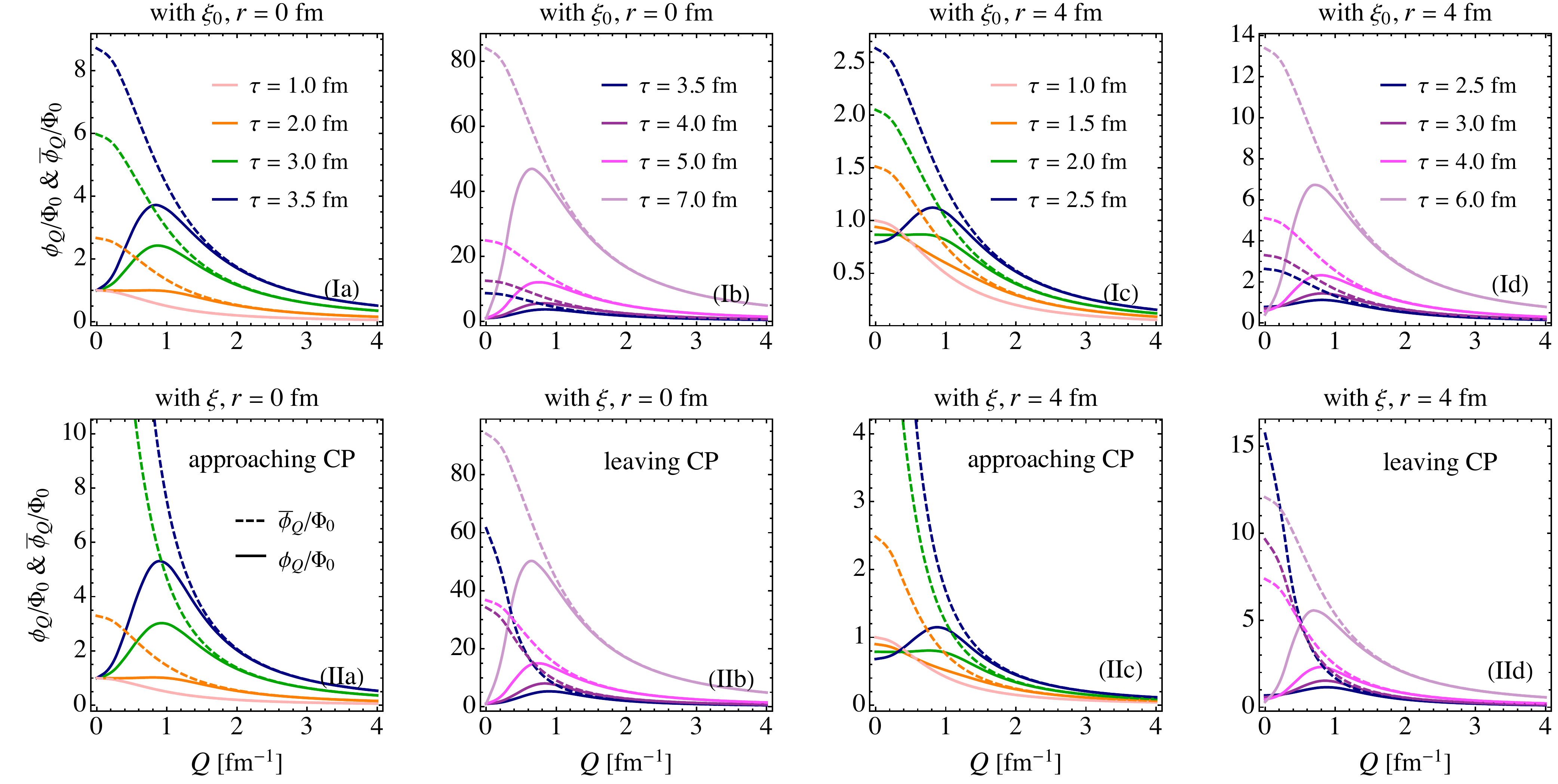}
    \caption{$Q$-dependence of the dynamics of the slow modes with $\xi = \xi_0$ (top row, I) and $\xi = \xi(T)$ (bottom row, II), at $r{\,=\,}0$\,fm (left two columns (a,b)) and $r{\,=\,}4$\,fm (right two columns (c,d)). At $r=0$\,fm (4 fm) $\xi$ approaches its maximum at $\tau{\,\approx\,}3.5$\,fm (2.5\,fm) (see Fig.~\ref{fig:gubser-setup}d). The dynamics at $r{\,=\,}0$\,fm (4\,fm) before $\tau{\,=\,}3.5$\,fm (2.5\,fm) (``approaching CP'') is shown in panels (a) (panels (c)); for later times (``leaving CP'') it is shown in panels (b) and (d), respectively.  [Note that the non-critical dynamics shown in the top row does not feel the critical point CP.] 
    In all eight panels dashed (solid) lines show the equilibrium (nonequilibrium) values of $\phi_Q$, both scaled by $\Phi_0(r)$. See text for discussion.\\
    \vspace{0.2cm}
    \label{fig:gubser-nearcp}
    \vspace*{-3mm}}
\end{figure*}

We close this subsection by showing in Fig.~\ref{fig:gubser-nearcp} for two radial positions ($r{\,=\,}0$ (a,b) and 4\,fm (c,d)) seven different time snapshots (as detailed in the legend) of the entire $Q$-spectrum of the slow modes, for both noncritical (top row) and critical dynamics (bottom row). Solid lines show the dynamically evolving slow modes spectra, dashed lines their corresponding equilibrium spectra.\footnote{%
    Note that at each $r$ we normalize $\bar\phi_Q$ and $\phi_Q$ by $\Phi_0(r)$, i.e. by the initial equilibrium value for $\xi=\xi_0$ at the same position.} 
For clarity, the dynamics is shown separately for the system approaching $T_c$ (columns a,c) and receding from $T_c$ (columns b,d); for $r=0$ (columns a,b) $T_c$ is reached at $\tau\approx3.5$\,fm, at $r{\,=\,}4$\,fm (columns c,d) this happens somewhat earlier at $\tau\approx 2.5$\,fm. In the bottom row, the equilibrium expectation for the magnitude of the slow modes (dashed curves) rises with time while approaching the critical point, and then begins to drop as the critical point is passed.  Note, however, that at very late times it rises again since, as the density keeps decreasing, the equilibrium $\bar\phi_Q$ grows like $1/n$, as we have discussed in Section~\ref{sec:advection}. In all plots, high-$Q$ modes are seen to closely follow their equilibrium values, hardly affected by advection. At $r{\,=\,}0$ (columns (a),(b)) low-$Q$ modes are basically frozen at their initial values while at $r{\,=\,}4$\,fm (columns (c), (d)) they are visibly affected by advection: even though $\Gamma_Q{\,=\,}0$ at $Q{\,=\,}0$, $\phi_0$ is seen to decrease with time instead of being frozen because advection moves the smaller value of $\phi_0$ at smaller $r$ to $r{\,=\,}4$\,fm.

The short summary of this subsection is that low-$Q$ slow modes are most strongly affected by the phenomenon of critical slowing down near a critical point.

%
\subsection{Space-time evolution of non-equilibrium slow mode effects and modified particle emission}
\label{sec:phenomenology}
%

In this subsection we will study, for the same setup as in the preceding subsection, the space-time structure of the non-equilibrium slow mode contribution to the entropy density, $\Delta s(\tau,r)$. Since this observable integrates over all slow-mode wave numbers $Q$, each of which evolves differently, it provides us with a global view of the interplay between off-equilibrium effects caused by expansion and advection before and after reaching the critical region, as well as their additional enhancement by critical slowing-down in the critical region.

\begin{figure*}[!htb]
    \hspace*{-3mm}
    \includegraphics[width= 1.02\textwidth]{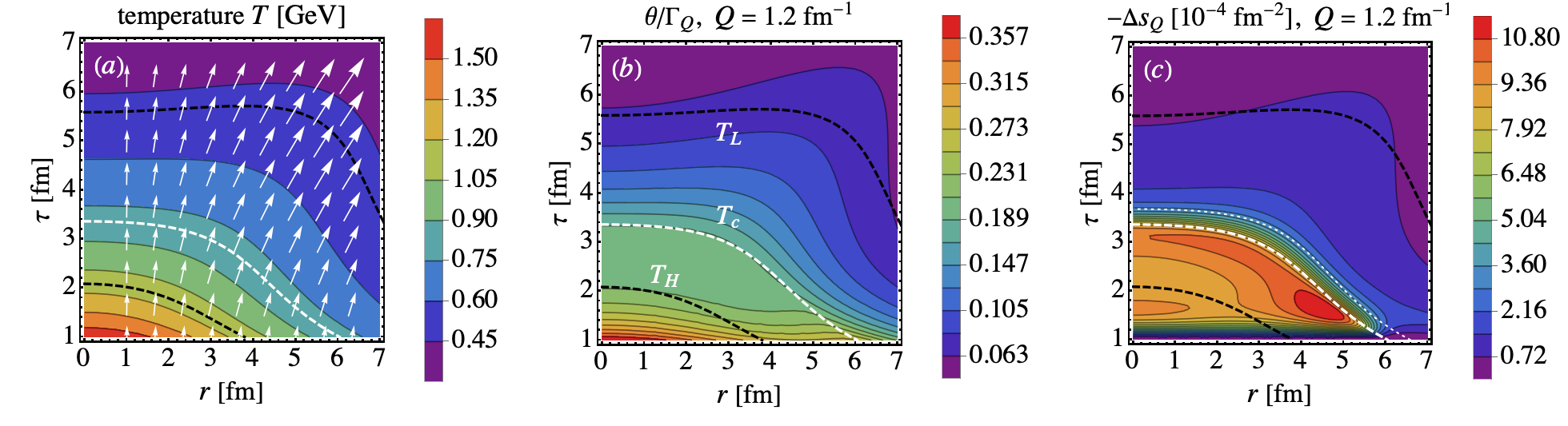}
    \caption{Full ideal Gubser space-time evolution of (a) temperature $T$ (contour lines) and flow velocity $(u^\tau, u^r)$ (white arrows), (b) critical Knudsen number $\theta/\Gamma_Q$ for $Q{\,=\,}1.2$\,\fm\ (contour lines), and (c) entropy modification $-\Delta s_Q$ contributed by this mode (contour lines), for a temperature dependent correlation length $\xi(T)$. The white dashed line in these panels denotes the $T(\tau,r){\,=\,}T_c{\,=\,}160$\,MeV isotherm; the black dashed lines show isotherms for  $T(\tau,r){\,=\,}T_H{\,=\,}224$\,MeV and $T(\tau,r){\,=\,}T_L{\,=\,}96$\,MeV which enclose the critical region. The dotted white line in panel (c) shows the ``freeze-out contour'' with temperature $T(\tau,r){\,=\,}T_f{\,=\,}148$\,MeV.
    \label{fig:gubser-pheno1}}
\end{figure*}

Figure~\ref{fig:gubser-pheno1} shows the space-time evolution of the background fluid (a) and of the critical Knudsen number (b) and non-equilibrium entropy density modification $-\Delta s_Q$ for a single representative slow mode with wave number $Q{\,=\,}1.2\,\mathrm{fm}^{-1}{\,\simeq\,}\xi^{-1}$ (c). The left panel (a) shows the evolution of the temperature contours and of the hydrodynamic flow, indicated by vectors. The middle panel (b)  demonstrates that critical Knudsen number is initially very large, due to the $1/\tau$ divergence of the longitudinal expansion rate at early times, and afterwards decays monotonically in time and also almost monotonically in radial direction. At later times ($\tau \gtrsim 4$ fm) growing radial flow causes the critical Knudsen number surface in $(\tau,r)$ to develop a weakly pronounced ridge along the direction pointing to the upper right corner. The right panel (c) shows that the slow-mode entropy correction $\Delta s_Q$ vanishes on the initial condition surface at $\tau_0{\,=\,}1$\,fm: this is a reflection of our (model-dependent) equilibrium initial conditions. Quickly thereafter, however, the large longitudinal expansion rate causes the slow mode to go out of equilibrium and generate a sizable amount of $|\Delta s_Q|$, which then does not, however, decrease with the critical Knudsen number as naively expected but, owing to the effects of critical slowing-down, remains high until the system has cooled below $T_c$ (denoted by the thick dashed white line), at which point it starts decreasing precipitously. On the $T_f{\,=\,}148$\,MeV ``freeze-out surface'' is already very small and will hardly affect the particle emission rate; the dominant phenomenological effects will likely be of second order, arising from the integrated effects of $\Delta s_Q(\tau,r)$ on the evolution history once the back-reaction onto the medium is taken into account (see Ref. \cite{Rajagopal:2019xwg}).    

\begin{figure}[!htb]
    \hspace*{-5mm}
    \centering
    \includegraphics[width=0.8\linewidth]{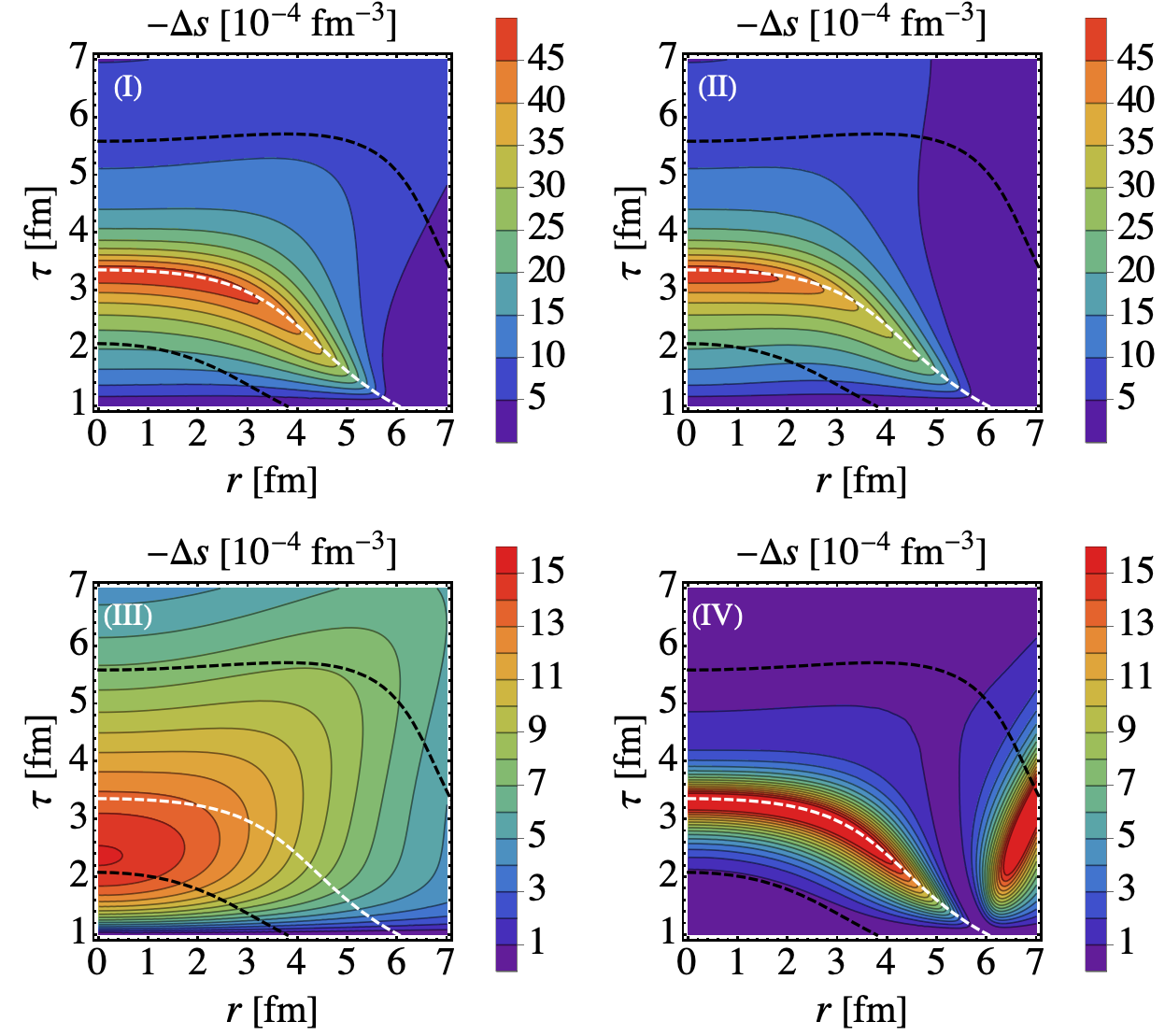}
    \caption{Full evolution of the $Q$-integrated entropy density modification $-\Delta s(\tau,r)$ for four different dynamic models: (I) full dynamics of the slow modes, (II) dynamics with the advection effects turned off by setting $u^r = 0$ in the slow-mode evolution equations, (III) dynamics with  constant correlation length $\xi = \xi_0$, and (IV) dynamics with $\Phi{\,=\,}\Phi_0$ and $\Gamma{\,=\,}\Gamma_0$, i.e. without accounting for the evolution of $\cpz/n^2$ and $\lambda_T/\cpz$. The black and white dashed lines are the same as in Fig.~\ref{fig:gubser-pheno1}.
    \label{fig:gubser-pheno2}}
\end{figure}

The $Q$-integrated entropy density modification effects from non-equilibrium slow-mode dynamics are illustrated in Fig.~\ref{fig:gubser-pheno2}. The four panels show four different dynamical models, as described in the caption. For the full dynamics shown in panel (I), the non-equilibrium entropy modification is strongly peaked near the critical isotherm. Turning off advection in panel (II) (which emulates the dynamics studied in \cite{Akamatsu:2018vjr}) pushes the contours of constant $\Delta s$ closer to the center, showing that conversely advection moves some of the non-equilibrium entropy effects to larger radii $r$. Removing the critical behavior of the correlation length $\xi(T)$ in panel (III), by setting $\xi{\,=\,}\xi_0{\,=\,}1$\,fm, strongly reduces and largely washes out the entropy modification effects; removing instead the temporal evolution of  $\cpz/n^2$ and $\lambda_T/\cpz$ from the calculation of the slow-mode damping rate and their equilibrium values in panel IV (emulating the dynamics studied in \cite{Rajagopal:2019xwg}) leads to both a reduction (by about a factor 3) and a tightening around $T_c$ of the non-equilibrium entropy modification effects. Interestingly, panel (IV) features a second branch of large entropy modification at $r{\,\gtrsim\,}6$\,fm, moving outward with the expanding fluid. This is because when the peak of $\bar\phi_Q/\bar\phi_{Q, \xi_0}$ moves inward with $T_c$, the peak of $\phi_Q/\bar\phi_{Q, \xi_0}$ from the initial condition moves outwards by advection (see Fig. \ref{fig:gubser-dynamics}, panel (Ic)). Since near the peak $\phi_Q$ is much larger than the local $\bar\phi_Q$, this yields a peak of $-\Delta s$. Something similar was also observed in Ref.~\cite{Rajagopal:2019xwg}.  

\begin{table}[!tp]
    \centering
    \begin{tabular}{c|c|c|c|c|c|c|c|c}
    \hline
    \hline
     & \multicolumn{2}{|c|}{Case I} & \multicolumn{2}{|c|}{Case II} & \multicolumn{2}{|c|}{Case III} & \multicolumn{2}{|c}{Case IV}\\
    \hline
    $|d\delta S/d\eta_s|$ & abs & $10^{-4}$ &  abs & $10^{-4}$ & abs & $10^{-4}$ & abs & $10^{-4}$ \\
    \hline
    \hline
     $T=160$\,MeV    & $1.230$  & $2.46$ & $1.019$ & $2.04$  & $0.319$ & $0.64$ & $0.446$ & $0.89$\\
     \hline
     $T=155$\,MeV  & $1.202$  & $2.40$ & $0.969$ & $1.94$  & $0.354$ & $0.71$ & $0.340$ & $0.78$\\
     \hline
     $T=148$\,MeV  & $1.056$  & $2.11$ & $0.804$ & $1.61$  & $0.409$ & $0.82$ & $0.254$ & $0.51$\\
    \hline
    \hline
    \end{tabular}
    \caption{The entropy modification $-d\delta S/d\eta_s$ from Eq.~(\ref{eq:fodsgubser}) at mid-rapidity on three isotherms with $T=160,\,155$, and 148\,MeV, for the four dynamical scenarios I-IV shown in Fig.~\ref{fig:gubser-pheno2}. ``abs'' stands for the absolute modification while the number next to it gives the relative modification (in units of $10^{-4}$), obtained by dividing by the initial total entropy content $dS/d\eta_s (\tau_0)\approx5000$ at mid-rapidity.
    \label{tab:modifiedS}}
\end{table}

Finally, to obtain a quantitative idea about how much, in the absence of back-reaction onto the medium, non-equilibrium slow mode dynamics might be able to affect particle emission from the freeze-out hypersurface, we use Eq.~(\ref{eq:fodsgubser}) to compute the total change in entropy per unit space-time rapidity, $d\delta S/d\eta_s$, integrated over the freeze-out surface. Since $\Delta s$ drops rather precipitously below $T_c$, we can ballpark the uncertainty of this prediction by working it out on three isotherms with $T=T_c=160,\,155$, and 148\,MeV. Table~\ref{tab:modifiedS} shows the entropy modification per unit rapidity on the three isotherms $T{\,=\,}160$, 155 and 148\,MeV. We see that in all cases the absolute modifications are tiny, of order $10^{-4}$ of the unmodified value. Using the variation among the results obtained on the three different isotherms we estimate the uncertainty in our calculation of the magnitude of this tiny effect to be at the few tens of percent level. We also note that the smallness of this effect is similar in magnitude to the changes in ellipticity caused by  non-equilibrium slow-mode effects studied in Sec.~\ref{sec:advection} and consistent with the rough estimate (\ref{eq:estdeltas}).

We note that although the slow-mode contribution to the entropy density is very small, its space-time evolution still provides an interesting reflection of the off-equilibrium slow mode dynamics: Since the magnitude of $\Delta s$ traces the magnitude of the expected critical point signatures (such as cumulants of fluctuations of produced particle yields \cite{Stephanov:1998dy, Stephanov:1999zu, Hatta:2003wn, Stephanov:2008qz, Stephanov:2011pb, Luo:2017faz}), Fig.~\ref{fig:gubser-pheno2} indicates which space-time regions of the fireball might contribute most prominently to such signals.

%
\subsection{Limits of the {\sc hydro+} framework}
%

In this final subsection we return to the groundwork of this study laid in Sec. \ref{sec:fluct}. The {\sc hydro+} framework is based on the assumption of a separation of scales, namely $\xi\ll\ell$ where $\xi$ is the correlation length and $\ell$ the hydrodynamic homogeneity length. For a quasi-1-dimensional expansion geometry such as Gubser flow there is only one macroscopic length scale parameter describing the (in-)homogeneity of the system, related to the scalar expansion rate: $\ell \sim 1/|\theta|$. The necessary scale separation (see Sec.~\ref{sec:hydrofluct}) thus requires $\xi\theta \ll \mathcal{O}(1)$. 

\begin{figure}[!tb]
    \hspace*{-5mm}\centering
    \includegraphics[width=0.8\linewidth]{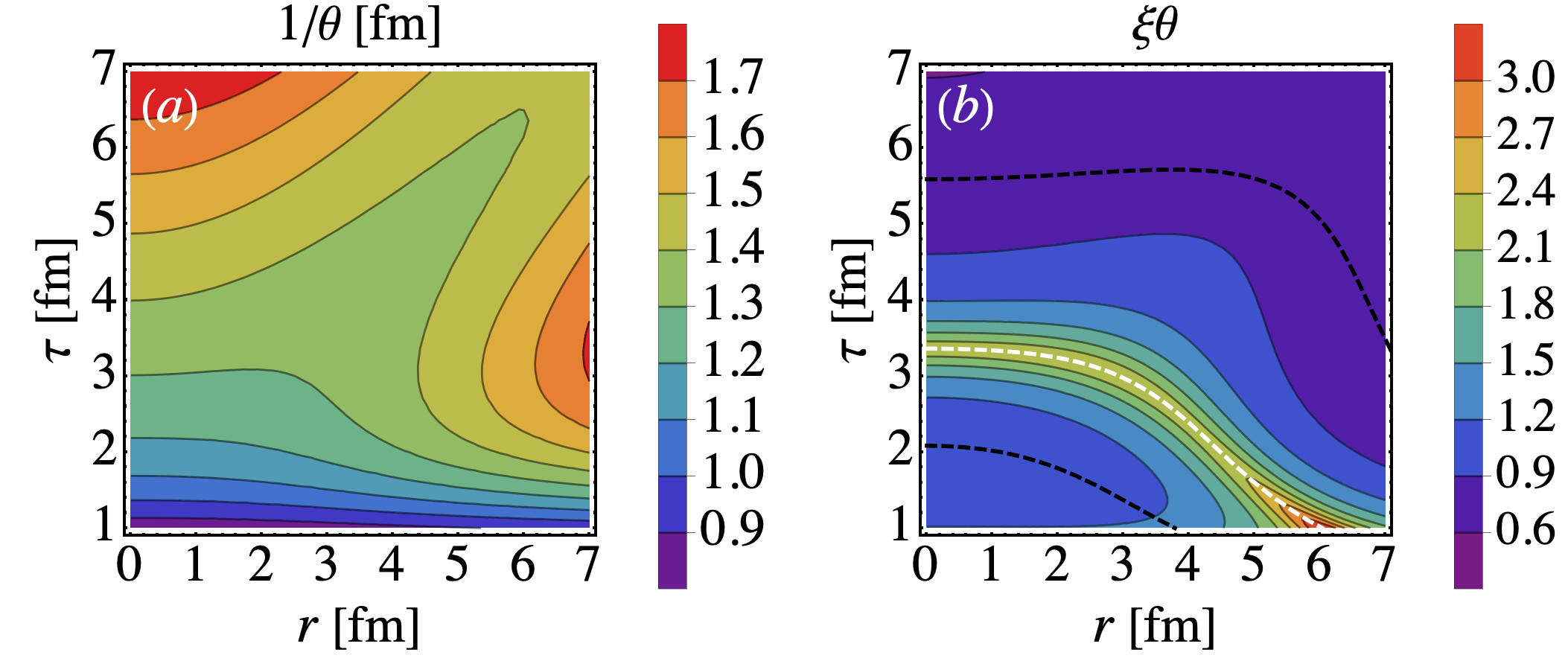}
    \caption{(a) Estimation of the hydrodynamic homogeneity length $\ell\sim 1/\theta$ for the ideal Gubser flow used in Secs. \ref{sec:corrlen} and \ref{sec:phenomenology}. (b) Estimation of the ratio $\xi/\ell \sim \xi\theta$. The black and white dashed lines are the same as in Fig.~\ref{fig:gubser-pheno1}.
    \label{fig:gubser-xitheta}}
\end{figure}

For ideal Gubser flow the scalar expansion rate (\ref{eq:expansion}) can be computed from the flow profile (\ref{eq-gubser-ut}-\ref{eq-gubser-uphi}). Its inverse, proportional to the hydrodynamic homogeneity length $\ell\sim 1/\theta$, is plotted in Fig.~\ref{fig:gubser-xitheta}a. The right panel, Fig.~\ref{fig:gubser-xitheta}b, shows the ratio  $\xi/\ell \sim \xi\theta$ which needs to be sufficiently small ($<\mathcal{O}(1)$) for the {\sc hydro+} framework to be valid. One sees that the framework gets stressed mostly in a narrow region around $T_c$ but elsewhere it works well, even at very early times where the homogeneity length $\ell$ is short. If we had used a more realistic parametrization of the correlation length $\xi$ as a function of both $T$ and $\mu$, which exhibits critical growth only near a critical point $(T_c,\mu_c)$ instead of on a critical isotherm $T_c$ as here, the $(\tau,r)$ region where {\sc hydro+} might break down would shrink correspondingly.

Real heavy-ion collisions exhibit an additional feature that is not shared by Gubser flow and therefore not reflected in Fig~\ref{fig:gubser-xitheta}: event-by-event quantum fluctuations in the initial spatial energy density profile (a.k.a. ``bumpiness''). This bumpiness arises from small ratios between the nucleon and nuclear radii and between the color correlation length inside a nucleon and the nucleon's radius (see Refs.~\cite{Shen:2017bsr, Du:2018mpf} for relevant recent studies at BES energies). Initial-state density fluctuations on subnucleonic length scales are of particular phenomenological importance in collisions involving small nuclei, such as proton-proton and proton-nucleus collisions \cite{Welsh:2016siu}. They have the potential of reducing the range of validity of the {\sc hydro+} framework in small collision systems below what is shown in Fig.~\ref{fig:gubser-xitheta}b, by locally shrinking the homogeneity length $\ell$ below the inverse expansion rate plotted in Fig.~\ref{fig:gubser-xitheta}a and thereby generating local bumps for the ratio $\xi/\ell$.

%
\section{Summary and conclusions}
\label{sec:conclusion}
%

In this chapter, we presented a systematic study of critical slow mode evolution in an expanding quark-gluon plasma (QGP) that passes close to a critical point in the QCD phase diagram. To achieve conceptual clarity of the mechanisms controlling the critical slow mode dynamics we used an analytical model, ideal Gubser flow discussed in Ch.~\ref{ch.gubser}, for the expansion of the QGP background fluid which qualitatively reproduces key features of the dynamics of the hot and dense medium created in relativistic heavy-ion collisions, in particular simultaneous and mutually coupled longitudinal and transverse flow. While the use of such a simplified expansion model robs us of the opportunity to make direct comparison with experimental data (this will be left for future work employing the (3+1)-dimensional numerical \bes+\ code developed for the study presented in this chapter and briefly described in the App.~\ref{app:hydroplus}), it provides us with the opportunity to selectively zoom in onto key mechanisms driving the critical slow mode evolution, by tuning the background flow analytically.

Just like other dissipative phenomena in a relativistic fluid, critical slow-mode dynamics is controlled by the competition between the rate of macroscopic hydrodynamic expansion (which drives the critical slow modes away from thermal equilibrium) and relaxation processes on length scales of order the correlation length $\xi$ and shorter, encoded in a wave number dependent relaxation rate $\Gamma_Q$, that help the slow modes to thermalize. Slow mode relaxation, as well as any other dissipative effects to which slow mode relaxation contributes, is affected by critical slowing down, i.e. by a dramatic reduction of the relaxation rate $\Gamma_Q$ close to the critical point where the correlation length $\xi$ for order parameter fluctuations becomes large.\footnote{%
    In fact, the critical slowing down of slow mode non-equilibrium dynamics is known to contribute, through its correction $\Delta p$ to the equilibrium pressure, to the critical slowing down of the relaxation of the bulk viscous pressure \cite{Stephanov:2017ghc, Rajagopal:2019xwg}.}
This comes in addition to a leading quadratic ($\sim Q^2$) wave number dependence of $\Gamma_Q$ which slows down the thermalization of long wavelength fluctuations already in the absence of a critical point. The competition between macroscopic expansion and microscopic relaxation is captured by the ($Q$-dependent) critical Knudsen number Kn$=\theta/\Gamma_Q$ which was shown in this chapter to be a good predictor for the (in-)ability of the critical slow modes to follow the dynamical evolution (via expansion of the background fluid) of their space-time dependent equilibrium value.

An important aspect of critical slow-mode dynamics in an expanding background is the phenomenon of advection, i.e. the outward transport of the slow mode with the expanding fluid by collective transverse flow which was ignored in some earlier work (e.g. \cite{Akamatsu:2018vjr}): As the system cools by longitudinal expansion, the critical surface $T(\tau,r)=T_c$ moves inward but, especially for small $Q$ where relaxation is anyhow suppressed, the critical maximum of the slow mode doesn't follow closely that inward motion of the critical surface but may instead even move outward, driven by outward radial flow transverse to the beam direction.

The present work is, to the best of our knowledge, the first one that studies {\it critical slow mode dynamics} in its full complexity\footnote{%
    Albeit not its back-reaction on the expanding fluid itself.} 
in a more or less realistic setting for relativistic heavy-ion collisions. We presented the space-time evolution of the full spectrum of wave numbers $Q$, as temporal profiles at fixed locations and as snapshots of spatial profiles at varying times, and we also computed their contribution to the overall entropy balance, in both space and time, in order to gauge the importance of feedback effects of critical slow mode dynamics on the hydrodynamical bulk evolution. While the critical slow modes $\phi_Q$ are expected to make a substantial contribution to fluctuation observables, we found that their corrections to the bulk entropy density and pressure, as well as to other macroscopic characteristics of the expanding fireball such as its elliptic geometric deformation $\epsilon_2$ in non-central heavy-ion collisions, are exceedingly small, of relative order $10^{-4}$. This confirms similar estimates presented in Ref.~\cite{Rajagopal:2019xwg} for a simpler dynamical setting for the slow modes. We therefore expect this feature to survive in upcoming fully realistic (3+1)-dimensional numerical simulations of the coupled macroscopic near-equilibrium expansion and the microscopic non-equilibrium kinetic slow mode dynamics that include all back-reaction effects; the tools for performing such a comprehensive study were developed and presented in this chapter. 

Confirming the smallness of back-reaction effects from non-equilibrium critical fluctuation dynamics onto the bulk properties of the fluid will be important for two reasons: it will firmly direct our attention away from bulk hydrodynamic features and towards direct fluctuation measurements when searching for critical point signatures, and it will simplify the description of the hydrodynamic fireball evolution by allowing us to ignore back-reaction effects without noticeable loss of accuracy.

\chapter{Conclusions and outlook}
\label{ch:concl}

To make experimental discoveries (e.g.~to locate or rule out the critical point in relativistic heavy-ion collisions), a well-constrained multistage theoretical framework is essential. In this thesis we advanced this cause in several directions. Considering the complexity of the relevant dynamics and the computational expense of realistic simulations, which are able to cover a vast range of collision energies, I personally think of a roadmap with the following necessary steps: (1) The (3+1)-dimensional evolution of collision fireballs with conserved charges should be benchmarked in the absence of critical effects. (2) Critical effects of phenomenological importance to the bulk evolution must be identified and exhaustively explored. (3) The thus identified ``essential critical effects'' must be added to the benchmarked non-critical dynamics, and the full model for bulk dynamics must be tuned via comprehensive model-to-data comparisons. (4) Iterations will be inevitable to reach a state of the art that allows for reliable and quantitatively precise calculations and predictions for observables that exhibit sensitivity to the critical point, in order to discover or rule out it using experimental data across relevant collision energies. In the meantime, observables that have constraining power on the bulk dynamics and criticality should be identified and explored so that they can be meaningfully integrated in the above process.

\begin{figure}[!htb]
\begin{center}
\includegraphics[width= \textwidth]{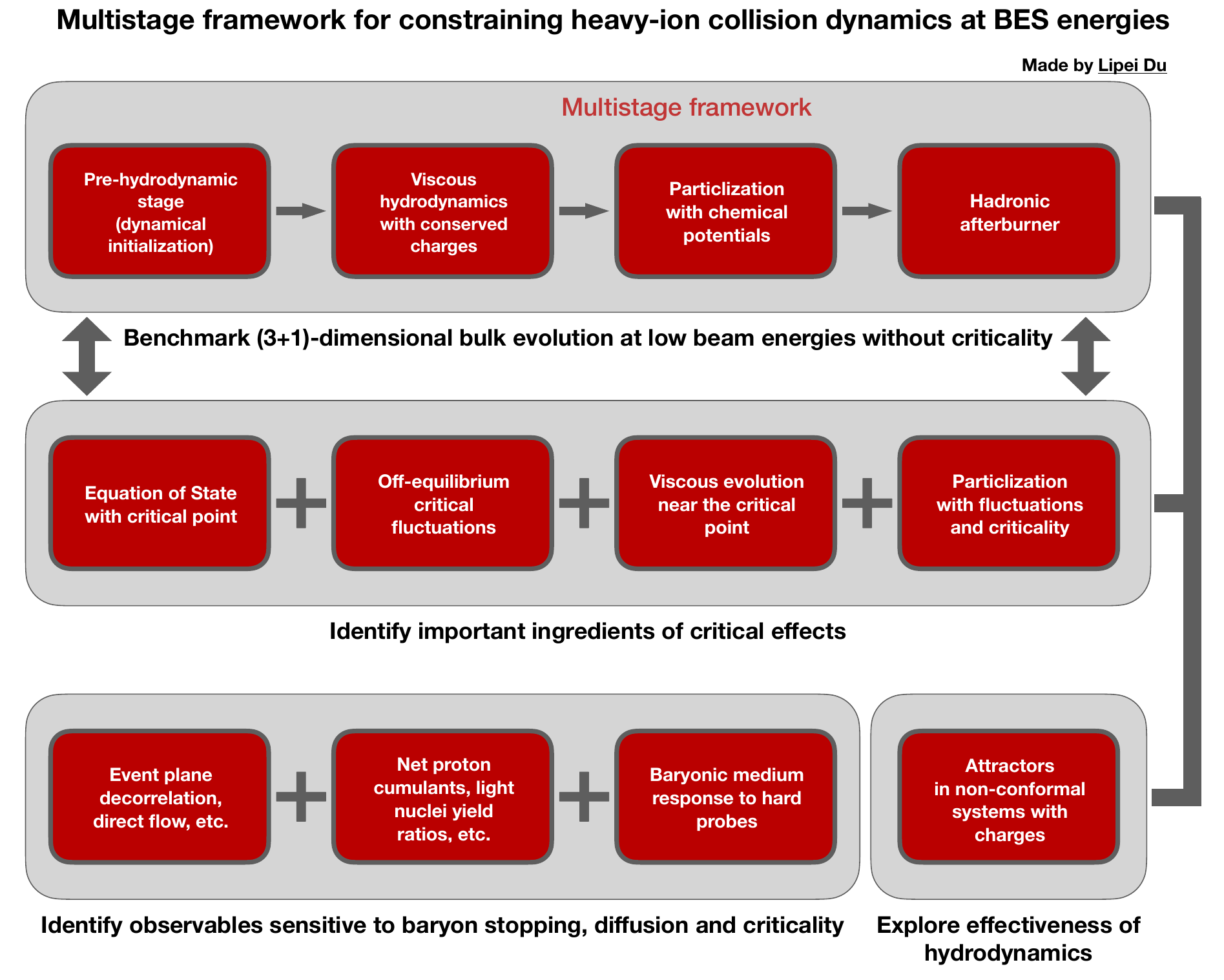}
\caption{A roadmap towards the multistage framework for heavy-ion collision dynamics at low beam energies.}
\label{fig:besdiagram}
\end{center}
\end{figure}

These steps are summarized in the roadmap illustrated by Fig.~\ref{fig:besdiagram}, which emerged from the studies presented in this thesis. In Ch.~\ref{ch:pheno}, a few observables which have constraining power on the bulk dynamics and which are sensitive to the criticality of heavy-ion collisions were discussed (bottom left box in Fig.~\ref{fig:besdiagram}). In Ch.~\ref{ch:multistage}, we illustrated the different stages and associated physics descriptions of the multistage framework for heavy-ion collisions at low beam energies (upper box in Fig.~\ref{fig:besdiagram}). In Chapters~\ref{ch:numerics} and~\ref{ch.gubser}, we presented and validated the numerical methods of \bes{}, which was developed to simulate the hydrodynamic evolution of systems produced in low energy nuclear collisions. A comprehensive validation protocol convinced us that the performance of \bes{} is exemplary. With these two chapters we contributed to the first step of the opening paragraph. 

In Ch.~\ref{ch:diffcp} we explored  baryon diffusion away from and close to the critical point and found that phase diagram trajectories of fluid cells at different space-time rapidities intersect each other, leading to strong reshuffling of baryon number in space-time by diffusion. We also found that because of a fast decay of the baryon diffusion current before the system enters the critical region, critical effects on bulk dynamics through baryon diffusion may be negligible. In Ch.~\ref{ch.fluctuations}, we studied the off-equilibrium dynamics of critical fluctuations and its back-reaction on the bulk evolution of the system which was found to be negligible. This implies that off-equilibrium critical effects may be safely neglected  when constraining the bulk evolution. On the other hand, the space-time evolution of the off-equilibrium critical fluctuations studied in Ch.~\ref{ch.fluctuations} provides valuable insights into a number of different mechanisms contributing to the final particle cumulants. Chapters~\ref{ch:diffcp} and~\ref{ch.fluctuations} are our contribution  to the second step of the roadmap and form part of the middle box in Fig.~\ref{fig:besdiagram}.

The exploratory studies presented in this thesis can be improved upon in the immediate future by including several missing ingredients. In the hydrodynamic sector, dynamical effects caused by non-zero values of conserved charges, the Equation of State at non-zero chemical potentials, and the related particlization with non-zero $(\mu_B, \mu_Q, \mu_S)$ should be investigated by replacing the model expansion scenarios used in this thesis by fully realistic (3+1) D simulations. More needs to be done in trying to identify phenomenologically important and robust critical effects. Areas to explore in greater detail include the critical slowing down of dissipative dynamics, adding critical singularities to the Equation of State, and the problem of particlizing critical fluctuations. A key challenge is to arrive at a robust conclusion whether  critical effects on bulk dynamics are phenomenological important, without knowing the location of the critical point and before the bulk dynamics gets well constrained. The framework developed in this thesis provides hope that we will be able to considerably improve our quantitive understanding of the dynamics of heavy-ion collisions at all collision energies in the near future.

\begin{appendices}
\chapter{Implementation of static critical behavior}
\label{app.corre_length}

\section{Parametrization of the correlation length}

\begin{figure}[!htb]
\begin{center}
    \includegraphics[width= 0.55\textwidth]{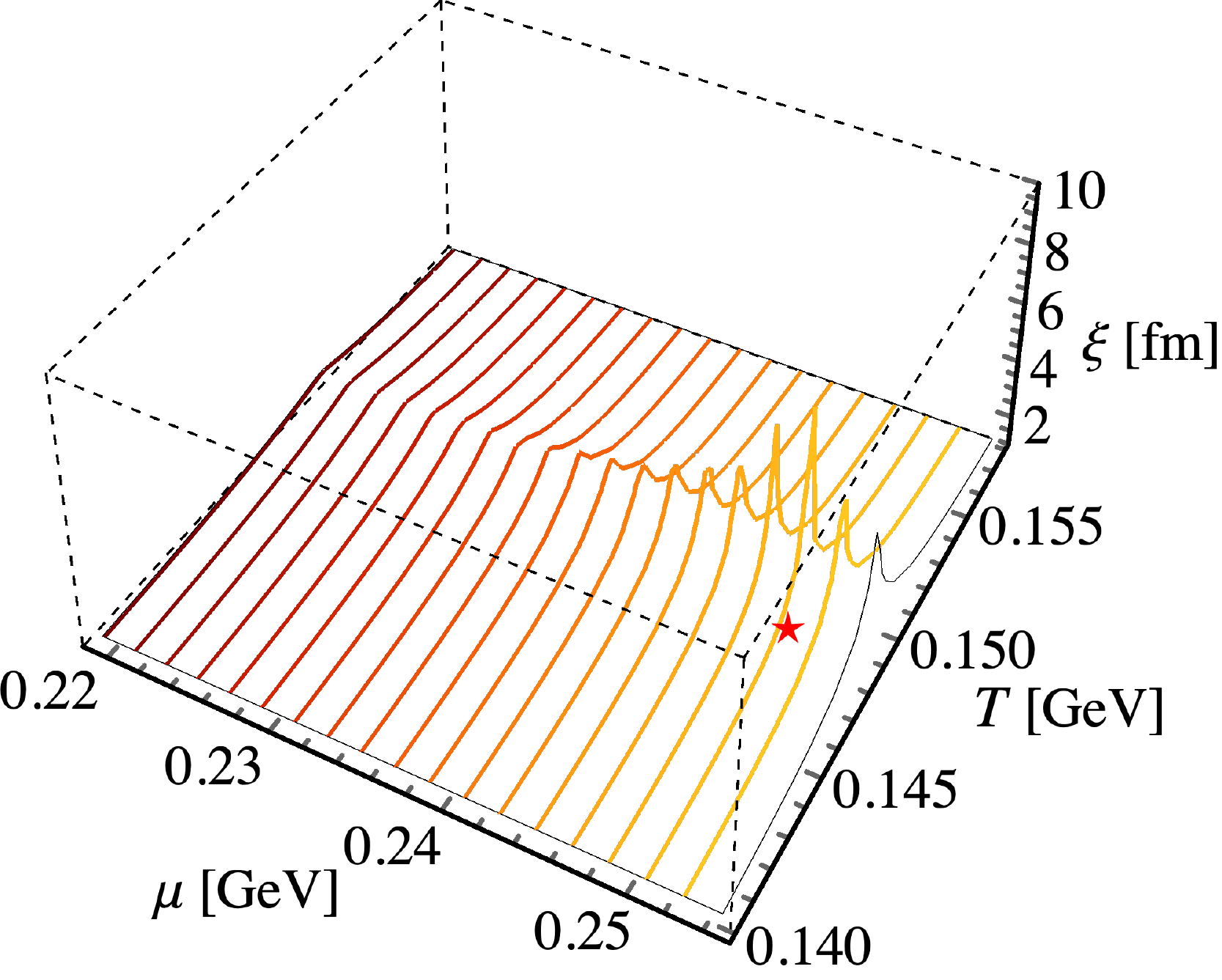}
\end{center}
\caption{%
    Distribution of the correlation length $\xi(\mu,T)$, with a critical point located in the $T$-$\mu$ plane at $T_c{\,=\,}0.149$\,GeV and ${\mu}_c{\,=\,}0.25$\,GeV (indicated by the red star), parametrized by Eq.~\eqref{eq:xi(mu,T)} with the parameters in Eq.~\eqref{eq:non-universal-parameters}.}
\label{fig:correlation_length}
\end{figure}

One significant feature of critical phenomena is that, when the system approaches a critical point adiabatically, the {\em equilibrium} correlation length, which is typically microscopically small, becomes macroscopically large and eventually diverges.
With the purpose of identifying qualitative signatures of a critical point, we characterize all equilibrium quantities exhibiting critical behavior in terms of their parametric dependence on the correlation length.\footnote{%
    If we knew the critical EoS explicitly, these dependencies would naturally follow from the thermodynamic identities relating these quantities to the thermal equilibrium partition function.}
We parametrize the correlation length as follows:
\begin{eqnarray}\label{eq:xi(mu,T)}
    \xi(\mu,T) =&\,\xi_0(\mu,T)\left\{\tanh [f(\mu,T)]\left(1-\left(\frac{\xi_0}{\xi_\mathrm{max}}\right)^{\frac{2}{\nu}}\right)+\left(\frac{\xi_0}{\xi_\mathrm{max}}\right)^{\frac{2}{\nu}}\right\}^{-\frac{\nu}{2}}\,.
\end{eqnarray}
Here $\xi_0(\mu,T)$ is the non-critical correlation length (measured far away from the critical point) while $\xi_\mathrm{max}$ is an infrared cutoff regulating the divergence at the critical point by implementing a maximum value for the correlation length. The crossover between the critical and non-critical regimes is mediated by the hyperbolic function $\tanh[f(\mu,T)]$, where
\begin{eqnarray}\label{eq:f(mu,T)}
    f(\mu,T)=\left|\frac{(\mu{-}\mu_c)\cos{\alpha_1}-(T{-}T_c)\sin{\alpha_1}}{\Delta \mu}\right|^2+\left|\frac{(\mu{-}\mu_c)\sin{\alpha_1}+(T{-}T_c)\cos{\alpha_1}}{\Delta T}\right|^{\frac{2}{\beta\delta}}.
\end{eqnarray}
In the above expression $(T_c,\mu_c)$ is the location of critical point, and $\Delta \mu$ and $\Delta T$ characterize the extent of the critical region along the $\mu$ and $T$ axes of the phase diagram; $\alpha_1$ is the angle between the crossover line ($h=0$ axis in the Ising model) and the negative $\mu$ axis (see Fig.~\ref{fig:phase_diagram_illu} below); $\nu=2/3$, $\beta=1/3$ and $\delta=5$ approximate the critical exponents of the 3-dimensional Ising universality class \cite{Berges:1998rc, Halasz:1998qr}. Eq.~\eqref{eq:xi(mu,T)} is designed to ensure the following properties:
\begin{itemize}
\setlength{\itemindent}{-1.3em}
    \item[] 1) $\xi=\xi_\mathrm{max}$ when $\mu=\mu_c$ and $T=T_c$\,;
    \item[] 2) $\xi\sim|T-T_c|^{-\frac{\nu}{\beta\delta}}$ when $\mu=\mu_c$ and $|T- T_c|\lesssim\Delta T$\,;
    \item[] 3) $\xi\sim|\mu-\mu_c|^{-\nu}$ when $T=T_c$ and $|\mu-\mu_c|\lesssim\Delta\mu$\,;
    \item[] 4) $\xi=\xi_0$ when $|\mu-\mu_c|\gg\Delta\mu$ and/or $|T-T_c|\gg\Delta T$\,.
\end{itemize}
To limit the number of free parameters, we ignore the $T$ and $\mu$ dependence of the non-critical correlation length $\xi_0$ and parametrize the crossover line as \cite{Bellwied:2015rza}
\begin{equation}
    \frac{T(\mu)}{T_0}=1-\kappa_2\left(\frac{\mu}{T_0}\right)^2+\mathcal{O}(\mu^4)\,,
\end{equation}
where $T_0=155$ MeV and $\kappa_2=0.0149$ are the transition temperature and the curvature of the transition line $T(\mu)$ at $\mu=0$. The location of the critical point $(T_c,\mu_c)$ is assumed to be on the crossover line \cite{Parotto:2018pwx}, and as a consequence
\begin{equation}
    \alpha_1=\arctan\left(\frac{2\kappa_2\mu_c}{T_0}\right).
\end{equation}
Thus $T_c$ and $\alpha_1$ are determined once $\mu_c$ is provided. Based on the above discussion we choose the following parameter values:
\begin{gather}
\xi_0 = 1\,\textrm{fm},\quad \xi_\mathrm{max} = 10\,\textrm{fm}, \nonumber\\
\mu_c=250\,\textrm{MeV}, \quad T_c = 149\,\textrm{MeV}, \quad \alpha_1=4.6^\circ\,, \label{eq:non-universal-parameters}\\
\Delta\mu = 92\,\textrm{MeV}, \quad \Delta T=18\,\textrm{MeV}. \nonumber
\end{gather}
Among these, $\Delta\mu$ and $\Delta T$ are determined by additional parameters provided in Eq.~\eqref{eq:setup_Delta_muT} of App.~\ref{sec:Delta_muT}. With those parameters, we visualize the correlation length as function of $(\mu,T)$ in Fig.~\ref{fig:correlation_length}.

Several comments are in order: First, our parametrization of the correlation length applies to the crossover region in the left part of the QCD $T$-$\mu$ phase diagram, at $\mu<\mu_c$, and not to the presumed first-order phase transition at $\mu>\mu_c$ where the theoretical description is complicated by possible phase coexistence and metastability \cite{Steinheimer:2012gc,An:2017brc}. This prompts us to choose the collision beam energy sufficiently high to avoid the latter situation, but not too high to be far from the critical point. Motivated by experimental hints \cite{Adam:2020unf} and earlier theoretical studies \cite{Monnai:2016kud,Denicol:2018wdp}
we here set $\snn\eq19.6\,$GeV. Second, although the correlation length diverges in the thermodynamic limit, heavy-ion collisions create small, rapidly expanding QGP droplets in which finite-size and finite-time effects as well as the critical slowing-down \cite{Stephanov:1999zu,Berdnikov:1999ph} prevent the correlation length from growing to infinity. A robust estimate for the largest correlation length the system might achieve in this dynamical environment is about 3\,fm \cite{Berdnikov:1999ph}. The system will thus never get close to our static infrared cutoff $\xi_{\rm max}=10$\,fm, and our final predictions turn out not to be sensitive to the precise value of this cutoff. 

Once all thermodynamic quantities and transport coefficients (introduced in the following subsection) are parametrized in terms of $\xi$ as given in Eq.~\eqref{eq:xi(mu,T)}, they are defined in both the non-critical and critical regions and thus ready for use in dynamical simulations describing the trajectory of the QGP fireball through the phase diagram. For economy we include in the following discussion not only the dynamic (transport) coefficients but also the thermal susceptibility $\chi$ and the specific heat $c_p$ which are static (thermodynamic) coefficients.

\section{Estimating the size of the critical region}
\label{sec:Delta_muT}

\begin{figure}[t]
\begin{center}
    \includegraphics[width= 0.55\textwidth]{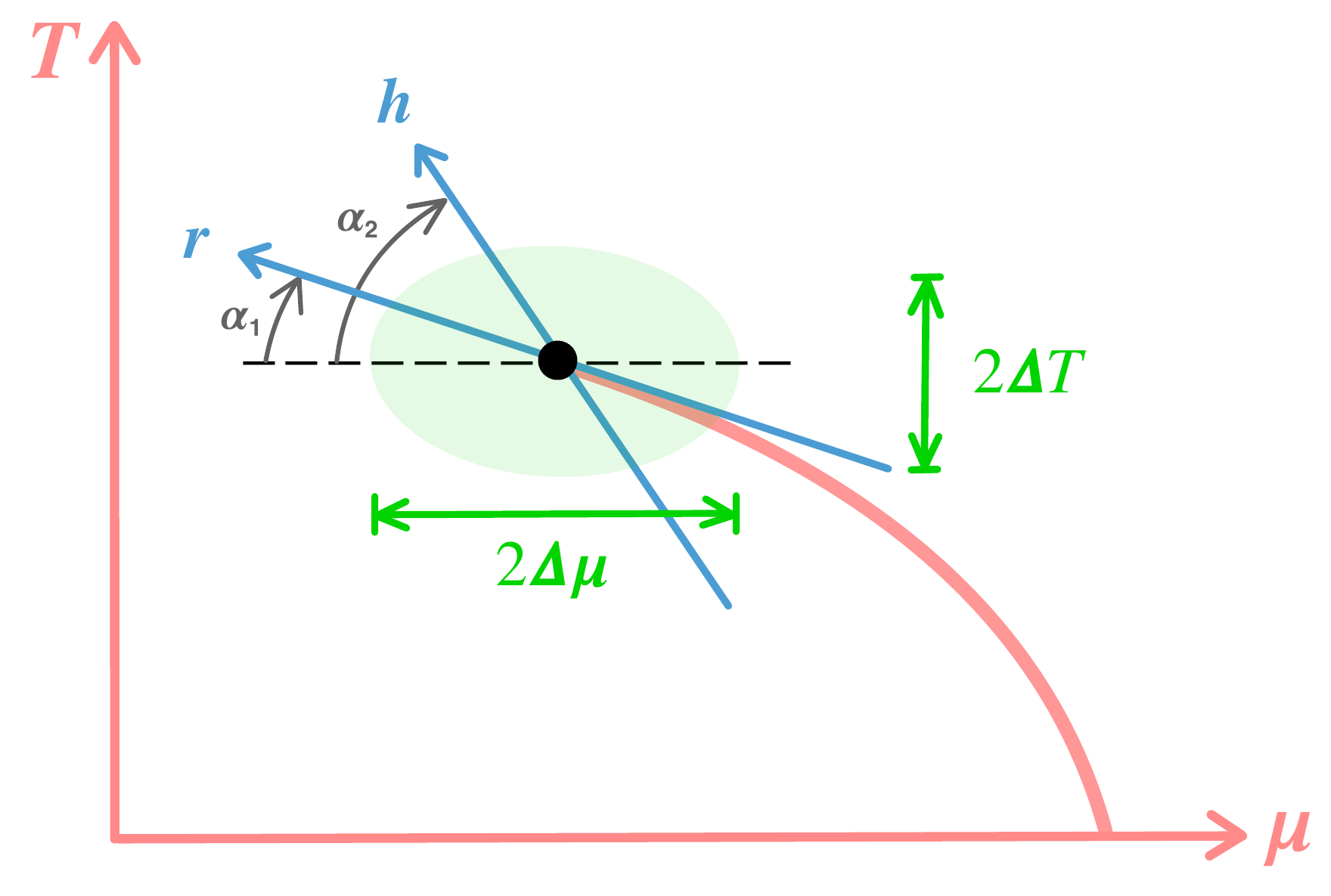}
    \caption{A schematic phase diagram for the mapping from Ising variables $(r, h)$ to the QCD phase diagram coordinate $(\mu,T)$. The black circle ($\mu_c, T_c$) the is QCD critical point, in the critical region with width $(2\Delta\mu, 2\Delta T)$.}
    \label{fig:phase_diagram_illu}
\end{center}
\end{figure}

In this Appendix we estimate the size of critical region characterize by $\Delta\mu$ and $\Delta T$ (cf. Fig.~\ref{fig:phase_diagram_illu}), as an input to Eq.~\eqref{eq:xi(mu,T)}. Following the analysis and notation convention of \cite{Parotto:2018pwx,Pradeep:2019ccv,Mroczek:2020rpm}, the linear mapping from 3-dimensional Ising variables (i.e., reduced Ising temperature $r$ and magnetic field $h$) to the coordinate variables of QCD phase diagram (i.e., temperature $T$ and baryon chemical potential $\mu$) reads
\begin{equation}\label{eq:mapping_variables1}
    \begin{aligned}
    \frac{T-T_c}{T_c}&=w(r\rho s_1+hs_2)\,,\\
    \frac{\mu-\mu_c}{\mu_c}&=w(-r\rho c_1-hc_2)\,,
    \end{aligned}
\end{equation}
where $(w,\rho)$ are scale factors for the Ising variables $r$ and $h$. For notational simplicity we let $s_i=\sin\alpha_i$, $c_i=\cos\alpha_i$, $i=1,2$, where $\alpha_1$ and $\alpha_2$ are the angles relative to the negative $\mu$ axis of the mapped $r$ and $h$ axes, respectively, defined in Fig.~\ref{fig:phase_diagram_illu}. From Eq.~\eqref{eq:mapping_variables1} one can also find the inverse mapping relations
\begin{equation}\label{eq:mapping_variables2}
    \begin{aligned}
    r&=\frac{\mu-\mu_c}{w\rho T_cc_1}\,,\\
    h&=\frac{c_1(T-T_c)}{wT_cs_{12}}\,,
    \end{aligned}
\end{equation}
where $s_{12}=\sin(\alpha_1-\alpha_2)=\sin\Delta\alpha$. With the map given by Eq.~\eqref{eq:mapping_variables2}, one can relate the leading singular contribution to the QCD pressure $p_{\rm crit}(\mu,T)$ and the Gibbs free energy in Ising theory $G(r,h)$ up to a constant of proportionality $A$:
\begin{equation}\label{eq:mapping_P-G}
    p_{\rm crit}(\mu,T)=-AG(r(\mu,T),h(\mu,T)).
\end{equation}
The coefficient $A$ can not be determined by universality, and thus we parametrize it as 
\begin{equation}
    A=aT_c^4
\end{equation}
where $a$ is an unknown parameter.

From Eq.~\eqref{eq:mapping_P-G} one can calculate the singular part of susceptibility, $\chi$, which, as our choice of measure for criticality, diverges near the critical point. Along the crossover ($h=0$) line, 
\begin{equation}
\begin{aligned}
    \chi^{\rm sing}_{h=0}&\sim AG_{\mu\mu}(r,0)\sim AG_{hh}(r,0)h_\mu^2\\
    &\sim A r^{\beta(1-\delta)}\left(\frac{s_1}{wT_cs_{12}}\right)^2\\
    &\sim A \left(\frac{\Delta\mu}{w\rho T_cc_1}\right)^{\beta(1-\delta)}\left(\frac{s_1}{wT_cs_{12}}\right)^2,
\end{aligned}
\end{equation}
while along the $h$ axis ($r=0$), 
\begin{equation}
\begin{aligned}
    \chi^{\rm sing}_{r=0}&\sim AG_{\mu\mu}(0,h)\sim AG_{hh}(0,h)h_\mu^2\\
    &\sim A h^{(1-\delta)/\delta}\left(\frac{s_1}{wT_cs_{12}}\right)^2\\
    &\sim A \left(\frac{c_1\Delta T}{w T_cs_{12}}\right)^{(1-\delta)/\delta}\left(\frac{s_1}{wT_cs_{12}}\right)^2.
\end{aligned}
\end{equation}
Here the subscripts $\mu$ or $h$ should be understood as a derivative in the corresponding direction with the other variable held fixed, for instance, $h_\mu\equiv(\partial h/\partial\mu)_T$. As the criterion for critical phenomena to be important we take that the singular part of the susceptibility is comparable to its regular part, i.e.,
\begin{subequations}\label{eq:Delta_mu_T1}
\begin{align}
    \xi^{\rm sing}_{h=0}&\sim\xi^{\rm reg}\,,\\
    \xi^{\rm sing}_{r=0}&\sim\xi^{\rm reg}\,.
\end{align}
\end{subequations}
Provided $\xi^{\rm reg}\sim T_c^2$, we find
\begin{subequations}\label{eq:Delta_mu_T2}
\begin{align}
    \Delta\mu&\sim T_c\rho w c_1\left(\frac{\sqrt{A}s_1}{w T_c^2s_{12}}\right)^{2/\gamma}, \\
    \Delta T&\sim\frac{(wT_cs_{12})^{-\beta(\delta+1)/\gamma}}{c_1}\left(\frac{\sqrt{A}s_1}{T_c}\right)^{2\beta\delta/\gamma},
\end{align}
\end{subequations}
where $\gamma=\beta(\delta-1)=4/3$.

In order to determine the values of $\Delta\mu$ and $\Delta T$, we adopt the following setup choices:
\begin{gather}\label{eq:setup_Delta_muT}
   \Delta\alpha=\alpha_1-\alpha_2=3^\circ\,, \quad  \rho=\frac{1}{2}\,, \quad w=\frac{1}{2}\,, \quad a=1\,,
\end{gather}
which is consistent with the findings from Ref.~\cite{Pradeep:2019ccv} in the small quark mass limit and the causality requirement from Ref.~\cite{Parotto:2018pwx}. Non-universal parameters given by Eq.~\eqref{eq:setup_Delta_muT} and \eqref{eq:non-universal-parameters} together determine the size of the critical region.

\chapter{Causality analysis near the critical point}\label{sec:causality}

In this Appendix we analysis how the causality (and stability) condition is satisfied near the critical point for Ch.~\ref{ch:diffcp}, focusing on the baryon diffusion only. A more complete analysis involving all non-hydrodynamic degrees of freedom (e.g., shear and bulk stress tensor) is plausible. We evoke small perturbations on top of a flat, homogeneous and static background (denoted by ``$\,\bar~\,$"),
\begin{gather}
    \ed=\bar\ed+\delta\ed(t,x)\,,\quad n=\bar n+\delta n(t,x)\,,\nonumber\\
    u^\mu=\bar u^\mu+\delta u^\mu(t,x)\,, \quad n^\mu=\delta n^\mu(t,x)\,,
\end{gather}
where the perturbations are assumed to be dependent on one spatial coordinate (i.e., $x$) only for simplicity. For a particular direction $x$ we linearize the conservation laws and \eqref{eq:IS_nmu} up to first order in gradient as
\begin{subequations}\label{eq:linearized}
\begin{align}
    &D\delta\ed+\bar w\nabla^x\delta u_x=0\,,\\
    &D\delta n+\bar n\nabla^x\delta u_x+\nabla^x\delta n_x=0\,,\\
    &\bar wD\delta u_x-p_\ed\nabla_x\delta\ed-p_n\nabla_x\delta n=0\,,\\
    &(1+\tau_nD)\delta n_x-\kappa_n(\alpha_\ed\nabla_x\delta\ed+\alpha_n\nabla_x\delta n)=0\,,
\end{align}
\end{subequations}
where
\begin{equation}
\begin{gathered}
    p_e=\left(\frac{\partial p}{\partial e}\right)_n, \quad p_n=\left(\frac{\partial p}{\partial n}\right)_e,\\ 
    \alpha_e=\left(\frac{\partial \alpha}{\partial e}\right)_n, \quad \alpha_n=\left(\frac{\partial \alpha}{\partial n}\right)_e.
\end{gathered}
\end{equation}
Introducing the Fourier component of the linearized variables collectively denoted by $\delta\phi=(\delta e, \delta n, \delta u_\mu, \delta n_\mu)^T$, i.e., 
\begin{equation}
    \delta\tilde\phi(\omega,k)=e^{i\omega t-ikx}\delta\phi(t,x)\,
\end{equation}
where $k=k_x=-k^x$, Eq.~\eqref{eq:linearized} can be transferred to
\begin{equation}
    M\delta\tilde\phi=0
\end{equation}
where
\begin{equation}
    M=\begin{pmatrix}
    i\omega & 0       & ik\bar w & 0 \\
    0       & i\omega & ik\bar n & ik \\
    ikp_e   & ikp_n   & i\omega\bar w & 0 \\
    ik\kappa_n\alpha_e &ik\kappa_n\alpha_n & 0 & i\omega\tau_n+1 \\
    \end{pmatrix}.
\end{equation}
The dispersion relations can be obtained by solving the determinant of the characteristic matrix $M$. For simplicity we assume at this moment $p=p(e)$, thus $p_n=0$ and $c_s^2=p_e$, we then find four eigenmodes:
\begin{equation}
    \omega_1^{\pm}=\pm c_s k\,, \quad \omega_2^{\pm}=i\frac{1\pm \sqrt{1-4\tau_nD_pk^2}}{2\tau_n}\,.
\end{equation}
$\omega_1^{\pm}$ are the modes propagating with the speed of sound; $\omega_2^{\pm}$ are the non-hydrodynamic modes that does not vanish at $k\to0$ (i.e., with finite decay time $\tau_n$), and turn to propagate at large $k$ limit, with the maximum group velocity 
\begin{equation}
v_\text{g}^{\text{max}}=\lim_{k\to\infty}\frac{d\text{Re}\omega}{dk}=\pm\sqrt{\frac{D_p}{\tau_n}}.
\end{equation}
In order to satisfy the causality condition $|v_{\text g}^{\text{max}}|\leq1$ near the critical point one must demand that $\tau_n$ grows at least as fast as $D_p$, which is obviously the case since $\tau_n\sim\xi^2\gg D_p\sim\xi^{-1}$. 

\chapter{Validation of \bes+}
\label{app:hydroplus}

%

Although in this work we studied the evolution of the slow modes only on top of a fixed ideal Gubser flow, with conformal EoS and without back-reaction, the {\sc hydro+} framework is embedded in the \bes~code \cite{Du:2019obx}, which can simulate the dissipative hydrodynamics at non-zero baryon density with realistic EoS for any expansion geometry; it also properly couples the evolution of the background fluid and the slow modes by including the back-reaction. In this Appendix we illustrate some numerical methods and the validation of \bes+.

At non-zero baryon density, one needs solve the ``root finding" problem \cite{Du:2019obx}, where one computes $(e, n)$ in LRF and $u^\mu$ from the energy-momentum tensor and net baryon current in the global computational frame
\begin{eqnarray}
    T^{\mu\nu} &=& e\, u^{\mu}u^{\nu}-p_\plus\Delta^{\mu\nu}+\pi^{\mu\nu}\;,\\
    N^{\mu} &=& n\, u^{\mu} +\V^\mu \;, 
\end{eqnarray}
where $p_\plus$ is given by,
\begin{equation}
    p_\plus(e, n) = p(e, n) + \D p(e, n, \phi(e, n))\,,
\end{equation}
which includes the back-reaction $\Delta p$ from Eq.~(\ref{eq:dp}). Here $p(e,n)$ implicitly includes the bulk viscous pressure, i.e. $p(e, n)=p_\mathrm{eq}(e, n)+\Pi$, and the equilibrium pressure $p_\mathrm{eq}(e, n)$ is given by the EoS.

The modified root finder in Ref.~\cite{Du:2019obx} can be extended to include contributions from the slow modes to solve the root finding problem in \bes+. We introduce
\begin{eqnarray}
  M^{\tau} &=& T^{\tau\tau}-\pi^{\tau\tau} = (e+p_\plus) \bigl(u^{\tau}\bigr)^{2} - p_\plus\,,
\\
  M^{i} &=& T^{\tau i}-\pi^{\tau i} = (e+p_\plus) u^{\tau} u^{i} \quad (i = x, y, \eta_s)\,,\quad
\\
  J^\tau &=& N^{\tau}-n^\tau =\n u^{\tau}\,,
\end{eqnarray}
and the flow speed, $v\equiv\sqrt{1{\,-\,}1/(u^\tau)^2}$. Then we solve iteratively
\begin{equation}
    v \equiv \frac{M}{M^\tau + p_\plus} = \frac{M}{M^\tau + p(e, n) + \D p(e, n, \phi(e, n))}\;,\label{root-v}
\end{equation}
where $M \equiv \sqrt{(M^x)^2+(M^y)^2+\tau^2(M^\eta)^2}$ and $(e, n)$ are both functions of $v$:
\begin{equation}
  \ed(v) = M^\tau - vM\,,\quad
  \n(v) = J^\tau\sqrt{1{\,-\,}v^2}\,.
\label{root-n-2}
\end{equation}
When $\Delta p$ is not added to Eq.~(\ref{root-v}) (as done in this work) the back-reaction is off and has no effects on the evolution of the background fluid.

%
\begin{figure}[!tb]
    \hspace*{-5mm}
   \centering
    \includegraphics[width= 0.85\textwidth]{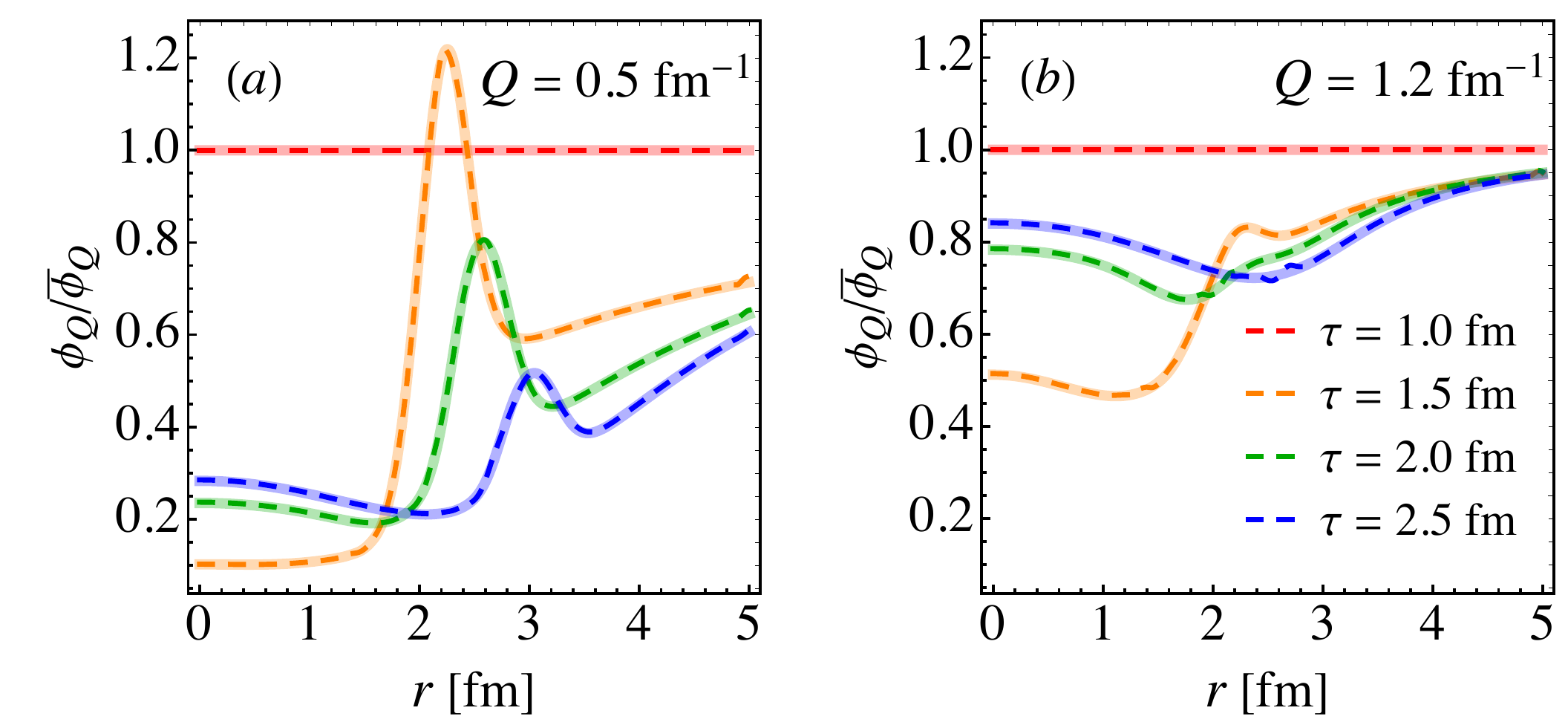}
    \caption{Comparison of $\phi_Q/\bar\phi_Q$ from semi-analytical solutions (solid lines) and numerical results from \bes+ (dashed lines) for two slow modes with (a) $Q{\,=\,}0.5$\,\fm\ and (b) $Q{\,=\,}1.2$\,\fm. For the comparison, the ideal Gubser flow provided by \bes \cite{Du:2019obx} was used with $\tau_0{\,=\,}1$\,fm, $q{\,=\,}1$\,\fm\ and $C{\,=\,}1.2$; for critical dynamics, the parametrization of the correlation length in Eq.~(\ref{eq:xipara2}) was used; otherwise the setup is the same as in Sec.~\ref{sec:corrlen}. The overall agreement is excellent, though some small wiggles can be observed in the numerical solution which originate from the numerical methods used to simulate the background fluid (see also Ref.~\cite{Du:2019obx}).
    \label{fig-validation}}
\end{figure}
%

Another numerical issue involves solving the equations of motion (\ref{eq:phi_scaling}) for the slow modes which share similarities with the evolution equations for the dissipative flows, including $\V^\mu$, $\pi^{\mu\nu}$ and especially $\Pi$. For the dissipative flows those equations are numerically solved by the Kurganov-Tadmor (KT) algorithm \cite{KURGANOV2000241}, with a second-order explicit Runge-Kutta (RK) ordinary differential equation solver \cite{leveque_2002} for the time integration step in \bes~\cite{Du:2019obx}, after being written in first-order flux-conserving form. Similarly, Eq.~(\ref{eq:phi_scaling}) can be rewritten in the same form as
\begin{eqnarray}
\!\!\!\!
    \partial_\tau \phi_Q + \partial_x (v^x \phi_Q) + \partial_y (v^y \phi_Q) + \partial_\eta (v^\eta \phi_Q) = S_Q,\quad 
\label{eq:eomre}
\end{eqnarray}
where $v^i \equiv u^i/u^\tau$ ($i = x, y, \eta_s$) is the 3-velocity of the fluid and $S_Q$ is the source term
\begin{equation}
    S_Q=-\frac{1}{u^\tau}\Gamma_Q\left(\phi_Q-\bar\phi_Q\right)+\phi_Q\partial_i v^i\;.
\end{equation}
Here we used $\partial_{i}v^{i} \equiv \partial_{x}v^{x} + \partial_{y}v^{y} + \partial_{\eta}v^{\eta}$. The slow mode equations can then be solved using the same KT-RK algorithm by straightforwardly extending \bes.

\bes~has been tested by comparing to semi-analytical solutions \cite{Du:2019obx}, and in the same spirit, we validate the numerical methods of the extened root finder (\ref{root-v}) and equations of motion (\ref{eq:eomre}) involving the slow modes, using the setup described in the caption of Fig. \ref{fig-validation}. As one can see from the figure, the agreement is excellent. Once such a numerically precise evolution of the slow modes has been achieved it is easy to derive the remaining off-equilibrium corrections, e.g. those to the pressure and entropy. The code is open source and can be freely downloaded from \url{https://github.com/LipeiDu/BEShydro}. Interested readers are encouraged to repeat the test with the setup described in the \texttt{HydroPlus} branch, especially after they make improvements to the code.

\end{appendices}


\backmatter
\bibliography{bibfile}

\bibliographystyle{elsarticle-num}

\end{document}